\documentclass[a4paper, 11pt]{article}
\usepackage{amsmath,amssymb,xcolor,comment}
\usepackage{jheppub} % for details on the use of the package, please
\usepackage[T1]{fontenc} % if needed
\usepackage{framed}

\title{Regulating Sommerfeld resonances for multi-state systems and higher partial waves}

\author[a]{Aditya Parikh,}
\author[b]{Ryosuke Sato,}
\author[c,d,e]{and Tracy R. Slatyer}

\affiliation[a\,]{C.\,N. Yang Institute for Theoretical Physics, Stony Brook University, Stony Brook, NY 11794}
\affiliation[b\,]{Department of Physics, Osaka University, Toyonaka, Osaka 560-0043, Japan}
\affiliation[c\,]{Center for Theoretical Physics, Massachusetts Institute of Technology, Cambridge, MA 02139, USA}
\affiliation[d\,]{Department of Physics, Harvard University, Cambridge, Massachusetts 02138, USA}
\affiliation[e\,]{Radcliffe Institute for Advanced Study at Harvard University, Cambridge,  Massachusetts 02138, USA}

\emailAdd{aditya.parikh@stonybrook.edu}
\emailAdd{rsato@het.phys.sci.osaka-u.ac.jp}
\emailAdd{tslatyer@mit.edu}

\abstract{Long-range attractive interactions between dark matter particles can significantly enhance their annihilation, particularly at low velocities. This ``Sommerfeld enhancement'' is typically computed by evaluating the deformation of the two-particle wavefunction due to the long-range potential, while ignoring the physics associated with the annihilation, and then scaling the appropriate annihilation matrix elements by factors that depend on the wavefunction in the limit where the particles approach zero relative separation. It has long been recognized that this approach is a valid approximation only in the limit where the annihilation rate is small, and breaks down in the regime where the enhanced annihilation rate approaches the unitarity bound, in which case ignoring the impact of the annihilation physics on the two-particle wavefunction cannot be justified and leads to apparent violations of unitarity. In the case where the physics relevant to annihilation occurs at a parametrically shorter distance scale (higher energy scale) compared with the long-range potential, we provide a simple prescription for correcting the Sommerfeld enhancement for the effects of the short-range physics, valid for all partial waves and for systems where multiple states are coupled by the long-range potential.}

\newcommand{\draftnote}[1]{} % added by RS. you can comment out this if \draftnote works in your environment.

\begin{document} 

\begin{flushright}
OU-HET-1243

MIT-CTP/5790
\end{flushright}
\maketitle
\flushbottom

\section{Introduction}

Stable particles may experience long-range interactions either mediated by known Standard Model (SM) particles or by new forces. It was pointed out as early as 2002 \cite{Hisano:2002fk, Hisano:2003ec, Hisano:2004ds} that such long-range interactions between annihilating dark matter (DM) particles could significantly enhance the annihilation rate, in analogy to the Sommerfeld enhancement of electron-positron annihilation via the Coulomb interaction \cite{Sommerfeld:1931qaf}. The original application of this insight was in the context of heavy DM in multiplets of the SM $SU(2)_L$ gauge group, where the force carriers are the electroweak gauge bosons and multiple states participate in the long-range interaction; subsequent studies explored the implications of Sommerfeld enhancement for more general DM scenarios (e.g. \cite{ArkaniHamed:2008qn, Iengo:2009ni, Cassel:2009wt, Slatyer:2009vg}).

The lowest-order prescription for calculating the Sommerfeld enhancement described in Refs.~\cite{Hisano:2002fk, Hisano:2003ec, Hisano:2004ds} gives the overall annihilation cross section in the factorized form:
\begin{equation}
\sigma v_\text{rel} = c_i \sum_{ab} A_a \Gamma_{ab} A_b^*,
\end{equation}
where $v_\text{rel}$ is the relative velocity between the two particles, $c_i$ is a symmetry coefficient dependent only on the initial state, $A$ is a vector obtained by solving the Schr\"odinger equation for the two-particle wavefunction in the presence of a long-range potential, and the matrix $\Gamma$ is built from the matrix elements for annihilation involving all the states coupled by the long-range potential. That long-range potential is characterized by a matrix, and includes diagonal terms corresponding to the mass splittings between the states coupled by the potential. 

As already noted in Ref.~\cite{Hisano:2004ds}, this factorization between the long-range potential effects and the annihilation matrix is only accurate to lowest order in $\Gamma$. More generally, the possibility for the particles to annihilate should be taken into account when working out the deformation of the wavefunction; ignoring this effect can lead to apparent violations of unitarity. Specifically, this effect becomes apparent (even for weakly-coupled interactions) when there is a near-zero-energy bound state in the spectrum, leading to a resonant enhancement to the annihilation cross section. The standard approach implies an enhanced $s$-wave cross section scaling as $\sigma v_\text{rel} \propto v_\text{rel}^{-2}$ at the center of the resonance (e.g.~\cite{Cassel:2009wt}), which at sufficiently low velocities contradicts the partial-wave unitarity bound $\sigma \le 2\pi/(\mu v_\text{rel})^2$ (for inelastic scattering of indistinguishable particles, and where $\mu$ is the reduced mass of the system).

Ref.~\cite{Blum:2016nrz} demonstrated how to capture this effect and restore manifest unitarity for a single-state system with only $s$-wave annihilation, by modeling the effect of annihilation (and possibly other short-range physics) as a $\delta$-function potential with a complex coefficient, and solving the Schr\"odinger equation to obtain the wavefunction in the presence of this added term\footnote{
This calculation treats both long-range and short-range effects in a non-perturbative way. Related works can be found in the context of nuclear physics, including the analysis of $p\bar p $ bound states \cite{Trueman:1961zza, Carbonell:1992wd}, $p\bar p$ annihilation \cite{Carbonell:1993, Carbonell:1996vd, Carbonell:1998ei, Protasov:1999ei}, and $pp$ scattering \cite{Kong:1998sx, Kong:1999tw}.
}. Refs.~\cite{Braaten:2017gpq, Braaten:2017kci, Braaten:2017dwq} developed a ``zero range effective field theory'' (ZREFT), allowing analytic study of Sommerfeld enhancement in the neighborhood of $s$-wave resonances, for DM that is a $SU(2)_L$ fermion triplet; the ZREFT parameterizes the behavior near the resonance in terms of a scattering length, and accounts for the annihilation effects naturally by allowing this scattering length to be complex. The ZREFT was restricted to $s$-wave resonances, but demonstrated the generalization of the unitarization prescription to a multi-state system. Ref.~\cite{Chu:2019awd} suggested a general parameterization for DM self-interaction cross sections in terms of the scattering length and effective range, and demonstrated how this prescription could be applied to the case of $s$-wave annihilation.
Ref.~\cite{Kamada:2023iol} showed the correlation between the resonant self-scattering cross section and the Sommerfeld factor for the annihilation by using Watson's theorem. 

The methodology of Ref.~\cite{Blum:2016nrz} did not generalize straightforwardly to higher partial waves, due to the lack of overlap between the higher-$\ell$ wavefunctions and a delta-function potential. In this work, we propose a new prescription for computing the corrected Sommerfeld enhancement, with conceptual similarities to the work of Refs.~\cite{Agrawal:2020lea,Parikh:2020ggm}. The resulting expression is manifestly unitarity-preserving, requires no additional calculations beyond solving the Schr\"odinger equation for the long-range potential and knowledge of any relevant short-distance interactions, and can be applied to all partial waves and to multi-state systems. Our work complements that of Ref.~\cite{Flores:2024sfy}, which showed how to extend the single-state unitarization calculation to higher partial waves via an approach based on the generalized optical theorem.

The central idea of our work is that if there is a separation of scales between annihilation and the interactions mediating the long-range potential, then we can treat the annihilation (and possibly other short-distance physics) as being localized to a small-$r$ region $r < a$ (where $r$ is the separation between the interacting particles). Outside this region, we can solve the Schr\"odinger equation with the long-range potential as usual, and the only possible effect of any short-range interactions on this calculation will be to modify the boundary conditions at the matching radius. The modified boundary conditions can be related to the $S$-matrix element for scattering in the {\it absence} of the long-range potential, accounting only for the short-distance interactions, and assuming the short-distance interactions are weakly coupled, this short-distance $S$-matrix can be computed in the Born approximation (or equivalently, using perturbative quantum field theory (QFT)). A similar approach was used in Ref.~\cite{Agrawal:2020lea} to study DM elastic self-scattering in potentials with singular short-distance behavior.

In Sec.~\ref{sec:singlestate} we introduce the method and apply it to the case where only a single state is relevant, obtain the regulated Sommerfeld enhancement for general $\ell$, demonstrate that we can recover the $\ell=0$ result of Ref.~\cite{Blum:2016nrz}, and work out some examples. In this section we also develop an analytic understanding of the momentum dependence of the enhancement in the low-momentum limit. In Sec.~\ref{sec:multistate} we extend this approach to the case where multiple states are coupled by the long-range potential and can participate in the short-range interactions, and show how to write the full regulated annihilation cross section in terms of the matrix elements for short-distance interactions and factors derived from the long-range Schr\"odinger equation. We work out the example of wino dark matter in Sec.~\ref{sec:wino}, and present our conclusions in Sec.~\ref{sec:conclusions}. 

Appendices \ref{app:analyticwavefn} and \ref{app:vpm} collect results that we use pertaining to solutions of the long-range Schr\"odinger equation; App.~\ref{app:analyticwavefn} discusses the analytic properties of the basis solutions to the long-range Schr\"odinger equation, while App.~\ref{app:vpm} presents a modification of the variable phase method for numerically solving the Schr\"odinger equation for the case with modified short-distance boundary conditions. App.~\ref{sec:comparison with 240502222} works out in detail the comparison between our single-state results and those of Ref.~\cite{Flores:2024sfy}, and shows that under a certain set of assumptions and approximations they are equivalent. Apps.~\ref{app:optical} and \ref{app:shortrangeinterp} discuss, respectively, the approach we use for computing the short-distance scattering amplitude and matching between the QFT and quantum mechanics calculations, and the interpretation of the short-distance amplitude when we use a convention where the long-range potential still has some support at $r < a$. We present some useful results on the partial-wave decomposition of the amplitude in App.~ \ref{app:partial_amps}; some supplemental results for the wino example in App.~\ref{app:winoextras}; and the generalization of our multi-state results to the case where the potential is Hermitian but not real in App.~\ref{app:Hermitian}.

\section{The single state case for general $\ell$}
\label{sec:singlestate}

Let us begin by working out the corrected Sommerfeld effect for the single-state case. Our discussion here will rely on techniques developed for the low energy nucleon-nucleon two-body problem,
originally due to Bethe \cite{Bethe:1949yr} and Chew-Goldberger \cite{Chew:1949zz} (this approach is also discussed in textbooks, e.g.~Section XIII of \cite{BetheMorrison} and Section 6.6 of \cite{GoldbergerWatson}).

\subsection{Basics of non-relativistic two-body scattering}
We first briefly review the basics of the two-body scattering problem in non-relativistic quantum mechanics.
The Schr\"odinger equation for the two-body scattering problem with a central potential $V(r)$ is
\begin{align}
-\frac{1}{2\mu} \nabla^2 \psi(\vec{r}) + V(r) \psi(\vec{r}) = \frac{p^2}{2\mu} \psi(\vec{r}),
\end{align}
where $\mu$ is the reduced mass of the two-body system, $\vec{r}$ is the separation vector between the particles, $r=|\vec{r}|$, and $p$ is the momentum of either particle in the center-of-momentum frame.
The asymptotic behavior of the wavefunction $\psi(\vec{r})$ as $r \to \infty$ is
\begin{align}
\psi(\vec r) = e^{ipz} + f(\theta) \frac{ e^{ipr} }{r}. \label{eq:asymptotic behavior psi}
\end{align}
where $\theta$ is the angle between $\vec r$ and the $z$-axis, and $z$ is the direction of the initial plane wave.
Let us take the following partial wave expansion:
\begin{align}
\psi(r,\theta) = \sum_\ell (2\ell+1) i^\ell P_\ell(\cos\theta) \frac{u_\ell(r)}{pr}.
\end{align}
The radial wavefunction $u_\ell(r)$ satisfies the reduced Schr\"odinger equation:
\begin{align}
\left( -\frac{d^2}{dr^2} + \frac{\ell(\ell+1)}{r^2} + 2\mu V(r) - p^2 \right) u_\ell(r) = 0.
\end{align}
$e^{ipz}$ can be expanded as $e^{ipz} = \sum_\ell (2\ell+1) i^\ell P_\ell(\cos\theta) j_\ell(pr)$. It will frequently be useful to work with the free wavefunctions:
\begin{align}
s_\ell(x) \equiv x j_\ell(x), \qquad
c_\ell(x) \equiv -	x y_\ell(x), \label{eq:def of sell and cell}
\end{align}
where $j_\ell(x)$ and $y_\ell(x)$ are the standard spherical Bessel functions. These wavefunctions have the small-$x$ asymptotic behavior:
\begin{align}
s_\ell(x) = \frac{1}{(2\ell+1)!!} x^{\ell+1} + \cdots, \qquad
c_\ell(x) = (2\ell-1)!! x^{-\ell} \cdots.
\label{eq:s and c at the origin}
\end{align}
At large $x$, their asymptotic behavior is:
\begin{align}
s_\ell(x) \rightarrow \sin(x - \pi \ell/2), \qquad
c_\ell(x) \rightarrow  \cos(x - \pi \ell/2).
\label{eq:s and c at large x}
\end{align}

Thus, we can read off the asymptotic behavior of $u_\ell(r)$ from the boundary condition of Eq.~\ref{eq:asymptotic behavior psi}, as:
\begin{align}
u_\ell(r) \to s_\ell(p r) + p f_\ell \left( c_\ell(p r) + i s_\ell(p r) \right) = 
 \frac{1}{2i}\left( (-i)^\ell S_\ell e^{ipr} - i^\ell e^{-ipr} \right), \quad r\rightarrow \infty, \label{eq:boundary condition at infinity}\end{align}
 where $f(\theta) = \sum_\ell (2\ell + 1) f_\ell P_\ell(\cos\theta)$ and $S_\ell = 1+ 2ipf_\ell$ is the $S$-matrix.

The elastic cross section $\sigma_{\text{sc}}$
and the inclusive annihilation cross section $\sigma_{\text{ann}}$ are given (for distinguishable particles) by
\begin{align}
\sigma_{\text{sc}} &= \frac{4\pi}{p^2} \sum_\ell (2\ell+1) \left| \frac{S_\ell-1}{2i}\right|^2, \\
\sigma_{\text{ann}} &= \frac{4\pi}{p^2} \sum_\ell (2\ell+1) \frac{1-|S_\ell|^2}{4}.
\end{align}
Here we take the inclusive annihilation cross section to include all inelastic processes that are not modeled by the Hermitian part of the potential $V(r)$, which thus manifest themselves as apparent non-unitarity of the $S$-matrix.

This relation assumes distinguishable particles in the initial state; for identical particles, the cross section will be zero for partial waves that do not have the correct symmetry properties, and enhanced by a factor of 2 otherwise (e.g. for identical fermions in a spin-singlet state, $\ell$ must be even; see Ref.~\cite{Asadi:2016ybp} for a more in-depth discussion). We will accordingly add a prefactor $c_i$ to cross sections, which is 1 for distinguishable particles and 2 for identical particles, while working with the scattering amplitudes appropriate to distinguishable particles.

We will often find it advantageous to expand the reduced wavefunction $u_\ell(r)$ in terms of the free wavefunctions $s_\ell(p r)$ and $c_\ell(p r)$; in particular, this facilitates numerical solution of the  Schr\"odinger equation via the variable phase method, which we review and justify in App.~\ref{app:vpm}. Let us define:
\begin{equation} f_\ell(r) \equiv s_\ell(p r)/\sqrt{p}, \qquad g_\ell(r) \equiv (c_\ell(p r) + i s_\ell(p r))/\sqrt{p}. \label{eq:vpmdefs} \end{equation}
Then for any reduced wavefunction $u_\ell(r)$ solving the Schr\"odinger equation, we can  decompose:
\begin{equation} u_\ell(r) = \alpha_\ell(r) f_\ell(r) - \beta_\ell(r) g_\ell(r),\end{equation}
and in order to uniquely define $\alpha_\ell(r)$ and $\beta_\ell(r)$ we can also impose the condition $u^\prime_\ell(r) = \alpha_\ell(r) f_\ell^\prime(r) - \beta_\ell(r) g^\prime_\ell(r)$ (see discussion in App.~\ref{app:vpm}). This condition ensures that matching the coefficients $\alpha_\ell(r)$ and $\beta_\ell(r)$ between two solutions (at a given choice of $r$) is sufficient to match both their values and their first derivatives for that $r$. Note also that $f_\ell^\prime(r) g_\ell(r) - f_\ell(r) g^\prime_\ell(r) =1$, for all $r$.

\subsection{$S$-matrix from the boundary condition}\label{sec:s matrix boundary condition}
We are interested in the annihilation cross section when the long-range force deforms the wavefunction from a plane wave.
We assume that the long-range force does not provide any annihilation effect directly, i.e. the corresponding potential is real.
Thus, it is useful to separate the short-range interactions (which may include inelastic/absorptive channels) and the long-range interactions (assumed to be well-described by a real potential) at some boundary $r=a$ as
\begin{align}
V(r) = \begin{cases}
V_{\rm short}(r) & (r<a) \\
V_{\rm long}(r) & (r\geq a)
\end{cases}.
\end{align}
Here $V_{\rm long}$ is real and $V_{\rm short}$ has an imaginary part which provides an effective description of the annihilation of particles.
We will primarily be interested in the case where the incoming momentum $p$ is much smaller than $1/a$; we will generally have in mind the case where $1/a$ is parametrically similar to (or larger than) the mass of the annihilating particles, so this low-momentum approximation is automatic if the particles are non-relativistic. As we will discuss, it is also not necessary that the short-range interactions be fully captured by a non-relativistic potential $V_{\rm short}$ or treated using non-relativistic quantum mechanics, as long as we can calculate the $S$-matrix associated with the short-range interactions.

In order to describe the wavefunction at $r>a$,
we introduce the functions $F_\ell(r)$ and $G_\ell(r)$, which are solutions of the Schr\"odinger equation with the long-range force:
\begin{align}
\left( -\frac{d^2}{dr^2} + \frac{\ell(\ell+1)}{r^2} + 2\mu V_{\rm long}(r) - p^2 \right)F_\ell(r) = 0 \label{eq:schroedinger eq F},\\
\left( -\frac{d^2}{dr^2} + \frac{\ell(\ell+1)}{r^2} + 2\mu V_{\rm long}(r) - p^2 \right)G_\ell(r) = 0 \label{eq:schroedinger eq G}.
\end{align}
$F_\ell(r)$ is regular at the origin and $G_\ell(r)$ is irregular,
and their asymptotic behavior at infinity is
\begin{align}
G_\ell(r) + i F_\ell(r) \to (-i)^\ell \exp\left( ipr + i\delta_\ell^{(L)} \right).
\label{eq:F and G at large r}
\end{align}
Here $\delta_\ell^{(L)}$ is the standard phase-shift induced by the long-range force.
Since we assume $V_{\rm long}(r)$ is real, $\delta_\ell^{(L)}$ is a real parameter.
For small $r$, $F_\ell(r)$ and $G_\ell(r)$ have the asymptotic behavior
\begin{align}
F_\ell(r) \simeq C_\ell s_\ell(p r) \simeq \frac{C_\ell}{(2\ell+1)!!} (pr)^{\ell+1}, \qquad
G_\ell(r) \simeq c_\ell(p r)/C_\ell \simeq \frac{(2\ell-1)!!}{C_\ell} (pr)^{-\ell}.
\label{eq:F and G at the origin}
\end{align}
Here $C_\ell$ is a function of $p$ and $\ell$, determined by $V_{\rm long}(r)$.
Note that
we can prove Eq.~\ref{eq:F and G at the origin}, relating $F_\ell(r)$ and $G_\ell(r)$ in the limit $r\rightarrow 0$, 
by using the fact that the Wronskian $F_\ell'(r) G_\ell(r) - F_\ell(r) G_\ell'(r)$ is independent of $r$, combined with the large-$r$ asymptotics given in Eq.~\ref{eq:F and G at large r}.

We obtain $F_\ell(r) \simeq s_\ell(p r)$ and $G_\ell(r) \simeq c_\ell(p r)$, for all $r$, if the long-range force is inefficient, and hence $C_\ell \simeq 1$ in this case. %
Comparing Eq.~\ref{eq:s and c at the origin} and Eq.~\ref{eq:F and G at the origin}, $C_\ell$ can be interpreted as the enhancement factor of the $\ell$-wave regular wavefunction near the origin, compared to the plane wave. As we will explicitly see later, $C_\ell^2$ is the conventional Sommerfeld factor.

The wavefunction which is consistent with the coefficient of $e^{-ipr}$ in Eq.~\ref{eq:boundary condition at infinity} is given by
\begin{align}
u_\ell(r) =
\begin{cases}
u_{<,\ell}(r) & (r<a) \\
u_{>,\ell}(r) = \exp\left( i \delta_\ell^{(L)} + i \delta_\ell^{(S)}  \right)
 \left[ F_\ell(r) \cos\delta_\ell^{(S)} + G_\ell(r) \sin\delta_\ell^{(S)} \right] & (r\geq a)
\end{cases}.\label{eq:matching}
\end{align}
From this expression, we can read off the $S$-matrix as $S_\ell = \exp\left( 2i\left(\delta_\ell^{(L)} + \delta_\ell^{(S)}\right) \right)$. 
We do not specify an explicit form for $u_{<,\ell}(r)$, and will only need its behavior at $r=a$ to obtain the full $S$-matrix.
Note that $\delta_\ell^{(S)}$ is a complex parameter in general,
and it can be determined from the boundary condition at $r=a$:
\begin{align}
\tan\delta_\ell^{(S)} \equiv - \left(\frac{F_\ell' - F_\ell (u'_{<,\ell} / u_{<,\ell})}{G_\ell' - G_\ell (u'_{<,\ell} / u_{<,\ell})}\right)_{r=a}. \label{eq:tan deltaS}
\end{align}
We will find it convenient to define a new parameter $k_\ell(p)$ as
\begin{align}
k_\ell(p) \equiv p^{2\ell+1} C_\ell^2 \left(\frac{G_\ell' - G_\ell (u'_{<,\ell} / u_{<,\ell})}{F_\ell' - F_\ell (u'_{<,\ell} / u_{<,\ell})}\right)\Bigg|_{r=a} = -p^{2\ell+1} C_\ell^2 / \tan\delta_\ell^{(S)} . \label{eq:kappa_def}
\end{align}
Then the $S$-matrix can be rewritten as
\begin{align}
S_\ell
= \exp\left( 2i \left(\delta_\ell^{(L)} + \delta_\ell^{(S)} \right) \right)
= \frac{k_\ell(p) - i p^{2\ell+1} C_\ell^2 }{k_\ell(p) + i p^{2\ell+1} C_\ell^2} \exp\left( 2i\delta_\ell^{(L)} \right).
\label{eq:S matrix before approximation}
\end{align}
In this expression, $C_\ell$ and $\delta_\ell^{(L)}$ are purely determined by the long-range potential.
On the other hand, $k_\ell(p)$ is affected by both the short-range and long-range interactions. 

To build some intuition for the meaning of $k_\ell(p)$, note that an equivalent expression has been employed in the nuclear physics literature, where the long-range potential is assumed to be a Coulomb interaction (e.g.~between protons or antiprotons) and the short-range effects correspond to a nuclear interaction. 
In this case, the $S$-matrix, from Ref.~\cite{Protasov:1999ei}, is given by:
\begin{align} S_\ell = S_{0,\ell} \frac{K^{-1} - 2 p^{2\ell+1} \eta h(\eta) g_\ell(\eta) +  i p^{2\ell+1} C_\ell^2}{K^{-1} - 2 p^{2\ell+1} \eta h(\eta) g_\ell(\eta) -  i p^{2\ell+1} C_\ell^2} \label{eq:KfromProtasov} \end{align}
where $S_{0,\ell} = e^{2i\delta_\ell^{(L)}}$ is the $S$-matrix present in the case of the attractive Coulomb potential alone, $\eta=-1/p a_B$ where $a_B$ is the Bohr radius of the Coulomb potential, $g_\ell(\eta) = \prod_{m=1}^\ell (1 + \eta^2/m^2)$, and $h(\eta) = \frac{1}{2} (\Psi(i \eta) + \Psi(-i\eta) - \ln(\eta^2))$.\footnote{Here $\Psi$ is the digamma function.} $K^{-1}$ in this case is a meromorphic function of $p^2$ whose leading-order term is $\mathcal{O}(p^0)$; the widely-used ``scattering length approximation'' consists of approximating $K^{-1}(p^2)$ by the lowest-order term in the small-$p$ expansion, i.e. $K^{-1}(p^2) \approx -1/a^{(cs)}_\ell$ where $a^{(cs)}_\ell$ is the (Coulomb-corrected) scattering length/volume/etc.~(with dimension $p^{-(2\ell+1)}$). This approximation can be systematically improved by including more terms in the expansion (e.g. the next-order term describes an ``effective range'' for the short-distance interaction). $K^{-1}$ thus fully captures the short-range nuclear interaction, although it is corrected even at leading order by the presence of the Coulomb interaction (an explicit example for a specific short-range potential is given in Ref.~\cite{PopovMur1985}). Our $k_\ell(p)$ function can be related in this case to $K^{-1}(p^2)$ by $k_\ell(p) = - \left[ K^{-1}(p^2) - 2 p^{2\ell+1} \eta h(\eta) g_\ell(\eta) \right]$. Based on this example, we might expect that $k_\ell(p)$ can be more generally separated into a term purely governed by the long-range interaction (akin to the $2 p^{2\ell+1} \eta h(\eta) g_\ell(\eta)$ term in the case of the Coulomb potential), and a term that mixes the long- and short-range interactions but has a well-characterized momentum dependence and becomes momentum-independent for sufficiently small $p$ (akin to the $K_\ell^{-1}(p^2)$ function). We will examine the momentum dependence of $k_\ell(p)$ in the general case (i.e.~not restricted to the Coulomb potential) in Sec.~\ref{eq:mom dependence kappa}.

\subsection{Obtaining the cross sections}

Since the inclusive annihilation cross section is determined by $|S_\ell|^2$, Eq.~\ref{eq:S matrix before approximation} tells us immediately that the cross section can be written solely in terms of $C_\ell^2 k_\ell^{-1}(p)$ (and the momentum $p$). The full cross sections for elastic scattering and inclusive annihilation are
\begin{align}
\sigma_{\text{sc},\ell} &= \frac{\pi}{p^2} (2\ell+1) c_i  \left| \frac{k_\ell(p) - i p^{2\ell+1} C_\ell^2}{k_\ell(p) + i p^{2\ell+1} C_\ell^2} \exp\left( 2i\delta_\ell^{(L)} \right) - 1 \right|^2, \label{eq:onestatesc}\\
\sigma_{\text{ann},\ell} &= 4\pi (2\ell+1) c_i \frac{ p^{2\ell-1} C_\ell^2 {\rm Im}( - k_\ell^{-1}(p))}{ |1 + i p^{2\ell+1} C_\ell^2 k_\ell^{-1}(p) |^2 }.\label{eq:onestateann}
\end{align}
Note that to lowest order in $k_\ell^{-1}$, the effect of the long-range force in the annihilation cross section is enhancement by a factor of $C_\ell^2$; this is the standard Sommerfeld enhancement. Note also that we can shift the definition of $k_\ell^{-1}(p)$ by a real number without modifying the numerator in the annihilation cross section; the effect of such a shift on the annihilation cross section would be to break the correction term in the denominator into two parts, which may be desirable in terms of separating the long-distance and short-distance behavior.

Although Eq.~\ref{eq:S matrix before approximation} is generic and exact, it is useful to consider the limit where $p$ is large and the long-range potential can be neglected. This assumes that such a regime exists consistent with the assumption of non-relativistic physics, but this should generally be true for weak coupling.  In this case, we expect $C_\ell^2 \rightarrow 1$, $e^{i\delta_\ell^{(L)}} \rightarrow 1$, and $S_\ell$ is given as
\begin{align}
S_\ell
&\simeq \frac{1 - i p^{2\ell+1} k_\ell^{-1}(p)}{1 + i p^{2\ell+1} k_\ell^{-1}(p)}.
\label{eq:S matrix s wave}
\end{align}
In this case, the cross section formulae (without a long-range force) at leading order in $k_\ell^{-1}$ are
\begin{align}
\sigma_{\text{sc},\ell}^{(0)} &\simeq 4\pi (2\ell+1) c_i p^{4\ell} |k_\ell(p)|^{-2}, \label{eq:sigma sc 0}\\
\sigma_{\text{ann},\ell}^{(0)} &\simeq 4\pi (2\ell+1) c_i p^{2\ell-1} \frac{{\rm Im} k_\ell(p)}{|k_\ell(p)|^2}. \label{eq:sigma ann 0}
\end{align}
This regime can be used for matching between the perturbative QFT calculation and the Schr\"odinger equation approach. The degree to which these results can be used to calibrate the corrected cross section at lower momenta depends on the momentum dependence of $k_\ell(p)$, so we will study this question next and present some examples.

Note that in defining $F_\ell(r)$ and $G_\ell(r)$ via Eqs.~\ref{eq:schroedinger eq F}, \ref{eq:schroedinger eq G}, we have the freedom to choose $V_{\rm long}(r)$ in the regime $r < a$. This choice does not affect the potential for the problem of interest, just the basis of solutions we use to study that problem; different choices lead to different properties for $F_\ell(r)$ and $G_\ell(r)$, and consequently to different coefficients for these functions in the $r > a$ regime, but the full wavefunction (and hence the $S$-matrix, cross sections, etc.~) will be the same in all cases. The various intermediate quantities calculated under the different conventions should also converge to each other in the limit where the short-distance interaction is described by a contact interaction and we take $a\rightarrow 0$; more generally, they will differ by terms that are suppressed by powers of $a$.

In the remainder of this section, we will define $V_\text{long}(r)$ as the $a$-independent long-range potential derived from the low-energy effective theory for all $r$, including $r < a$. This implies that the $F_\ell(r)$ and $G_\ell(r)$ functions, and consequently the $C_\ell$ factors, are formally independent of $a$. This choice will simplify our examples by allowing the use of well-known results from the literature for the solutions for commonly-used potentials.

As an alternative to calibrating the short-range physics by measuring at a high momentum and predicting its momentum dependence, in some cases we may be able to calculate the $S$-matrix element for the short-range physics directly (e.g.~because we know the full relativistic theory). Since $k_\ell(p)$ is defined in terms of the boundary conditions at $r=a$ (Eq.~\ref{eq:kappa_def}), we can in principle rewrite it in terms of the $S$-matrix element arising solely from the physics at $r < a$. However, we defer a discussion of this approach to Sec.~\ref{sec:multistate}, where we will work out the relation between (the generalization of) $k_\ell(p)$ and the short-range amplitude for the general multi-state problem. In that section, we will also address the question of the ambiguity of $V_\text{long}(r)$ in the regime $r < a$.

\subsection{Momentum dependence of the $k_\ell(p)$ coefficient} \label{eq:mom dependence kappa}

The $k_\ell(p)$ function is affected by both short-range and long-range interactions. In this section, we will discuss the momentum dependence of $k_\ell(p)$, with the goal of separating these effects. The results in this section are not in general exact, but instead focus on identifying terms that are potentially large at low momentum. 

\subsubsection{Behavior of $u_{<,\ell}$, $F_\ell$, and $G_\ell$ at $r=a$}
As defined in Eq.~\ref{eq:kappa_def}, $k_\ell(p)$ is determined by $u_{<,\ell}$, $F_\ell$, and $G_\ell$. Before discussing $k_\ell(p)$, let us introduce some relevant properties of $u_{<,\ell}$, $F_\ell$, and $G_\ell$. For details, see also App.~\ref{sec:detail on Fell and Gell}.

$u_{<,\ell}(r)$ is determined by the short-range potential $V_{\rm short}(r)$. If this potential has a characteristic length scale of $a$, then while we do not specify an explicit form of $u_{<,\ell}$, we can assume $u'_{<,\ell}/u_{<,\ell}$ can be expanded as
\begin{align}
\frac{u'_{<,\ell}}{u_{<,\ell}} = \frac{1}{a}\left( c_{0,\ell} + c_{1,\ell} (pa)^2 + c_{2,\ell} (pa)^4 + \cdots  \right), \label{eq:bc at r=a}
\end{align}
where the $c_{i,\ell}$ coefficients are ${\cal O}(1)$ dimensionless numbers.  As long as $p \ll a^{-1}$, $u'_{<,\ell}/u_{<,\ell}$ can be approximated as a momentum-independent constant.

$F_\ell(r)$ is a regular solution of the Schr\"odinger equation with long-range potential $V_{\rm long}(r)$, and it can be expanded as a series in $r$ at small $r$. As we have seen in Eq.~\ref{eq:F and G at the origin}, the leading term of $F_\ell(r)$ around $r=a$ is expected to scale as $r^{\ell + 1}$, and $F_\ell(r)$ behaves as
\begin{align}
F_\ell(r) \simeq \frac{C_\ell}{(2\ell + 1)!!} (pr)^{\ell + 1} \label{eq:F at r=a}
\end{align}
The Schr\"odinger equation at small $r$ provides recurrence relations among the coefficients of higher powers of $r$ as shown in Eqs.~\ref{eq:f recurrence 1}-\ref{eq:f recurrence 3}. Thus, the form of $F_\ell(r)$ (except for its overall prefactor) is uniquely determined by the Schr\"odinger equation at small $r$.

$G_\ell(r)$ is an irregular solution of the Schr\"odinger equation with long-range potential $V_{\rm long}(r)$.
The recurrence relations for $G_\ell(r)$ are given in Eqs.~\ref{eq:g recurrence 4}-\ref{eq:g recurrence 8}. The coefficients of $r^\ell$, $r^{\ell-1}$, $\cdots$, $r^{-\ell+1}$ can be derived from those recurrence relations, and we can see that those terms are subdominant at $r=a$, compared to the leading $r^{-\ell}$ term. In more detail, we expect the coefficients $g_{\ell,k}$ (which are normalized so that the coefficient for the $r^{-\ell}$ term is dimensionless and $\mathcal{O}(1)$, see Eq.~\ref{eq:g recurrence 4}) to take the form of polynomials in the momentum $p$ and the coefficients $v_k$ defined from the expansion of the potential in Eq.~\ref{eq:vk}. These coefficients do not have any singular behavior near the resonance points, and have positive mass dimension $k+2$. If we write $v_k \sim p_k^{k+2}$, where $p_k$ is a momentum scale associated with the potential, we expect $p_k \ll \mu \sim 1/a$ by our assumption that the potential is long-range (note that for the Coulomb potential, $v_{-1}$ is just set by the inverse Bohr radius). We thus expect the coefficients to be built from positive powers of these $p_k$ factors and the small momentum $p$, and the resulting terms to be suppressed (relative to the $r^{-\ell}$ term) for $r \sim a$ by products of factors of the form $p_k a \ll 1$, $p a \ll 1$. 

However, this argument breaks down for the $r^{\ell+1}$ term, because it is impossible to get the relation between the coefficient of $r^{-\ell}$ and $r^{\ell + 1}$ only from the recurrence relations near the origin.
This is because the sum of an irregular solution and a regular solution is another irregular solution,  satisfying the same Schr\"odinger equation.  This means that the $r^{\ell+1}$ term can in principle have a coefficient that is parametrically enhanced, and a value much larger than one would naively expect from dimensional analysis + the assumption that the only relevant momentum scales are small compared to $1/a$.

The coefficient of $r^{\ell+1}$ is determined by imposing the boundary condition at infinity (Eq.~\ref{eq:F and G at large r}).
Thus, although $r^{\ell}$, $r^{\ell-1}$, $\cdots$, $r^{-\ell+1}$ are subdominant at $r=a$, we cannot conclude whether the $r^{\ell+1}$ term is subdominant or not.
A general expansion of $G_\ell$ is given in Eq.~\ref{eq:G expansion}.
For now, we just keep both the $r^{-\ell}$ term and $r^{\ell+1}$ term and write $G_\ell$ at $r=a$ as
\begin{align}
G_\ell(r) \simeq \frac{(2\ell - 1)!!}{C_\ell} (pr)^{-\ell} + \frac{1}{ (2\ell + 1)!! C_\ell p^{2\ell+1}} \left[ z_\ell(p) (pr)^{\ell+1} + x_\ell(p)  (pr)^{\ell+1} \log\frac{r}{r_0} \right].
\label{eq:G at r=a}
\end{align}
Note that we also keep track of the term scaling as $(pr)^{\ell+1} \log r/r_0$, because it causes a subtle issue which we will discuss later.
Again, $z_\ell(p)$ cannot be determined by the recurrence relations at the origin. On the other hand, $x_\ell(p)$ can be determined by Eq.~\ref{eq:g recurrence 7}.
Here $r_0$ is a parameter which has been introduced to make the argument of the log dimensionless. $r_0$ can be taken to be any value because the difference of $r_0$ can be absorbed by redefining $z_\ell(p)$; however, we will find it useful to take $r_0 \sim a$.
We will discuss the behavior of $z_\ell(p)$ for some examples of $V(r)$ in Sec.~\ref{sec:example one state}.

\subsubsection{Momentum scaling of terms in $k_\ell(p)$} \label{sec:mom dep small k}
Now let us discuss the momentum dependence of the coefficient $k_\ell(p)$.
We only keep the leading term and drop terms of ${\cal O}(p^2 a^2)$ in Eq.~\ref{eq:bc at r=a}.
Then, the boundary condition from the short-range physics can be expressed by two parameters; $c_{0,\ell}$ and $a$.
By using Eqs.~\ref{eq:bc at r=a}-\ref{eq:G at r=a}, we obtain
\begin{align}
k_\ell(p) \simeq \tilde k_\ell + z_\ell(p) + \tilde c_\ell x_\ell(p).\label{eq:kell}
\end{align}
where $\tilde k_\ell$ and $\tilde c_\ell$ are momentum-independent constants which are defined as
\begin{align}
\tilde k_\ell &\equiv -\frac{(2\ell+1)!!(2\ell-1)!!}{a^{2\ell+1}} \frac{\ell + c_{0,\ell}}{\ell + 1 - c_{0,\ell}}, \\
\tilde c_\ell &\equiv \log \frac{a}{r_0} + \frac{1}{\ell+1-c_{0,\ell}}.
\end{align}
Instead of $c_{0,\ell}$ and $a$, we can use these two parameters $\tilde k_\ell$ and $\tilde c_\ell$ to parametrize the effects of the short-range physics.
$\tilde k_\ell$ is ${\cal O}(a^{-2\ell-1})$ and the leading term in $k_\ell(p)$ in most of the cases. As we will see in explicit examples, $z_\ell(p)$ can be large at small $p$ if there exists a zero-energy resonance.
$x_\ell(p)$ can be determined by a recurrence relation given in Eq.~\ref{eq:g recurrence 7}, and it is a polynomial of momentum $p$ in general.

We can extract the momentum-dependent term in $k_\ell(p)$ from the behavior of $G_\ell(r)$.
Let us define $\tilde z_\ell(p)$ as
\begin{align}
\tilde z_\ell(p;a)
\equiv \left[ \frac{p^\ell C_\ell(p)}{2^\ell \ell!}  \frac{d^{2\ell+1}}{dr^{2\ell+1}} r^{\ell} G_\ell(p;r)\right]\Bigg|_{r=a}
\simeq z_\ell(p) + x_\ell(p) \log \frac{a}{r_0} + c'_{2\ell+1} x_\ell(p). \label{eq: def ztilde}
\end{align}
Note that $c'_n$ is an integer $\sim n! \log n$.\footnote{
$c'_n$ is defined as $\frac{d^n}{dx^n}(x^n \log x) = n! + c'_n \log x$.
The recurrence relation is $c'_1 = 1$ and $c'_{n+1} = (n+1) c'_n + n!$.
%The solution is $c'_n = n!(\gamma_E + \Psi(n))$ where $\gamma_E$ is Euler's gamma.}
The solution is $c'_n = n!(\gamma_E + \Psi(n+1))$ where $\gamma_E$ is Euler's gamma.}
The difference between $\tilde z_\ell(p;a)$ and $z_\ell(p) + \tilde c_\ell x_\ell(p)$ is evaluated as
\begin{align}
\tilde z_\ell(p;a) - \left[ z_\ell(p) + \tilde c_\ell x_\ell(p) \right]
\simeq
\left( c'_{2\ell+1} - \frac{1}{\ell + 1 - c_{0,\ell}} \right) x_\ell(p).
\end{align}
Since $x_\ell(p)$ is determined by the long-range force, the RHS of the above equation is at most ${\cal O}( {\rm max}(p^{2\ell+1}, R^{-2\ell-1}  )  )$ where $R$ is the typical length scale of the long-range force (e.g., the Bohr radius).
For example, this will be explicitly seen in Eq.~\ref{eq:zell coulomb approximation} for the Coulomb potential case.
On the other hand, $\tilde k_\ell$ is ${\cal O}(a^{-2\ell-1})$ and the difference between $\tilde z_\ell(p;a)$ and $z_\ell(p) + \tilde c_\ell x_\ell(p)$ is negligible compared to $\tilde k_\ell$.
Thus, we can evaluate $k_\ell(p)$ by replacing $z_\ell(p) + \tilde c_\ell x_\ell(p)$ in Eq.~\ref{eq:kell} with $\tilde z_\ell(p;a)$ as
\begin{align}
	k_\ell(p) \simeq {\tilde k}_\ell + \tilde z_\ell(p;a)
\end{align}
For practical purposes, it is useful to evaluate $k_\ell$ as
\begin{align}
k_\ell(p) \simeq k_\ell(p_0) + \Delta z_\ell(p,p_0;a),
\label{eq:k parameter}
\end{align}
where we define $\Delta z_\ell(p,p_0;a)$ as
\begin{align}
\Delta z_\ell(p,p_0;a) \equiv& \tilde z_\ell(p;a) - \tilde z_\ell(p_0;a).
\end{align}

In Eq.~\ref{eq:k parameter}, $k_\ell(p_0)$ and $\Delta z_\ell(p,p_0;a)$ parametrize the effect of short-range physics and long-range physics, respectively. $\Delta z_\ell(p,p_0;a)$ has a dependence on $a$ and so it seems that the separation between long-range physics and short-range physics is not complete.
However, $\Delta z_\ell(p,p_0;a)$ depends on $a$ via a term $[x_\ell(p) - x_\ell(p_0)] \log(a/r_0)$ and this term is always subdominant compared to the $k_\ell(p_0)$ term. Therefore, we can safely ignore the $a$ dependence in $\Delta z_\ell(p,p_0;a)$.
Note that, in some special cases, $x_\ell(p)$ does not depend on momentum. For example, this happens for the case with $\ell = 0$, or when the potential $V(r)$ has no $1/r$ behavior near the origin. In this case, the $a$-dependence completely disappears and the single parameter $k_\ell(p_0)$ parameterizes the information of the short-range physics. However, for $\ell \geq 1$ and $V(r)$ containing a $1/r$ term, $x_\ell(p)$ in general does depend on $p$. For details, see App.~\ref{sec:mom dep xp}.

To summarize, as shown in Eq.~\ref{eq:k parameter}, the long-range and short-range contributions to $k_\ell(p)$ can be separated into $k_\ell (p_0)$ (short-range effect) and $\Delta z_\ell (p,p_0;a)$ (long-range effect).

\subsubsection{Implications for the $S$-matrix and cross sections}

By using Eq.~\ref{eq:k parameter}, $S_\ell$ can be written as
\begin{align}
S_\ell
&= \frac{k_\ell(p_0) + \Delta z_\ell(p,p_0;a) - i p^{2\ell+1} C_\ell^2}{k_\ell(p_0) + \Delta z_\ell(p,p_0;a) + i p^{2\ell+1} C_\ell^2} \exp\left(2i\delta_\ell^{(L)}\right). \label{eq:Smatrix single state}
\end{align}
Substituting into Eqs.~\ref{eq:onestatesc}, \ref{eq:onestateann}, we have:
\begin{align}
\sigma_{\text{sc},\ell} &= \frac{\pi}{p^2} (2\ell+1) c_i \left| \frac{k_\ell(p_0) + \Delta z_\ell(p,p_0;a) - i p^{2\ell+1} C_\ell^2}{k_\ell(p_0) + \Delta z_\ell(p,p_0;a) + i p^{2\ell+1} C_\ell^2} \exp\left( 2i\delta_\ell^{(L)} \right) - 1 \right|^2, \label{eq:onestatescfinal}\\
\sigma_{\text{ann},\ell} &= 4\pi (2\ell+1) c_i \frac{ p^{2\ell-1} C_\ell^2 {\rm Im}k_\ell(p_0)}{ |k_\ell(p_0) +  \Delta z_\ell(p,p_0;a) + i p^{2\ell+1} C_\ell^2 |^2 }.\label{eq:onestateannfinal}
\end{align}
Choosing a large reference momentum $p_0$ such that the long-range force can be neglected, and applying Eqs.~\ref{eq:sigma sc 0}, \ref{eq:sigma ann 0}, the full annihilation cross section can be written as
\begin{align}
\sigma_{\text{ann},\ell} &= \sigma_{\text{ann},\ell}^{(0)} C_\ell^2
\times \left| 1 + \frac{1}{k_\ell^2(p_0) } \left( \Delta z_\ell(p;p_0) + i p^{2\ell+1} C_\ell^2 \right) \right|^{-2}
%\times \left[ 1 + \frac{1}{k_\ell^2(p_0) } \left( \Delta z_\ell(p;p_0) + i p^{2\ell+1} C_\ell^2 \right) \right]^{-1}
\label{eq:annihilation cross section}, \\
\frac{1}{k_\ell(p_0)} &\simeq \pm \sqrt{ \frac{ \sigma_{\text{sc},\ell}^{(0)}(p_0) }{ 4\pi (2\ell+1) p_0^{4\ell} }  -\left( \frac{ \sigma_{\text{ann},\ell}^{(0)}(p_0) }{ 4\pi (2\ell+1) p_0^{2\ell-1} }\right)^2 } - i \frac{ \sigma_{\text{ann},\ell}^{(0)}(p_0) }{ 4\pi (2\ell+1) p_0^{2\ell-1} }.
\end{align}
Note that the sign of the real part of $k_\ell^{-1}(p_0)$ cannot be determined directly from Eq.~\ref{eq:sigma sc 0}; it depends on whether the short-range force is repulsive or attractive.

\subsection{Comparison with literature}
Let us compare our formula with the literature.
If the short-range cross section is sufficiently small, we can assume $k_\ell(p_0) \gg \Delta z_\ell(p,p_0;a),~p^{2\ell+1} C_\ell^2$.
Then, we can take  $1 + \displaystyle\frac{1}{k^2_\ell(p_0)} (\Delta z_\ell(p) + i p^{2\ell+1} C_\ell^2) \simeq 1$ in Eq.~\ref{eq:annihilation cross section} and obtain 
\begin{align}
\sigma_{\text{ann},\ell} &\simeq \sigma_{\text{ann},\ell}^{(0)} \times C_\ell^2. \label{eq:conventional SE formula}
\end{align}
As mentioned previously, we can see that our formula in this limit is consistent with the conventional formula with Sommerfeld enhancement (e.g., Ref.~\cite{ArkaniHamed:2008qn} for the $s$-wave case and Refs.~\cite{Iengo:2009ni, Cassel:2009wt, Slatyer:2009vg} for higher-$\ell$ cases).

Next, let us compare our results with the $s$-wave formulae given in  Ref.~\cite{Blum:2016nrz}.
As we have discussed, $x_0(p)$ is momentum-independent for the $s$-wave case and we can define
\begin{align}
\Delta z_0(p,p_0) \equiv \lim_{a\to0} \Delta z_0(p,p_0;a).
\end{align}
The mapping between our notation and that of \cite{Blum:2016nrz} can then be written as:
\begin{align}
\exp\left( 2i\delta_0^{(L)} \right) &\to \frac{d_p}{d_p^*}, \\
k_0(p_0) &\to k_{p_0}, \\
\Delta z_0(p,p_0) &\to {\rm Re} g_p'(0) - {\rm Re} g_{p_0}'(0), \\
p C_0^2 &\to {\rm Im} g_p'(0).
\end{align}
With this mapping, we can see that Eqs.~\ref{eq:Smatrix single state}, \ref{eq:onestatescfinal}, \ref{eq:onestateannfinal} are equivalent to Eqs.~(25,~27,~28) in Ref.~\cite{Blum:2016nrz}, respectively.

Ref.~\cite{Flores:2024sfy} studies the preservation of unitarity in the presence of an effective imaginary potential\footnote{A similar attempt at unitarization, to ensure consistency with the optical theorem, has been discussed in Ref.~\cite{Kamada:2022zwb}.}. Whereas our method separates short-distance from long-distance physics, that work's approach separates the real potential from the imaginary part, and does not require the imaginary part to be short-distance; in fact, the assumption needed in that work to obtain a simple expression for the corrected cross section fails for absorptive contact interactions. Ref.~\cite{Flores:2024sfy} defines parameters $x_{\ell,\text{(un)reg}} = \sigma^{\text{elas}}_{\ell,\text{(un)reg}}/\sigma_\ell^U$, $y_{\ell,\text{(un)reg}} = \sigma^{\text{inel}}_{\ell,\text{(un)reg}}/\sigma_\ell^U$, where in our language $\sigma^{\text{inel}}_{\ell,\text{(un)reg}}$ is the inclusive (un)corrected annihilation cross section, $\sigma^{\text{elas}}_{\ell,\text{(un)reg}}$ is the (un)corrected elastic scattering cross section due to the long-range potential, and $\sigma_\ell^U = 2^\delta 4 \pi (2\ell+1)/p^2$ is the partial wave unitarity bound, where $\delta$ is 0 or 1 depending on whether the incoming state contains distinguishable or identical particles (so $c_i=2^\delta$ in our notation).  Our Eqs.~\ref{eq:onestatesc}, \ref{eq:onestateann} thus predict:
\begin{align}
% y_{\ell,\text{reg}} & = 4 p^{2\ell+1} C_\ell^2 \text{Im} (-k_\ell^{-1}(p)) / |1 + i p^{2\ell+1} C_\ell^2 k_\ell^{-1}(p)|^2, \nonumber \\
 y_{\ell,\text{reg}} & =  p^{2\ell+1} C_\ell^2 \text{Im} (-k_\ell^{-1}(p)) / |1 + i p^{2\ell+1} C_\ell^2 k_\ell^{-1}(p)|^2, \nonumber \\
x_{\ell,\text{reg}} & = \frac{\left|(1 - i p^{2\ell+1} C_\ell^2 k_\ell^{-1}(p)) e^{2i\delta_\ell^{(L)}} - (1 +  i p^{2\ell+1} C_\ell^2 k_\ell^{-1}(p))   \right|^2} {4 |1 + i p^{2\ell+1} C_\ell^2 k_\ell^{-1}(p)|^2} . 
 \end{align}
If the ``unregulated'' cross sections are obtained by expanding to lowest order in $k_\ell^{-1}(p)$, and if we could assume $k_\ell^{-1}(p)$ to be purely imaginary, we would then obtain:
\begin{align}
%y_{\ell,\text{unreg}} & = - 4 p^{2\ell+1} C_\ell^2 \text{Im} k_\ell^{-1}(p), \nonumber \\
y_{\ell,\text{unreg}} & = -  p^{2\ell+1} C_\ell^2 \text{Im} k_\ell^{-1}(p), \nonumber \\
x_{\ell,\text{unreg}} & =  \frac{1}{4} \left| e^{2i\delta_\ell^{(L)}} - 1  \right|^2 = \frac{1}{2} (1 - \cos(2\delta_\ell^{(L)})) \nonumber \\
%y_{\ell,\text{reg}} & = y_{\ell,\text{unreg}}  / (1 + y_{\ell,\text{unreg}} /4)^2,\nonumber \\
y_{\ell,\text{reg}} & = y_{\ell,\text{unreg}}  / (1 + y_{\ell,\text{unreg}})^2,\nonumber \\
 x_{\ell,\text{reg}} 
% & = \left[ y_{\ell,\text{unreg}}^2/4  + x_{\ell,\text{unreg}} (1 -  y_{\ell,\text{unreg}}^2 /16)  \right]/ (1 + y_{\ell,\text{unreg}} /4)^2 .
 & = \left[ y_{\ell,\text{unreg}}^2  + x_{\ell,\text{unreg}} (1 -  y_{\ell,\text{unreg}}^2)  \right]/ (1 + y_{\ell,\text{unreg}})^2 .
\end{align}
 These results are identical to the expressions given in Eq.~60 of Ref.~\cite{Flores:2024sfy}. For a detailed comparison of our wave function $u_\ell(r)$ and Ref.~\cite{Flores:2024sfy}, see App.~\ref{sec:comparison with 240502222}.
 In general our $k_\ell^{-1}(p)$ will not be purely imaginary and is not governed purely by the short-range physics, receiving irreducible contributions from the long-range real potential; however, it may often be true that the term arising from the short-range physics is the dominant one.

\subsection{Examples} \label{sec:example one state}
As shown in Eq.~\ref{eq:Smatrix single state}, the $S$-matrix can be written in terms of $\delta_\ell^{(L)}$, $C_\ell^2$, and $\Delta z_\ell(p,p_0;a)$. In order to obtain an explicit form for these variables, we have to solve the Schr\"odinger equation. Let us discuss several explicit examples.

\subsubsection{Spherical well potential}\label{sec:example spherical well}
Let us take the following attractive spherical well potential:
\begin{align}
V(r) = -\frac{p_V^2}{2\mu} \theta(R-r).
\end{align}
We assume that the effective range of the long-range force, $R$, satisfies $R>a$.
The solution of the Schr\"odinger equation (Eqs.~\ref{eq:schroedinger eq F}, \ref{eq:schroedinger eq G}) with the boundary conditions given by Eqs.~\ref{eq:F and G at large r}, \ref{eq:F and G at the origin} can be written as
\begin{align}
G_\ell + iF_\ell = 
\begin{cases}
c_G c_\ell(\tilde p r) + i c_F s_\ell(\tilde p r) & (r<R) \\
\exp\left( i \delta_\ell^{(L)} \right) [c_\ell(pr) + i s_\ell(pr)] & (r\geq R)
\end{cases},
\end{align}
where $\tilde p \equiv \sqrt{p^2 + p_V^2}$.
By using the large-$r$ boundary conditions of Eqs.~\ref{eq:F and G at large r}, \ref{eq:F and G at the origin}, and the requirement of continuity of the wavefunction and its first derivative at $r=R$, we obtain $c_F$ and $c_G$ as
\begin{align}
c_F &= \frac{1}{\tilde p} \exp\left( i\delta_\ell^{(L)} \right) \biggl\{ i\tilde p c'_\ell(\tilde pR) \Bigl[ c_\ell(pR) + is_\ell(pR) \Bigr] -i p c_\ell(\tilde pR) \Bigl[c'_\ell(pR) + is'_\ell(pR) \Bigr] \biggr\}, \\
c_G &= \frac{1}{\tilde p} \exp\left( i\delta_\ell^{(L)} \right) \biggl\{  \tilde p s'_\ell(\tilde pR) \Bigl[c_\ell(pR) + is_\ell(pR) \Bigr] - p s_\ell(\tilde pR) \Bigl[c'_\ell(pR) + is'_\ell(pR) \Bigr] \biggr\}.
\end{align}
where $c_\ell'(x) = dc_\ell(x)/dx$ and $s_\ell'(x) = ds_\ell(x)/dx$.
$\delta_\ell^{(L)}$ is given as
\begin{align}
\delta_\ell^{(L)} &= -{\rm arg}\biggl( \tilde p s'_\ell(\tilde p R) \Bigl[ c_\ell(pR)+is_\ell(pR) \Bigr] - p s_\ell(\tilde p R) \Bigl[ c_\ell'(pR)+is_\ell'(pR) \Bigr] \biggr)
\end{align}
Note that ${\rm Im}c_G = 0$ and ${\rm Re}c_F = p / \tilde p c_G$.

Next we turn to evaluating the Sommerfeld factor.
$C_\ell$ defined in Eq.~\ref{eq:F and G at the origin} is given by
\begin{align}
C_\ell = \frac{\tilde p^{\ell+1}}{p^{\ell+1}} {\rm Re}c_F = \frac{\tilde p^\ell}{p^\ell} \frac{1}{c_G}. \label{eq:Cell square well}
\end{align}
Thus the Sommerfeld enhancement to annihilation, $C_\ell^2$, can be calculated as
\begin{align}
C_\ell^2
= \frac{\tilde p^{2\ell+2}/p^{2\ell}}{ \left[ p s_\ell(\tilde p R) c_\ell'(p R) - \tilde p c_\ell(p R) s_\ell'(\tilde p R) \right]^2 + \left[ p s_\ell(\tilde p R) s_\ell'(p R) - \tilde p s_\ell(p R) s_\ell'(\tilde p R) \right]^2  }
\end{align}

\begin{figure}
	\centering
	\includegraphics[width=0.32\hsize]{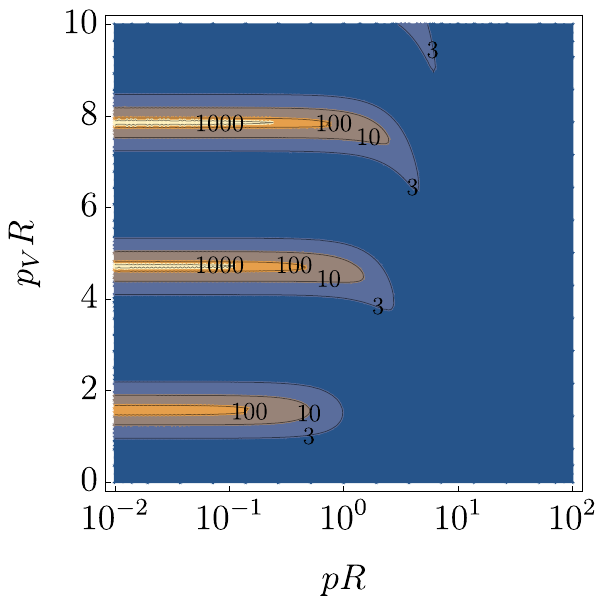}
	\includegraphics[width=0.32\hsize]{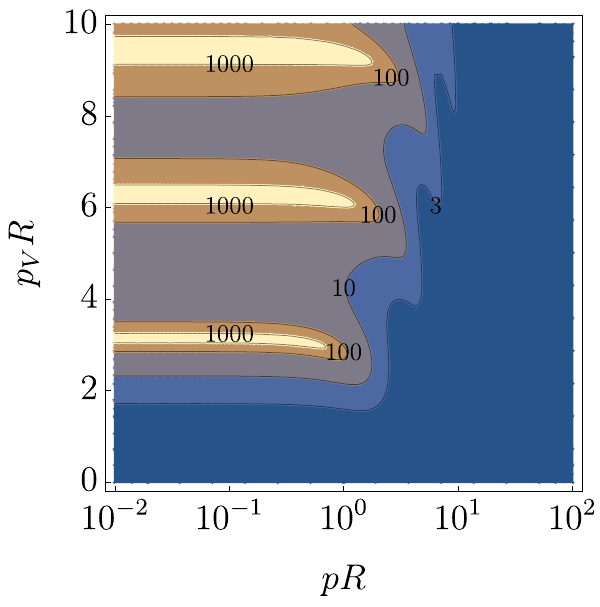}
	\includegraphics[width=0.32\hsize]{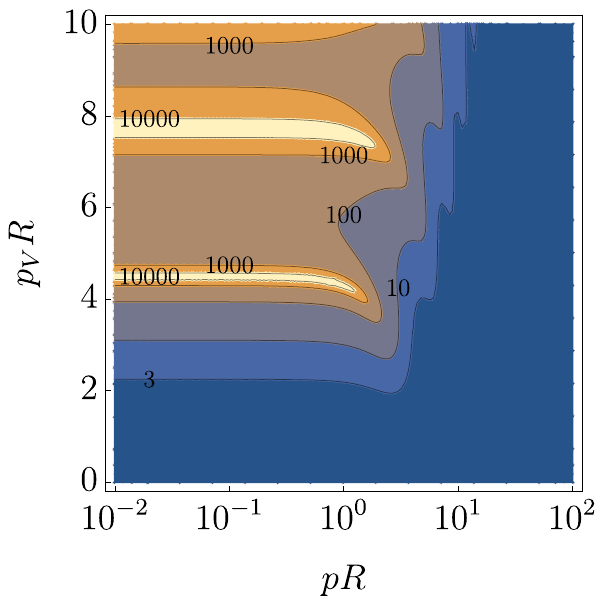}
	\caption{The uncorrected Sommerfeld enhancement factor $C_\ell^2$ for a finite square-well potential with $\ell = 0$ ({\it left panel}), $\ell = 1$ ({\it middle panel}), and $\ell = 2$ ({\it right panel}).
}\label{fig:sfactor squarewell}
\end{figure}

The numerical value of $C_\ell^2$ is plotted in Fig.~\ref{fig:sfactor squarewell}. We can see that $C_\ell^2$ is strongly enhanced at small velocity if $p_V R$ takes some specific values. This behavior comes from the existence of a zero-energy resonance.
The wavefunction of the zero-energy state behaves as $u_\ell \propto s_\ell(p_V r)$ at $r<R$ and $u_\ell \propto r^{-\ell}$ at $r>R$. Then, by requiring continuity of $u_\ell' / u_\ell$ at $r=R$, we obtain
\begin{align}
J_{\ell-1/2}(p_V R) = 0.
\end{align}
By using the Bessel function zeros $j_{n,k}$, we obtain $p_V R = j_{\ell-1/2,k}$ as the condition to have a zero-energy resonance.
For example,
\begin{align}
p_V R &= \frac{\pi}{2},~ \frac{3\pi}{2},~ \frac{5\pi}{2}, \cdots (\ell = 0), \\
p_V R &= \pi,~ 2\pi,~ 3\pi, \cdots (\ell = 1), \\
p_V R &= 4.49,~7.73,~10.9,\cdots (\ell = 2).
\end{align}
If this condition is satisfied, the low energy behavior of $C_\ell^2$ is given as follows:
\begin{align}
	C_\ell^2 \propto
	\begin{cases}
	p^{-2} & (\ell = 0 ) \\
	p^{-4} & (\ell \geq 1 ) 
	\end{cases}.
\end{align}
This behavior is consistent with the result obtained by using Levinson's theorem and Omn\`es solution in Ref.~\cite{Kamada:2023iol}.

\begin{figure}
	\centering
	\includegraphics[width=0.46\hsize]{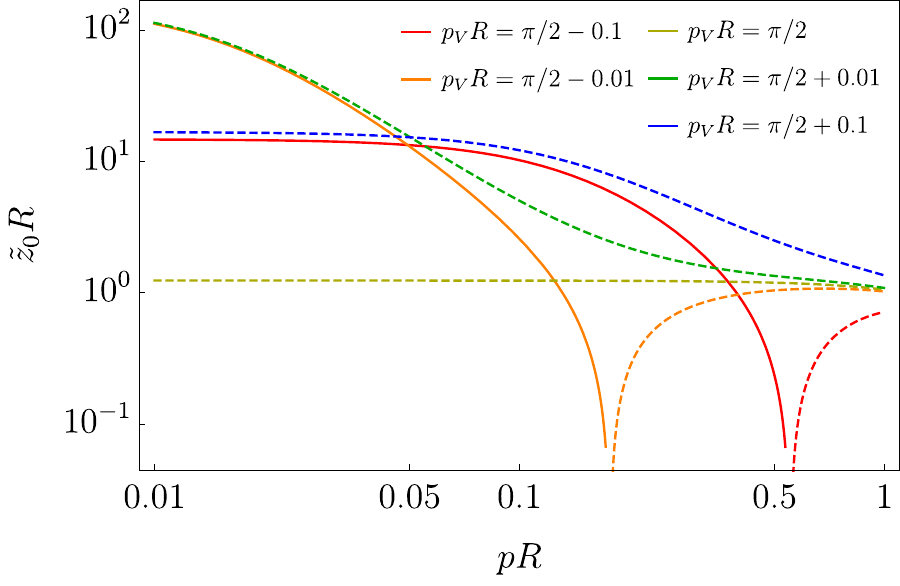}
	\includegraphics[width=0.46\hsize]{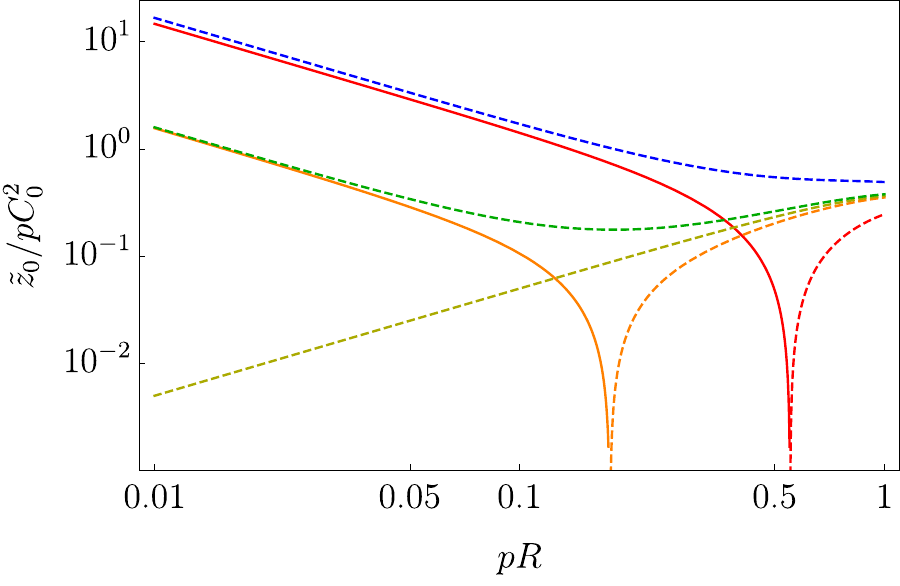}\\[5mm]
	\includegraphics[width=0.46\hsize]{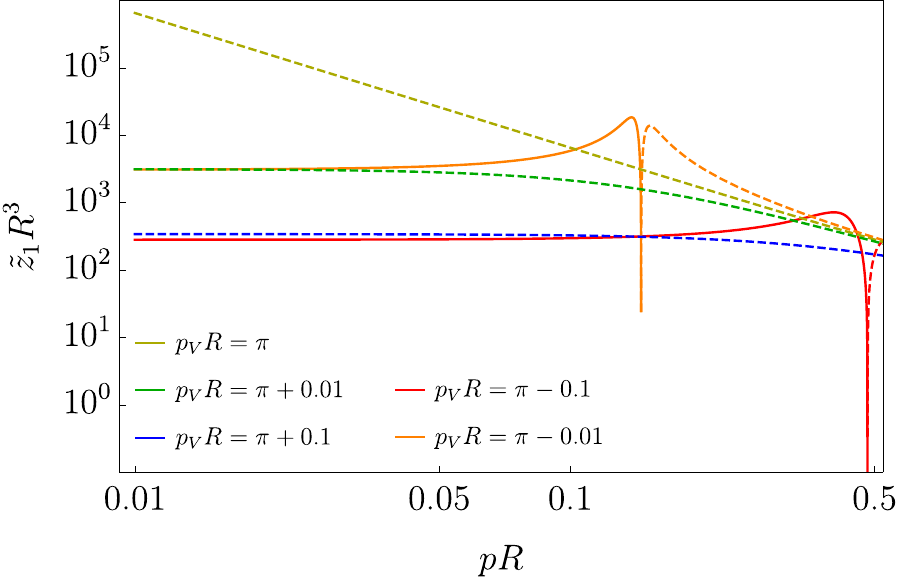}
	\includegraphics[width=0.46\hsize]{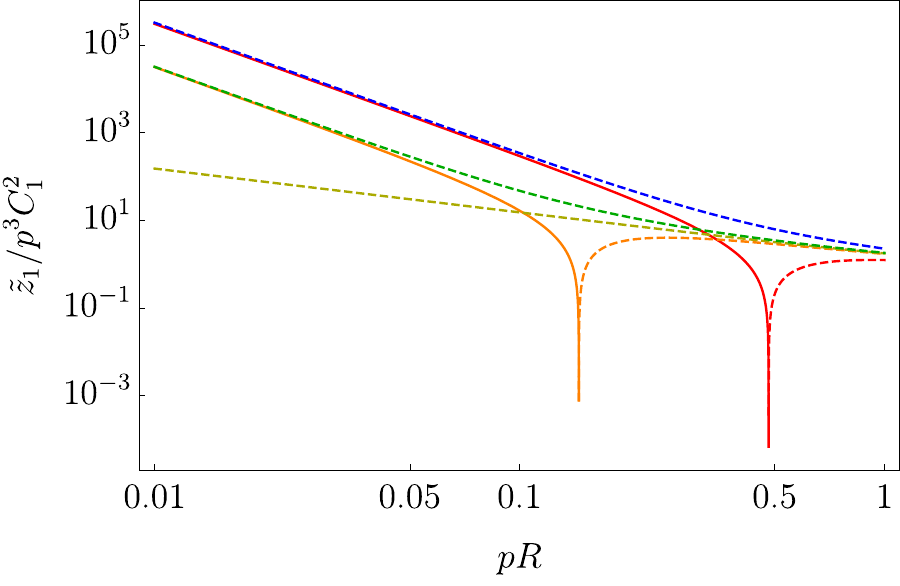}\\[5mm]
	\includegraphics[width=0.46\hsize]{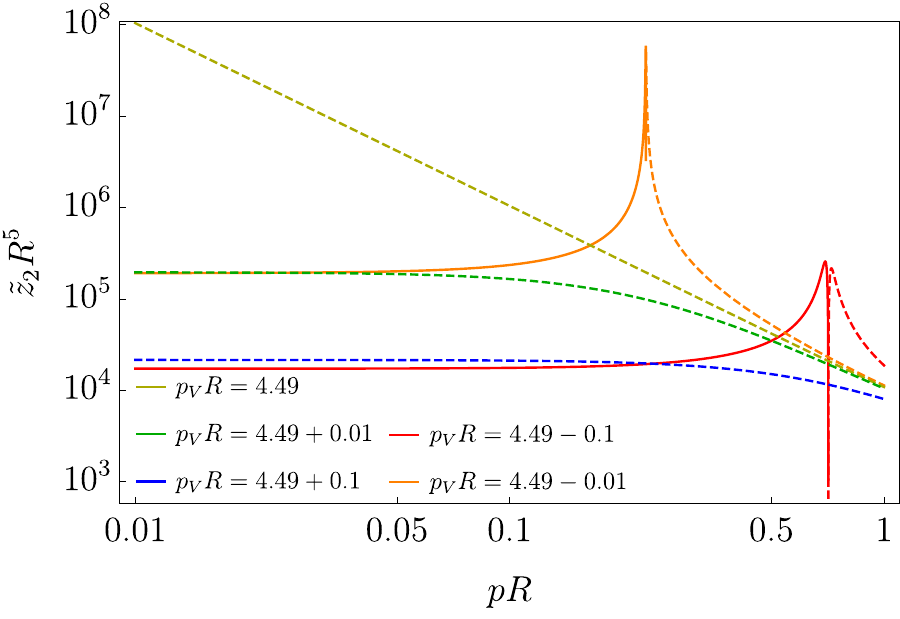}
	\includegraphics[width=0.46\hsize]{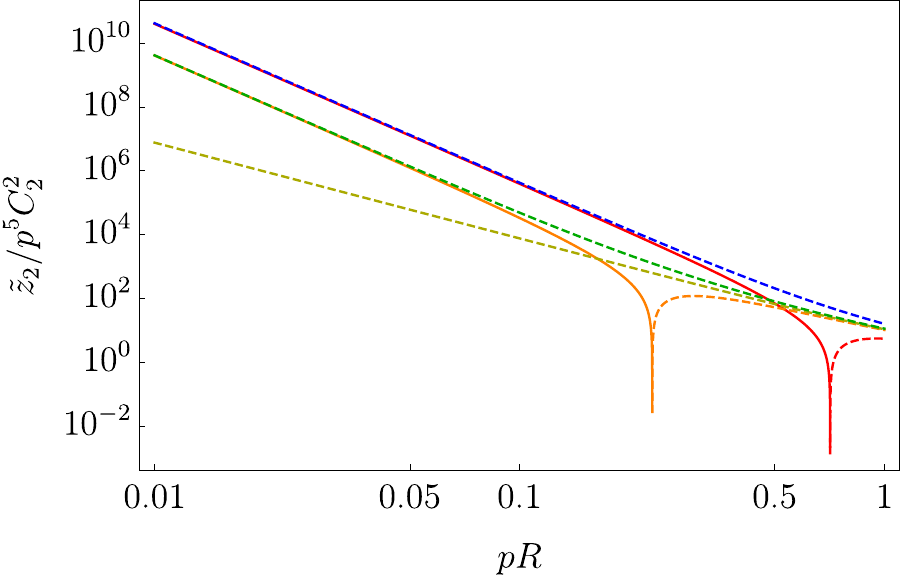}\\[5mm]
	\caption{\emph{Left:} $\tilde z_\ell(p;0) R^{2\ell+1}$ for the finite square-well potential for $\ell=0$ ({\it upper}), $\ell = 1$ ({\it middle}), and $\ell = 2$ ({\it lower}). \emph{Right:} $\tilde z_\ell(p;0) / p^{2\ell+1} C_\ell^2$ for the finite square-well potential for $\ell=0$ ({\it upper}), $\ell = 1$ ({\it middle}), and $\ell = 2$ ({\it lower}). Solid (dashed) curves indicate positive (negative) $\tilde z_\ell(p;0)$. }
	\label{fig:z squarewell}
\end{figure}

\begin{figure}
	\centering
	\includegraphics[width=0.32\hsize]{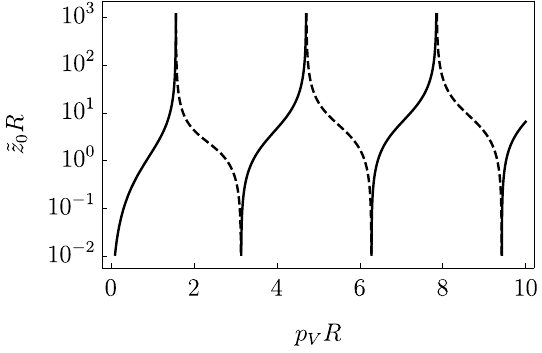}
	\includegraphics[width=0.32\hsize]{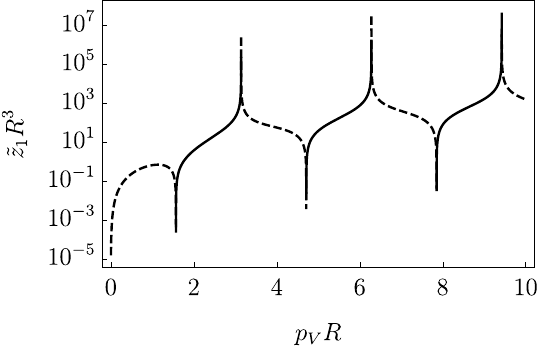}
	\includegraphics[width=0.32\hsize]{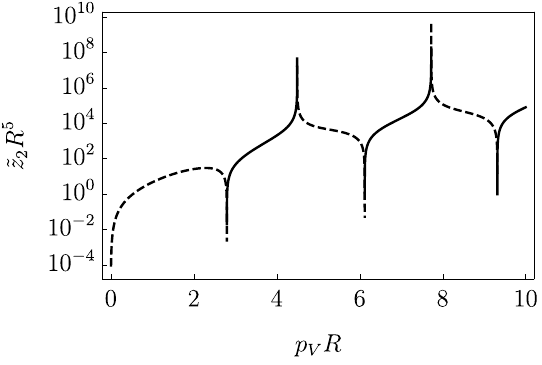}
	\caption{$|\tilde z_\ell(0;0) R^{2\ell+1}|$ for the finite square-well potential with $\ell = 0$ ({\it left}), $\ell = 1$ ({\it middle}), and $\ell = 2$ ({\it right}). Solid (dashed) curves indicate positive (negative) $\tilde z_\ell(0;0)$.}\label{fig:zlowv squarewell}
\end{figure}

Now let us discuss the $\tilde z_\ell$ function. Since the potential $V(r)$ around $r=0$ is flat and there is no $1/r$ term, $x_\ell$ is equal to $0$ in this case. Thus, $\tilde z_\ell$ is finite in the limit $a \to 0$. By using Eq.~\ref{eq: def ztilde}, we obtain
\begin{align}
\lim_{a\to 0} \tilde z_\ell(p;a) = - p^\ell {\tilde p}^{\ell+1} C_\ell {\rm Im} c_F.
\end{align}
The momentum dependence of $\tilde z_\ell$ around a zero-energy resonance is shown in the left panels of Fig.~\ref{fig:z squarewell}, and $\tilde z_\ell$ in the zero-momentum limit is shown in Fig.~\ref{fig:zlowv squarewell}.
The left panels of Fig.~\ref{fig:z squarewell} show, if there exists a zero-energy resonance, i.e., $p_V R = j_{\ell-1/2,k}$ is satisfied, the low energy behavior of $\tilde z_\ell(p;0)$ is given as follows:
\begin{align}
\tilde z_\ell(p;0) \propto
\begin{cases}
p^0 & (\ell = 0) \\
p^{-2} & (\ell \geq 1)
\end{cases}
\end{align}

We can see that $\tilde z_\ell(p;0)$ diverges in the limit of $p\to 0$ for $\ell \geq 1$. For the $\ell = 0$ case, $\tilde z_\ell(p;0)$ does not diverge at $p=0$. However, $\tilde z_\ell(p;0)$ at small $p$ is enhanced when the parameter is slightly displaced from the zero-energy resonance.
For $p_V R \simeq (2k-1)\pi$, $\tilde z_0(0;0)$ in the limit of $p\to 0$ behaves as
\begin{align}
	\tilde z_0(0;0) \simeq \frac{(2k-1)\pi}{2R}\frac{1}{p_V R - (2k-1)\pi}
\end{align}
Thus, $\tilde z_\ell$ is not negligible if there is a zero-energy resonance. We can define $\Delta z_\ell(p,p_0) \equiv \tilde z_\ell(p) - \tilde z_\ell(p_0)$ where $p_0$ is a reference momentum. Obviously, $\Delta z_\ell(p,p_0)$ becomes large at low velocity if there exists a zero-energy resonance.

\begin{figure}
	\centering
	\includegraphics[width=0.32\hsize]{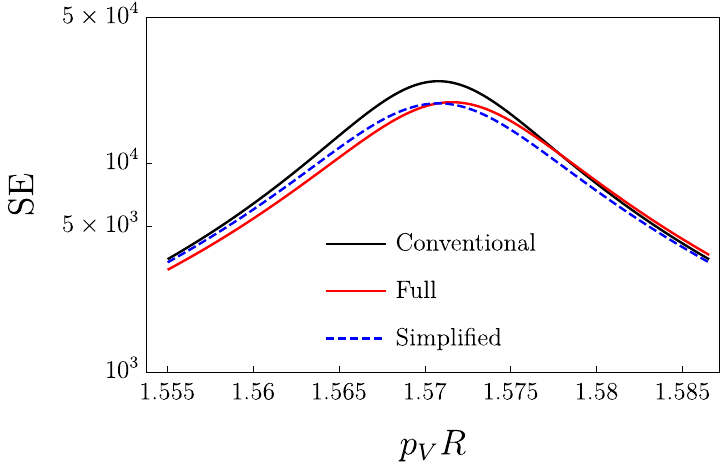}
	\includegraphics[width=0.32\hsize]{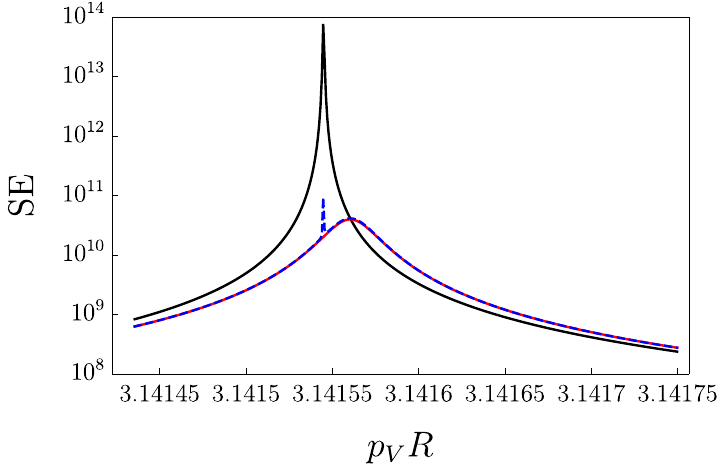}
	\includegraphics[width=0.32\hsize]{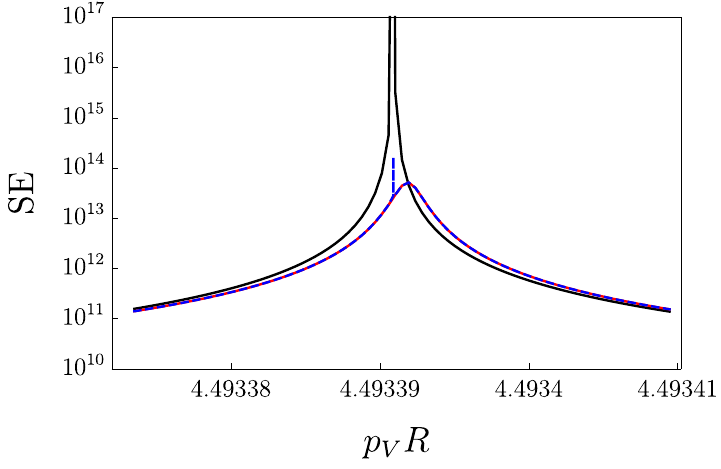}
	\caption{The ratio between the annihilation cross sections with and without the Sommerfeld effect (i.e.~the Sommerfeld enhancement factor) for the finite square-well potential, for the $\ell=0$ ({\it left}), $\ell=1$ ({\it middle}) and $\ell=2$ ({\it right}) cases. The black solid curves shows the results computed using the conventional formula in Eq.~\ref{eq:conventional SE formula}, the red solid curves correspond to the full corrected result given in Eq.~\ref{eq:onestateannfinal}, and the blue dashed curves correspond to the simplified formula of Eqs.~\ref{eq:sigma simplified swave}, \ref{eq:sigma simplified higherell}. We take $k(p_0) = (10^3 + 10^3 i)R^{-1}$ for $s$-wave, $(10^6 + 10^6 i)R^{-3}$ for $p$-wave, and $(10^9 + 10^9 i)R^{-5}$ for $d$-wave, and choose $p = 10^{-2} R^{-1}$ in all cases.}\label{fig:annihilation cross section square well}
\end{figure}

At the location of a zero-energy resonance with $\ell = 0$, $|\tilde z_0(p;0)|$ tends to be smaller than $p C_0^2$ at low velocity; the upper right panel of Fig.~\ref{fig:z squarewell} displays the numerical value of the ratio between $\tilde z_0$ and $p C_0^2$. We can approximate the annihilation cross section as
\begin{align}
	\sigma_{{\rm ann},\ell} \simeq \sigma_{{\rm ann},\ell}^{(0)} \times C_\ell^2 \left|1 + \frac{ipC_0^2 }{k_\ell(p_0)}\right|^{-2}. \label{eq:sigma simplified swave}
\end{align}
As pointed out in Ref.~\cite{Blum:2016nrz}, although the conventional formula $\sigma_{{\rm ann},\ell} \simeq \sigma_{{\rm ann},\ell}^{(0)} \times C_\ell^2$ could violate the unitarity bound, the factor $\left|1 + ipC_0^2 /k_\ell(p_0)\right|^{-2}$ in the above formula regulates the divergent behavior of the annihilation cross section. We see that $\sigma_{{\rm ann},\ell} / \sigma_{{\rm ann},\ell}^{(0)}$ becomes a momentum-independent constant in the limit $p\to 0$.
On the other hand, in proximity to a zero-energy resonance with $\ell \geq 1$, $|\tilde z_\ell(p;0)| \gg p^{2\ell+1}C_\ell^2$ is satisfied at low velocity; again the ratio between $\tilde z_\ell$ and $p^{2\ell+1} C_\ell^2$ is plotted in the right panels of Fig.~\ref{fig:z squarewell}. We can approximate the annihilation cross section as
\begin{align}
	\sigma_{{\rm ann},\ell} \simeq \sigma_{{\rm ann},\ell}^{(0)} \times C_\ell^2 \left|1 + \frac{\tilde z_\ell(p;0) }{k_\ell(p_0)}\right|^{-2}. \label{eq:sigma simplified higherell}
\end{align}
We can see that the factor $\left|1 + \tilde z_\ell(p;0) /k_\ell(p_0)\right|^{-2}$ cancels the divergent behavior of $C_\ell$ at low velocity,
and $\sigma_{{\rm ann},\ell} / \sigma_{{\rm ann},\ell}^{(0)}$ becomes a momentum-independent constant in the limit of $p\to 0$.
In Fig.~\ref{fig:annihilation cross section square well}, we show the Sommerfeld enhancement computed by the conventional formula Eq.~\ref{eq:conventional SE formula}, the full formula Eq.~\ref{eq:onestateannfinal}, and the simplified formula Eqs.~\ref{eq:sigma simplified swave}, \ref{eq:sigma simplified higherell}. We can see that the divergent behavior in the conventional formula (which can lead to unitarity violation) is regularized in the full formula given in Eq.~\ref{eq:conventional SE formula}.
For the $s$-wave case, the behavior of the full formula is well captured by the simplified formula in Eq.~\ref{eq:sigma simplified swave}. For higher partial waves, the simplified formula in Eq.~\ref{eq:sigma simplified higherell} describes the behavior of the full formula fairly well except for the points where $\tilde z_\ell$ crosses $0$ and flips sign.

%\begin{figure}
%	\centering
%	\includegraphics[width=0.32\hsize]{fig/sfactor_squarewell_swave.pdf}
%	\includegraphics[width=0.32\hsize]{fig/sfactor_squarewell_pwave.pdf}
%	\includegraphics[width=0.32\hsize]{fig/sfactor_squarewell_dwave.pdf}
%	\caption{The uncorrected Sommerfeld enhancement factor $C_\ell^2$ for a finite square-well potential with $\ell = 0$ ({\it left panel}), $\ell = 1$ ({\it middle panel}), and $\ell = 2$ ({\it right panel}).
%}\label{fig:sfactor squarewell}
%\end{figure}

%\begin{figure}
%	\centering
%	\includegraphics[width=0.46\hsize]{fig/squarewell_z_swave.pdf}
%	\includegraphics[width=0.46\hsize]{fig/squarewell_zovercsq_swave.pdf}\\[5mm]
%	\includegraphics[width=0.46\hsize]{fig/squarewell_z_pwave.pdf}
%	\includegraphics[width=0.46\hsize]{fig/squarewell_zovercsq_pwave.pdf}\\[5mm]
%	\includegraphics[width=0.46\hsize]{fig/squarewell_z_dwave.pdf}
%	\includegraphics[width=0.46\hsize]{fig/squarewell_zovercsq_dwave.pdf}\\[5mm]
%	\caption{\emph{Left:} $\tilde z_\ell(p;0) R^{2\ell+1}$ for the finite square-well potential for $\ell=0$ ({\it upper}), $\ell = 1$ ({\it middle}), and $\ell = 2$ ({\it lower}). \emph{Right:} $\tilde z_\ell(p;0) / p^{2\ell+1} C_\ell^2$ for the finite square-well potential for $\ell=0$ ({\it upper}), $\ell = 1$ ({\it middle}), and $\ell = 2$ ({\it lower}). Solid (dashed) curves indicate positive (negative) $\tilde z_\ell(p;0)$. }
%	\label{fig:z squarewell}
%\end{figure}

\subsubsection{Coulomb potential} \label{sec:example coulomb}
Let us take the attractive Coulomb potential:
\begin{align}
V(r) = -\frac{1}{\mu a_B r}.
\end{align}
Note that $a_B = (\alpha\mu)^{-1}$ is the Bohr radius of this system.
It is useful to define the Coulomb Hankel function:
\begin{align}
H^\pm_\ell(\eta,\rho) &= \frac{(\mp i)^\ell}{\ell!} \frac{1}{\sqrt{g_\ell(\eta)} C_0(\eta)} \Gamma(\ell+1 \pm i\eta) W_{\mp i\eta,\ell+1/2}(\mp 2 i\rho).
\end{align}
where $W$ is the Whittaker function, and $C_0$ and $g_\ell$ are defined as
\begin{align}
C_0(\eta) = \sqrt{\frac{2\pi\eta}{\exp(2\pi\eta)-1}}, \qquad
g_\ell(\eta) = \prod_{k=1}^\ell \left( 1+ \frac{\eta^2}{k^2} \right).
\end{align}
Note that $C_0^2(\eta = -1/pa_B)$ is the conventional $s$-wave Sommerfeld factor. For details, see, e.g., Ref.~\cite{Gaspard:2018sqy}.

The solutions of the Schr\"odinger equation described in Eqs.~\ref{eq:schroedinger eq F}, \ref{eq:schroedinger eq G},
with boundary conditions given by Eqs.~\ref{eq:F and G at large r}, \ref{eq:F and G at the origin}, can be written as
\begin{align}
F_\ell(r) = {\rm Im} H_\ell^+ \left( -\frac{1}{pa_B}, pr \right), \qquad
G_\ell(r) = {\rm Re} H_\ell^+ \left( -\frac{1}{pa_B}, pr \right).
\end{align}
and we can show that the conventional Sommerfeld factor for the $\ell$-wave case is
\begin{align}
C_\ell = \sqrt{g_\ell\left( -\frac{1}{pa_B}\right)} C_0\left( - \frac{1}{pa_B}\right).
\end{align}
Since the Coulomb potential does not have a zero-energy resonance, $C_\ell$ does not have a singular behavior.
For $p \ll 1/a_B$, $C_\ell^2$ behaves as
\begin{align}
	C_\ell^2 \simeq \frac{2\pi}{(\ell!)^2 (pa_B)^{2\ell+1}}.
\end{align}
We calculate $x_\ell(p)$ in Eq.~\ref{eq:G at r=a} as
\begin{align}
%x_\ell(p) = -2 p^{\ell-1} a_B^{-\ell-2} g_\ell\left( -\frac{1}{pa_B}\right).
x_\ell(p) = -2 p^{2\ell} a_B^{-1} g_\ell\left( -\frac{1}{pa_B}\right).
\end{align}
and we can also calculate $\tilde z_\ell(p)$ as
\begin{align}
\tilde z_0 = \frac{p^0 C_0}{2^0 0!} \frac{d}{dr} G_0 \biggr|_{r=a}
=&
%p a_B
x_0(p) \left[ \log\left(\frac{2a}{a_B}\right)  + 2\gamma_E + h\left(\frac{1}{pa_B}\right) \right], \label{eq:z0 coulomb}\\
\tilde z_1 = \frac{p^1 C_1}{2^1 1!} \frac{d^3}{dr^3} r^1 G_1 \biggr|_{r=a}
=&
%p^2 a_B^2
x_1(p) \left[ \log\left(\frac{2a}{a_B}\right)  + 2\gamma_E + h\left(\frac{1}{pa_B}\right) \right] %\nonumber \\ 
%& 
- \frac{p^2}{2a_B}, \label{eq:z1 coulomb}\\
\tilde z_2 = \frac{p^2 C_2}{2^2 2!} \frac{d^5}{dr^5} r^2 G_2 \biggr|_{r=a}
=&
%p^3 a_B^3
x_2(p) \left[ \log\left(\frac{2a}{a_B}\right)  + 2\gamma_E + h\left(\frac{1}{pa_B}\right) \right] %\nonumber \\
%& 
- \frac{p^2}{16a_B^3} (4 + 13 p^2 a_B^2), \label{eq:z2 coulomb}\\
\tilde z_3 = \frac{p^3 C_3}{2^3 3!} \frac{d^7}{dr^7} r^3 G_3 \biggr|_{r=a}
=&
%p^4 a_B^4
x_3(p) \left[ \log\left(\frac{2a}{a_B}\right)  + 2\gamma_E + h\left(\frac{1}{pa_B}\right) \right] \nonumber \\
& \qquad - \frac{p^2}{48a_B^5} (2 + 23 p^2 a_B^2 + 50 p^4 a_B^4), \label{eq:z3 coulomb}%, \\
\end{align}

\begin{figure}
	\centering
	\includegraphics[width=0.32\hsize]{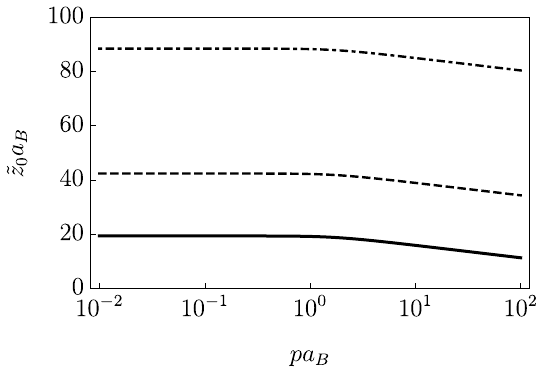}
	\includegraphics[width=0.32\hsize]{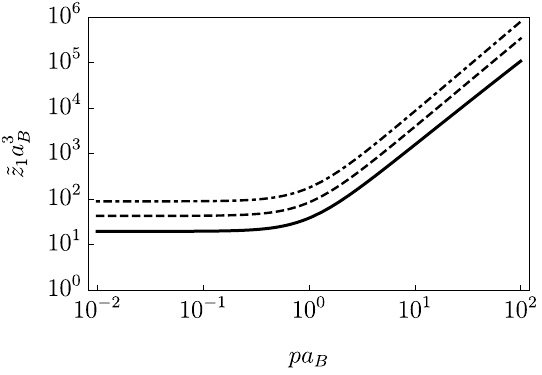}
	\includegraphics[width=0.32\hsize]{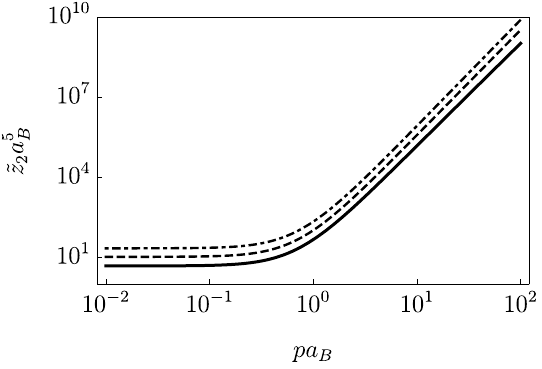}
	\caption{$\tilde z_\ell a_B^{2\ell+1}$ for the Coulomb potential with $\ell = 0$ ({\it left}), $\ell = 1$ ({\it middle}), and $\ell = 2$ ({\it right}). $a$ is taken to be $10^{-5} a_B$ (solid curves), $10^{-10} a_B$ (dashed curves), and $10^{-20} a_B$ (dot-dashed curves).}\label{fig:z coulomb}
\end{figure}

Here $h(\eta)$ is defined as below Eq.~\ref{eq:KfromProtasov}.
Eq.~\ref{eq:z0 coulomb} shows that the $\log a$ term in $G_\ell$ has a momentum-independent coefficient for $\ell = 0$. On the other hand, Eqs.~\ref{eq:z1 coulomb}--\ref{eq:z3 coulomb} show the coefficient is momentum-dependent for $\ell \geq 1$ (although at least in this example, the coefficient becomes momentum-independent for $p \ll 1/a_B$ for general $\ell$, as $g_\ell(\eta) \propto \eta^{2\ell}$ for large $\eta$).
Thus we cannot take the $a\to 0$ limit (and absorb the log term into the high-momentum amplitude) naively for $\ell \geq 1$ as discussed in Sec.~\ref{eq:mom dependence kappa}.
$\tilde z_\ell$ is constructed from a polynomial in $p$ and the function $h(1/pa_B)$.
Note that $h(\eta)$ does not have a singularity for $\eta>0$ and its asymptotic behavior is $h(\eta) \simeq -\log\eta - \gamma_E$ for $\eta\ll 1$ and $h(\eta) \simeq 1/12\eta^2$ for $\eta\gg 1$.
Thus, $\tilde z_\ell$ also does not have divergent behavior at small $p$. By using dimensional analysis, $|\tilde z_\ell|$ should be bounded by $[{\rm max}(p,a_B^{-1})]^{2\ell+1}$. Thus, as long as the short-range effect $k_\ell^{-1}$ is perturbative, $|k_\ell| \gg |z_\ell|$ should be satisfied and the effect of $z_\ell$ is negligible.
The numerical value of $\tilde z_\ell$ is shown in Fig.~\ref{fig:z coulomb}, which shows that a rough behavior of $\tilde z_\ell$ is given as
\begin{align}
\tilde z_\ell \sim a_B^{-2\ell-1} {\rm max}[1, (p a_B)^{2\ell}]. \label{eq:zell coulomb approximation}
\end{align}
In contrast to the case of a finite square well potential, $C_\ell$ and $\tilde z_\ell$ do not demonstrate resonant behavior. This is the expected result because there is no zero-energy resonance in the Coulomb potential.
Since $C_\ell$ and $\tilde z_\ell$ do not have any singular behavior, we can guarantee $|k_\ell(p_0)| \gg |\Delta z(p,p_0;a)|,~p^{2\ell+1} C_\ell^2$, provided the short-range physics effect can be treated perturbatively. Then we can approximate the annihilation cross section as
\begin{align}
	\sigma_{{\rm ann},\ell} \simeq \sigma_{{\rm ann},\ell}^{(0)} \times C_\ell^2.
\end{align}
This formula is the same as the conventional annihilation cross section formula with the Sommerfeld effect.

The scattering problem with a short-range effect and Coulomb force has been discussed in depth in the context of nuclear physics \cite{PopovMur1985, Carbonell:1993, Carbonell:1996vd, Carbonell:1998ei, Protasov:1999ei}.
In order to compare our result with those references, let us rewrite Eq.~\ref{eq:onestateann} by using $C_0$ and $g_\ell$ as
\begin{align}
\sigma_{\text{ann},\ell}
=
4 \pi (2\ell + 1) \frac{ p^{2\ell-1} g_\ell C_0^2 {\rm Im}[-k_\ell^{-1}]}{ |1 + i p^{2\ell+1} g_\ell ( C_0 - i p^{-2\ell-1} g_\ell^{-1} \Delta z )k_\ell^{-1} |^2}. \label{eq:ann2}
\end{align}
Note that
\begin{align}
p^{-2\ell-1} g_\ell^{-1} \Delta z
=
\frac{2}{pa_B} h\left(\frac{1}{pa_B}\right) + \cdots. \label{eq:denominator}
\end{align}
In the $s$-wave case, 
the second term of Eq.~\ref{eq:denominator} does not depend on the momentum $p$. Thus, we can absorb this residual term by redefining $k_\ell$ and we can see that Eq.~\ref{eq:ann2} is consistent with, e.g., Eq.~(1) of \cite{Carbonell:1998ei}.
On the other hand, the second term of Eq.~\ref{eq:denominator} is a polynomial in $p$ and it \textit{does} depend on the momentum $p$ for $\ell \geq 1$.
Thus, we cannot absorb this term by redefining a single momentum independent parameter (although we could absorb it order-by-order in $p^2$, by increasing the number of parameters to describe the short-range physics). This residual polynomial term is not explicitly given in Eq.~(2) of \cite{Carbonell:1998ei} and Eq.~(2) of \cite{Protasov:1999ei}, but they absorb this residual term into their definition of the short-range matrix element $K$, as discussed in Sec.~\ref{sec:s matrix boundary condition}. Thus, our formulation is consistent with the nuclear physics literature \cite{PopovMur1985, Carbonell:1993, Carbonell:1996vd, Carbonell:1998ei, Protasov:1999ei}.

%\begin{figure}
%	\centering
%	\includegraphics[width=0.32\hsize]{fig/z0_coulomb.pdf}
%	\includegraphics[width=0.32\hsize]{fig/z1_coulomb.pdf}
%	\includegraphics[width=0.32\hsize]{fig/z2_coulomb.pdf}
%	\caption{$\tilde z_\ell a_B^{2\ell+1}$ for the Coulomb potential with $\ell = 0$ ({\it left}), $\ell = 1$ ({\it middle}), and $\ell = 2$ ({\it right}). $a$ is taken to be $10^{-5} a_B$ (solid curves), $10^{-10} a_B$ (dashed curves), and $10^{-20} a_B$ (dot-dashed curves).}\label{fig:z coulomb}
%\end{figure}

\subsubsection{Finite range Coulomb potential} \label{example finite coulomb}
Let us take the following finite range Coulomb potential.
\begin{align}
V(r) = -\frac{1}{\mu a_B r} \theta(R-r).
\end{align}
We assume that the effective range of the long-range force, $R$, satisfies $R>a$.
This potential can be regarded as a solvable approximate description of the Yukawa potential.
The solutions of the Schr\"odinger equation described in Eqs.~\ref{eq:schroedinger eq F}, \ref{eq:schroedinger eq G},
with boundary conditions given by Eqs.~\ref{eq:F and G at large r}, \ref{eq:F and G at the origin}, can be written as
\begin{align}
G_\ell + i F_\ell
=
\begin{cases}
c_G G_\ell^{(C)} \left( -1/pa_B, pr \right) + i c_F F_\ell^{(C)} \left( -1/pa_B, pr \right) & (r < R) \\
\exp\left( i\delta_\ell^{(L)} \right) [ c_\ell(pr) + i s_\ell(pr) ] & (r\geq R)
\end{cases}.
\end{align}
Here we use the $(C)$ superscript to indicate results from the Coulomb potential without a cutoff.
Note that ${\rm Im}c_G = 0$ and ${\rm Re}c_F = 1/c_G$ can be shown by using the boundary conditions Eq.~\ref{eq:F and G at large r} and Eq.~\ref{eq:F and G at the origin}, respectively.
The continuity of $G_\ell + iF_\ell$ and $G'_\ell + iF'_\ell$ at $r=R$ imposes:
\begin{align}
c_F &= \exp\left( i\delta_\ell^{(L)} \right) \left( i\frac{1}{p} \frac{dG_\ell^{(C)}}{dr} (c_\ell + is_\ell) -i G_\ell^{(C)} (c'_\ell + is'_\ell) \right) \biggr|_{r=R}, \\
c_G &= \exp\left( i\delta_\ell^{(L)} \right) \left(  \frac{1}{p} \frac{dF_\ell^{(C)}}{dr} (c_\ell + is_\ell) - F_\ell^{(C)} (c'_\ell + is'_\ell) \right) \biggr|_{r=R}.
\end{align}

\begin{figure}
\centering
\includegraphics[width=0.32\hsize]{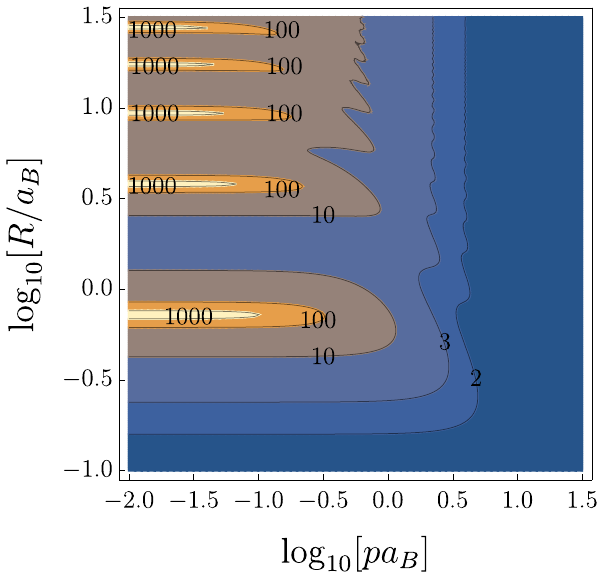}
\includegraphics[width=0.32\hsize]{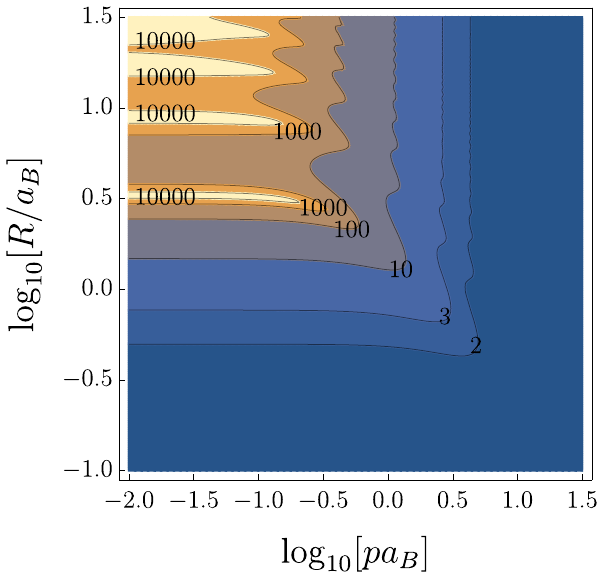}
\includegraphics[width=0.32\hsize]{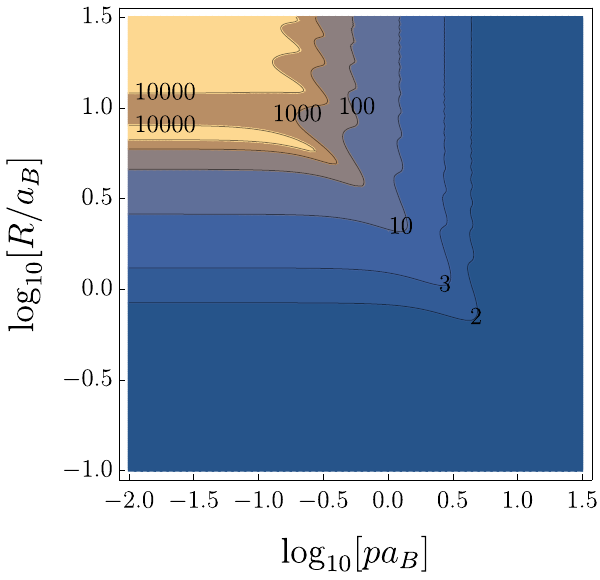}
\caption{The conventional Sommerfeld factor $C_\ell^2$ for a finite-range Coulomb potential (set to zero for $r>R$) for $\ell = 0$ ({\it left}), $\ell = 1$ ({\it middle}), and $\ell = 2$ ({\it right}).}\label{fig:sfactor finitecoulomb}
\end{figure}

The phase shift $\delta_\ell^{(L)}$ and Sommerfeld factor $C_\ell^2$ are calculated as
\begin{align}
\delta_\ell^{(L)} &= -{\rm arg}\left( \frac{1}{p} \frac{dF_\ell^{(C)}}{dr} (c_\ell + is_\ell) - F_\ell^{(C)} (c'_\ell + is'_\ell) \right) \biggr|_{r=R}, \\
C_\ell^2 &= \frac{|C_\ell^{(C)}|^2}{|c_G|^2} =  |C_\ell^{(C)}|^2\left| \frac{1}{p} \frac{dF_\ell^{(C)}}{dr} (c_\ell + is_\ell) - F_\ell^{(C)} (c'_\ell + is'_\ell) \right|^{-2}_{r=R}.
\end{align}
The numerical values of the unregulated Sommerfeld enhancement $C_\ell^2$ are shown in Fig.~\ref{fig:sfactor finitecoulomb}.
Similarly to the spherical well potential discussed in Sec.~\ref{sec:example spherical well}, $C_\ell$ is strongly enhanced at small velocity if $R/a_B$ takes  specific values; this is the standard zero-energy resonance behavior.

The wavefunction of a zero-energy state behaves as $u_\ell \propto \sqrt{r} J_{2\ell+1}(2\sqrt{2r/a_B})$ at $r<R$ and $u_\ell \propto r^{-\ell}$ at $r>R$. Then, by requiring continuity of $u'_\ell/u_\ell$ at $r=R$, we obtain
\begin{align}
\,{}_0F_1\!\left(2\ell+1, -\frac{2R}{a_B} \right) = 0
\end{align}
where $\,{}_0F_1(a,b)$ is the confluent hypergeometric function.
A few numerical solutions of this condition are
\begin{align}
\frac{R}{a_B} &= 0.723,~3.81,~9.36,~\cdots (\ell=0), \\
\frac{R}{a_B} &= 3.30,~8.86,~16.9,~\cdots (\ell=1), \\
\frac{R}{a_B} &= 7.20,~15.3,~25.8,~\cdots (\ell=2).
\end{align}

\begin{figure}[h]
	\centering
	\includegraphics[width=0.46\hsize]{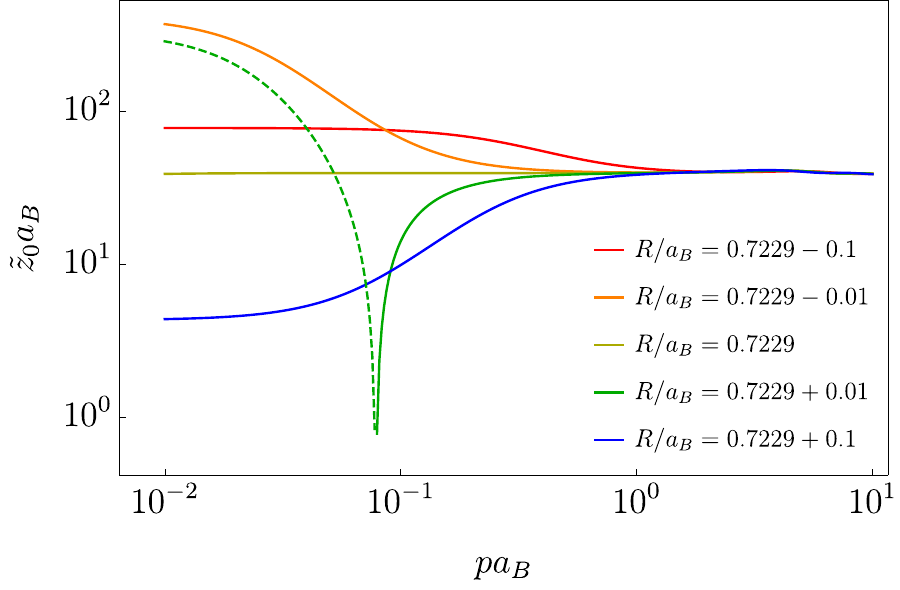}
	\includegraphics[width=0.46\hsize]{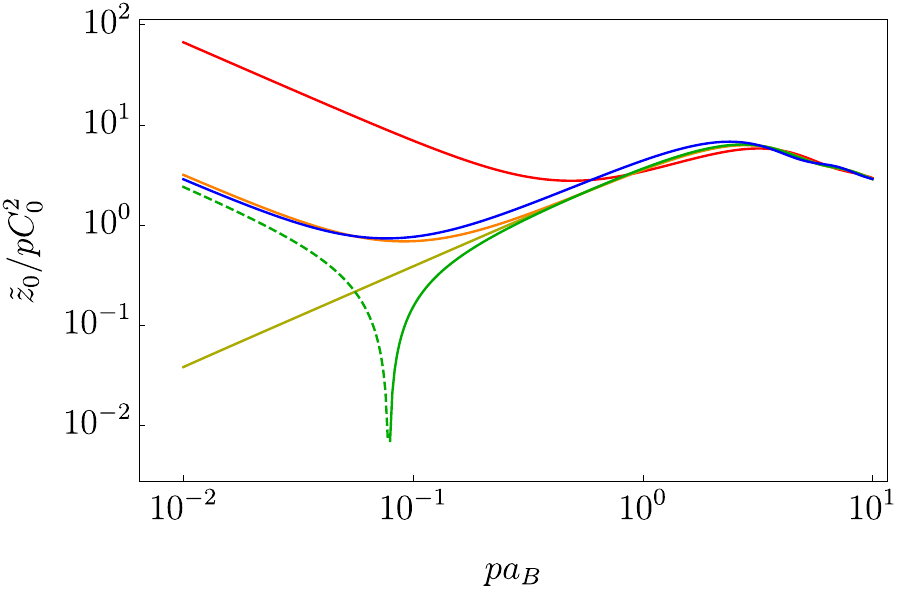}\\[5mm]
	\includegraphics[width=0.46\hsize]{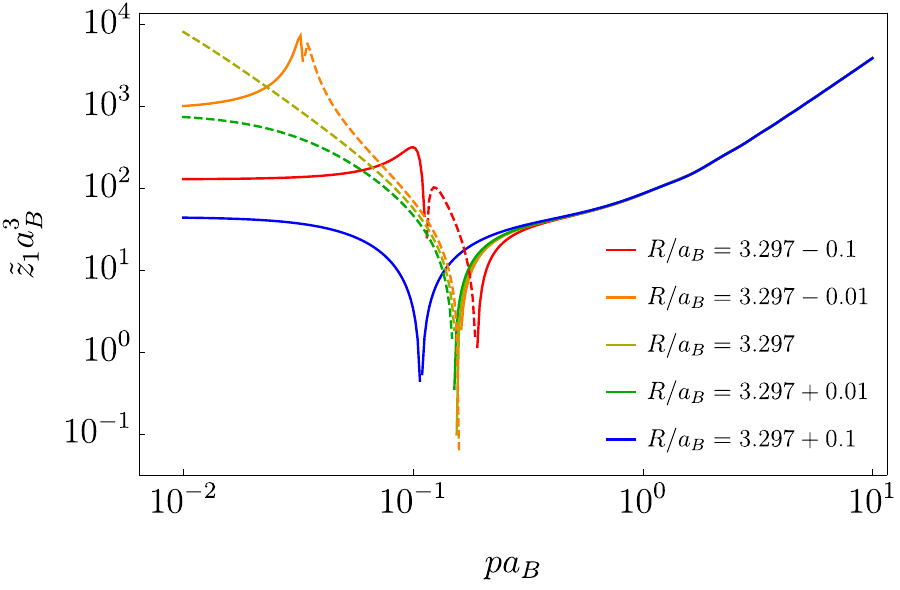}
	\includegraphics[width=0.46\hsize]{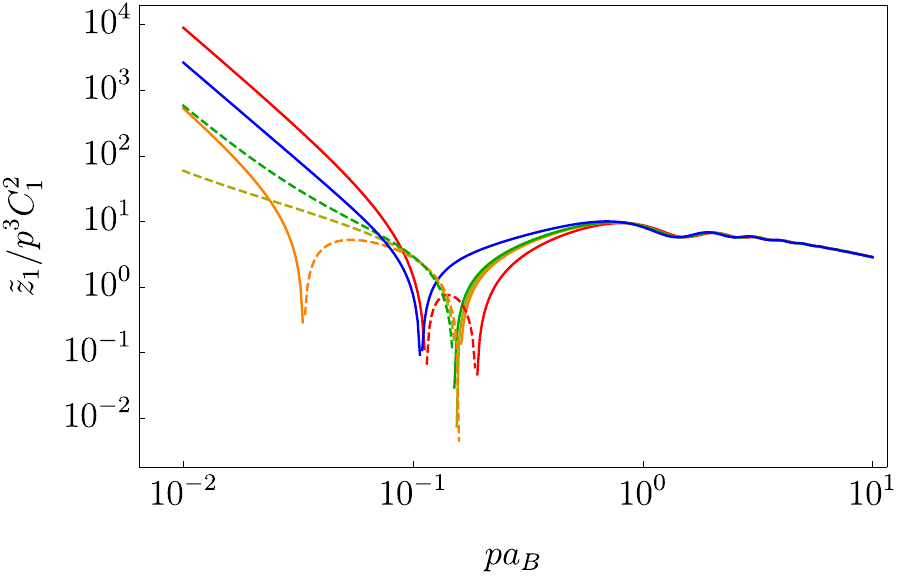}\\[5mm]
	\includegraphics[width=0.46\hsize]{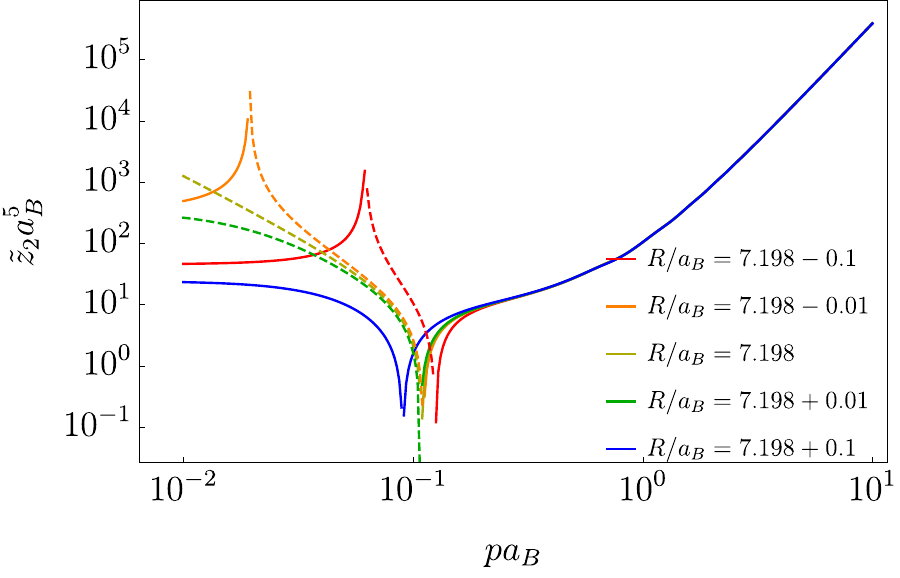}
	\includegraphics[width=0.46\hsize]{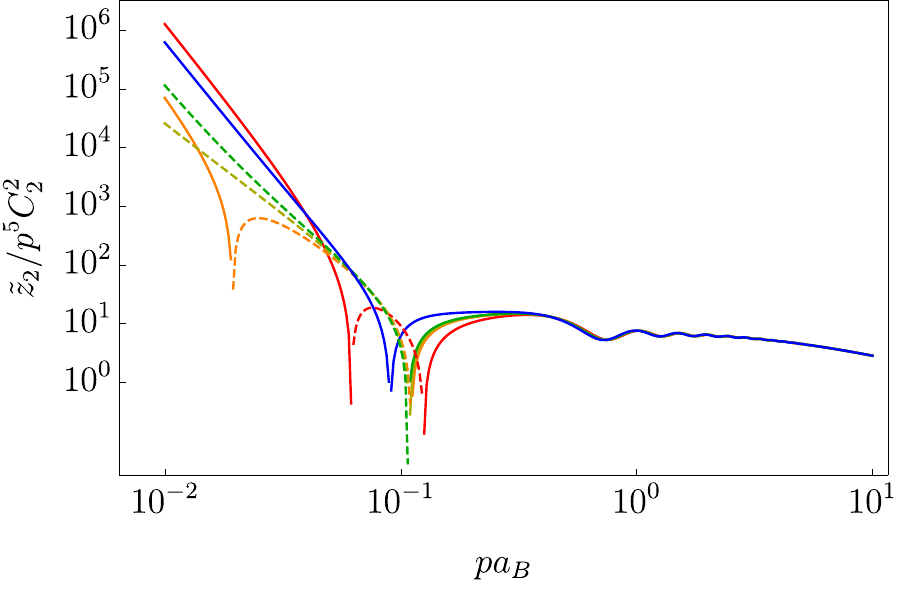}\\[5mm]
	\caption{\emph{Left:} $\tilde z_\ell a_B^{2\ell+1}$ as a function of $p$ for the finite-range Coulomb potential for $\ell=0$ ({\it upper}), $\ell = 1$ ({\it middle}), and $\ell = 2$ ({\it lower}). We take $a = 10^{-10}a_B$. \emph{Right:} $\tilde z_\ell(p;0) / p^{2\ell+1} C_\ell^2$ for the finite-range Coulomb potential for $\ell=0$ ({\it upper}), $\ell = 1$ ({\it middle}), and $\ell = 2$ ({\it lower}). We take $a=10^{-10} a_B$. Solid (dashed) curves indicate positive (negative) values of $\tilde z_\ell$.}\label{fig:z lowv finitecoulomb}
%~\\[1cm]
\end{figure}
If this condition is satisfied, the low energy behavior of $C_\ell^2$ is given as follows:
\begin{align}
	C_\ell^2 \propto
	\begin{cases}
	p^{-2} & (\ell = 0 ) \\
	p^{-4} & (\ell \geq 1 ) 
	\end{cases}.
\end{align}
As in the square-well potential case, this behavior is consistent with the result obtained by using Levinson's theorem and Omn\`es solution in Ref.~\cite{Kamada:2023iol}.

%\begin{figure}
%\centering
%\includegraphics[width=0.32\hsize]{fig/sfactor_finitecoulomb_swave.pdf}
%\includegraphics[width=0.32\hsize]{fig/sfactor_finitecoulomb_pwave.pdf}
%\includegraphics[width=0.32\hsize]{fig/sfactor_finitecoulomb_dwave.pdf}
%\caption{The conventional Sommerfeld factor $C_\ell^2$ for a finite-range Coulomb potential (set to zero for $r>R$) for $\ell = 0$ ({\it left}), $\ell = 1$ ({\it middle}), and $\ell = 2$ ({\it right}).}\label{fig:sfactor finitecoulomb}
%\end{figure}

An explicit calculation shows
\begin{align}
x_\ell(p) = \frac{c_G C_\ell}{C_\ell^{(C)} }  x_\ell^{(C)}(p) = x_\ell^{(C)}(p).
\end{align}
We can see that $x_\ell(p)$ is the same as the Coulomb potential because $x_\ell(p)$ is determined by the recursion relations at the origin, as shown in Eq.~\ref{eq:xell formula}.
As shown in Eq.~\ref{eq: def ztilde}, the $a$-dependence in $\tilde z_\ell$ is determined by $x_\ell$.
Thus, we can expect that $z_\ell$ has a similar term to $z_\ell^{(C)}$ and can explicitly show that
\begin{align}
\tilde z_\ell = 
\frac{p^\ell C_\ell}{2^\ell \ell!} \frac{d^{2\ell+1}}{dr^{2\ell+1}} G_\ell \biggr|_{r=a}
&= -{\rm Im}c_F  C_\ell C_\ell^{(C)} p^{2\ell+1} + c_G \frac{p^\ell C_\ell}{2^\ell \ell!} \frac{d^{2\ell+1}}{dr^{2\ell+1}} G_\ell^{(C)} \biggr|_{r=a} \nonumber\\
&= -{\rm Im}c_F  C_\ell C_\ell^{(C)} p^{2\ell+1} + \tilde z_\ell^{(C)}.
\end{align}

\begin{figure}
	\centering
	\includegraphics[width=0.32\hsize]{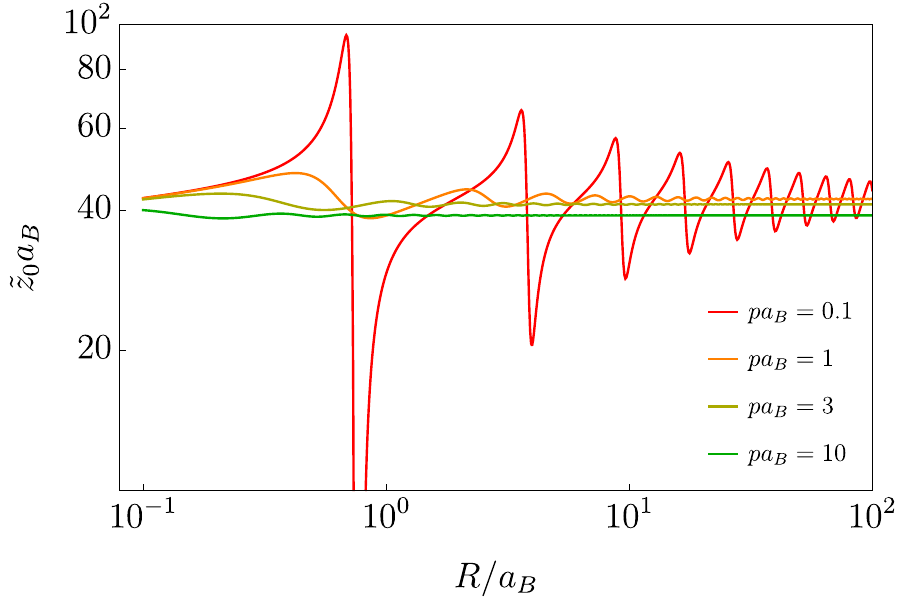}
	\includegraphics[width=0.32\hsize]{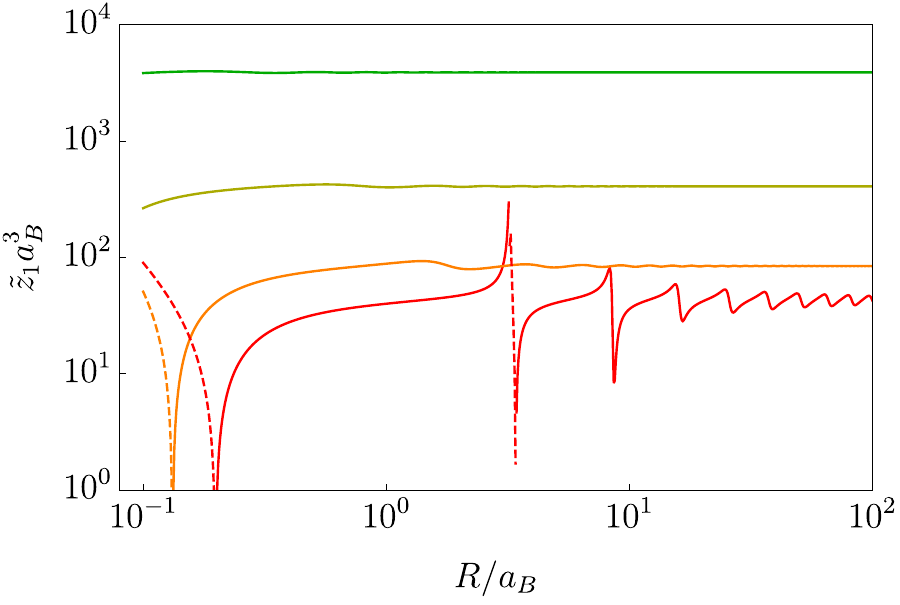}
	\includegraphics[width=0.32\hsize]{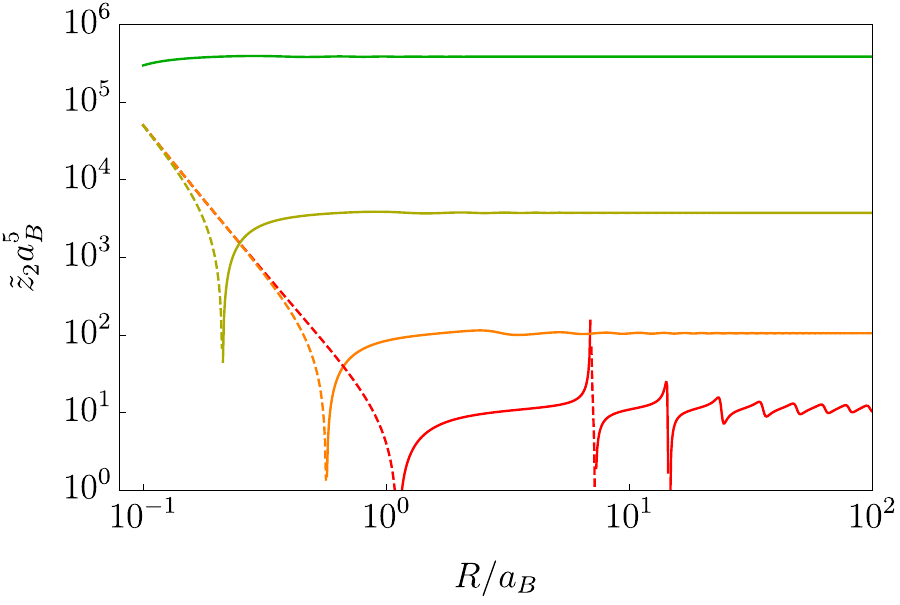}
	\caption{$\tilde z_\ell a_B^{2\ell+1}$ as a function of $R$ for the finite-range Coulomb potential with $\ell=0$ ({\it left}), $\ell = 1$ ({\it middle}), and $\ell = 2$ ({\it right}). We show results for several different values of the momentum: $p = 0.1a_B^{-1}$ (red curves), $a_B^{-1}$ (orange curves), $3 a_B^{-1}$ (yellow curves), and $10 a_B^{-1}$ (green curves). We take $a = 10^{-10} a_B$. Solid (dashed) curves indicate positive (negative) values of $\tilde z_\ell$.}
\label{fig:z finitecoulomb}
\end{figure}

Note that we have used $c_G C_\ell = C_\ell^{(C)}$.
The $a$-dependent term in $\tilde z_\ell$ is the second term of the RHS, $\tilde z_\ell^{(C)}$.
Since the coefficient of the $\log a$ term depends on $p$ for $\ell \geq 1$ (at least for $p\gtrsim 1/a_B$), we cannot absorb $\log a$ dependent terms into a momentum independent constant $k_\ell$ for $\ell \geq 1$.
Fig.~\ref{fig:z lowv finitecoulomb} and Fig.~\ref{fig:z finitecoulomb} show that, if there exists a zero-energy resonance, i.e., $\,{}_0F_1\!\left(2\ell+1, -2R/a_B \right) = 0$ is satisfied, the low energy behavior of $\tilde z_\ell$ is given as follows:
\begin{align}
\tilde z_\ell \propto
\begin{cases}
p^0 & (\ell = 0 ) \\
p^{-2} & (\ell \geq 1 ) 
\end{cases}.
\end{align}
We can see $\tilde z_\ell$ diverges in the limit of $p\to 0$ for $\ell \geq 1$. For the $\ell = 0$ case, $\tilde z_\ell$ does not diverge at $p=0$, however, $\tilde z_\ell$ at small $p$ is enhanced when the parameter is slightly displaced from the zero-energy resonance.
Thus, $\tilde z_\ell$ is not negligible if there is a zero-energy resonance.
We can define $\Delta z_\ell(p,p_0) \equiv \tilde z_\ell(p) - \tilde z_\ell(p_0)$ where $p_0$ is a reference momentum. Obviously, $\Delta z_\ell(p,p_0)$ becomes large at low velocity if there exists a zero-energy resonance.

\begin{figure}
	\centering
	\includegraphics[width=0.32\hsize]{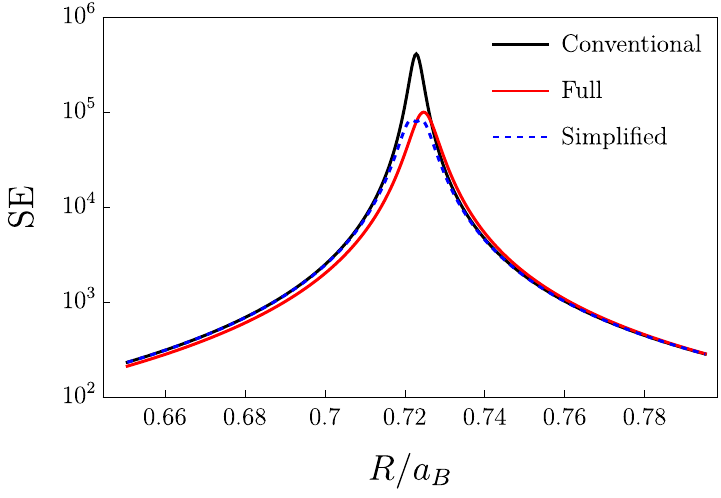}
	\includegraphics[width=0.32\hsize]{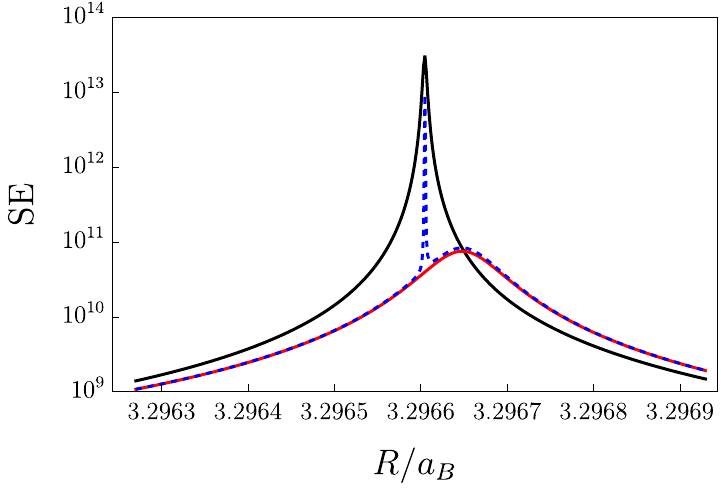}
	\includegraphics[width=0.32\hsize]{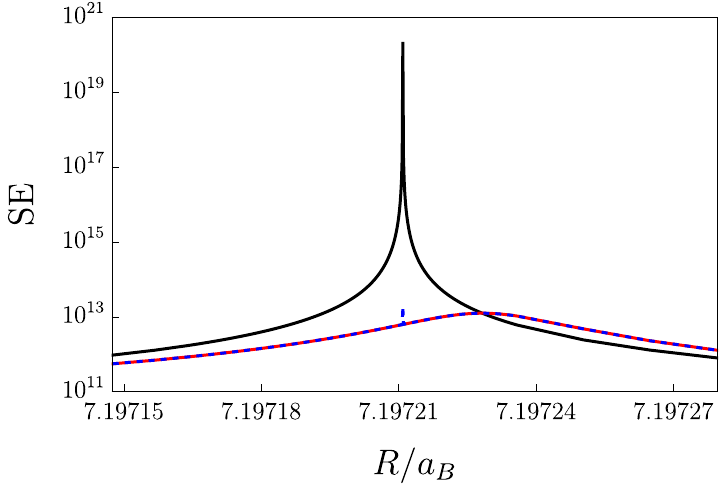}
	\caption{The ratio between the annihilation cross sections with and without the Sommerfeld effect for the finite-range Coulomb potential, for the $\ell=0$ ({\it left}), $\ell=1$ ({\it middle}) and $\ell=2$ ({\it right}) cases. The black solid curves shows the results computed using the conventional formula in Eq.~\ref{eq:conventional SE formula}, the red solid curves correspond to the full corrected result given in Eq.~\ref{eq:onestateannfinal}, and the blue dashed curves correspond to the simplified formula of
 Eqs.~\ref{eq:sigma simplified swave finitecoulomb}, \ref{eq:sigma simplified higherell finitecoulomb}. We take $k(p_0) = (10^3 + 10^3 i)a_B^{-1}$ for the $s$-wave case, $(10^5 + 10^5 i)a_B^{-3}$ for the $p$-wave case, and $(10^5 + 10^5 i)a_B^{-5}$ for the $d$-wave case. We choose $p = 5 \times 10^{-3} a_B^{-1}$ and $a = 10^{-10} a_B$ in all cases.}\label{fig:annihilation cross section finitecoulomb}
\end{figure}

The behavior of the annihilation cross section in proximity to a zero-energy resonance is similar to the square-well potential case. For $\ell = 0$, we can assume $|\tilde z_0| \ll p C_0^2$ at low velocity; see the upper right panel of Fig.~\ref{fig:z lowv finitecoulomb} for the numerical value of the ratio between $\tilde z_0$ and $p C_0^2$. We obtain
\begin{align}
	\sigma_{{\rm ann},\ell} \simeq \sigma_{{\rm ann},\ell}^{(0)} \times C_\ell^2 \left|1 + \frac{ipC_0^2 }{k_\ell(p_0)}\right|^{-2}. \label{eq:sigma simplified swave finitecoulomb}
\end{align}
For $\ell \geq 1$, we can assume $|\tilde z_\ell| \gg p^{2\ell+1} C_\ell^2$ at low velocity. See the right panels of Fig.~\ref{fig:z lowv finitecoulomb} for numerical value of the ratio between $\tilde z_\ell$ and $p^{2\ell+1} C_\ell^2$. We obtain
\begin{align}
	\sigma_{{\rm ann},\ell} \simeq \sigma_{{\rm ann},\ell}^{(0)} \times C_\ell^2 \left|1 + \frac{\Delta z_\ell(p;p_0) }{k_\ell(p_0)}\right|^{-2}. \label{eq:sigma simplified higherell finitecoulomb}
\end{align}
In both the $\ell=0$ and $\ell \geq 1$ cases, the divergent behavior of $C_\ell^2$ as $p\rightarrow 0$ is suppressed by the last factor on the right-hand side of this equation.
In Fig.~\ref{fig:annihilation cross section finitecoulomb}, we show the Sommerfeld enhancement as calculated by the conventional formula Eq.~\ref{eq:conventional SE formula}, the full formula Eq.~\ref{eq:onestateannfinal}, and the simplified formula Eqs.~\ref{eq:sigma simplified swave finitecoulomb}, \ref{eq:sigma simplified higherell finitecoulomb}.
For the $s$-wave case, the behavior of the full formula is well captured by the simplified formula Eq.~\ref{eq:sigma simplified swave finitecoulomb}. For higher partial waves, the simplified formula Eq.~\ref{eq:sigma simplified higherell finitecoulomb} describes the behavior of the full formula fairly well, except for the points where $\tilde z_\ell$ crosses $0$ and flips its sign; these behaviors are similar to those for the finite square well.

\newpage
\section{The multi-state case for general $\ell$}
\label{sec:multistate}

In general, the long-range potential may couple a number of two-body states, and the short-range annihilation may proceed through any of these states. Scenarios with multiple interacting nearly-degenerate states were already considered in the first works on Sommerfeld enhancement as applied to dark matter annihilation \cite{Hisano:2002fk, Hisano:2003ec, Hisano:2004ds}, and methods for the standard (uncorrected) Sommerfeld calculation are well-developed for multi-state systems.

Assuming the energy associated with the potential to be small compared to the mass, we will generally treat all states that can be excited due to the potential as having approximately equal masses $M_\chi$, and treat the mass differences between them as terms in the potential.  In this case, $V_\text{long}(r)$ is a matrix describing the potential, and its diagonal elements include terms for the mass splittings between states, as in e.g. \cite{Hisano:2004ds}. 

 The states coupled by the potential may or may not be kinematically accessible at large distances, depending on the velocity of the interacting particles; we will consider the general case where there are $N$ coupled two-body states, and $M$ of these are kinematically accessible. If the 3-momentum of a single particle in the lowest-mass state is defined to be $p$, and the reduced mass of the system is $\mu$, we can define the single-particle momentum in the $n$th two-body state as follows, by energy conservation:
\begin{equation}
p_n \equiv \sqrt{p^2 - 2 \mu \mathrm{lim}_{r \rightarrow\infty} V_{\text{long},nn}(r)},\label{eq:multistatep}
\end{equation}
where now $p_n$ is real for the $M$ kinematically accessible states and imaginary otherwise.

We will assume throughout this section that the potential is spherically symmetric and ignore couplings between states of different $\ell$. Then the Schr\"odinger equation satisfied by the $\ell$th partial wave of the component of the radial reduced wavefunction in the $n$th state can be written:
\begin{equation}
\left[ \left(-\frac{d^2}{dr^2} + \frac{\ell(\ell+1)}{r^2} - p_a^2 \right) \delta_{ab} + 2\mu (\hat{V}_\text{long}(r))_{ab} \right] u_b(r) = 0, \label{eq:SEmultistate}
\end{equation}
where $\hat{V}_\text{long}(r)$ is a $N\times N$ matrix that goes to zero as $r\rightarrow \infty$. We will assume throughout that $V_\text{long}(r)$ (and hence $\hat{V}_\text{long}(r)$) is real and symmetric for all $r$.\footnote{The generalization to a complex Hermitian $V_\text{long}(r)$ is not difficult; see App.~\ref{app:Hermitian}.}

\subsection{The two-body scattering problem for multiple states}

The contribution to the wavefunction corresponding to scattering from such a potential is given asymptotically by $(\psi_n({\bf x}))_\mathrm{scattered} \rightarrow f_n(\theta) e^{i p_n r}/r$, for $n=1, \ldots, M$ (again, only the first $M$ states are kinematically allowed at large $r$). Performing the partial wave expansion for a plane wave $e^{ipz}$ as previously, and defining $f_n(\theta) = \sum_\ell (2\ell+1) f_{nl} P_\ell(\cos\theta)$, we can write, for $n=1, \ldots, M$:
\begin{equation} \psi_n(r,\theta) \rightarrow (2\ell+1) P_l(\cos\theta) \left[ a_n \frac{e^{i p_n r} - (-1)^\ell e^{- i p_n r}}{2 i p_n r} + f_{n\ell} \frac{e^{i p_n r}}{r} \right]. \label{eq:partialwave}, \end{equation}
where the $a_n$ coefficients describe the two-body state for the incoming particles at large $r$. For states with $n > M$, where $p_n$ is imaginary, we can use a similar expansion, but with $a_n=0$ (corresponding to the fact that there is no incoming plane wave in this state); the purely-outgoing $f_{nl}$ component then translates into an exponentially suppressed component, as required to have a physical solution.

Consider the solution of the Schr\"odinger equation where the initial unscattered plane wave corresponds entirely to a single two-body state (labeled $i$, $i=1, \ldots, M$), i.e. we can write $a_{n} = \delta_{ni}$ where $i$ runs from 1 to $M$, and so the incoming particles each carry momentum $p_i$. Let $f_{ni\ell}$ be defined as $f_{n\ell}$ when the initial state is purely the $i$th state; when we write $f_\ell$, we will henceforth mean the $N\times M$ matrix with elements $f_{ni\ell}$. The scattering amplitude from the $i$th kinematically allowed state to the $n$th kinematically allowed state will be given by the truncation of $f_\ell$ to $M\times M$.

The form of the $n$-component of the wavefunction at large $r$, for an initial plane-wave in component $i$, will be:
\begin{align} \psi_{ni}(r,\theta) & \rightarrow (2\ell+1) P_l(\cos\theta) \left[ \delta_{ni} \frac{e^{i p_n r} - (-1)^l e^{- i p_n r}}{2 i p_n r} + f_{ni\ell} \frac{e^{i p_n r}}{r} \right] \nonumber \\
& \sim (2\ell +1) P_l(\cos\theta)e^{i \pi \ell/2} \frac{1}{p_n r} \left[ \left(\delta_{ni} +i p_n f_{ni\ell} \right) s_\ell(p_n r)  + f_{ni\ell} p_n c_\ell(p_n r) \right]. \end{align}

We can define a reduced wavefunction vector, with $n$-component $u_{ni\ell}(r)$, that is directly proportional to $r \psi_{ni}(r,\theta)$ with a proportionality factor that depends only on $\ell, \theta$ and $p_i$; for consistency with the one-state case, we will choose the reduced wavefunction to have the large-$r$ asymptotics:
\begin{equation} u_{ni\ell}(r) \rightarrow \frac{p_i}{p_n}  \left[ \left(\delta_{ni} +i p_n f_{ni\ell} \right) s_\ell(p_n r)  + f_{ni\ell} p_n c_\ell (p_n r) \right] =   \left(\delta_{ni} +i p_i f_{ni\ell} \right) s_\ell(p_n r)  + f_{ni\ell} p_i c_\ell (p_n r) \label{eq:largermultistate}. \end{equation}
In the unperturbed case we see that $u_{ni\ell}(r) \rightarrow \delta_{ni}s_\ell(p_n r)$ at large $r$.

In this case, the incident probability flux is $\langle J_\mathrm{in} \rangle = p_i/\mu$.
The scattered flux into a (kinematically allowed) state with momentum $p_f$, where $p_f$ is real ($f = 1, \ldots, M$), is given by $\langle J_\mathrm{out} \rangle = \frac{p_f}{\mu} \frac{1}{r^2} |f_{fi}(\theta)|^2$, where $f_{fi}(\theta) = \sum_l (2\ell + 1) f_{fi\ell} P_l(\cos\theta)$. The cross section for this scattering process is given by:
\begin{equation} \frac{d\sigma_{i\rightarrow f}}{d\Omega} = r^2 \frac{\langle J_\mathrm{out} \rangle}{\langle J_\mathrm{in} \rangle} =  \frac{p_f}{p_i} \left|\sum_l (2l + 1) f_{fi\ell} P_l(\cos\theta) \right|^2.  \end{equation}

Performing the angular integral using the orthogonality of the Legendre polynomials, we obtain the scattering cross sections:
\begin{equation} \sigma_{i\rightarrow f} = 4\pi \frac{p_f}{p_i}  \sum_l (2l+1)  \left| f_{fi\ell} \right|^2 = \pi  \frac{1}{p_i^2} \sum_l (2l+1) \left| S_{fi\ell} - \delta_{fi} \right|^2, \label{eq:multistatescattering}  \end{equation}
where we have defined the normalized $S$-matrix $S_{fi\ell} \equiv  \delta_{fi} + 2 i f_{fi\ell} \sqrt{p_i p_f}$. This choice ensures that the probability conservation condition remains $S_\ell^\dagger S_\ell=1$, in the absence of absorptive processes such as annihilation. Considering $S_{fi\ell}$ and $f_{fi\ell}$ as the $(f,i)$ components of matrices $S_\ell$ and $f_\ell$, we observe that $S_\ell$ is a $M\times M$ matrix, describing scattering from one kinematically allowed two-body state to another kinematically allowed two-body state. 

The annihilation cross section, for the $i$th initial state, can be derived from the apparent non-conservation of the probability current and is given by:
\begin{equation} \sigma_{i,\mathrm{ann}} = \frac{\pi}{p_i^2} \sum_l (2\ell +1) \left(1 - \sum_f |S_{fi\ell}|^2  \right) =  \frac{\pi}{p_i^2} \sum_\ell (2\ell +1) \left(1 - S_\ell^\dagger S_\ell  \right)_{ii}.\label{eq:multistatexsec} \end{equation}
As in Sec.~\ref{sec:singlestate}, this relation assumes distinguishable particles in the initial state, and should be corrected by a prefactor $c_i$ which is 1 for distinguishable particles and 2 for identical particles. We will generally work with scattering amplitudes appropriate for distinguishable particles unless noted otherwise (see App.~\ref{app:optical} for a discussion of how to obtain the correctly rescaled matrix elements from QFT for short-distance scattering amplitudes involving identical particles).

For the remainder of this section, we will generally suppress $\ell$ subscripts when writing out the components of objects that carry state indices, in an attempt to simplify the notation. All wavefunctions, $S$-matrix elements, cross sections, etc., are separated into partial waves (and thus carry an implicit $\ell$ index) unless explicitly noted otherwise. We will make the $\ell$ indices explicit when dealing with the objects themselves rather than their components, e.g. we will write $S_\ell$ for the $S$-matrix but $S_{fi}$ for its components.

\subsection{Constructing the wavefunction with modified short-range physics}
\label{sec:multistate_decomposition}

Our goal now is to relate $f_\ell$ and $S_\ell$ to the Sommerfeld enhancement arising from the long-range potential for $r > a$ (obtained from the regular solution to the long-range Schr\"odinger equation), and the modification of the boundary condition at $r=a$ due to short-range physics. The results of this section will generalize our one-state results from Sec.~\ref{sec:singlestate}.

Let us decompose the reduced wavefunction for $r > a$, whose components we denote $u_{ni}(r)$, into the regular solution to the long-range Schr\"odinger equation, denoted $w_{ni}(r)$ (with the standard boundary conditions as $r\rightarrow 0$), and a correction associated with the short-range physics. As previously, $i$ indexes the initial state and $n$ indexes the wavefunction component. We choose $w_{ni}(r)$ and $u_{ni}(r)$ to correspond to the same ingoing plane wave (i.e. in the $i$th state and with standard normalization), so that $u_{ni}(r)\rightarrow w_{ni}(r)$ for $r > a$ in the limit where the long-range potential is the only relevant physics, and in general $u_{ni}(r) - w_{ni}(r)$ is purely outgoing at large $r$. We can write this purely-outgoing component as a linear combination of basis solutions,
\begin{equation} u_{ni}(r) = w_{ni}(r) + \tilde{w}_{nj}(r) R_{ji}, \label{eq:decomposition} \end{equation}
where we define the basis wavefunctions $\tilde{w}_{nj}(r)$ as the unique irregular solutions to the long-range Schr\"odinger equation that are purely outgoing (or exponentially suppressed) at infinity, and whose leading-order short-distance behavior is: 
\begin{equation} \tilde{w}_{nj}(r) \rightarrow \delta_{nj} (2\ell - 1)!! r^{-\ell} \,  \, \text{as} \, r\rightarrow 0.\end{equation}
The constant coefficients $R_{ji}$ (which collectively inhabit a matrix $R_\ell$) should be chosen to fix the desired boundary conditions at $r=a$, and thus will be determined by the short-distance physics combined with the properties of the $w_{ni}(r)$ and $\tilde{w}_{ni}(r)$ solutions as $r\rightarrow a$. Note that we must allow $j$ to take values from $1$ to $N$, not solely $1$ to $M$, to allow for cases where there is a non-trivial short-distance boundary condition in a component that is kinematically suppressed at large $r$.

\subsection{Properties of the basis solutions}
\label{sec:multistate_longrange}

Solutions similar to $\tilde{w}_{nj}(r)$ have been studied previously (e.g.~in Ref.~\cite{Slatyer:2009vg}); because the boundary conditions here are slightly modified, and for completeness, we repeat some calculations of key properties here. For this section, we will only need to consider the standard long-range Schr\"odinger equation, in the range $0 < r < \infty$, without any short-distance perturbations. Because we treat the potential as fully general (beyond the assumption that it is real and symmetric), this covers our full range of conventions for $V_\text{long}(r)$ at $r < a$.

Assuming the long-distance potential is real, the real and imaginary parts of $\tilde{w}_{ni}(r)$ are independently (real) solutions of the Schr\"odinger equation (with the long-distance potential). Let us write $\tilde{w}_{ni}(r) = \tilde{G}_{ni}(r) + i \tilde{F}_{ni}(r)$, with $\tilde{G}_\ell$ and $\tilde{F}_\ell$ being real functions with components $\tilde{G}_{ni}$, $\tilde{F}_{ni}$ respectively (similarly to the one-state case, but note the normalization convention is different here, to avoid having to deal with the question of whether factors of the momentum are real or imaginary). Note also that by our choice of boundary condition, $\tilde{F}_{ni}(r)$ is a regular solution to the long-range Schr\"odinger equation (for the reduced wavefunction), i.e. it vanishes at the origin. Then the boundary conditions on $\tilde{F}_\ell$ and $\tilde{G}_\ell$ as $r\rightarrow 0$ can be written:
\begin{align}  \tilde{G}_{ni}(r)  & \rightarrow \delta_{ni} (2\ell - 1)!! r^{-\ell}, \quad \tilde{F}_{ni}(r) \rightarrow( z_{ni}/p_n^{\ell+1}) s_\ell(p_n r), \quad r\rightarrow 0. \label{eq:GshortdistBC} \end{align}
where $z_{ni}$ denotes the components of a real $N\times N$ matrix.

For the boundary condition at large $r$, since $\tilde{w}_{nj}(r)$ is purely outgoing at large $r$, we can define the components of a constant matrix $f_{1,\ell}$ by:
\begin{equation} \tilde{w}_{nj}(r) \rightarrow p_j f_{1,nj} (c_\ell(p_n r) + i s_\ell(p_n r)), \quad r\rightarrow \infty. \label{eq:f1def} \end{equation}

Thus we obtain the following conditions on $\tilde{G}_\ell$ and $\tilde{F}_\ell$ at large $r$:
\begin{align} \tilde{G}_{ni}(r) & \rightarrow  \text{Re} (p_i f_{1,ni} c_\ell(p_n r)) - \text{Im} (p_i f_{1,ni} s_\ell(p_n r)),\nonumber \\
\tilde{F}_{ni}(r) & \rightarrow \text{Re} (p_i f_{1,ni} s_\ell(p_n r)) + \text{Im} (p_i f_{1,ni} c_\ell(p_n r)), \quad r\rightarrow \infty.\end{align}
Let $f_{0,\ell}$ (with components denoted $f_{0,ni}$) be the scattering amplitude for the $w_{ni}(r)$ solution (i.e. the scattering amplitude from the long-range potential with no short-distance physics), and let the short-distance behavior of $w_{ni}(r)$ be given by:
\begin{align} w_{ni}(r) & \rightarrow Q_{ni} s_\ell(p_n r). \label{eq:wshortdistBC} \end{align}

Let us first consider the Wronskian computed from the regular solution $w_\ell$ (with components $w_{ni}$) and the irregular solution $\tilde{w}_{\ell} = \tilde{G}_\ell + i \tilde{F}_\ell$, in the form $W_{ij} = (w_\ell^T \tilde{w}^\prime_\ell - w^{T \prime}_\ell \tilde{w}_\ell)_{ij} = \sum_n w_{ni} \tilde{w}_{nj}^\prime - \sum_n w_{ni}^\prime \tilde{w}_{nj}$. Using the small-distance asymptotic behavior of $s_\ell(x)$ and $c_\ell(x)$, we observe:
\begin{align} W_{ij}(r\rightarrow 0) & = \sum_n Q_{ni} \frac{1}{(2\ell + 1)!!} (p_n r)^{\ell+1} \delta_{nj} (2\ell-1)!! (-\ell) r^{-\ell-1} \nonumber \\
& - \sum_n Q_{ni} \frac{1}{(2\ell + 1)!!} p_n (\ell+1) (p_n r)^{\ell}  \delta_{nj} (2\ell-1)!!  r^{-\ell} \nonumber \\
& = - Q_{ji} p_j^{\ell+1} . \label{eq:wronskian1} \end{align}

Then for the Wronskian we obtain:
\begin{align} W_{ij}(r\rightarrow \infty) & = \sum_{n\le M}  (\delta_{ni} s_\ell(p_n r) + p_i f_{0,ni} (c_\ell(p_n r) + i s_\ell(p_n r)))  f_{1,nj} p_j p_n (-s_\ell(p_n r) + i c_\ell(p_n r)) \nonumber \\
& - (\delta_{ni} p_n c_\ell(p_n r) + p_i p_n f_{0,ni} (-s_\ell(p_n r) + i c_\ell(p_n r))) f_{1,nj} p_j (c_\ell(p_n r) + i s_\ell(p_n r)) \nonumber \\
& = - f_{1,ij} p_j p_i  . \label{eq:f1wronskian} \end{align}
Thus, constancy of the Wronskian (which follows from the Schr\"odinger equation) implies $Q_{ji} p_j^{\ell+1} = f_{1,ij} p_j p_i$, or $f_{1,ij} = (p_j^\ell/p_i) Q_{ji}$.

We can also compute the large-$r$ and small-$r$ limits of the Wronskian derived from treating $\tilde{G}_\ell$ and $\tilde{F}_\ell$ as independent solutions, $\tilde{W}_{ij} = (\tilde{G}_\ell^T \tilde{F}_\ell^\prime - \tilde{G}^{T\prime}_\ell \tilde{F}_\ell)_{ij}$, and from invariance of the Wronskian. We obtain:
\begin{align}\tilde{W}_{ij}(r\rightarrow 0) & = z_{ij}, \nonumber \\
\tilde{W}_{ij}(r\rightarrow \infty) & = \sum_{n\le M} p_n ( \text{Re} (p_i f_{1,ni}) \text{Re} (p_j f_{1,nj}) + \text{Im} (p_i f_{1,ni}) \text{Im} (p_j f_{1,nj})), \nonumber \\
\Rightarrow z_{ij} & =
 \sum_{n \le M} \frac{1}{p_n} ( \text{Re} (p_i^{\ell+1} Q_{in}) \text{Re} (p_j^{\ell+1} Q_{jn}) + \text{Im} (p_i^{\ell+1} Q_{in}) \text{Im} (p_j^{\ell+1}  Q_{jn})). \end{align} 
where only the states with $n \le M$ contribute in the calculation of $\tilde{W}_{ij}(r \rightarrow \infty)$ as the others are exponentially suppressed as $r\rightarrow \infty$.

For convenience, we can define $\bar{Q}_\ell$ as a $N\times M$ matrix with elements $\bar{Q}_{in} = p_i^{\ell+1} Q_{in} / \sqrt{p_n}$, and then write:
\begin{align} z_{ij} & = \sum_{n \le M}  (\text{Re} \bar{Q}_{in}) (\text{Re} \bar{Q}_{jn}) + (\text{Im} \bar{Q}_{in}) (\text{Im} \bar{Q}_{jn})  = \text{Re} (\bar{Q}_\ell \bar{Q}_\ell^\dagger)_{ij}. \label{eq:QZrelation} \end{align}

We can do one further Wronskian calculation to relate $f_{0,\ell}$ to $Q_\ell$, this time computing $\hat{W}_{ij} = (w^\dagger_\ell \tilde{w}^\prime_\ell - w^{\dagger \prime}_\ell \tilde{w}_\ell)_{ij} = \sum_n w_{ni}^* \tilde{w}_{nj}^\prime - \sum_n w_{ni}^{* \prime} \tilde{w}_{nj}$, where $i$ runs from 1 to $M$ and $j$ runs from 1 to $N$. Now at short distances we obtain:
\begin{align} \hat{W}_{ij}(r\rightarrow 0) & = \sum_n Q_{ni}^* \frac{1}{(2\ell + 1)!!} (p_n^* r)^{\ell+1} \delta_{nj} (2\ell-1)!! (-\ell) r^{-\ell-1} \nonumber \\
& - \sum_n Q_{ni}^* \frac{1}{(2\ell + 1)!!} p_n^* (\ell+1) (p_n^* r)^{\ell}  \delta_{nj} (2\ell-1)!!  r^{-\ell} \nonumber \\
& = - Q_{ji}^* (p_j^*)^{\ell+1} .\end{align}
At large distances, we have:
\begin{align} \hat{W}_{ij}(r\rightarrow \infty) & = \sum_{n\le M}  (\delta_{ni} s_\ell(p_n r) + p_i f^*_{0,ni} (c_\ell(p_n r) - i s_\ell(p_n r)))  i f_{1,nj} p_j p_n ( c_\ell(p_n r) + i s_\ell(p_n r)) \nonumber \\
& - (\delta_{ni} p_n c_\ell(p_n r) - i p_i p_n f^*_{0,ni} (c_\ell(p_n r) - i s_\ell(p_n r)) f_{1,nj} p_j (c_\ell(p_n r) + i s_\ell(p_n r)) \nonumber \\
& =- f_{1,ij} p_j p_i +2 i  p_i p_j \sum_{n \le M} f^*_{0,nil} p_n f_{1,nj}.\end{align}
From our previous result that $Q_{ji} p_j^{\ell+1} = f_{1,ij} p_j p_i$, constancy of the Wronskian implies:
\begin{align} - Q_{ji}^* (p_j^*)^{\ell+1}  & = - Q_{ji} p_j^{\ell+1} +2 i  p_i p_j^{\ell+1} \sum_{n \le M} f^*_{0,ni} Q_{jn}, \nonumber \\
\Rightarrow \frac{\bar{Q}_{ji} - \bar{Q}_{ji}^*}{2i} & = \sum_{n \le M} \sqrt{p_i p_n} f^*_{0,ni} \bar{Q}_{jn}. \label{eq:Qf0relation} \end{align}

\subsection{Extracting the modified $S$-matrix element}

Now we can use the properties derived in the previous subsection to write down expressions for the $S$-matrix element in the presence of additional short-distance physics. We can read off the full scattering amplitude, $f_\ell$ (with components denoted $f_{ni}$), from the long-range behavior of $u_{ni}(r)$, obtaining a contribution from $w_{ni}(r)$ which is identical to that in the case with no short-range physics, and then an additional contribution from the purely-outgoing $\tilde{w}$ term:
\begin{equation} f_{ni} = f_{0,ni} + f_{1,nj} (p_j/p_i) R_{ji} \label{eq:fdecomposition}. \end{equation}

In turn, we can write the full $S$-matrix element as:
\begin{equation} S_{ni} = S_{0,ni} + 2i f_{1,nj} (p_j/p_i) R_{ji} \sqrt{p_i p_n} = S_{0,ni} + 2i (p_j^{\ell+1}) Q_{jn} R_{ji}/ \sqrt{p_n p_i}, \label{eq:sdecomposition} \end{equation}
where $S_{0,ni }= \delta_{ni} + 2 i f_{0,ni} \sqrt{p_i p_n}$ is the $S$-matrix corresponding to scattering through the long-range potential only. $n$ and $i$ take values from 1 to $M$, whereas $j$ runs from 1 to $N$. 

It is also helpful to define a complex $N\times N$ diagonal matrix $\tilde{P}$ with elements $\tilde{P}_{ij} = \sqrt{p_i} \delta_{ij}$; we denote by $P$ the (real) truncation of $\tilde{P}$ to the $M\times M$ subspace. Then in matrix form (and noting that $R$ is $N\times M$), we can write:
\begin{equation} S_\ell = S_{0,\ell} + 2i \bar{Q}^T_\ell R_\ell P^{-1}.\end{equation}

Thus we have reduced the problem of computing the full $S$-matrix element to that of computing $S_{0,\ell}$, $\bar{Q}_\ell$, and the coefficients in the matrix $R_\ell$. $S_{0,\ell}$ and $\bar{Q}_\ell$ can be computed using standard methods from the Schr\"odinger equation with only the long-distance potential. We can also use Eq.~\ref{eq:Qf0relation} to relate these quantities:
\begin{equation} \frac{\bar{Q}_{ji} - \bar{Q}_{ji}^*}{2i} = \sum_{n \le M} \left(\frac{S_{0,ni} - \delta_{ni}}{2i} \right)^* \bar{Q}_{jn}, \end{equation}
or in matrix form,
\begin{align} \bar{Q}^T_\ell - \bar{Q}^\dagger_\ell & = - (S_{0,\ell} - 1)^\dagger \bar{Q}_\ell^T \Rightarrow \bar{Q}_\ell^* S_{0,\ell} = \bar{Q}_\ell. \label{eq:S0Qrelation} \end{align}
Consequently, we also have $\bar{Q}^\dagger_\ell = S_{0,\ell}^\dagger \bar{Q}^T_\ell$. Provided the long-range physics has no absorptive component (as we have assumed to date), $S_{0,\ell}^\dagger S_{0,\ell} = S_{0,\ell} S_{0,\ell}^\dagger = 1$, so multiplying both sides by $S_{0,\ell}$ gives $\bar{Q}^T = S_{0,\ell} \bar{Q}^\dagger$. Thus we can rewrite the full $S$-matrix element in the form:
\begin{equation} S_\ell = S_{0,\ell} (1 + 2i \bar{Q}^\dagger_\ell R_\ell P^{-1}) \label{eq:modifiedS}.\end{equation}
This form will be convenient for computing $S_\ell^\dagger S_\ell = (1 + 2i \bar{Q}_\ell^\dagger R_\ell P^{-1})^\dagger (1 + 2i \bar{Q}_\ell^\dagger R_\ell P^{-1})$, since $S_{0,\ell}^\dagger S_{0,\ell} =1$. The unitarity of $S_{0,\ell}$ also has implications for the properties of $Q_\ell$, as we can write:
\begin{align} \bar{Q}_\ell \bar{Q}^\dagger_\ell & = \bar{Q}^*_\ell S_{0,\ell} S_{0,\ell}^\dagger \bar{Q}^T_\ell = \bar{Q}^*_\ell \bar{Q}^T_\ell = (\bar{Q}_\ell \bar{Q}^\dagger_\ell)^*. \label{eq:QQdagger} \end{align}
Thus all elements of $\bar{Q}_\ell \bar{Q}^\dagger_\ell$ are real.

\subsection{Characterizing the effects from short-range physics}
\label{subsec:shortrange}

Finally, we wish to determine $R_\ell$ and separate out the effects of the short-range and long-range physics; this is analogous to determining $k_\ell(p)$ in the one-state case. Let us first observe that we can relate the regular solutions $w_{ni}$ and $\tilde{F}_{ni}$. For fixed $i$, these solutions each have $N$ complex boundary conditions fixed by regularity at the origin, and a further $N$ boundary conditions at $r=0$ determined by the coefficients of $s_\ell(p_n r)$ in  $w_{ni}(r)$ and $\tilde{F}_{ni}(r)$; thus if the coefficients of $s_\ell(p_n r)$ agree for all $n$, these solutions must coincide. We observe that as $r\rightarrow 0$:
\begin{align} \tilde{F}_{ni}(r) & \rightarrow (\bar{Q}_\ell \bar{Q}^\dagger_\ell)_{ni}/p_n^{\ell+1} s_\ell(p_n r) \nonumber \\
& = \sum_{j \le M} \bar{Q}^*_{ij} Q_{nj} \frac{p_n^{\ell+1}}{\sqrt{p_j} p_n^{\ell+1}} s_\ell(p_n r) \nonumber \\
& =  \sum_{j \le M} \frac{\bar{Q}^*_{ij} }{\sqrt{p_j}} (Q_{nj} s_\ell(p_n r)),\end{align}
and thus we can identify $\tilde{F}_{ni}(r) = \sum_{j \le M} \frac{\bar{Q}^*_{ij} }{\sqrt{p_j}} w_{nj}(r)$ from Eq.~\ref{eq:wshortdistBC}. Then the expression for the full wavefunction for $r > a$ (Eq.~\ref{eq:decomposition}) can be expanded as:
\begin{align} u_{ni}(r) & = w_{ni}(r) + \left[ \tilde{G}_{nj}(r) + i \sum_{k \le M} \frac{\bar{Q}^*_{jk} }{\sqrt{p_k}} w_{nk}(r) \right] R_{ji}, \nonumber \\
\Rightarrow u_\ell(r) &= \tilde{G}_\ell(r) R_\ell + w_\ell(r) \left[1 +  i P^{-1} \bar{Q}^\dagger_\ell R_\ell \right], \label{eq:multistatewavefn} \end{align}
where we have used matrix notation in the second line. This expression is analogous to Eq.~\ref{eq:matching} in the one-state case.

We will find it helpful to decompose the solutions to the long-range Schr\"odinger equation according to the variable phase method. In the multi-state case,  we can write any solution of the Schr\"odinger equation (with the long-range potential) in the form
\begin{equation} u_n(r) = \alpha_n(r) f_n(r) - \beta_n(r) g_n(r), \label{eq:multistatevpm} \end{equation}
where without loss of generality we also impose $u^\prime_n(r) = \alpha_n(r) f_n^\prime(r) - \beta_n(r) g^\prime_n(r)$, and where $f_n(r) = s_\ell(p_n r)/\sqrt{p_n}$ and $g_n(r) = (c_\ell(p_n r) + i s_\ell(p_n r))/\sqrt{p_n}$. Here $n$ indexes the components of the solution. Let the matrices $\alpha_{ni}(r)$ and $\beta_{ni}(r)$ denote the $\alpha_n(r)$ and $\beta_n(r)$ functions for the $u_{ni}(r)$ solution with an incoming plane wave in the $i$th state. Similarly, we can define matrices $\alpha_{\tilde{G}_\ell}(r)$, $\beta_{\tilde{G}_\ell}(r)$, $\alpha_{w_\ell}(r)$ and $\beta_{w_\ell}(r)$ to denote the $\alpha_{ni}$ and $\beta_{ni}$ matrices corresponding to the $\tilde{G}_\ell(r)$ and $w_\ell(r)$ solutions.

The boundary conditions at $r=0$, given in Eq.~\ref{eq:GshortdistBC} and Eq.~\ref{eq:wshortdistBC}, can then be expressed in the form:
\begin{align} \alpha_{w_\ell}(0) = \tilde{P} Q_\ell, \quad \beta_{w_\ell}(0) = 0, \quad \beta_{\tilde{G}_\ell}(0) = -  \tilde{P}^{2\ell+1}. \label{eq:smallralphabeta} \end{align}
To translate these boundary conditions to $r=a$, we must specify how $V_\text{long}(r)$ is defined for $r < a$, as this affects the functions $w_\ell(r)$ and $\tilde{G}_\ell(r)$ (this choice also affects $S_{0,\ell}$ and $Q_\ell$). 

For the moment, let us leave $V_\text{long}(r)$ unspecified for $r < a$. 
To solve for the $R_\ell$ coefficients, we need to match $u_\ell(r)$ given by Eq.~\ref{eq:multistatewavefn} onto the solution for $r < a$, $u_{<,\ell}(r)$. We can characterize the short-distance solution at $r=a$ (up to its first derivatives, which is all we need for the matching) in terms of the short-distance scattering amplitude as:
\begin{equation} u_{<,ni}(r) \simeq \gamma_{ni} s_\ell(p_n r) + \sum_{j \le N} \gamma_{ji} f_{s,nj} p_j (c_\ell(p_n r) + i s_\ell(p_n r)), \quad r = a, \label{eq:multistatematching} \end{equation}
where $f_{s,\ell}$ is the scattering amplitude describing the short-distance physics.
%(according to our convention, this has no contribution from $V_\text{long}(r)$, but $f_{s,\ell}$ may include physical effects that we might usually absorb into the long-range potential).
We see that the matching conditions then correspond to choosing $\alpha_{ni}(a)/\sqrt{p_n} = \gamma_{ni}$ and $-\beta_{ni}(a)/\sqrt{p_n} = \sum_{j \le N} \gamma_{ji} f_{s,nj} p_j$, and obtain
\begin{align}
	\gamma_\ell  = \tilde{P}^{-1} \alpha_\ell(a), \quad f_{s,\ell} \tilde{P}^2 \gamma_\ell  = -\tilde{P}^{-1} \beta_\ell(a) \label{eq:multistatematchsol}
\end{align}
Then we obtain $\tilde P f_{s,\ell} \tilde P \alpha_\ell(a) = - \beta_\ell(a)$.
Working in matrix notation and decomposing $u_\ell(r)$ into its $\tilde{G}_\ell(r)$ and $w(r)$ components as in Eq.~\ref{eq:multistatewavefn}, we find that:
\begin{align}
	\tilde{P} f_{s,\ell} \tilde{P}   \left(\alpha_{\tilde{G}_\ell}(a) R_\ell + \alpha_{w_\ell}(a) \left[ 1 + i P^{-1}\bar{Q}_\ell^\dagger R_\ell \right]  \right)   = - \left(\beta_{\tilde{G}_\ell}(a) R_\ell + \beta_{w_\ell}(a) \left[ 1 + i P^{-1}\bar{Q}_\ell^\dagger R_\ell \right]  \right)
\end{align}
Thus, $R_\ell$ can be written as
\begin{align}
	R_\ell =&  - \left[\beta_{\tilde{G}_\ell}(a) + i \beta_{w_\ell}(a)  P^{-1}\bar{Q}_\ell^\dagger +  \tilde{P} f_{s,\ell} \tilde{P}   \left(\alpha_{\tilde{G}_\ell}(a)  + i  \alpha_{w_\ell}(a) P^{-1}\bar{Q}_\ell^\dagger \right) \right]^{-1}  \nonumber \\
& \times \left(\tilde{P} f_{s,\ell} \tilde{P}  \alpha_{w_\ell}(a)  +  \beta_{w_\ell}(a)  \right).
\end{align}
%%%%%%%%%%%%%%%%%%

Let $f_{b,\ell}$ be the short-distance scattering amplitude corresponding only to the part of $V_\text{long}(r)$ with $r < a$ (i.e. it is a contribution to $f_{s,\ell}$, and may be zero if we choose to set $V_\text{long}(r) = 0$ for $r < a$). To read off $f_{b,\ell}$, consider a regular solution $(u_{b,\ell})_{ni}$ to the Schr\"odinger equation with potential $V_\text{long}(r)$ for $r < a$ (and matching onto a free-particle solution for $r > a$). Let us denote the $\alpha$ and $\beta$ coefficients associated with this solution by $\alpha_{b,\ell}$, $\beta_{b,\ell}$, and choose the boundary condition at $r=a$ to be $(\alpha_{b,\ell}(a))_{ni} =\sqrt{p_i} \delta_{ni}$, or $\alpha_{b,\ell}(a) = \tilde{P}$. This corresponds to sending in a unit-normalized plane wave with component $\delta_{ni} s_\ell(p_n r)$ in the $n$th component. Then the coefficient of the outgoing $s_\ell(p_n r) + i c_\ell(p_n r)$ term at $r=a$ is $-(\beta_{b,\ell}(a))_{ni}/\sqrt{p_n}$. For the free-particle propagation at $r > a$, $\alpha_n(r)$ and $\beta_n(r)$ remain constant, so we can identify the coefficient of this outgoing term with $p_i f_{b,ni}$, and $ f_{b,\ell} = - \tilde{P}^{-1} \beta_{b,\ell}(a) \tilde{P}^{-2}$, or equivalently $\beta_{b,\ell}(a) = -\tilde{P}  f_{b,\ell} \tilde{P}^2$.

Now any regular solution can be written as a linear combination of the components of $u_{b,\ell}$ with different $i$ (corresponding to an arbitrary combination of ingoing plane waves at $r=a$). In particular, for the $(w_\ell(r))_{nj}$ regular solution, the coefficients of the $i$th such contribution are given by $(\alpha_{w_\ell}(a))_{ij}/\sqrt{p_i}$, so we can write $(\beta_{w_\ell}(a))_{nj} = (-\tilde{P}  f_{b,\ell} \tilde{P}^2)_{ni} (\alpha_{w_\ell}(a))_{ij}/\sqrt{p_i}$, i.e. $\beta_{w_\ell}(a) = -\tilde{P} f_{b,\ell} \tilde{P} \alpha_{w_\ell}(a)$.

Furthermore, if we measure $\alpha_{b,\ell}(0)$, we can similarly write any other regular solution as a linear combination of the components of $u_{b,\ell}$ to determine $\alpha_{w_\ell}(0)$, via $(\alpha_{w_\ell}(0))_{nj} = (\alpha_{b,\ell}(0))_{ni} (\alpha_{w_\ell}(a))_{ij}/\sqrt{p_i}$, i.e. $\tilde{P} Q_\ell = \alpha_{w_\ell}(0) = \alpha_{b,\ell}(0) \tilde{P}^{-1} \alpha_{w_\ell}(a)$.

Using the result for $\beta_{w_\ell}(a)$, and defining $\hat{f}_{s,\ell} = f_{s,\ell} - f_{b,\ell}$, we can write: 
\begin{align} R_\ell  &=  - \left[\beta_{\tilde{G}_\ell}(a) - i \tilde{P} f_{b,\ell} \tilde{P} \alpha_{w_\ell}(a) P^{-1}\bar{Q}_\ell^\dagger +  \tilde{P} (\hat{f}_{s,\ell} + f_{b,\ell}) \tilde{P}   \left(\alpha_{\tilde{G}_\ell}(a)  + i  \alpha_{w_\ell}(a) P^{-1}\bar{Q}_\ell^\dagger \right) \right]^{-1}  \nonumber \\
& \times \left(\tilde{P} \hat{f}_{s,\ell} \tilde{P}  \alpha_{w_\ell}(a)  \right) \nonumber \\
& = - \left[\beta_{\tilde{G}_\ell}(a) - \beta_{b,\ell}(a) \tilde{P}^{-1} \alpha_{\tilde{G}_\ell}(a) + \tilde{P} \hat{f}_{s,\ell} \tilde{P}   \left(\alpha_{\tilde{G}_\ell}(a)  + i  \alpha_{w_\ell}(a) P^{-1}\bar{Q}_\ell^\dagger \right) \right]^{-1}\left(\tilde{P} \hat{f}_{s,\ell} \tilde{P}  \alpha_{w_\ell}(a)  \right)  \label{eq:Rell} \end{align}
Now consider the Wronskian between the wavefunctions $\tilde{G}_\ell(r)$ and $u_{b,\ell}(r) \tilde{P}^{-1}$ (note this is still a valid solution matrix as the $n$th component of the $i$th solution is $(u_{b,\ell})_{ni}(r) /\sqrt{p_i}$, i.e. the $\sqrt{p_i}$ rescaling factor is the same constant for all components of any given solution), which are both solutions to the same Schr\"odinger equation for $r<a$. From Eq.~\ref{eq:vpmwronskian} we find that $(\alpha_{b,\ell}(r) \tilde{P}^{-1})^T \beta_{\tilde{G}_\ell}(r) - (\beta_{b,\ell}(r) \tilde{P}^{-1})^T \alpha_{\tilde{G}_\ell}(r)$ is a $r$-independent constant. Since $\alpha_{b,\ell}(a) = \tilde{P}$, we have:
\begin{align} \beta_{\tilde{G}_\ell}(a) - (\beta_{b,\ell}(a) \tilde{P}^{-1})^T \alpha_{\tilde{G}_\ell}(a) & = (\alpha_{b,\ell}(0) \tilde{P}^{-1})^T \beta_{\tilde{G}_\ell}(0) - (\beta_{b,\ell}(0) \tilde{P}^{-1})^T \alpha_{\tilde{G}_\ell}(0) \nonumber \\
& = - (\alpha_{b,\ell}(0) \tilde{P}^{-1})^T \tilde{P}^{2\ell+1} \label{eq:Wronskiangenconvention} \end{align}
where we have used the regularity of $u_{b,\ell}$ at the origin to infer $\beta_{b,\ell}(0) =0$. Furthermore, we can show that $\beta_{b,\ell}(a)\tilde{P}^{-1}$ is symmetric using the version of the variable phase method detailed in Appendix A of Ref.~\cite{Asadi:2016ybp}; in the notation of that work (which differs from our App.~\ref{app:vpm}), $\beta_{b,\ell}(r) \tilde{P}^{-1} = O(r) \alpha_{b,\ell}(r) \tilde{P}^{-1}$, where $O(r)$ is a matrix function. It is readily proven that $O(r)$ in Ref.~\cite{Asadi:2016ybp} is symmetric for all $r$, using an identical argument to that given in App.~\ref{app:vpm}. Since $\alpha_{b,\ell}(a) \tilde{P}^{-1}=1$, it follows that $\beta_{b,\ell}(a) \tilde{P}^{-1}$ is symmetric, as required. Thus we can write:
\begin{align} R_\ell  
& = \left[(\alpha_{b,\ell}(0) \tilde{P}^{-1})^T \tilde{P}^{2\ell+1} - \tilde{P} \hat{f}_{s,\ell} \tilde{P}   \left(\alpha_{\tilde{G}_\ell}(a)  + i  \alpha_{w_\ell}(a) P^{-1}\bar{Q}_\ell^\dagger \right) \right]^{-1}\left(\tilde{P} \hat{f}_{s,\ell} \tilde{P}  \alpha_{w_\ell}(a)  \right) \nonumber \\
& =  \left[( \tilde{P} \hat{f}_{s,\ell} \tilde{P} )^{-1} (\alpha_{b,\ell}(0) \tilde{P}^{-1})^T \tilde{P}^{2\ell+1} -  \left(\alpha_{\tilde{G}_\ell}(a)  + i  (\alpha_{b,\ell}(0)\tilde{P}^{-1})^{-1} \tilde{P} Q_\ell P^{-1}\bar{Q}_\ell^\dagger \right) \right]^{-1} \nonumber \\
&\times (\alpha_{b,\ell}(0)\tilde{P}^{-1})^{-1} \tilde{P} Q_\ell \nonumber \\
& = \left[\alpha_{b,\ell}(0)\tilde{P}^{-2} \hat{f}_{s,\ell}^{-1} \tilde{P}^{-2} (\alpha_{b,\ell}(0))^T \tilde{P}^{2\ell+1} - \alpha_{b,\ell}(0) \tilde{P}^{-1}  \alpha_{\tilde{G}_\ell}(a)  - i  \tilde{P} Q_\ell P^{-1}\bar{Q}_\ell^\dagger  \right]^{-1}  \tilde{P} Q_\ell .  \end{align}
In going from the first to the second line we have used the relation $\tilde{P} Q_\ell = \alpha_{b,\ell}(0) \tilde{P}^{-1} \alpha_{w_\ell}(a)$, and moved the $\tilde{P}\hat{f}_{s,\ell}\tilde{P}$ factor inside the inverse.

Now we can write down the $S$-matrix from Eq.~\ref{eq:modifiedS} as:
\begin{align} S_\ell & = S_{0,\ell} \left(1 + 2 i \bar{Q}^\dagger_\ell  \left[\alpha_{b,\ell}(0)\tilde{P}^{-2} \hat{f}_{s,\ell}^{-1} \tilde{P}^{-2} (\alpha_{b,\ell}(0))^T \tilde{P}^{2\ell+1} - \alpha_{b,\ell}(0) \tilde{P}^{-1}  \alpha_{\tilde{G}_\ell}(a) \right. \right. \nonumber \\
& \left. \left. - i  \tilde{P} Q_\ell P^{-1}\bar{Q}_\ell^\dagger  \right]^{-1}  \tilde{P} Q_\ell P^{-1} \right) \end{align}

Recalling $\bar{Q}_\ell = \tilde{P}^{2(\ell+1)} Q_\ell P^{-1}$, let us define $\Sigma_{0,\ell} \equiv \tilde{P}^2 Q_\ell P^{-2}$ (we will see later that this is the standard Sommerfeld matrix). Note that the identity $\bar{Q}_\ell^T = S_{0,\ell} \bar{Q}_\ell^\dagger$ now translates to:
\begin{equation} P \Sigma_{0,\ell}^T = S_{0,\ell} P \Sigma_{0,\ell}^\dagger (\tilde{P}^\dagger \tilde{P}^{-1})^{2\ell} \label{eq:sommerfeldidentity} \end{equation}

We can then write the $S$-matrix element in the simple form:
\begin{align} S_\ell & = S_{0,\ell} \left(1 + 2 i P \Sigma_{0,\ell}^\dagger \left[\kappa_\ell^{-1} - i \Sigma_{0,\ell} P^2 \Sigma_{0,\ell}^\dagger \right]^{-1} \Sigma_{0,\ell} P \right), \nonumber \\
\kappa_\ell^{-1} & \equiv \left[ \tilde{P} \alpha_{b,\ell}(0)\tilde{P}^{-2} \right] \left( \hat{f}_{s,\ell}^{-1} \left[\tilde{P}  \alpha_{b,\ell}(0) \tilde{P}^{-2}\right]^T   \tilde{P}^{2\ell} - \tilde{P}  \alpha_{\tilde{G}_\ell}(a) \right) (\tilde{P}^\dagger)^{-2 \ell} . \label{eq:multistateSmatrix}  \end{align}

Note that $\alpha_{b,\ell}(0)$ describes the normalization of a regular solution at the origin, so it can be related to the (standard calculation of the) Sommerfeld enhancement if we were to restrict $V_\text{long}$ to the region $r < a$ only. Let us write the analogue of $\Sigma_{0,\ell}$ when we set $V_\text{long}(r)$ to zero outside $r=a$ as $\Sigma^a_{0,\ell}$ (noting however that $\Sigma^a_{0,\ell}$ is $N\times N$ rather than $N\times M$). Then using the analogue of Eq.~\ref{eq:smallralphabeta} for this case, we can write $\alpha_{b,\ell}(0) = \tilde{P}^{-1} \Sigma^a_{0,\ell} \tilde{P}^2$, and we can write $\kappa_\ell^{-1}$ in the form:
\begin{align} \kappa_\ell^{-1} & \equiv \bar{f}_{s,\ell}^{-1} - \bar{Z}_\ell, \quad 
\bar{f}_{s,\ell}  \equiv (\tilde{P}^\dagger \tilde{P}^{-1})^{2\ell} ( \Sigma^{a T}_{0,\ell})^{-1} \hat{f}_{s,\ell} ( \Sigma^a_{0,\ell})^{-1}, \quad \bar{Z}_\ell  \equiv \Sigma^a_{0,\ell} \tilde{P}  \alpha_{\tilde{G}_\ell}(a) (\tilde{P}^\dagger)^{-2 \ell}.  \label{eq:multistatekappa} \end{align}
In general, we see that $\bar{f}_{s,\ell}$ encodes the short-range physics, potentially including effects from evolution under $V_\text{long}(r)$ at $r < a$. In contrast, $\bar{Z}_\ell$ depends solely on $V_\text{long}(r)$.

We show in App.~\ref{app:shortrangeinterp} that if we were to model the physics at $r < a$ as a contact interaction (potentially including both elastic and inelastic interactions) combined with whatever physics is already captured in $V_\text{long}(r)$ for $r < a$, and work to lowest order in the contact interaction, then $\bar{f}_{s,\ell}$ would be fixed by the scattering amplitude associated with the contact interaction, and hence would be $a$-independent. Roughly speaking, the factor of $(\Sigma^a_{0,\ell})^{-1}$ and its transpose take the full scattering amplitude and factor out the Sommerfeld factors arising from the already-modeled $V_\text{long}(r)$ at $r<a$, leaving only the raw contact-interaction contribution. In a similar way, to obtain $\hat{f}_s$ we already subtracted off the scattering amplitude arising solely from interactions involving $V_\text{long}(r)$ at $r < a$. Thus we may understand $\bar{f}_{s,\ell}$ as capturing the $r < a$ physics but stripping off the leading additive and multiplicative terms associated with $V_\text{long}(r)$ at $r < a$.

We see that our choice of how to define $V_\text{long}(r)$ for $r < a$ will in general affect the $S$-matrix via the $\Sigma_{0,\ell}$, $S_{0,\ell}$, $\bar{f}_{s,\ell}$ and $\bar{Z}_\ell$ terms (although the $S_{0,\ell}$ dependence will drop out in terms depending only on $S_\ell^\dagger S_\ell$). There are two particularly simple conventions one might choose.

One choice would be to fix $V_\text{long}(r) = 0$ for $r < a$, so that $\bar{f}_{s,\ell}$ explicitly includes all contributions from the short-range part of the non-relativistic potential, and so will be $a$-dependent. In this case, there is no potential for $r < a$, so $\Sigma^a_{0,\ell}=1$, $f_{b,\ell}=0$, and consequently $\bar{f}_{s,\ell} = (\tilde{P}^\dagger \tilde{P}^{-1})^{2\ell} f_{s,\ell}$. The $\Sigma_{0,\ell}$, $S_{0,\ell}$ and $\bar{Z}_\ell$ factors are also explicitly $a$-dependent. This convention choice corresponds to the simplest definition of what it means to separate ``short-range'' ($r < a$) and ``long-range'' ($r > a$) physics. We will find it helpful for proving a number of results in the following sections, as it gives the simplest expressions for $\bar{f}_{s,\ell}$ and $\bar{Z}_\ell$.

An alternative choice is to set $V_\text{long}(r)$ to be equal to the non-relativistic long-range potential for all $r$, including $r < a$, such that the definition of $V_\text{long}(r)$ does not change when we vary $a$. In this case $S_{0,\ell}$ and $\Sigma_{0,\ell}$ are $a$-independent, and $\alpha_{\tilde{G}_\ell}(r)$ is independent of $a$ as a function (i.e. $\alpha_{\tilde{G}_\ell}(a)$ depends on $a$ only via the evaluation point of the function; the boundary conditions defining the function do not vary as we change $a$). In this convention choice, if the short-range physics {\it not} captured in $V_\text{long}(r)$ were modeled as a contact interaction, then (as discussed above) $\bar{f}_{s,\ell}$ would be determined solely by the amplitude for the contact interaction (at least at leading order), and hence would be $a$-independent. However, higher-order terms involving both $V_\text{long}(r < a)$, and the short-range physics not captured in $V_\text{long}(r)$, could induce an $a$-dependence.

To the degree that we model the additional (non-potential) short-range physics as a contact interaction computed through perturbative quantum field theory (i.e. by taking the low-momentum limit), we can match the resulting amplitude directly onto $\bar{f}_{s,\ell}$ in the 2nd convention choice (with a subtlety due to the $(\tilde{P}^\dagger \tilde{P}^{-1})^{2\ell}$ term in Eq.~\ref{eq:multistatekappa} that is discussed in App.~\ref{app:optical}). If we switch to the 1st convention choice, $\bar{f}_{s,\ell}$ should also include additional terms from scattering in the potential at $r < a$ and the short-range Sommerfeld factors, as described in App.~\ref{app:shortrangeinterp}. These factors can be computed perturbatively in QFT, or by solving the wavefunction evolution in quantum mechanics for the truncated potential.

In either convention, there is also a choice as to the order at which one computes the non-relativistic potential $V_\text{long}(r)$. Provided the higher-order corrections to $V_\text{long}(r)$ are real and symmetric (or Hermitian, if the more general results of App.~\ref{app:Hermitian} are used), there is no difficulty in including them in $V_\text{long}(r)$; alternatively, if they have suppressed support outside $r=a$, they can be folded into $f_{s,\ell}$. If these higher-order corrections are omitted entirely, the main effect is to slightly shift the positions of the resonance peaks (see e.g.~\cite{Beneke:2019qaa, Urban:2021cdu}) in parameter space. This shift can drastically change the annihilation cross section at a specified parameter point if that point is sufficiently close to a resonance, but is not expected to change the overall range of possibilities (we demonstrate this point with an example in Sec.~\ref{sec:wino}).

Relatedly, we will see that setting $V_\text{long}(r)=0$ for $r < a$ changes the pattern of resonance peaks in $\Sigma_{0,\ell}$, which is analogous to the standard Sommerfeld factor; by construction, the additional $a$-dependent terms cancel this shift and restore the correct resonance positions in the full $S$-matrix. This provides one practical reason to choose a convention in which $V_\text{long}(r)$ is $a$-independent, in order to avoid numerical error in the delicate interplay of $a$-dependent factors that controls the correct resonance positions, and consequently in the evaluation of the cross section close to resonance.

In any case, we can extract the partial-wave annihilation cross section as:
\begin{align} (\sigma_{i,\text{ann}})_\ell & = c_i \frac{\pi}{p_i^2} (2\ell+1) \left[ 1 - S_\ell^\dagger S_\ell \right]_{ii} \nonumber \\
& = c_i \frac{2 \pi i}{p_i} (2\ell+1) \left[ \Sigma_{0,\ell}^\dagger \left( 1 + i  \kappa_\ell^\dagger \Sigma_{0,\ell} P^2 \Sigma_{0,\ell}^\dagger \right)^{-1} \left(\kappa_\ell^\dagger - \kappa_\ell   \right) \left( 1 - i  \Sigma_{0,\ell} P^2 \Sigma_{0,\ell}^\dagger \kappa_\ell \right)^{-1} \Sigma_{0,\ell} \right]_{ii} 
\label{eq:annxseccalc} \end{align}
 where we have used the unitarity of $S_{0,\ell}$ to set $S_{0,\ell}^\dagger S_{0,\ell}=1$. We can write this result in a more compact form by defining:
 \begin{align}\Sigma_\ell & \equiv \left[ 1 - i  \Sigma_{0,\ell} P^2 \Sigma_{0,\ell}^\dagger \kappa_\ell \right]^{-1} \Sigma_{0,\ell}, \label{eq:symboldefsnew} \end{align}
 so that if $\mu$ is the reduced mass of the system and $v_\text{rel}$ is the relative velocity of the incoming particles in state $i$, we have:
 \begin{align} (\sigma_{i,\text{ann}} v_\text{rel})_\ell 
& = c_i  \frac{2 \pi i}{\mu}  (2\ell+1) \left[ \Sigma_{\ell}^\dagger (\kappa_\ell^\dagger - \kappa_\ell) \Sigma_{\ell} \right]_{ii}. \label{eq:annxsecfinalnew} \end{align} 
We can view $\kappa_\ell$ as a corrected short-distance scattering amplitude -- it depends on the long-range potential via the $\bar{Z}_\ell$ term, but asymptotes to $\bar{f}_{s,\ell}$ in the limit where $\bar{f}_{s,\ell}$ is small. Similarly, we can regard $\Sigma_\ell$ as a corrected Sommerfeld enhancement. As we will discuss in Sec.~\ref{sec:singlestatecomparison}, the inverse of $\kappa_\ell$ is related to $k_\ell$ by a simple power of the momentum matrix, so much of our intuition for $k_\ell$ from the single-state case can be carried over.

We have some freedom to shift terms between the corrected short-distance matrix element $\kappa_\ell$ and the corrected Sommerfeld enhancement $\Sigma_\ell$, while maintaining a similar form for the overall cross section. Suppose we choose $\tilde{\kappa}_\ell^{-1} \equiv \kappa_\ell^{-1} + \Delta_\ell$, where $\Delta_\ell$ is Hermitian, then we can write:
\begin{align} \tilde{\Sigma}_\ell & \equiv \left[ 1 - (\Delta_\ell + i  \Sigma_{0,\ell} P^2 \Sigma_{0,\ell}^\dagger) \tilde{\kappa}_\ell \right]^{-1} \Sigma_{0,\ell} , \nonumber \\
 (\sigma_{i,\text{ann}} v_\text{rel})_\ell 
& = c_i  \frac{2 \pi i}{\mu}  (2\ell+1) \left[ \tilde{\Sigma}_{\ell}^\dagger (\tilde{\kappa}_\ell^\dagger - \tilde{\kappa}_\ell) \tilde{\Sigma}_{\ell} \right]_{ii}. \label{eq:deltashift} \end{align} 
For example, we might choose to subtract terms from $\kappa_\ell^{-1}$ that have a non-analytic dependence on $p$, and capture them explicitly with $\Delta_\ell$ instead, to guarantee desired properties for $\tilde{\kappa}_\ell^{-1}$.

If we wish to write the cross section directly in terms of $\bar{f}_{s,\ell}$ rather than $\kappa_\ell$, we can expand Eq.~\ref{eq:annxsecfinalnew} as:
\begin{align} (\sigma_{i,\text{ann}} v_\text{rel})_\ell 
&  = c_i \frac{2 \pi i}{\mu} (2\ell+1) \left[ \Sigma_{0,\ell}^\dagger \left[ (\bar{f}_{s,\ell}^\dagger)^{-1} -\bar{Z}_\ell^\dagger + i  \Sigma_{0,\ell} P^2 \Sigma_{0,\ell}^\dagger \right]^{-1} \left(\bar{f}_{s,\ell}^{-1} - (\bar{f}_{s,\ell}^\dagger)^{-1} +\bar{Z}_\ell^\dagger - \bar{Z}_\ell \right) \right. \nonumber \\
&\left.  \left[ \bar{f}_{s,\ell}^{-1} - \bar{Z}_\ell - i  \Sigma_{0,\ell} P^2 \Sigma_{0,\ell}^\dagger \right]^{-1} \Sigma_{0,\ell} \right]_{ii}  \nonumber \\
& = c_i \frac{2 \pi i}{\mu} (2\ell+1) \left[ \bar{\Sigma}_\ell^\dagger \left(\bar{f}_{s,\ell}^\dagger - \bar{f}_{s,\ell} + \bar{f}_{s,\ell}^\dagger (\bar{Z}_\ell^\dagger - \bar{Z}_\ell) \bar{f}_{s,\ell} \right) \bar{\Sigma}_\ell \right]_{ii}, \label{eq:explicitfxsec} \nonumber \\
\bar{\Sigma}_\ell & = \left[ 1 - (\bar{Z}_\ell + i  \Sigma_{0,\ell} P^2 \Sigma_{0,\ell}^\dagger) \bar{f}_{s,\ell} \right]^{-1} \Sigma_{0,\ell}. \end{align}

However, because $\bar{Z}_\ell$ is in general non-Hermitian, the numerator in this case has a more complicated form than in Eq.~\ref{eq:deltashift}. It is easiest to interpret the additional $\bar{f}_{s,\ell}^\dagger (\bar{Z}_\ell^\dagger - \bar{Z}_\ell) \bar{f}_{s,\ell}$ numerator term if we use the convention where we set $V_\text{long}(r)=0$ for $r < a$, so that $\bar{f}_{s,\ell} = (\tilde{P}^\dagger \tilde{P}^{-1})^{2\ell} f_{s,\ell}$ contains all the short-distance physics including scattering in the potential at $r < a$, and $\bar{Z}_\ell = \tilde{P}  \alpha_{\tilde{G}_\ell}(a) (\tilde{P}^\dagger)^{-2 \ell}$.

%To maintain the form of the cross section given in Eq.~\ref{eq:deltashift} but make $\tilde{\kappa}_\ell$ independent of the long-range physics, we could choose a Hermitian $\Delta_\ell$ to absorb the dependence on the long-range potential, i.e.~$\Delta_\ell = \frac{1}{2} (\bar{Z}_\ell + \bar{Z}_\ell^\dagger)$ (the Hermitian part of $\bar{Z}_\ell$). Then $\tilde{\kappa}_\ell^{-1} = \bar{f}_{s,\ell}^{-1} - \frac{1}{2} (\bar{Z}_\ell - \bar{Z}_\ell^\dagger)$. It turns out that the anti-Hermitian part of $\bar{Z}_\ell$ holds no information on the long-range potential, so this choice moves all the terms dependent on the long-range potential into $\Delta_\ell$.

In this convention, $\bar{Z}_\ell^\dagger - \bar{Z}_\ell = i (\tilde{P}^{-2\ell} (\tilde{P}^\dagger)^{2(\ell+1)} + \tilde{P}^{2(\ell+1)} (\tilde{P}^\dagger)^{-2\ell})$. To prove this, note that $\tilde{G}_{ni}(a)$ is real by construction (it was defined as a real wavefunction), and we have
 \begin{align} \tilde{G}_{ni}(a) & = f_n(a) (\alpha_{\tilde{G}_\ell}(a))_{ni} + g_n(a) (\tilde{P}^{2\ell+1})_{ni} \nonumber \\
 & = s_\ell(p_n a) p_n^{-(\ell+1)} (\tilde{P}^{2\ell + 1} \alpha_{\tilde{G}_\ell}(a) + i \tilde{P}^{2(2 \ell + 1)} )_{ni} + c_\ell(p_n a)  (\tilde{P}^{2\ell})_{ni}.\end{align}
Since $s_\ell(p_n a) p_n^{-(\ell+1)}$, $c_\ell(p_n a) p_n^\ell$, and $G_{ni}(a)$ are all real, it follows that  $\tilde{P}^{2\ell + 1} (\alpha_{\tilde{G}_\ell}(a) + i \tilde{P}^{2 \ell + 1})$ must also be real. Furthermore, we show in App.~\ref{app:vpm} that $\tilde{P}^{2\ell+1} \alpha_{\tilde{G}_\ell}(0) = \tilde{P}^{2\ell+1} \alpha_{\tilde{G}_\ell}(a)$ must also be symmetric, by direct construction. Thus $\tilde{P}^{2\ell+1} (\alpha_{\tilde{G}_\ell}(a) + i \tilde{P}^{2 \ell + 1})$ is Hermitian, and so is $\tilde{P}^{-2\ell} \tilde{P}^{2\ell+1} (\alpha_{\tilde{G}_\ell}(a) + i \tilde{P}^{2 \ell + 1}) (\tilde{P}^\dagger)^{-2\ell} = \bar{Z}_\ell + i \tilde{P}^{2(\ell+1)} (\tilde{P}^\dagger)^{-2\ell}$. The required relationship follows directly. 

Consequently, in this convention $\bar{Z}_\ell^\dagger - \bar{Z}_\ell$ is a diagonal matrix with the $n$th diagonal component being $2 i p_n$ for $n \le M$, and zero for $n > M$. When we write out the non-Hermitian part of the short-range scattering amplitude, according to the optical theorem it will include a contribution corresponding to the cross section for scattering into kinematically allowed states (i.e. $n \le M$), but this contribution is not considered part of our inclusive annihilation cross section and must be subtracted. This subtraction is implemented by the $\bar{f}_{s,\ell}^\dagger (\bar{Z}_\ell^\dagger - \bar{Z}_\ell) \bar{f}_{s,\ell}$ term in Eq.~\ref{eq:explicitfxsec}. We discuss this issue more explicitly in App.~\ref{app:optical}.

  \subsection{Limiting cases}
  
  \subsubsection{Case with no long-range potential}

Let us first consider the case where we turn off the long-range potential completely, setting $V_\text{long}(r)=0$ (both inside and outside $r=a$). In this case the convention choice for $r < a$ is trivial. Then we have $N=M$, $Q_\ell=\Sigma_{0,\ell}=1$, and $f_{s.\ell} = \bar{f}_{s,\ell}$. The wavefunctions for the long-range potential are just the free-particle wavefunctions, and $\tilde{G}_{ni}(r) = \delta_{ni} p_n^\ell c_\ell(p_n r) = \delta_{ni} p_n^{\ell + 1/2} (g_n(r) - i f_n(r))$ (note that in this case $p_n$ is always real as $M=N$). Thus we can read off $\alpha_{\tilde{G}_\ell}(a) = - i P^{2\ell+1}$, and
\begin{equation} \kappa_\ell^{-1} = f_{s,\ell}^{-1} + i P^2. \end{equation}
Then $\Sigma_\ell = (1 - i P^2 \kappa_\ell)^{-1} = \kappa_\ell^{-1} (\kappa_\ell - i P^2)^{-1} = \kappa_\ell^{-1} f_{s,\ell}$, and so we can write:
\begin{align} (\sigma_{i,\text{ann}} v_\text{rel})_{\ell,\text{bare}} & = c_i \frac{2 \pi i}{\mu} (2\ell+1) \left[f_{s,\ell}^\dagger (\kappa_\ell^{-1})^\dagger  (\kappa_\ell^\dagger - \kappa_\ell ) \kappa_\ell^{-1} f_{s,\ell} \right]_{ii} \nonumber \\
& =  c_i \frac{2 \pi i}{\mu} (2\ell+1) \left[ f_{s,\ell}^\dagger - f_{s,\ell} + 2 i  f_{s,\ell}^\dagger P^2 f_{s,\ell} \right]_{ii}. \end{align} 
This is the correct expression if the full scattering amplitude is replaced with $f_{s,\ell}$ (e.g. in Eq.~\ref{eq:multistatexsec}). Explicitly, in our matrix notation we can write the relationship between the $S$-matrix and the scattering amplitude as $S_\ell = 1 + 2 i P f_\ell P$, and so $1 - S_\ell^\dagger S_\ell = 2 i P \left[f_\ell^\dagger - f_\ell + 2 i f_\ell^\dagger P^2 f_\ell\right] P$, from which the desired result follows immediately. Note that from the perspective of the optical theorem, the $f_{s,\ell}^\dagger P^2 f_{s,\ell}$ term has the effect of subtracting the cross section for scattering into dark matter states from the inclusive annihilation cross section.

\subsubsection{Single-state case}
\label{sec:singlestatecomparison}

In the case $M=N=1$, let us check we can recover the results of Sec.~\ref{sec:singlestate}. In the notation of Sec.~\ref{sec:singlestate}, we have $\Sigma_{0,\ell}=Q_\ell=C_\ell e^{i\delta_\ell}$  (where $C_\ell$ is real). Our expression for the $S$-matrix from Eq.~\ref{eq:multistateSmatrix} then takes the form:
\begin{align} S_\ell = S_{0,\ell} \left( 1 + \frac{2 i p C_\ell^2}{\kappa_\ell^{-1} - i p C_\ell^2} \right) = S_{0,\ell} \left( \frac{\kappa_\ell^{-1} +  i p C_\ell^2}{\kappa_\ell^{-1} - i p C_\ell^2} \right).  \end{align}
This form will be consistent with the result of Eq.~\ref{eq:S matrix before approximation} if we can identify $k_\ell(p) = -p^{2\ell} \kappa_\ell^{-1}$. In the one-state case our equation for $\kappa_\ell^{-1}$ from Eq.~\ref{eq:multistateSmatrix} becomes:
\begin{align}\kappa_\ell^{-1} = \alpha_{b,\ell}(0) (\alpha_{b,\ell}(0) \hat{f}_{s,\ell}^{-1}/p - \alpha_{\tilde{G}_\ell}(a)p^{-\ell}). \label{eq:kappa_singlestate} \end{align}

To check the guess that $k_\ell(p) = -p^{2\ell} \kappa_\ell^{-1}$, let us go back to the definition of $k_\ell$ given in Eq.~\ref{eq:kappa_def}, which is constructed from the $u_{<,\ell}(r)$, $G_\ell(r)$ and $F_\ell(r)$ functions and their derivatives. Taking the one-state version of Eq.~\ref{eq:multistatematching} for $u_{<,\ell}(r)$, we observe that (where $f_\ell$ and $g_\ell$ are the basis functions in the variable phase method, see App.~\ref{app:vpm}):
\begin{equation} (u'_{<,\ell} / u_{<,\ell})\Big|_{r=a} = \frac{f^\prime_\ell(a) + f_{s,\ell} p g^\prime_\ell(a)}{f_\ell(a) + f_{s,\ell} p g_\ell(a)}.\end{equation}
Now decomposing $F_\ell(r) = \alpha_{F_\ell}(r) f_\ell(r) - \beta_{F_\ell}(r) g_\ell(r)$, $G_\ell(r) = \alpha_{G_\ell}(r) f_\ell(r) - \beta_{G_\ell}(r) g_\ell(r)$, employing the variable phase method as usual, substituting into Eq.~\ref{eq:kappa_def} gives:
\begin{align}k_\ell(p) 
& = p^{2\ell+1} C_\ell^2 \frac{\beta_{G_\ell}(a) + \alpha_{G_\ell}(a) f_{s,\ell} p}{\beta_{F_\ell}(a) + \alpha_{F_\ell}(a) f_{s,\ell} p} .\end{align}

Now since $F_\ell(r)$ is a regular solution, we can write $\beta_{F_\ell}(a) = - p f_{b,\ell} \alpha_{F_\ell}(a)$, yielding:
\begin{align}k_\ell(p) 
& = p^{2\ell+1} C_\ell^2 \frac{\beta_{G_\ell}(a) + \alpha_{G_\ell}(a) f_{s,\ell} p}{\alpha_{F_\ell}(a) p (f_{s,\ell} -f_{b,\ell})} = p^{2\ell+1} C_\ell^2 \left[ \frac{\beta_{G_\ell}(a) + p \alpha_{G_\ell}(a) f_{b,\ell}}{\alpha_{F_\ell}(a) p \hat{f}_{s,\ell}} + \frac{\alpha_{G_\ell}(a)}{\alpha_{F_\ell}(a)} \right] .\end{align}
Noting that the boundary condition on $\tilde{G}_{\ell}(r)$ from Eq.~\ref{eq:GshortdistBC} differs from that for $G_{\ell}(r)$ from Eq.~\ref{eq:F and G at the origin} by a factor of $p^\ell C_\ell$, we have $ \alpha_{\tilde{G}_\ell}(a) = C_\ell p^\ell \alpha_{G_\ell}(a)$, $ \beta_{\tilde{G}_\ell}(a) = C_\ell p^\ell \beta_{G_\ell}(a)$. Similarly, the boundary condition at $r=0$ on $w_\ell(r)$ from Eq.~\ref{eq:wshortdistBC} matches that for $F_\ell(r)$ from Eq.~\ref{eq:F and G at the origin} up to a factor of $e^{i\delta_\ell}$, so we have $\alpha_{F_\ell}(a) = e^{-i\delta} \alpha_{w_\ell}(a) = e^{-i\delta} p Q_\ell / \alpha_{b,\ell}(0)$. Furthermore, from Eq.~\ref{eq:Wronskiangenconvention} applied to the single-state case, we have that $ \beta_{\tilde{G}_\ell}(a) + p f_{b,\ell} \alpha_{\tilde{G}_\ell}(a) =  - \alpha_{b,\ell}(0) p^\ell$. Thus we obtain:
\begin{align}k_\ell(p) = - p^{2 \ell} \alpha_{b,\ell}(0) \left[ \frac{\alpha_{b,\ell}(0)}{p \hat{f}_{s,\ell}} - \alpha_{\tilde{G}_\ell}(a) p^{-\ell} \right] = -p^{2\ell} \kappa_\ell^{-1}, \end{align}
as required. Note that $k_\ell$ has dimensions of $p^{2\ell+1}$ but is expected to be near-momentum-independent where the $\hat{f}_{s,\ell}$ term dominates; it can be viewed as an effective inverse scattering length for $\ell=0$, an inverse scattering volume for $\ell=1$, etc. $\kappa_\ell$ has dimensions of $p^{-1}$ (i.e. it is a length) but is expected to scale as $p^{2\ell}$ when the $\hat{f}_{s,\ell}$ term dominates, similar to $\hat{f}_{s,\ell}$ itself.

Inserting the decomposition for $\kappa_\ell^{-1}$ given in Eq.~\ref{eq:multistatekappa}, we obtain:
\begin{equation} k_\ell(p) = p^{2\ell} \left[- \bar{f}_{s,\ell}^{-1} + \bar{Z}_\ell  \right]. \label{eq:kellSmatch} \end{equation}
This resembles the decomposition discussed in Sec.~\ref{sec:singlestate} into a short-range term ($\bar{f}_{s,\ell}^{-1}$) and a term calculable from the long-range potential only ($\bar{Z}_\ell$). In cases where the $\bar{Z}_\ell$ term diverges as $a\rightarrow 0$, but with a low-$p$ scaling that is equal or subleading to that of the $f_{s,\ell}^{-1}$ term, the divergent piece can be absorbed into the effective scattering length/volume/etc.~defined by $f_{s,\ell}^{-1}$ (or into terms in the series that are higher-order in momentum), corresponding to renormalization of the short-distance scattering amplitude. A similar renormalization step was done for the case of a true contact interaction and $s$-wave scattering in Ref.~\cite{Blum:2016nrz}.
%

%If we choose the convention where $V_\text{long}(r)=0$ for $r < a$, then $\alpha_{b,\ell}(0)=\sqrt{p}$, $f_s=\hat{f}_s$, and we obtain the simplified expression:
%\begin{equation} k_\ell(p) = p^{2\ell} \left[- f_{s,\ell}^{-1} + p^{-\ell+1/2} \alpha_{\tilde{G}_\ell}(a)  \right]. \label{eq:kellSmatch} \end{equation}
%This resembles the decomposition discussed in Sec.~\ref{sec:singlestate} into a short-range term ($f_{s,\ell}^{-1}$) and a term calculable from the long-range potential only ($\alpha_{\tilde{G}_\ell}(a)$). The $\alpha_{\tilde{G}_\ell}(a)$ term does depend on the choice of matching radius $a$ (as does $C_\ell$). In cases where this term diverges as $a\rightarrow 0$, but with a low-$p$ scaling that is equal or subleading to that of the $f_{s,\ell}^{-1}$ term, the divergent piece can be absorbed into the effective scattering length/volume/etc.~defined by $f_{s,\ell}^{-1}$ (or into terms in the series that are higher-order in momentum), corresponding to renormalization of the short-distance scattering amplitude. This renormalization step was done for the case of a true contact interaction and $s$-wave scattering in Ref.~\cite{Blum:2016nrz}.
%

\subsubsection{Comparison with un-corrected multi-state enhancement}

We can expand to lowest order in the short-range physics (controlled by $\bar{f}_{s,\ell}$) in the general form of Eq.~\ref{eq:deltashift},  obtaining 
$\tilde{\Sigma}_\ell \approx \Sigma_{0,\ell}$, $\tilde{\kappa}_\ell \approx \bar{f}_{s,\ell}$ (for any choice of $\Delta_\ell$). We see in this case the long-range physics encoded in $\Sigma_{0,\ell}$ factorizes from the purely short-range physics encoded in $\bar{f}_{s,\ell}^\dagger - \bar{f}_{s,\ell}$ (we could also start from Eq.~\ref{eq:explicitfxsec}, but would drop the $(\bar{f}_{s,\ell}^\dagger (\bar{Z}_\ell^\dagger - \bar{Z}_\ell) \bar{f}_{s,\ell})$ term as higher-order in $\bar{f}_{s,\ell}$). 

We thus expect $\Sigma_{0,\ell}$ to yield the standard Sommerfeld enhancement for the multi-state case, while $\bar{f}_{s,\ell}^\dagger - \bar{f}_{s,\ell}$ gives rise to the standard short-range annihilation matrix. To check this, let us compare our result for $\Sigma_{0,\ell}$ to the standard prescription for the multi-state Sommerfeld enhancement. In Ref.~\cite{Slatyer:2009vg}, the bare cross section for $\ell$-wave annihilation with an initial state labeled by $i$ is given by $(\sigma_{i,\text{ann}} v_\text{rel})_\ell  = \alpha_\ell p_i^{2\ell} \Gamma_{ii}$, where $\alpha_\ell$ is an overall proportionality factor and $\Gamma$ is a $N\times N$ ``annihilation matrix'' characterizing the short-distance physics. The enhanced annihilation cross section is then given by:
\begin{equation} (\sigma_{i,\text{ann}} v_\text{rel})_\ell  = \alpha_\ell ((2\ell-1)!!)^2 (T^* \Gamma T^T)_{ii},\end{equation}
where $T_{ij}$ is defined in that work as the coefficient of $e^{i k_i r}$ in the $i$th component of a purely-outgoing large-$r$ solution $\rho_{j}(r)$, which has the short-distance boundary condition $\rho_{ij}(r) \approx r^{-\ell} \delta_{ij}$.

Now in terms of the wavefunctions we have worked with, $\tilde{w}_{ij}(r) = (2\ell-1)!! \rho_{ij}(r)$, and we previously defined $f_{1,ij} p_j$ to be the coefficient of $c_\ell(p_i r) + i s_\ell(p_i r)$ in $\tilde{w}_{ij}(r)$ at large $r$, so we have $f_{1,ij} p_j = (2\ell - 1)!! T_{ij}$ up to a phase factor (which will cancel out in the final expression). As we previously proved, $f_{1,ij} = (p_j^\ell/p_i) Q_{ji}$, so we can write $T_{ij} (2\ell - 1)!! = (p_j^{\ell+1}/p_i) Q_{ji}$; in other words, computations of the $T$ factor from Ref.~\cite{Slatyer:2009vg} can be translated into our $Q_\ell$ factor via:
\begin{equation} Q_{ji} = (p_i / p_j^{\ell+1}) (2\ell-1)!! T_{ij}.\end{equation}

Now our prescription for the Sommerfeld enhancement to lowest-order in $f_{s,\ell}$ can be written as:
\begin{align} (\Sigma_{0,\ell})_{ij} = (p_i/p_j) Q_{ij} = (p_i/p_j) (p_j/p_i^{\ell+1}) (2\ell-1)!! T_{ji} = T_{ji} (2\ell-1)!!/p_i^\ell,\end{align}

Consequently, our prescription for the enhanced cross section in Eq.~\ref{eq:explicitfxsec} can be written (to lowest order in $f_{s,\ell}$) in terms of $T$ as:
\begin{align} (\sigma_{i,\text{ann}} v_\text{rel})_\ell  & \approx  c_i \frac{2 \pi i}{\mu}  (2\ell+1) ((2\ell-1)!!)^2 (T^*)_{ij} \frac{1}{p_j^\ell} (\bar{f}_{s,\ell}^\dagger  - \bar{f}_{s,\ell})_{jk} \frac{1}{p_k^\ell} T_{ik},\end{align} 

This is consistent with Ref.~\cite{Slatyer:2009vg} provided the matrices describing the short-distance physics are related by $\alpha_\ell \Gamma_{jk} \leftrightarrow c_i \frac{2\pi i}{\mu} (2\ell+1) (\bar{f}_{s,\ell}^\dagger  - \bar{f}_{s,\ell})_{jk}/(p_j p_k)^\ell$. As a cross-check, in the case with no enhancement, our results imply that the cross section becomes:
\begin{align} (\sigma_{i,\text{ann}} v_\text{rel})_\ell  & = c_i  \frac{2 \pi i}{\mu}  (2\ell+1) (\bar{f}_{s,\ell}^\dagger  - \bar{f}_{s,\ell})_{ii} \leftrightarrow \alpha_\ell p_i^{2\ell} \Gamma_{ii},\end{align} 
consistent with the bare cross section employed in Ref.~\cite{Slatyer:2009vg}. A similar framework is employed in Ref.~\cite{Beneke:2014gja}, with the matrices describing the short-range physics written in terms of Wilson coefficients.

\subsection{Summary of multi-state results}

In this section we aim to summarize our results, for readers who are only interested in the final expression. If the system involves $N$ coupled two-particle states of which $M$ are kinematically accessible at large separations, we can write the full Sommerfeld-enhanced annihilation cross section for the $\ell$th partial wave in the form:
\begin{align} (\sigma_{i,\text{ann}} v_\text{rel})_\ell 
& = c_i  \frac{2 \pi i}{\mu}  (2\ell+1) \left[ \tilde{\Sigma}_{\ell}^\dagger (\tilde{\kappa}^\dagger_\ell - \tilde{\kappa}_\ell) \tilde{\Sigma}_{\ell} \right]_{ii}, \nonumber \\
\tilde{\kappa}_\ell^{-1} & \equiv \bar{f}_{s,\ell}^{-1} - \bar{Z}_\ell + \Delta_\ell, \nonumber \\
\tilde{\Sigma}_\ell & \equiv \left[ 1 - (\Delta_\ell + i  \Sigma_{0,\ell} P^2 \Sigma_{0,\ell}^\dagger) \tilde{\kappa}_\ell \right]^{-1} \Sigma_{0,\ell}. \end{align} 
Here $\mu$ is the reduced mass of the system, $i$ indexes the initial state, $c_i=2$ for identical particles in the initial state and 1 otherwise, $\Sigma_{0,\ell}$ is the standard $N\times M$ Sommerfeld enhancement matrix calculated from the long-range real potential, $\Delta_\ell$ is an arbitrary Hermitian $N\times N$ matrix (which can be chosen freely to simplify the form of $\tilde{\kappa}_\ell$), and $\bar{f}_{s,\ell}$ is the short-range $N\times N$ scattering amplitude when the interactions included in $V_\text{long}(r)$ for $r < a$ are factored out (see Eq.~\ref{eq:multistatekappa} and subsequent discussion). In the case $\Delta_\ell=0$, we denote $\tilde{\kappa}_\ell$ instead as $\kappa_\ell$, and $\tilde{\Sigma}_\ell$ instead as $\Sigma_\ell$. $\tilde{P}$ is the diagonal $N\times N$ matrix whose entries are $\sqrt{p_i}$ if $p_i$ is the momentum associated with the $i$th two-particle state, and $P$ is the $M\times M$ truncated version of the same matrix, restricted to the kinematically allowed states.

The $\bar{Z}_\ell$ coefficient is fully determined by the long-range potential and can be obtained by solving the long-range matrix Schr\"odinger equation with potential $V_\text{long}(r)$. Specifically, it can be written (Eq.~\ref{eq:multistatekappa}) in terms of the coefficients $ \alpha_{\tilde{G}_\ell}(a)$ and $\Sigma^a_{0,\ell}$; the latter of these is the conventional Sommerfeld-enhancement matrix that would be derived if $V_\text{long}(r)$ is set to zero outside $r=a$ (considering all $N$ possible incoming states, so the matrix is $N\times N$). The $ \alpha_{\tilde{G}_\ell}(a)$ coefficient is fully determined by the long-range potential and can be obtained by solving the long-range matrix Schr\"odinger equation with potential $V_\text{long}(r)$, for the real part $\tilde{G}_\ell(r)$ of a solution that is purely outgoing at large $r$ and satisfies Eq.~\ref{eq:GshortdistBC} at small $r$. This can be done using the variable phase method outlined in App.~\ref{app:vpm} that casts the solution in terms of $\alpha(r)$ and $\beta(r)$ coefficients for free-particle basis wavefunctions; the value of the $\alpha$ coefficient at $r=a$ gives the desired coefficient $ \alpha_{\tilde{G}_\ell}(a)$.

We can choose $V_\text{long}(r)$ freely for $r < a$ so long as it is real and so long as we self-consistently recalculate $\bar{f}_{s,\ell}$, $\Sigma_{0,\ell}$, and $\bar{Z}_\ell$. The corrections applied to the raw short-distance amplitude $f_{s,\ell}$ in calculating $\bar{f}_{s,\ell}$ will automatically remove contributions already included in $V_\text{long}(r)$ for $r < a$, and so any such contributions also need to be removed when matching $\bar{f}_s$ to a perturbative QFT calculation.

Our multi-state formalism could also be applied to cases where the final state of the annihilation process is also non-relativistic and can be treated in the framework of non-relativistic quantum mechanics, as studied in e.g.~Ref.~\cite{Cui:2020ppc}; in this case the ``annihilation'' could be treated as a scattering process, and captured either in the long-range potential (as an off-diagonal term) or the short-range amplitude.

\subsection{Inclusive vs exclusive cross sections}

We have chosen above to focus on the inclusive annihilation cross section, as this is the quantity that controls the apparent non-unitarity in the quantum mechanics scattering problem. However, for observational searches for annihilation, the exclusive cross sections for annihilation to particular channels are of course very relevant. One might therefore ask to what degree the corrections we find are universal across different SM final states.~\footnote{We thank the anonymous referee for raising this interesting question.} 

%Ref.~\cite{Flores:2024sfy} uses an alternate formalism, capable of studying individual annihilation channels; under specific circumstances (involving the momentum dependence of the amplitudes), they obtain a simplified expression for the correction that is universal across all final states, depending only on the inclusive cross section.

In the single-state case where the absorptive physics is localized at the origin (as in e.g.~Ref.~\cite{Blum:2016nrz}), it is easy to see that all final states will obtain an equivalent (multiplicative) correction to the annihilation rates; the enhancement to annihilation can be written in terms of the (corrected) wavefunction at the origin, and the (initial state) wavefunction is the same for all final states. This is less obvious in the cases studied by  Ref.~\cite{Flores:2024sfy} where the absorptive physics need not be localized, as the corrected matrix element involves a radial overlap integral between the modified wavefunction and the (position space) amplitude for annihilation. If the modification to the wavefunction is $r$-dependent {\it and} if annihilation to different final states can have quite different momentum dependence (corresponding to different $r$-dependence in the position-space effective potential), the effect of the correction may in principle differ; nonetheless, Ref.~\cite{Flores:2024sfy} find a universal correction in the case where their simplified expression for the corrected cross section is valid, depending only on the inclusive cross section.

To address this question in our formalism, we can consider the case where there are additional states that are part of the Hilbert space, but not coupled by $V_\text{long}(r)$; they can be populated from the incoming states only for $r < a$. Our short-distance scattering amplitude $f_{s,\ell}$ could then be extended to explicitly include scattering into these states, and would source additional outgoing terms in Eq.~\ref{eq:multistatematching}. If we can model the final state as a plane wave characterized by momentum $p_n$ (with $n > N$), produced from state $j$ at short distances with amplitude $f_{s,nj}$, then the corresponding term in Eq.~\ref{eq:multistatematching} will have the same form as previously:
\begin{align} u_{ni}(r) = \sum_{j \le N} \gamma_{ji} f_{s,nj} p_j (c_\ell(p_n r) + i s_\ell(p_n r)), \quad n > N.\end{align}
Furthermore, while previously we described Eq.~\ref{eq:multistatematching} as giving only the wavefunction at the matching radius $r=a$, for a state where the propagation outside $r=a$ is entirely free, we can read off this component of the full scattering amplitude directly at $r=a$, using Eq.~\ref{eq:largermultistate}:
\begin{align} f_{ni}  = \sum_{j \le N} \frac{p_j}{p_i} \gamma_{ji} f_{s,nj}, \quad n > N . \label{eq:annconversion} \end{align}
(If the state has extra degrees of freedom, e.g. a high-multiplicity final state, we would need to refine our description of how to extract the scattering amplitude, but we would expect a similar logic to carry over.)

Thus we can directly read off the full amplitude from the short-range amplitude once we know the matrix $\gamma_{ji}$, which has no dependence on these additional states. From Eq.~\ref{eq:multistatematchsol} we have:
\begin{equation} \gamma_\ell = \tilde{P}^{-1} \alpha_\ell(a) = \tilde{P}^{-1} \left( \alpha_{w_\ell}(a)  + \left[ \alpha_{\tilde{G}_\ell}(a) + i \alpha_{w_\ell}(a) P^{-1}\bar{Q}_\ell^{\dagger}\right]  R_\ell \right). \end{equation}

From the results earlier in this section we can write down a number of relevant identities, in particular:
\begin{align}
\alpha_{w_\ell}(a) & = \tilde{P}^{-1}  (\Sigma^a_{0,\ell})^{-1} \Sigma_{0,\ell} P^2 \nonumber \\
\alpha_{\tilde{G}_\ell}(a) & =\tilde{P}^{-1} (\Sigma^a_{0,\ell})^{-1} \bar{Z}_\ell  (\tilde{P}^\dagger)^{2\ell}  \nonumber \\
R_\ell & = (\tilde{P}^\dagger)^{-2\ell}\left[ ( \kappa_\ell^{-1} - i \Sigma_{0,\ell} P^2 \Sigma_{0,\ell}^\dagger)  \right]^{-1} \Sigma_{0,\ell} P^2 \end{align}

Putting these pieces together we find:
\begin{align} \gamma_\ell 
& =    \tilde{P}^{-2}  (\Sigma^a_{0,\ell})^{-1}   \left[ 1 - \bar{Z}_\ell \bar{f}_{s,\ell} - i \Sigma_{0,\ell} P^2 \Sigma_{0,\ell}^\dagger   \bar{f}_{s,\ell} \right]^{-1} \Sigma_{0,\ell} P^2 \nonumber \\
\Rightarrow \gamma_{ji} & = \frac{p_i}{p_j} \sum_{k \le N} ((\Sigma^a_{0,\ell})^{-1} )_{jk} (\bar{\Sigma}_\ell)_{ki},  \end{align}
where $\bar{\Sigma}_\ell$ is defined as in Eq.~\ref{eq:explicitfxsec}. Then Eq.~\ref{eq:annconversion} becomes:
\begin{align} f_{ni}  = \sum_{k \le N}  \left[ \sum_{j \le N} f_{s,nj} ((\Sigma^a_{0,\ell})^{-1} )_{jk} \right] (\bar{\Sigma}_\ell)_{ki} . \end{align}

The term in square brackets can be interpreted as the bare annihilation amplitude to the desired final state, with the enhancement factor $\Sigma^a_{0,\ell}$ from the non-zero potential $V_\text{long}(r)$ at $r<a$ stripped off (note that because this is an annihilation amplitude, not a scattering amplitude or an annihilation cross section, there is only one factor of the Sommerfeld matrix rather than two). Let us denote this amplitude by the (row) vector $\bar{f}_\ell(X)$ for a given SM final state $X$, i.e.~if $X$ corresponds to the index $n$, then $(\bar{f}_\ell(X))_i =  \sum_{j \le N} f_{s,nj} ((\Sigma^a_{0,\ell})^{-1} )_{ji}$. Then the effect of the long-range potential would be to apply the corrected Sommerfeld factor $\bar{\Sigma}_\ell$ to this bare amplitude, equivalent to contracting the bare annihilation matrix (built from annihilation amplitudes from the different initial states) with factors of $\bar{\Sigma}_\ell^\dagger$ and $\bar{\Sigma}_\ell$ to obtain the cross section. Specifically, applying Eq.~\ref{eq:multistatescattering} we would obtain:
\begin{align} \sigma(i\rightarrow X) & = 4\pi c_i \frac{p_X}{p_i} \sum_{\ell} (2\ell+1) (\bar{\Sigma}_\ell^\dagger \bar{f}(X)^\dagger \bar{f}(X) \bar{\Sigma}_\ell)_{ii} \nonumber \\
\sigma(i\rightarrow X) v_\text{rel} & = c_i \frac{2 \pi i}{\mu} \sum_{\ell} (2\ell+1)\left[ \bar{\Sigma}_\ell^\dagger \left(  -2 i  p_X \bar{f}_\ell(X)^\dagger \bar{f}_\ell(X) \right) \bar{\Sigma}_\ell\right]_{ii}.\end{align}

Comparing the sum of exclusive cross sections to the inclusive result in Eq.~\ref{eq:explicitfxsec}, we see that they share the same structure -- having a ``bare'' cross section matrix contracted with factors of $\bar{\Sigma}_\ell^\dagger$ and $\bar{\Sigma}_\ell$ -- and will agree provided:
\begin{align} \bar{f}_{s,\ell}^\dagger - \bar{f}_{s,\ell} & = - \bar{f}_{s,\ell}^\dagger (\bar{Z}_\ell^\dagger - \bar{Z}_\ell) \bar{f}_{s,\ell} + \sum_X  \left(  -2 i  p_X \bar{f}_\ell(X)^\dagger \bar{f}_\ell(X) \right) . \end{align} 
The form of this relation is familiar from the optical theorem. As discussed previously, the first term on the right-hand side corresponds to transitions into the kinematically available DM states, whereas the sum over $X$ corresponds to annihilation into SM states.

In using Eq.~\ref{eq:multistatescattering} we tacitly assumed that the final state is a two-particle state characterized by momentum $p_X$. For more general final states, the appropriate generalization replaces the $p_X$ factor with a phase-space integral over the final state, as in Eq.~\ref{eq:gamma0}:
\begin{align} -2 i p_X  \bar{f}(X)^\dagger \bar{f}(X) \rightarrow \bar{\Gamma}_{0,\ell}(X), \quad (\bar{\Gamma}_{0,\ell}(X))_{ba}  \equiv \frac{1}{32 i \pi \mu} \int d\Pi_X \mathcal{M}_\ell^*(b \rightarrow X) \mathcal{M}_\ell(a \rightarrow X),  \end{align}
where $\mathcal{M}_\ell(a\rightarrow X)$ is the matrix element for annihilation from the $a$ initial state to the $X$ final state, ignoring any Sommerfeld factors, computed using the QFT normalization convention discussed in App.~\ref{app:optical}. (To check the normalization, note that in the case where the final state has two particles and center-of-mass energy $E_\text{CM}\approx 4\mu$, $\int d\Pi_X = \int d\Omega \frac{1}{16\pi^2} p_X/E_\text{CM} = p_X/(4\pi \times 4\mu)$.) Thus we can write the exclusive cross section as:
\begin{align} \sigma(i\rightarrow X) v_\text{rel} & = c_i \frac{2 \pi i}{\mu} \sum_{\ell} (2\ell+1)\left[ \bar{\Sigma}_\ell^\dagger \bar{\Gamma}_{0,\ell}(X) \bar{\Sigma}_\ell\right]_{ii},\end{align}
compute $\bar{\Gamma}_{0,\ell}(X)$ from perturbative quantum field theory, and capture the correction to the Sommerfeld enhancement purely in $\bar{\Sigma}_\ell$ as defined in Eq.~\ref{eq:explicitfxsec}. Note that $\bar{\Sigma}_\ell$ depends on the short-distance physics only through the inclusive short-distance scattering amplitude; it is universal with respect to the SM final state. 

%Note that this is structurally very similar to Eq.~\ref{eq:explicitfxsec}. The optical theorem (or equivalently, unitarity of the full expanded $S$-matrix) for the short-range physics will relate the sum of the bare cross sections for the various annihilation channels to the numerator term $\bar{f}_{s,\ell}^\dagger - \bar{f}_{s,\ell} + \bar{f}_{s,\ell}^\dagger (\bar{Z}_\ell^\dagger - \bar{Z}_\ell) \bar{f}_{s,\ell}$ (at least up to higher-order corrections in our matching prescription) in Eq.~\ref{eq:explicitfxsec}, which is then contracted with $\bar{\Sigma}_\ell^\dagger$ and $\bar{\Sigma}_\ell$.

As a practical matter, however, while the correction to the Sommerfeld matrix $\bar{\Sigma}_\ell$ is universal, the magnitude of the correction to the cross section may differ between final states. Different final states may be sensitive to different components of the (initial-state) wavefunction: for example, in the case where the DM is neutral but the potential couples it to slightly higher-mass chargino states, the charginos can annihilate at tree level to photons, whereas the DM-DM state cannot. Thus a correction to $\bar{\Sigma}_\ell$ that is concentrated in the rows/columns relevant to the DM-DM state might have minimal effect on the cross section for annihilation to photons, but a significant effect on the cross section for annihilation to $W$ bosons.

\subsection{Simplified algorithm for multi-state calculations}

Finally, we provide a summary of the algorithm for calculating the corrected annihilation cross section in the multi-state case, for the convenience of the reader. As explained above, there are a number of self-consistent choices for how to perform the calculation (e.g. the behavior of $V_\text{long}(r)$ for $r < a$ and the choice of the Hermitian matrix $\Delta_\ell$). For this summary, we choose $V_\text{long}(r)$ to be $a$-independent, i.e. we use the conventional non-relativistic result for $V_\text{long}(r)$ even for $r < a$, and take $\Delta_\ell=0$. We show in App.~\ref{app:optical} that this convention facilitates a particularly simple matching between $\bar{f}_{s,\ell}$ and the leading-order annihilation amplitude. 
We note that the matching prescription here omits corrections to scattering amongst the DM states that are not captured by the potential $V_\text{long}(r)$ (although this is improvable by including higher-order corrections in the potential), as well as corrections to the short-range annihilation amplitude that are not captured as a Sommerfeld correction from the potential in the $r < a$ range (e.g.~non-ladder loop diagrams). Developing a systematically improvable matching procedure is a possible direction for future work.

\begin{framed}
\begin{enumerate}
\item Compute the bare annihilation matrix $\bar{\Gamma}_{0,\ell}(X)$ for the desired final state $X$, and also for the inclusive sum over all available SM final states (not including scattering to other states that are coupled by the potential $V_\text{long}(r)$, as this process is already captured by the Schr\"odinger equation). Specifically, we define this matrix as (App.~\ref{app:optical}) $(\bar{\Gamma}_{0,\ell})_{dc}(X) \equiv \frac{1}{32 i \pi \mu} \int d\Pi_X \mathcal{M}^*(d\rightarrow X)  \mathcal{M}(c\rightarrow X)$, where $c$ and $d$ index the initial two-body DM states. The inclusive result is $\bar{\Gamma}_{0,\ell} \equiv \sum_X \bar{\Gamma}_{0,\ell}(X)$. The matrix elements $\mathcal{M}$ are obtained from standard perturbative QFT methods, divided by a factor of $\sqrt{2}$ if either the initial or final states are comprised of two identical particles, or a factor of 2 if both initial and final states have that property. An example is given for the wino in App.~\ref{app:winomatrix}.
\item Obtain the inclusive short-distance scattering amplitude at leading order via the matching relation: $\bar{f}_{s,\ell} = -\frac{1}{2} \bar{\Gamma}_{0,\ell}$ (Eq.~\ref{eq:improvedmatchingcon2}).
\item Solve the Schr\"odinger equation to obtain the standard/unregulated Sommerfeld enhancement $\Sigma_{0,\ell}$ and the correction matrix $\bar{Z}_\ell = \Sigma^a_{0,\ell} \tilde{P} \alpha_{\tilde{G}_\ell}(a) (\tilde{P}^\dagger)^{-2\ell}$. $\tilde{P}$ is the diagonal matrix whose entries are $\sqrt{p_j}$ for the $j$th two-particle state, where $p_j$ is the single-particle momentum in the $j$th state (see Eq.~\ref{eq:multistatep}). There are several ways to obtain $\Sigma^a_{0,\ell}$, $\Sigma_{0,\ell}$ and $\alpha_{\tilde{G}_\ell}(a)$. Using the variable phase method (App.~\ref{app:vpm}), with each solution characterized by matrices $\alpha(r)$ and $\beta(r)$, one recipe is:
\begin{itemize}
\item Solve for the regular solution $w_\ell(r)$ by imposing boundary conditions $\alpha_{w_\ell}(r) \rightarrow \tilde{P}$ as $r\rightarrow \infty$, $\beta_{w_\ell}(0)=0$, for the two cases where the potential is (i) $V_\text{long}(r)$ for all $r$, (ii) $V_\text{long}(r)$ for $r < a$, zero otherwise. In each case, read off the uncorrected Sommerfeld enhancement matrix ($\Sigma_{0,\ell}$ in case (i) and $\Sigma^a_{0,\ell}$ in case (ii)) as $\tilde{P} \alpha_{w_\ell}(0) \tilde{P}^{-2}$. For case (i), in $\alpha_{w_\ell}(r)$ and $\Sigma_{0,\ell}$, only columns corresponding to kinematically allowed states at infinity are retained (but all rows must be kept; rows index the component while columns index the initial state).
\item Solve for the irregular solution $\tilde{w}_\ell(r)$ by imposing boundary conditions $\beta_{\tilde{w}_\ell}(0)=-\tilde{P}^{2\ell+1}$, $\alpha_{\tilde{w}}(r) \rightarrow 0$ as $r\rightarrow \infty$. Read off $\alpha_{\tilde{G}_\ell}(a) = \alpha_{\tilde{w}_\ell}(a) - i \alpha_{w_\ell}(a) \Sigma_{0,\ell}^\dagger (\tilde{P}^\dagger)^{2\ell}$ (Eq.~\ref{eq:alphaGextraction}).
\end{itemize}
\item Let $P$ be the truncation of $\tilde{P}$ to include only kinematically allowed states (i.e.~$p_j$ real); let $\mu$ be the reduced mass, and $c_i=2$ if the initial particles in the $i$th state are identical (otherwise 1). Compute the corrected Sommerfeld enhancement and cross section (from the $i$th two-particle state, to the final state $X$) via:
\begin{align} 
\bar{\Sigma}_\ell \equiv \left[ 1 -  (\bar{Z}_\ell + i  \Sigma_{0,\ell} P^2 \Sigma_{0,\ell}^\dagger) \bar{f}_{s,\ell} \right]^{-1} \Sigma_{0,\ell} \nonumber \\
(\sigma_{i,\text{ann}}(X) v_\text{rel})_\ell = c_i  \frac{2 \pi i}{\mu}  (2\ell+1) \left[ \bar{\Sigma}_{\ell}^\dagger \bar{\Gamma}_{0,\ell}(X) \bar{\Sigma}_{\ell} \right]_{ii}.
 \end{align} 
\end{enumerate}\end{framed}

\section{Applying the multi-state result to wino dark matter}
\label{sec:wino}

The Sommerfeld-enhanced annihilation of wino DM has been studied extensively in the literature (e.g.~\cite{Hisano:2003ec, Hisano:2004ds, Cirelli:2007xd, Beneke:2014gja, Braaten:2017kci}). The wino multiplet consists of a neutral Majorana fermion $\chi^0$, which is the DM candidate, and a slightly heavier Dirac fermion $\chi^{\pm}$. An initial state consisting of two DM particles, $\chi^0\chi^0$, is coupled to the $\chi^+\chi^-$ two-particle state through $W$ boson exchange. We will employ the framework of Ref.~\cite{Beneke:2014gja}, using the ``method-2 basis'' discussed in that work, where the $\chi^+\chi^-$ and $\chi^-\chi^+$ two-particle states are treated as components of a single (appropriately symmetrized) state. This means we can treat the wino as a $N=2$ system (with the states corresponding to $\chi^0\chi^0$ and $\chi^+\chi^-$), with $M=1$ or $M=2$ depending on the kinetic energy of the particles; in the Milky Way halo, typically we expect $M=1$. Note that while two-particle states such as $\chi^0 \chi^+$ can exist, they are not coupled to $\chi^0 \chi^0$ through the long-range potential.

The leading-order (LO) potential in the wino case (and the method-2 basis) can be written as \cite{Hisano:2004ds}:
\begin{equation} V_\text{long}(\vec{r}) = \begin{pmatrix} 0 & -\sqrt{2} \alpha_W \frac{e^{-m_W r}}{r}  \\ -\sqrt{2} \frac{\alpha_W}{r} e^{-m_W r} & 2\Delta - \frac{\alpha}{r} - \frac{\alpha_W c_W^2}{r} e^{-m_Z r},\end{pmatrix} \label{eq:winopot} \end{equation}
where the first row/column corresponds to the $\chi^0 \chi^0$ state, the second row/column corresponds to the $\chi^+\chi^-$ state, and $\Delta$ denotes the mass splitting between $\chi^+$ and $\chi^0$, and $c_W=\cos\theta_W$. Note that when we perform perturbative calculations using this potential, we include the $2\Delta$ term in the unperturbed Hamiltonian. Note that because the $\chi^0 \chi^0$ state consists of two identical fermions, its wavefunction must be antisymmetric overall; this means that spin-singlet configurations must have even $\ell$, and spin-triplet configurations must have odd $\ell$. It also means that we must take $c_i=2$ in our calculations. 

The next-to-leading-order (NLO) corrections to this potential have been computed in Refs.~\cite{Beneke:2019qaa, Urban:2021cdu}; for our numerical results, we employ the analytic fitting functions derived in those works. Specifically, we make the following replacements in Eq.~\ref{eq:winopot}, where $L\equiv \ln(m_W r)$ and $s_W=\sin\theta_W$:
\begin{align}
     &e^{-m_W r}  \rightarrow   e^{-m_W r} \nonumber \\
     &+ \frac{2595}{\pi} \alpha_W 
     \begin{cases} (-1) \exp\left[-\frac{79 \left(L - \frac{787}{12}\right)\left(L - \frac{736}{373}\right) \left(L - \frac{116}{65}\right)\left(L^2 - \frac{286 L}{59} + \frac{533}{77}\right) }{34 \left(L - \frac{512}{19}\right) \left(L - \frac{339}{176}\right) \left(L - \frac{501}{281}\right) \left(L^2 -  \frac{268 L}{61} + \frac{38}{7}\right)} \right]\!, & m_W r < 555/94 \\
     \exp\left[-\frac{13267 \left(L - \frac{76}{43}\right)\left(L - \frac{28}{17}\right) \left(L + \frac{37}{30}\right)\left(L^2 - \frac{389 L}{88} + \frac{676}{129}\right) }{5 \left(L - \frac{191}{108}\right) \left(L - \frac{256}{153}\right) \left(L + \frac{8412}{13}\right) \left(L^2 -  \frac{457 L}{103} + \frac{773}{146}\right)} \right] & m_W r > 555/94, \end{cases} 
     \nonumber \\
     &c_W^2 e^{-m_Z r}  \rightarrow c_W^2 e^{-m_Z r} \nonumber \\
     &+ \alpha_W \left[ -\frac{80}{9} \frac{s_W^4 \left(\ln(m_Z r) + \gamma_E\right)}{2\pi (1 + (32/11) (m_W r)^{-22/9})}   + \frac{\left( \frac{19}{6}  \ln(m_Z r) - \frac{1960}{433} \right)}{2\pi(1 + (7/59) (m_W r)^{61/29})} \right. \nonumber \\
     & \left. - \frac{s_W^2 \left(-\frac{1}{30} + \frac{4}{135} \ln(m_W r) \right)}{1 + (58/79) (m_W r)^{-17/15} + (1/30) (m_W r)^{119/120} + (8/177) (m_W r)^{17/8}}\right]\!.
\label{eq:vnlo}     
\end{align}

For simplicity, in this example we will focus on the dominant $\ell=0$ annihilation channel, and the case where the kinetic energy is insufficient to excite the $\chi^+\chi^-$ state.

\subsection{General results for $M=1$, $N=2$}

In this low-momentum wino case we have $M=1$, $N=2$, and the relevant momenta are $p_2=-p_2^*$, $p_1=p_1^*$. We will work with the form of the cross section given in Eq.~\ref{eq:symboldefsnew}-\ref{eq:annxsecfinalnew}.

The baseline Sommerfeld enhancement is encapsulated in a $2\times 1$ vector $\Sigma_{0,\ell} = \begin{pmatrix} \sigma_{0,\ell} \\ \sigma_{\pm,\ell}\end{pmatrix}$. We can show that the relative phase between $\sigma_{0,\ell}$ and $\sigma_{\pm,\ell}$ is predetermined. From our discussion in the main text, our prescription for the ordinary Sommerfeld enhancement states that $\sigma_{0,\ell} = Q_{11}$, $\sigma_{\pm,\ell} = (\Sigma_{0,\ell})_{21} = (p_2/p_1) Q_{21}$, and all elements of $\bar{Q} \bar{Q}^\dagger$ are real.  Now $\bar{Q}$ is a $2\times 1$ matrix with elements $\bar{Q}_{in} = p_i^{\ell+1}/\sqrt{p_n} Q_{in}$, i.e. $\bar{Q}_{11} = p_1^{\ell+1/2} Q_{11}$, $\bar{Q}_{21} = p_2^{\ell+1}/\sqrt{p_1} Q_{21}$. The matrix $\bar{Q}\bar{Q}^\dagger$ being real requires that $\bar{Q}_{11} \bar{Q}_{21}^*$  is real, i.e. $Q_{11} p_1^{\ell}  (p_2^*)^{\ell+1}  Q_{21}^*$ is real, and so in turn $\sigma_{0,\ell} p_1^{\ell}  (p_2^*)^{\ell+1} (p_1/p_2^*) \sigma_{\pm,\ell}^*= p_1 (p_1 p_2^*)^\ell \sigma_{0,\ell}  \sigma_{\pm,\ell}^*$ must be real. Since $p_1$ is real and $p_2$ is imaginary, in this case we see that $\sigma_{0,\ell}  \sigma_{\pm,\ell}^*$ is purely real for even $\ell$, and purely imaginary for odd $\ell$; thus $\sigma_{0,\ell}$, $\sigma_{\pm,\ell}$ must share the same phase for even $\ell$, and have a phase offset of $\pi/2$ for odd $\ell$. 

We observe that in the expression $\Sigma_{0,\ell} P^2\Sigma_{0,\ell}^\dagger$ any overall phase factor in $\Sigma_{0,\ell}$ will cancel out. Thus in Eq.~\ref{eq:symboldefsnew}, the effect of introducing an overall phase factor into $\Sigma_{0,\ell}$ will just be to rescale $\Sigma_\ell$ by the same phase factor. This overall phase factor then cancels out in the expression for the annihilation rate (which depends on $\Sigma_\ell$ via $\Sigma_\ell^\dagger (\kappa_\ell^\dagger - \kappa_\ell) \Sigma_\ell$). Thus we have the freedom to rescale $\Sigma_{0,\ell}$ by a phase factor, allowing us to assume without loss of generality that either $\sigma_0, \sigma_\pm$ are both real (for even $\ell$) or that $\sigma_0$ is real and $\sigma_\pm$ purely imaginary (for odd $\ell$). 

Now let us choose $\ell=0$ and remove the $\ell$ subscript on the various quantities. The reduced mass $\mu=M_\chi/2$ where $M_\chi$ is the wino mass, so that the $s$-wave annihilation cross section is given by:
\begin{align} \sigma_{\text{ann}} v_\text{rel} & = \frac{8 \pi i}{M_\chi} \begin{pmatrix} \sigma_0 & \sigma_\pm \end{pmatrix}  \left[1 + i  p_1 \kappa^\dagger \begin{pmatrix} \sigma_0^2 & \sigma_0 \sigma_\pm \\ \sigma_0 \sigma_\pm & \sigma_\pm^2 \end{pmatrix} \right]^{-1} (\kappa^\dagger - \kappa) \nonumber \\
& \times \left[1 - i p_1 \begin{pmatrix} \sigma_0^2 & \sigma_0 \sigma_\pm \\ \sigma_0 \sigma_\pm & \sigma_\pm^2 \end{pmatrix}  \kappa \right]^{-1} \begin{pmatrix} \sigma_0 \\ \sigma_\pm \end{pmatrix}.\label{eq:winoxsec} 
\end{align}
Writing $\kappa=\begin{pmatrix} \kappa_{11} & \kappa_{12} \\ \kappa_{21} & \kappa_{22}\end{pmatrix}$ and explicitly performing the matrix inversion yields:
\begin{align} \sigma_{\text{ann}} v_\text{rel} & = -\frac{8 \pi i}{M_\chi} \frac{\left[(\kappa_{11} -\kappa_{11}^*) \sigma_0^2 + (\kappa_{22} - \kappa_{22}^*) \sigma_\pm^2 + (\kappa_{12} + \kappa_{21} - \kappa_{21}^* - \kappa_{12}^*) \sigma_0 \sigma_\pm \right]}{\left|i + p_1 \left(\sigma_0^2 \kappa_{11} + \sigma_\pm^2 \kappa_{22} + \sigma_0 \sigma_\pm (\kappa_{12} + \kappa_{21})\right) \right|^2} \nonumber \\
& =  \frac{16 \pi}{M_\chi} \frac{\left[ \sigma_0^2 \text{Im} \kappa_{11}  + \sigma_\pm^2 \text{Im} \kappa_{22}  + \sigma_0 \sigma_\pm \text{Im}(\kappa_{12} + \kappa_{21})  \right]}{\left|i + p_1 \left(\sigma_0^2 \kappa_{11} + \sigma_\pm^2 \kappa_{22} + \sigma_0 \sigma_\pm (\kappa_{12} + \kappa_{21})\right) \right|^2} 
\end{align}

Let us define
\begin{equation} \gamma_0 =  p_1^2 \left(\sigma_0^2 \kappa_{11} + \sigma_\pm^2 \kappa_{22} + \sigma_0 \sigma_\pm (\kappa_{12} + \kappa_{21})\right),\end{equation}
then we can write the regulated annihilation cross section as:
\begin{align} \sigma_{\text{ann}} v_\text{rel} 
& =  \frac{16 \pi}{M_\chi} \frac{\text{Im} \gamma_0 / p_1^2}{\left|i + \gamma_0  / p_1 \right|^2}  =  \frac{16 \pi}{M_\chi} \frac{\text{Im} \gamma_0}{( p_1 + \text{Im} \gamma_0 )^2 + (\text{Re}\gamma_0)^2}. 
\end{align}
This is exactly the form of the universal near-resonance $s$-wave cross section described in Ref.~\cite{Braaten:2017dwq}, if $\gamma_0$ is approximately momentum-independent. For $s$-wave resonances we expect $\sigma_0, \sigma_\pm \propto 1/p_1$, with the components of $\kappa$ being close to momentum-independent for $\ell=0$ (at least if the short-range contribution dominates), and so $\gamma_0$ as defined above can indeed be approximately momentum-independent.

Note that the unitarity bound on this $s$-wave cross section corresponds to $\sigma_{\text{ann}} \le 2 \pi / p_1^2$, i.e. $\sigma_{\text{ann}} v_\text{rel} \le 2 \pi / (p_1 \mu) = 4 \pi / (M_\chi p_1)$. This bound is manifestly satisfied since $\displaystyle{\frac{\text{Im} \gamma_0}{( p_1 + \text{Im} \gamma_0 )^2 + (\text{Re}\gamma_0)^2} \le \frac{\text{Im} \gamma_0}{( p_1 + \text{Im} \gamma_0 )^2} \le \frac{1}{4 p_1}}$.
 
\subsection{Calculation of the corrected enhancement}

We now compute the unregulated $s$-wave Sommerfeld enhancement factors $(\sigma_0,\sigma_\pm)$, and the coefficients $\Sigma^a_{0,\ell}$, using the variable phase method~\cite{PhysRevC.84.064308}, as recast in Refs.~\cite{Beneke:2014gja, Asadi:2016ybp}. We use the same method, adapted as discussed in App.~\ref{app:vpm}, to solve for the irregular solution $\tilde{G}_\ell(r)$ and to extract the coefficient $\alpha_{\tilde{G}_\ell}(a)$ and hence $\bar{Z}_\ell$ (as defined in Eq.~\ref{eq:multistatekappa}).

In the main text we will choose $V_\text{long}(r)$ to be $a$-independent and employ the NLO potential derived in Refs.~\cite{Beneke:2019qaa,Urban:2021cdu} for our numerical calculations. In App.~\ref{app:winoextras} we show the result of including the leading real correction to the $11$ term of the potential (which is not included in the NLO potential), the result of instead using the LO potential, and the result of instead using the convention where $V_\text{long}(r)$ is set to zero for $r < a$.

We choose a mass splitting of $\Delta = 0.1644$ GeV~\cite{Ibe:2012sx}, and for the electroweak parameters, we employ: $m_W = 80.3692$ GeV, $m_Z = 91.1880$ GeV, and $\sin^{2}\theta_W = 0.23129$ \cite{ParticleDataGroup:2024cfk}. For the electroweak coupling we use $G_F = 1.1663788 \times 10^{-5}$ GeV$^{-2}$ \cite{ParticleDataGroup:2024cfk}, combined with the relation $G_F = \pi \alpha_W/(\sqrt{2} m_W^2)$.

For all calculations we set the matching radius at $a=0.5/M_\chi$, $1/M_\chi$ and $2/M_\chi$ and compare results, in order to test the sensitivity to $a$.

We use Eq.~\ref{eq:multistatekappa} to determine $\kappa$. To compute $\bar{f}_{s,\ell}$, for our main calculation we work in the convention where $V_\text{long}(r)$ is $a$-independent and is given by Eqs.~\ref{eq:winopot}-\ref{eq:vnlo}. As derived in App.~\ref{app:optical}, we match $\bar{f}_{s,\ell}$ in this convention directly to $-(1/2)\bar{\Gamma}_{0,\ell}$ computed from the tree-level annihilation amplitudes.

Specifically, for $s$-wave wino annihilation we have:
\begin{equation} \bar{f}_{s,0} = i\frac{ \alpha_W^2}{2 (2 \mu)} \begin{pmatrix} 1 & \frac{1}{\sqrt{2}} \\ \frac{1}{\sqrt{2}} & \frac{3}{2} \end{pmatrix} . \end{equation}

\subsection{Numerical results}

\begin{figure}
\centering
\includegraphics[width=0.49\hsize]{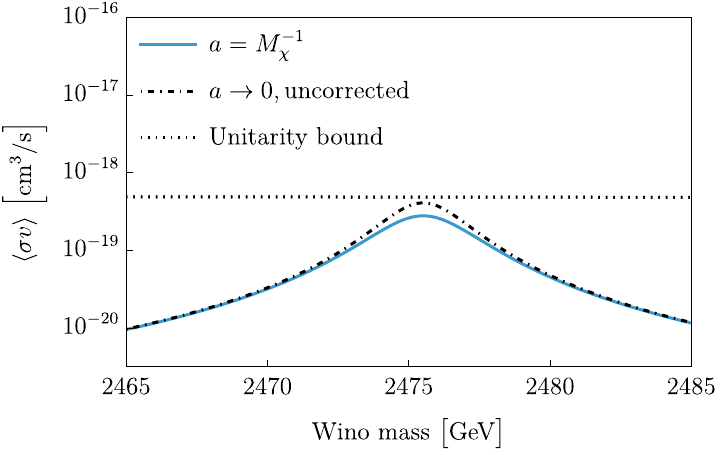}
\includegraphics[width=0.49\hsize]{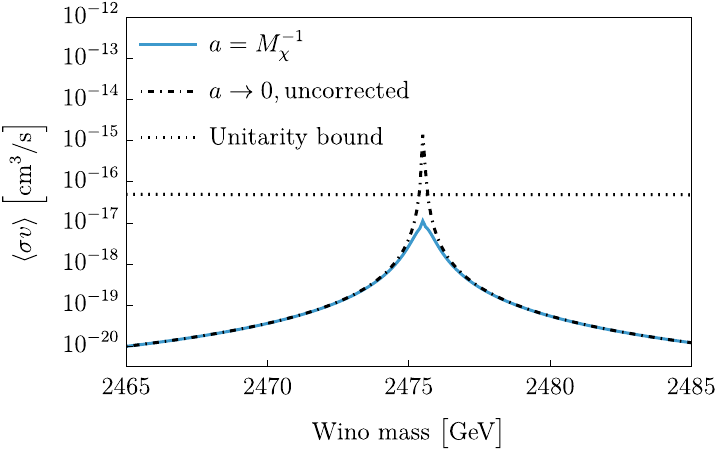}
\caption{The blue lines show the Sommerfeld-enhanced annihilation cross section for wino dark matter, using the corrected expression derived in this work, for a matching radius of $a = M_{\chi}^{-1}$. The relative velocity between DM particles is taken to be 
$v=10^{-4}$ (\emph{left}) or $v=10^{-6}$ (\emph{right}). We additionally computed the Sommerfeld-enhanced annihilation cross section using matching radii of $a = 0.5 M_{\chi}^{-1}$ and $a = 2M_{\chi}^{-1}$ and find that they are identical to the $a = M_{\chi}^{-1}$ results to one part in $10^5$ for $v=10^{-4}$ and one part in $10^4$ for $v=10^{-6}$. 
The dot-dashed black line shows the result of computing the uncorrected Sommerfeld-enhanced annihilation cross section. The dotted black line indicates the unitarity bound on the annihilation cross section.}\label{fig:xsec_wino_NLO}
\end{figure}

We plot in Fig.~\ref{fig:xsec_wino_NLO} both the full annihilation cross section, and the cross section obtained using the uncorrected Sommerfeld enhancement $\sigma_{\ell,0}$, as a function of the wino mass $M_\chi$. We zoom in on the resonance region for $v_\text{rel}=10^{-4}$ and $v_\text{rel}=10^{-6}$ in order to demonstrate the regulation of the resonances. We observe that the results are nearly $a$-independent, as expected (there is a more dramatic $a$-dependence in the intermediate results when we choose $V_\text{long}(r)$ to be explicitly $a$-dependent, which cancels out in the full expression, as we demonstrate in App.~\ref{app:winoextras}). We also checked the results for $v_\text{rel}=10^{-3}$, which is comparable to the typical velocity of dark matter particles in the Milky Way, and found that there the uncorrected cross section was always well below the unitarity bound and a good approximation to the corrected result.

At $v_\text{rel}=10^{-4}$, the uncorrected cross section approaches the unitarity bound and the correction induces a non-negligible suppression. This suppression is more pronounced for $v_\text{rel}=10^{-6}$, where the uncorrected result significantly exceeds the unitarity bound on resonance. The corrected results all respect the unitarity bound, as we expect by construction. Note that in the case of $v_\text{rel}=10^{-6}$, the corrected cross section does not come close to saturating the unitarity limit; the effect of the correction is not simply to provide an upper limit on the cross section at a given momentum.

\begin{figure*}
\centering
\includegraphics[width=0.6\hsize]{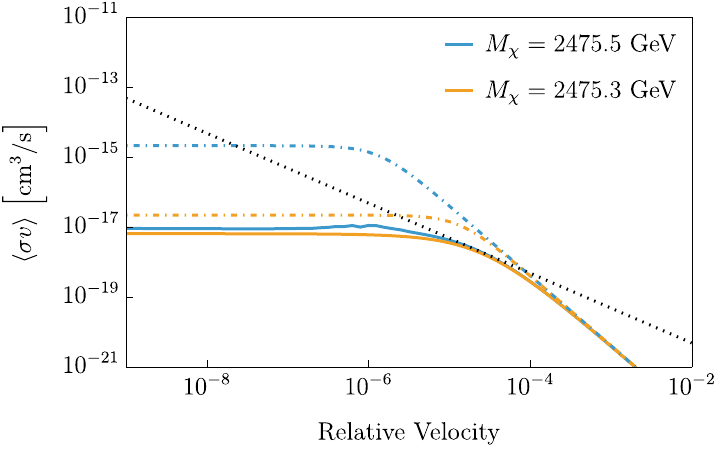}
%xsec_velocity_dependence_NLO.pdf}
\caption{Colored lines show the Sommerfeld-enhanced annihilation cross section for wino dark matter, using the corrected expression derived in this work and taking $a=M_\chi^{-1}$, for different choices of the wino mass in proximity to the first resonance. For each mass, the solid line shows the full result and the dot-dashed line shows the uncorrected Sommerfeld-enhanced result. The black dotted line shows the unitarity bound at the mass corresponding to the center of the resonance (the mass variation between the different lines in this figure changes the unitarity bound at the sub-percent level).}\label{fig:vdep_NLO}
\end{figure*}

We demonstrate the velocity dependence of the corrected and uncorrected cross section near the resonance in Fig.~\ref{fig:vdep_NLO}, for two different masses corresponding to the cases where the uncorrected cross section exceeds the unitarity bound, in one case only slightly and for a small range of momenta, and in the other case by a large factor (with the second case being closer to the resonance peak). We observe that as $v_\text{rel}$ is decreased, in both cases the corrected cross section saturates at a roughly $v_\text{rel}$-independent value soon after approaching the unitarity bound. In the first case, this induces only a modest shift in the saturation velocity and saturated cross section relative to the uncorrected calculation, but in the second case, this leads to earlier saturation and a much lower cross section (again relative to the uncorrected results).

 The low-velocity saturated cross section near the resonance can be compared to the results of the zero-range effective field theory of Ref.~\cite{Braaten:2017dwq}, which provides a unitarized result for the inclusive wino annihilation cross section that is valid in proximity to the resonance. That work found that the resonance peak was at 2.39 TeV, offset from ours by a few percent, so we should not expect agreement in the cross sections  better than this level; we note that the resonance position is sensitive to the exact choice of electroweak parameters and to higher-order corrections to the potential (see App.~\ref{app:winoextras}). That work finds that at the resonance, $\gamma_0 = 9.59 \times 10^{-4} i \times 28.5 \text{GeV}$, corresponding to a cross section in the low-velocity limit of $\sigma_\text{ann} v_\text{rel} = 16\pi/(M_\chi \text{Im} \gamma_0) = 0.769 \text{GeV}^{-2} = 8.98 \times 10^{-18}$ cm$^3$/s, which is very similar to the result shown in Fig.~\ref{fig:vdep_NLO} ($\sigma_\text{ann} v_\text{rel} = 1.02 \times 10^{-17}$ cm$^3$/s for the nearest-resonance mass and lowest velocity we tested). Our methods are quite different so this correspondence is encouraging.

\section{Conclusions}
\label{sec:conclusions}

We have derived a general expression for the regulated Sommerfeld enhancement, valid for arbitrary partial waves and where multiple states are coupled by the long-range potential. We assume a hierarchy between the momentum scale associated with the long-range potential (parametrically $\alpha M_\chi$) and the momentum scale associated with the short-range physics (parametrically $M_\chi$ for annihilation), and further assume that the $S$-matrix element associated with the long-range potential $S_{0,\ell}$ satisfies $S_{0,\ell}^\dagger S_{0,\ell} = 1$. Under these assumptions, we have demonstrated that the full cross section, which manifestly preserves unitarity, can be written in terms of the conventional Sommerfeld enhancement and a corrected scattering amplitude $\kappa_{\ell}$, for any partial wave $\ell$ and for arbitrary numbers of states coupled by the potential. The corrected amplitude $\kappa_\ell$ can be constructed in general from the short-distance amplitude $f_{s,\ell}$ and two additional factors obtained by solving the Schr\"odinger equation with the long-range potential, and we have explored its generic properties for the case where only a single two-particle state is involved.

 We have provided expressions for the corrected $S$-matrix and cross section valid when any number of the states are kinematically allowed at asymptotically large distances, as well as simplified expressions valid for the one-state case. We have confirmed we can reproduce earlier expressions in the literature for the leading-order multi-state Sommerfeld enhancement for arbitrary $\ell$, as well as the manifestly unitarity-preserving cross section for the $s$-wave single-state case, and the near-resonance behavior previously studied using zero range effective field theory in the case of the wino. Our expressions nominally depend on the matching radius, but we have shown (analytically in the single-state case and numerically in a multi-state example) that the leading dependence on the matching radius can be absorbed/canceled, and the residual dependence is small. We expect that a more complete effective field theory treatment, including higher-order corrections to both the potential and the short-range physics, would further reduce the residual dependence on the matching radius, which can be thought of as an auxiliary parameter akin to the renormalization scale.
 
One point of interest is that we have found that even when the uncorrected cross section is well below the unitarity bound at a given velocity, there can be significant corrections to the standard expression in cases where the uncorrected cross section at higher velocities exceeds the unitarity bound. Our method helps clarify the origin of these corrections, which can be cast as modifications to the effective short-range matrix element from the long-range potential.

\section*{Acknowledgements} We thank the anonymous referee for helpful comments that improved the clarity of the paper.
TRS thanks Tobias Binder, Kalliopi Petraki, and Marcos Flores for useful discussions.
RS thanks Tomohiro Abe, Ayuki Kamada, and Takumu Yamanaka for useful discussions.
We also thank Barry Cimring for pointing out errors in equations in an earlier version of this manuscript.
The work of RS is supported in part by JSPS KAKENHI No.~23K03415, No.~24H02236, and No.~24H02244.
The work of AP is supported in part by the US National Science Foundation Grant PHY2210533 and the Simons Foundation Grant Nos.~623940 and MPS-SIP-00010469.
TRS was supported in part by a Guggenheim Fellowship; the Edward, Frances, and Shirley B. Daniels Fellowship of the Harvard Radcliffe Institute; the Bershadsky Distinguished Fellowship of the Harvard Physics Department; and the Simons Foundation (Grant Number 929255, T.R.S). TRS and AP thank the Kavli Institute for Theoretical Physics (KITP) and the Mainz Institute for Theoretical Physics for their hospitality during the completion of this work; this research was supported in part by grant no.~NSF PHY-2309135 to KITP. TRS thanks the Aspen Center for Physics for their hospitality during the completion of this work, which was performed in part at the Aspen Center for Physics, which is supported by National Science Foundation grant PHY-2210452. TRS' work was supported by the U.S. Department of Energy, Office of Science, Office of High Energy Physics of U.S. Department of Energy under grant Contract Number DE-SC0012567. 

\begin{appendix}
\section{Analytic properties of $F_\ell$ and $G_\ell$} \label{sec:detail on Fell and Gell}
\label{app:analyticwavefn} 
In this appendix, we discuss the analytic properties of the $F_\ell$ and $G_\ell$
basis solutions defined in Sec.~\ref{sec:s matrix boundary condition}, following the discussion of \textit{e.g.,} Ref.~\cite{teichmann1951interpretation}.
We assume the most singular term of the potential at the origin is $1/r$ and the potential can be expanded as a series in $r$:
\begin{align}
2\mu V(r) = \sum_{k=-1}^\infty v_k r^k.\label{eq:vk}
\end{align}

\subsection{Regular solution : $F_\ell$}
The regular solution $F_\ell$ can be expanded as a power series in $r$:
\begin{align}
F_\ell(r) = C_\ell p^{\ell+1} \sum_{k=\ell+1}^\infty f_{\ell,k} r^k. \label{eq:F expansion}
\end{align}
In order to be consistent with the boundary condition of Eq.~\ref{eq:F and G at the origin}, the coefficient of the $r^{\ell+1}$ term is determined to be
\begin{align}
f_{\ell,\ell+1} &= \frac{1}{(2\ell+1)!!}. \label{eq:f recurrence 1}
\end{align}
The Schr\"odinger equation Eq.~\ref{eq:schroedinger eq F} around $r=0$ tells us the recurrence relation for $f_{\ell,k}$ with $k \geq \ell+2$:
\begin{align}
&-(2\ell+2) f_{\ell,\ell+2} + v_{-1} f_{\ell,\ell+1} = 0, \label{eq:f recurrence 2}\\
&[-k(k-1)+\ell(\ell+1)] f_{\ell,k} + \left(\sum_{m=-1}^{k-\ell-3} v_m f_{\ell,k-2-m} \right) - p^2 f_{\ell,k-2} = 0 \qquad (k\geq \ell+3) \label{eq:f recurrence 3}.
\end{align}
We can determine $f_{\ell,k}$ by applying this recurrence relation repeatedly starting from small $k$, and see that all of the $f_{\ell,k}$ coefficients are analytic functions of $p^2$.

\subsection{Irregular solution : $G_\ell$}
On the other hand, the irregular solution $G_\ell(r)$ can be expanded as
\begin{align}
G_\ell(r) &=
	\frac{1}{C_\ell p^\ell} \left[ \sum_{k=-\ell}^\infty g_{\ell,k} r^k
	+ x_\ell \log \frac{r}{r_0} \times \frac{1}{C_\ell p^{\ell+1}} F_\ell(r)\right]
	\nonumber\\
	&=
	\frac{1}{C_\ell p^\ell} \left[ \sum_{k=-\ell}^\infty g_{\ell,k} r^k
	+ x_\ell \sum_{k=\ell+1}^\infty f_{\ell,k} r^k \log \frac{r}{r_0} \right]. \label{eq:G expansion}
\end{align}
Here $r_0$ is introduced to make the argument of the logarithm dimensionless.
Shifting $r_0$ can be absorbed by a redefinition of $g_{\ell,\ell+1},~g_{\ell,\ell+2},~\cdots$.
In order to be consistent with the boundary condition of Eq.~\ref{eq:F and G at the origin}, the coefficient of $r^{-\ell}$ term can be determined as
\begin{align}
g_{\ell,-\ell} &= (2\ell-1)!!. \label{eq:g recurrence 4}
\end{align}

The Schr\"odinger equation Eq.~\ref{eq:schroedinger eq F} around $r=0$ tells us the recurrence relation for $g_{\ell,k}$ with $k \geq \ell+2$.
For $\ell = 0$,
\begin{align}
&v_{-1} g_{0,0} + x_0 f_{0,1} = 0, \label{eq:g recurrence 2}\\
&-k(k-1) g_{0,k} + \left(\sum_{m=-1}^{k-2} v_m g_{0,k-2-m} \right) - p^2 g_{0,k-2} + x_0 k f_{0,k} = 0 \qquad (k \geq 2), \label{eq:g recurrence 3}
\end{align}
For $\ell \geq 1$,
\begin{align}
&2 \ell g_{\ell,-\ell+1} + v_{-1} g_{\ell,-\ell} = 0, \label{eq:g recurrence 5}\\
&[-k(k-1)+\ell(\ell+1)] g_{\ell,k} + \left(\sum_{m=-1}^{k-2+\ell} v_m g_{\ell,k-2-m} \right) - p^2 g_{\ell,k-2} = 0 \qquad (-\ell+2 \leq k \leq \ell), \label{eq:g recurrence 6}\\
%&\left(\sum_{m=-1}^{2\ell-1} v_m g_{\ell,\ell-1-m} \right) - p^2 g_{\ell,\ell-1} + x_\ell (\ell+1) f_{\ell,\ell+1} = 0, \label{eq:g recurrence 7}\\
&\left(\sum_{m=-1}^{2\ell-1} v_m g_{\ell,\ell-1-m} \right) - p^2 g_{\ell,\ell-1} + x_\ell (2\ell+1) f_{\ell,\ell+1} = 0, \label{eq:g recurrence 7}\\
%&[-k(k-1)+\ell(\ell+1)] g_{\ell,k} + \left(\sum_{m=-1}^{k-2+\ell} v_m g_{\ell,k-2-m} \right) - p^2 g_{\ell,k-2} + x_\ell k f_{\ell,k} = 0 \qquad (k \geq \ell+2).\label{eq:g recurrence 8}
&[-k(k-1)+\ell(\ell+1)] g_{\ell,k} + \left(\sum_{m=-1}^{k-2+\ell} v_m g_{\ell,k-2-m} \right) - p^2 g_{\ell,k-2} + x_\ell (2k-1) f_{\ell,k} = 0 \qquad (k \geq \ell+2).\label{eq:g recurrence 8}
\end{align}
We can determine many of the coefficients by applying the recurrence relations in the following order: $g_{\ell,-\ell}, g_{\ell,-\ell+1}, \ldots, g_{\ell,\ell}, x_\ell, g_{\ell,\ell+2}, g_{\ell,\ell+3}, \ldots$.
On the other hand, we cannot determine $g_{\ell,\ell+1}$ from this procedure.
$g_{\ell,\ell+1}$ can be determined by 
imposing the boundary condition on $G_\ell$ at $r\to\infty$ given in Eq.~\ref{eq:F and G at large r}.

\subsection{Momentum dependence of $x_\ell(p)$} \label{sec:mom dep xp}
Let us discuss the momentum dependence of $x_\ell(p)$ in the expansion of $G_\ell(r)$ given in Eq.~\ref{eq:G expansion}.
This property is important for the renormalization discussion given in Sec.~\ref{sec:mom dep small k}.

For the $s$-wave case ($\ell = 0$),
Eqs.~\ref{eq:f recurrence 1}, \ref{eq:g recurrence 4}, \ref{eq:g recurrence 2} tell us that
\begin{align}
x_0 = -v_{-1}.
\end{align}
Thus, we can see $x_0$ is constant ($p$-independent) for the $s$-wave case. On the other hand, $x_\ell$ with $\ell \geq 1$ can be determined from Eqs.~\ref{eq:f recurrence 1}, \ref{eq:g recurrence 7} as:
\begin{align}
%x_\ell = -\frac{(2\ell+1)!!}{\ell(\ell+1)} \left[ \left(\sum_{m=-1}^{2\ell-1} v_m g_{\ell,\ell-1-m} \right) - p^2 g_{\ell,\ell-1}  \right]. \label{eq:xell formula}
x_\ell = -\frac{(2\ell+1)!!}{2\ell+1} \left[ \left(\sum_{m=-1}^{2\ell-1} v_m g_{\ell,\ell-1-m} \right) - p^2 g_{\ell,\ell-1}  \right]. \label{eq:xell formula}
\end{align}
Thus, for higher partial waves ($\ell \ge 1$) $x_\ell$ is momentum-dependent except in special cases.

As one example of such a special case, let us discuss a potential $V(r)$ which does not contain any odd powers of $r$, \textit{i.e.},
\begin{align}
v_{-1} = v_1 = v_3 = \cdots = 0.
\end{align}
This is the case when the potential $V(r)$ is an analytic function of $x$, $y$, and $z$ around the origin.
In this case, by using the recurrence relations, we can show that
\begin{align}
g_{\ell,-\ell+1} = g_{\ell,-\ell+3} = g_{\ell,-\ell+5} = \cdots = 0.
\end{align}
Then, we obtain
\begin{align}
x_\ell = 0.
\end{align}

\section{Variable phase method for irregular solutions}
\label{app:vpm}

It can be helpful to directly numerically compute the solutions to the multi-state Schr\"odinger equation with perturbed short-distance boundary conditions (corresponding to irregular solutions of the Schr\"odinger equation with the usual long-range potential). For such cases, we can adapt the variable phase method proposed in Ref.~\cite{PhysRevC.84.064308} as recast in Refs.~\cite{Beneke:2014gja, Asadi:2016ybp}. Following the notation of Ref.~\cite{Asadi:2016ybp}, we can decompose a reduced radial wavefunction $u(r)$ as:
\begin{equation} u_{n}(r) = \alpha_{n}(r) f_n(r) -\beta_{n}(r) g_n(r),\end{equation}
where $n$ indexes the component and there is no summation over $n$ on the right-hand side, and as previously (e.g. Eq.~\ref{eq:vpmdefs}), $f_n(r) = s_\ell(p_n r)/\sqrt{p_n}$ and $g_n(r) = \left[ c_\ell(p_n r) + i s_\ell(p_n r) \right]/\sqrt{p_n}$. In particular, at large $r$, $g_n(r) \propto e^{i p_n r}$, the purely outgoing mode, and $g_n(r)$ is irregular at the origin, while $f_n(r)$ is regular. We may add a second index to describe an ensemble of solutions, as in the discussion below Eq.~\ref{eq:multistatevpm}. For the Schr\"odinger equation with no potential, and an incoming wave in the $i$th state, $u_{ni}(r) = \delta_{ni} f_n(r)$ and $u_{ni}(r) = \delta_{ni} g_n(r)$ are independent solutions for each choice of $i$, and so their Wronskian is constant: by considering the large-$r$ regime, it then follows straightforwardly that $f_n^\prime g_n - f_n g_n^\prime = 1$ for each choice of $n$ (this can also be checked directly). 

As in Ref.~\cite{Asadi:2016ybp}, to account for the extra degrees of freedom induced by solving for two functions rather than one, we will impose the additional condition:
\begin{equation} \alpha_{a}^\prime(r) f_a(r) -\beta_{a}^\prime(r) g_a(r) = 0,\end{equation}
for each individual choice of $n$ and $i$.

The setup in Ref.~\cite{Asadi:2016ybp} is appropriate for determining the regular solution. Suppose we are instead interested in the irregular outgoing solutions, e.g. those labeled $\tilde{w}_{ni}(r)$ in Sec.~\ref{sec:multistate_decomposition}, with non-zero $\beta_n(r)$ at small $r$, and $\alpha_n(r) \rightarrow 0$ at large $r$. Let us write $\alpha_{a}(r) = O_{ab}(r) \beta_{b}(r)$; we will solve for $O_{ab}(r)$ and $\beta_a(r)$.

These functions can be obtained by solving the first-order differential equations:
\begin{align} O_{ab}^\prime(r) & = \sum_{c,d} (g_a\delta_{ac} - O_{ac} f_c) \hat{U}_{cd} (g_b \delta_{bd} - f_d O_{db}) \nonumber \\
\beta_a^\prime & = \sum_{b,c} f_a \hat{U}_{ac} (g_b \delta_{cb} - f_c O_{cb}) \beta_b, \label{eq:vpmdes} \end{align}
where $\hat{U}_{ac}$ can be read off from the potential matrix for the multi-state system.

To check this, note that if these equations are satisfied, then we obtain:
\begin{align} &\alpha_{a}^\prime(r) f_a(r) -\beta_{a}^\prime(r) g_a(r) \\
& = O_{ab}^\prime \beta_b f_a + O_{ab} \beta_b^\prime f_a - \beta_a^\prime g_a \nonumber \\
& = (g_a\delta_{ac} - O_{ac} f_c) \hat{U}_{cd} (g_b \delta_{bd} - f_d O_{db}) \beta_b f_a + f_a O_{ab} f_b \hat{U}_{bc} (g_d \delta_{cd} - f_c O_{cd}) \beta_d \nonumber \\
& - (f_a \hat{U}_{ac} g_c \beta_c - f_a \hat{U}_{ac} f_c O_{cb} \beta_b) g_a \nonumber \\
& =  g_a g_b \hat{U}_{ab} \beta_b f_a + O_{ac} f_c \hat{U}_{cd} f_d O_{db} \beta_b f_a - g_a \hat{U}_{ad} f_d O_{db} f_a \beta_b - O_{ac} f_c \hat{U}_{cb} g_b \beta_b f_a \nonumber \\
& + f_a O_{ab} f_b \hat{U}_{bc} g_c \beta_c - f_a O_{ab} f_b \hat{U}_{bc} f_c O_{cd} \beta_d - (f_a \hat{U}_{ac} g_c \beta_c - f_a \hat{U}_{ac} f_c O_{cb} \beta_b) g_a \nonumber \\
& = 0, \end{align}
as required. Furthermore,
\begin{align} u_n^{\prime \prime}(r) & = \left[\alpha_n(r) f_n^\prime(r) -\beta_{n}(r) g^\prime_n(r) \right]^\prime \nonumber \\
& = \alpha^\prime_n(r) f_n^\prime(r) -\beta^\prime_{n}(r) g^\prime_n(r) + \alpha_n(r) f_n^{\prime \prime}(r) -\beta_{n}(r) g^{\prime \prime}_n(r) \nonumber \\
& =  \beta^\prime_n(r) g_n(r) f_n^\prime(r)/f_n(r) -\beta^\prime_{n}(r) g^\prime_n(r) + \alpha_n(r) f_n^{\prime \prime}(r) -\beta_{n}(r) g^{\prime \prime}_n(r) \nonumber \\
& = \frac{\beta^\prime_n(r)}{f_n(r)} \left[ g_n(r) f_n^\prime(r) - g^\prime_n(r) f_n(r) \right] + \alpha_n(r) f_n^{\prime \prime}(r) -\beta_{n}(r) g^{\prime \prime}_n(r) \nonumber \\
& =  \hat{U}_{nc} (g_b(r) \delta_{cb} - f_c O_{cb}(r))\beta_b(r) + \alpha_n(r) f_n^{\prime \prime}(r) -\beta_{n}(r) g^{\prime \prime}_n(r)  \nonumber \\
& =  - \hat{U}_{nb} u_b(r) + \alpha_n(r) f_n^{\prime \prime}(r) -\beta_{n}(r) g^{\prime \prime}_n(r)  \end{align}

Since $f_n$ and $g_n$ are solutions to the Schr\"odinger equation with no potential and momentum $p_n$, we can write $f_n^{\prime \prime}(r) = (\ell(\ell+1)/r^2 -p_n^2) f_n(r)$, $g_n^{\prime \prime}(r) = (\ell(\ell+1)/r^2 -p_n^2) g_n(r)$, and thus we obtain:
\begin{align} u_n^{\prime \prime}(r)
& =  - \hat{U}_{nb} u_b(r) + (\ell(\ell+1)/r^2 -p_n^2) \left[ \alpha_n(r) f_n(r) -\beta_{n}(r) g_n(r) \right] \nonumber \\
& =  \left[\frac{\ell(\ell+1)}{r^2} \delta_{nb} - p_n^2 \delta_{nb} - \hat{U}_{nb}  \right] u_b(r) \end{align}
By comparison to Eq.~\ref{eq:SEmultistate}, we see that we should identify $\hat{U} = -2\mu \hat{V}_\text{long}$.

We will often be interested in purely outgoing solutions, in which case we will require $O_{ab} \rightarrow 0$ at large $r$.\footnote{If we wish to study irregular solutions that are not purely outgoing, we can add regular solutions with an ingoing component to match the desired large-$r$ boundary conditions; the variable phase method used in Refs.~\cite{Beneke:2014gja, Asadi:2016ybp} is sufficient to compute these solutions.} Note that the equation for $O_{ab}$ is nonlinear and in particular $O_{ab}=0$ is not a valid solution in the presence of a non-zero long-range potential, so imposing this boundary condition will not automatically lead to a trivial solution. Our first step in studying a purely outgoing solution is thus to solve the first line of Eq.~\ref{eq:vpmdes} for $O_{ab}(r)$ with the boundary condition $O_{ab}(r\rightarrow \infty)=0$.

In particular, consider the $\tilde{w}_\ell(r)$ solutions introduced in Eq.~\ref{eq:decomposition}, with components $\tilde{w}_{ni}$; here $n$ indexes the component and $i$ describes the small-$r$ boundary condition. As shown in Eq.~\ref{eq:smallralphabeta}, the small-$r$ boundary conditions for these solutions can be written in matrix form as $\beta_{\tilde{w}_\ell}(0) = \beta_{\tilde{G}_\ell}(0) + i \beta_{\tilde{F}_\ell}(0) = -\tilde{P}^{2\ell+1}$. This fully specifies the boundary conditions needed to solve for $\tilde{w}_\ell(r)$.

In order to compute the corrected cross sections in this work, for general choices of the convention for $V_\text{long}(r)$ at $r < a$, we need to compute the $\Sigma^a_{0,\ell}$ matrices derived from regular solutions to the Schr\"odinger equation, the standard uncorrected Sommerfeld enhancement $\Sigma_{0,\ell}$, and the matrix $\alpha_{\tilde{G}_\ell}(a)$. For regular solutions we employ the method described in Appendix A of Ref.~\cite{Asadi:2016ybp}. In the notation of that work, we solve for $\tilde{\alpha}_{ni}(r) \equiv \alpha_{ni}(r)/g_n(r)$. The prescription defined in the text for $\alpha_{b,\ell}(0)$ is to set boundary conditions of $\alpha_{b,\ell}(a) = \tilde{P}$, and then read off $\alpha_{b,\ell}(0)$ (and hence $\Sigma^a_{0,\ell}= \tilde{P} \alpha_{b,\ell}(0) \tilde{P}^{-2}$) for a regular solution. This amounts to setting boundary conditions on $\tilde{\alpha}_{ni}(a) \equiv \delta_{ni}\sqrt{p_n}/g_n(a)$, and then extracting $\tilde{\alpha}_{ni}(0) g_n(0)$.\footnote{It is also viable to set the boundary condition $\alpha(0) =\tilde{P}$ at $r=0$ and read off $\tilde{P}^{-1} \alpha(a)$, which is then related to $\tilde{P}^{-1} \alpha_{b,\ell}(0)$ (and hence $\Sigma^a_{0,\ell}$) by matrix inversion. This is somewhat more efficient if one wishes to scan over a range of values for the matching radius $a$.}

The methods of Appendix A of Ref.~\cite{Asadi:2016ybp} can likewise be applied to directly solve for the regular solution $w_\ell(r)$. Once (the $\alpha$ functions for) both $\tilde{w}_\ell(r)$ and $w_\ell(r)$ are in hand, the matrix $\alpha_{\tilde{G}_\ell}(a)$ satisfies:
\begin{align}\alpha_{\tilde{w}_\ell}(r) & = \alpha_{\tilde{G}_\ell}(r) + i \alpha_{w_\ell}(r) P^{-1}\bar{Q}^\dagger_\ell \nonumber \\
\Rightarrow   \alpha_{\tilde{G}_\ell}(a)  & = \alpha_{\tilde{w}_\ell}(a) - i \alpha_{w_\ell}(a) \Sigma_{0,\ell}^\dagger (\tilde{P}^\dagger)^{2\ell} \label{eq:alphaGextraction} \end{align}

The standard Sommerfeld factor $\Sigma_{0,\ell}$ can be extracted from the regular solution $w_\ell(r)$, computed as in Ref.~\cite{Asadi:2016ybp}, or from the irregular solution $\tilde{w}_\ell$ (which is equivalent to the widely-used approach of extracting the Sommerfeld enhancement from the behavior of the irregular solution at large $r$), by solving the second line of Eq.~\ref{eq:vpmdes} for $\beta_{\tilde{w}_\ell}(r)$. We observe that $- \lim_{r\rightarrow \infty} (\beta_{\tilde{w}_\ell}(r))_{ni} / \sqrt{p_n}$ is the coefficient of $c_\ell(p_n r) + i s_\ell(p_n r)$ for the $n$th component, and consequently $\lim_{r\rightarrow \infty} (\beta_{\tilde{w}_\ell}(r))_{nj}  = - \sqrt{p_n} p_j f_{1,nj}$ (Eq.~\ref{eq:f1def}). Since $\Sigma_{0,\ell} = \tilde{P}^2 Q_\ell P^{-2}$ and $f_{1,ij} = (p_j^\ell/p_i) Q_{ji}$ for $i \le M$ (as derived below Eq.~\ref{eq:f1wronskian}), we have $- \lim_{r\rightarrow \infty} (\beta_{\tilde{w}_\ell}(r))_{nj}  = (p_j^{\ell+1}/\sqrt{p_n}) Q_{jn}$, and $\Sigma_{0,ij} = p_i Q_{ij} / p_j$, so finally we obtain:
\begin{equation} \Sigma_{0,ij} = - \frac{1}{p_i^\ell \sqrt{p_j}}  \lim_{r\rightarrow \infty} (\beta_{\tilde{w}_\ell}(r))_{ji}, \quad \Sigma_{0,\ell} = - \lim_{r\rightarrow \infty} \left(  P^{-1}  \beta_{\tilde{w}_\ell}(r) \tilde{P}^{-2\ell} \right)^T ,\end{equation}
where in the final matrix expression we truncate $\beta_{\tilde{w}_\ell}(r)$ to $M\times N$, i.e. we only need to read off the components for the states that are kinematically accessible. 

We can use this construction to show that $O_{ab}(r)$ is always symmetric, provided the long-range potential matrix is symmetric and the boundary condition on $O_{ab}$ is symmetric. Let us consider the evolution equations for $O_{ab}(r)$ and $O_{ba}(r)$.  We have:
\begin{align} O_{ab}^\prime(r) & = \sum_{c,d} (g_a\delta_{ac} - O_{ac} f_c) \hat{U}_{cd} (g_b \delta_{bd} - f_d O_{db}) \nonumber \\
 O_{ba}^\prime(r) & = \sum_{c,d} (g_b\delta_{bc} - O_{bc} f_c) \hat{U}_{cd} (g_a \delta_{ad} - f_d O_{da}).  \end{align}
Relabeling dummy variables ($c \leftrightarrow d$) in the second equation,  and using $\hat{U}_{cd} = \hat{U}_{dc}$ by symmetry of the potential, we have:
\begin{align}  O_{ba}^\prime(r) & = \sum_{c,d}(g_a \delta_{ac} - O_{ca} f_c )  \hat{U}_{cd}  (g_b\delta_{bd} -  f_d O_{bd}).  \end{align}
We observe that if $O=O^T$ at a point (so  $O_{ca}=O_{ac}$ and $O_{bd}=O_{db}$), then the derivatives of $O_{ba}$ and $O_{ab}$ will also coincide at that point, and thus the evolution equation will preserve the relation $O=O^T$. Thus provided we have a symmetric initial/boundary condition, $O$ will remain symmetric for all $r$.  

Now for $r\rightarrow 0$, for the $\tilde{w}_{ni}$ solutions, $p_a^{\ell + 1/2} \alpha_{ai} = p_a^{\ell + 1/2} O_{ab} \beta_{bi} \rightarrow - O_{ab} p_a^{\ell + 1/2} p_b^{\ell + 1/2} \delta_{bi} = - O_{ai} p_a^{\ell + 1/2} p_i^{\ell + 1/2}$, which is invariant under $a\leftrightarrow i$. Thus $\tilde{P}^{2\ell+1} \alpha$ is symmetric for the $\tilde{G}_\ell(r)$ and $\tilde{F}_\ell(r)$ solutions (corresponding to the real and imaginary parts of $\tilde{w}_\ell$).

Finally, we note that given two solutions $u$ and $v$ with corresponding $\alpha$ and $\beta$ functions $\alpha_u, \alpha_v, \beta_u$ and $\beta_v$, we can compute the Wronskian as (again assuming a real symmetric potential):
\begin{align} (u^\prime)^T v - u^T v^\prime & = \sum_a \left[ (\alpha_{u a}(r) f_a^\prime(r) -\beta_{u a}(r) g_a^\prime(r)) (\alpha_{va}(r) f_a(r) - \beta_{va}(r) g_a(r)) \right. \nonumber \\
& \left. - (\alpha_{u a}(r) f_a(r) -\beta_{u a}(r) g_a(r)) (\alpha_{va}(r) f_a^\prime(r) - \beta_{va}(r) g^\prime_a(r))\right] \nonumber \\
& = \sum_a (\alpha_{ua}(r) \beta_{va}(r) - \beta_{ua}(r) \alpha_{va}(r)) (- f_a^\prime(r) g_a(r) + f_a(r) g^\prime_a(r)), \nonumber \\
& = - (\alpha_u^T \beta_v - \beta_u^T \alpha_v). \label{eq:vpmwronskian} \end{align}

\section{Comparison with Flores-Petraki wavefunction}\label{sec:comparison with 240502222}
\begin{table}[ht]
	\centering
	\begin{tabular}{c||c|c}
		& Flores-Petraki \cite{Flores:2024sfy} & This paper \\\hline\hline
		Wave function & $u_{p,\ell}(r)$ & $i^\ell u_\ell(r)$\\
		Long-range potential & $V(r)$ & $V_{\rm long}(r)$ \\
		Phase-shift by long-range potential & $\theta_\ell(p)$ & $\delta_\ell^{(L)}$ \\
		Regular wave function & ${\cal F}_{p,\ell}(r)$ & $i^\ell \exp\left( i \delta_\ell^{(L)} \right) F_\ell(r)$ \\
		Irregular wave function & ${\cal G}_{p,\ell}(r)$ & $-i^\ell \exp\left( i \delta_\ell^{(L)} \right) G_\ell(r)$ \\
		Out-going wave function & ${\cal H}_{p,\ell}^{(+)}(r)$ & $-i^{\ell+1} \exp\left( i \delta_\ell^{(L)} \right) [G_\ell(r)+iF_\ell(r)]$ \\
	\end{tabular}
	\caption{Notation in Flores-Petraki \cite{Flores:2024sfy}. The rightmost column shows the corresponding expression in the notation of this paper.} \label{tab:notation in FP}
\end{table}

\subsection{Comparison to results of Section~\ref{sec:singlestate}}

In this Appendix, we show a detailed comparison with our wavefunction $u_\ell(r)$ given in Eq.~\ref{eq:matching} and the wavefunction $u_{p,\ell}(r)$ given in Eq.~38 in Ref.~\cite{Flores:2024sfy}. The translation between the notation in Ref.~\cite{Flores:2024sfy} and this paper is summarized in Tab.~\ref{tab:notation in FP}.
To describe annihilation, Ref.~\cite{Flores:2024sfy} takes an imaginary non-local term in the potential; ${\cal V_\ell}(r,r') = - i\sum_j v_\ell^{j*}(r) v_\ell^j(r')$, where $v_\ell^j(r)$ is defined in Eq.~24 in Ref.~\cite{Flores:2024sfy} as
\begin{align}
	v_\ell^j(r) = \sqrt{ \frac{2^{1+\delta}}{\pi^2\mu} } \int_0^\infty dp p^{3/2} j_\ell(pr) a_\ell^j(s,p).
\end{align}
Here $a_\ell^j(s,p)$ is the annihilation amplitude. Since $a_\ell^j(s,p)$ scales like $p^\ell$ in the small-$p$ limit, we can show that $v_\ell^j(r) = \tilde v_\ell^j r^{-\ell-3/2}$ where $\tilde v_\ell^j$ is a $r$-independent constant. Because of this universal $r$-dependence in the small-momentum limit, we can write ${\cal V_\ell}(r,r') = -i v_\ell(r) v_\ell(r')$ where $v_\ell(r) \equiv \sqrt{\sum_j |\tilde v_\ell^j|^2} \times r^{-\ell-3/2}$. Alternatively, we can reach the same expression without knowing the momentum-dependence of the amplitudes by focusing on the case where only one annihilation channel is available, for ease of comparison to our calculation where we do not distinguish final states. The Schr\"odinger equation in Ref.~\cite{Flores:2024sfy} is written as
\begin{align}
	\left[ -\frac{1}{2\mu}\frac{d^2}{dr^2} + \frac{\ell(\ell+1)}{r^2} + V(r) - \frac{p^2}{2\mu}\right] u_{p,\ell}(r)
	- i r v_\ell(r) \int_0^\infty dr' r' v_\ell(r') u_{p,\ell}(r') = 0.
\end{align}
Since $v_\ell(r)$ drops rapidly for large $r$, we treat $v_\ell(r)\simeq 0$ for $r$ greater than a threshold distance $a$. Then, by using Eqs.~35b, 37, 38, 50 in Ref.~\cite{Flores:2024sfy}, the wavefunction $u_{p,\ell}(r)$ for $r>a$ in Ref.~\cite{Flores:2024sfy} can be written as
\begin{align}
    u_{p,\ell}(r)
    =
    {\cal F}_{p,\ell}(r)
    + \frac{i|\int_0^a r dr {\cal F}_{p,\ell}(r) v_\ell(r) |^2}{1 - i\int_0^a dr r \int_0^a dr' r' v_\ell(r) G_{p,\ell}(r,r') v_\ell(r')}  \frac{2\mu i}{2^\delta p}{\cal H}_{p,\ell}^{(+)}(r) \label{eq:u FP}
\end{align}
To compare this $u_{p,\ell}(r)$ and our $u_\ell(r)$,
let us define the $S$-matrix $S_\ell^{\rm (FP)}$, and the phase-shift from short-range physics, $\delta_\ell^{(S,{\rm FP})}$, using Eq.~\ref{eq:u FP}.
By using Eqs.~33a, 48a in Ref.~\cite{Flores:2024sfy}, we obtain the asymptotic behavior of $u_{p,\ell}(r)$ at $r\to\infty$ as
\begin{align}
u_{p,\ell}(r) \to \frac{1}{2i}\left( S_{\ell}^{\rm (FP)} e^{ipr} - (-1)^\ell e^{-ipr} \right)
\end{align}
where $S_\ell^{\rm (FP)}$ is defined as
\begin{align}
	S_\ell^{\rm (FP)} \equiv e^{2i\theta_\ell(p)} \left[ 1 + \frac{2i|\int_0^a r dr {\cal F}_{p,\ell}(r) v_\ell(r) |^2}{1 - i\int_0^a dr r \int_0^a dr' r' v_\ell(r) G_{p,\ell}(r,r') v_\ell(r')}  \frac{2\mu i}{2^\delta p} \right]. \label{eq:S FP}
\end{align}
$\delta_\ell^{(S)}$ is defined from $S_\ell^{\rm (FP)} = \exp\left( 2i\theta_\ell(p) + 2i\delta_\ell^{(S,{\rm FP})} \right)$. From Eq.~\ref{eq:S FP}, we obtain
\begin{align}
	\exp\left( 2i\delta_\ell^{(S,{\rm FP})} \right)
	=
	\left[ 1 + \frac{2i|\int_0^a r dr {\cal F}_{p,\ell}(r) v_\ell(r) |^2}{1 - i\int_0^a dr r \int_0^a dr' r' v_\ell(r) G_{p,\ell}(r,r') v_\ell(r')}  \frac{2\mu i}{2^\delta p} \right] \label{eq:exp 2ideltaS from FP}
\end{align}
By using Eq.~\ref{eq:exp 2ideltaS from FP}, we can write $\cot\delta_\ell^{(S,{\rm FP})}$ as
\begin{align}
    \cot\delta_\ell^{(S,{\rm FP})} %&= -i \frac{ \exp\left( 2i\delta_\ell^{(S,{\rm FP})} \right) + 1  }{ \exp\left( 2i\delta_\ell^{(S,{\rm FP})} \right) - 1 }\nonumber\\
% \biggr|_{\cite{Parikh:2024mwa}}
%    &=
%    i \left[ 1 + \left( \frac{i|\int_0^a r{\cal F}_{k,\ell}(r) v_\ell(r) |^2}{1 - i\int_0^a dr r \int_0^a dr' r' v_\ell(r) G_{k,\ell}(r,r') v_\ell^*(r')}  \frac{2\mu i}{2^\delta k} \right)^{-1} \right] \biggr|_{\cite{Flores:2024sfy}} \nonumber\\
%    &=
%    i \left[ 1 + \left( \frac{i|\int_0^a r{\cal F}_{k,\ell}(r) v_\ell(r) |^2}{2^\delta k/2\mu i - i\int_0^a dr r \int_0^a dr' r' v_\ell(r) {\cal F}_{k,\ell}^*(r_<) {\cal H}_{k,\ell}^{(+)}(r_>) v_\ell^*(r')}   \right)^{-1} \right] \biggr|_{\cite{Flores:2024sfy}} \nonumber\\
%    &=
%    i \left[ 1 + \frac{2^\delta k/2\mu i - i\int_0^a dr r \int_0^a dr' r' v_\ell(r) {\cal F}_{k,\ell}^*(r_<) {\cal H}_{k,\ell}^{(+)}(r_>) v_\ell^*(r')}{i|\int_0^a r{\cal F}_{k,\ell}(r) v_\ell(r) |^2} \right] \biggr|_{\cite{Flores:2024sfy}} \nonumber\\
    &=
	\frac{1}{|\int_0^a r dr {\cal F}_{p,\ell}(r) v_\ell(r) |^2}
    \left[ \frac{2^\delta p}{2\mu i} - \int_0^a dr r \int_0^a dr' r' v_\ell(r) {\cal F}_{p,\ell}^*(r_<) {\cal G}_{p,\ell}(r_>) v_\ell(r')\right]. %\biggr|_{\cite{Flores:2024sfy}}
	\label{eq:cotdelta}
\end{align}
Here we have used Eq.~50 in Ref.~\cite{Flores:2024sfy} as
\begin{align}
	G_{p,\ell}(r,r') &= \frac{2\mu i}{2^\delta p} {\cal F}_{p,\ell}^*(r_<) {\cal H}_{p,\ell}^{(+)}(r_>) \nonumber\\
	&= \frac{2\mu i}{2^\delta p} {\cal F}_{p,\ell}^*(r_<) {\cal F}_{p,\ell}(r_>) - \frac{2\mu}{2^\delta p} {\cal F}_{p,\ell}^*(r_<) {\cal G}_{p,\ell}(r_>) \label{eq:Gpl}
\end{align}
where $r_< \equiv {\rm min}\{r,r'\}$ and $r_> \equiv {\rm max}\{r,r'\}$.

Let us assume $p \ll 1/a$. In this case, we can take the leading terms of ${\cal F}_{p,\ell}(r)$ and ${\cal G}_{p,\ell}(r)$ in the small-$r$ expansion. By using Eq.~\ref{eq:F at r=a}, \ref{eq:G at r=a} and Tab.~\ref{tab:notation in FP}, ${\cal F}_{p,\ell}(r)$ and ${\cal G}_{p,\ell}(r)$ at short-distance can be evaluated as
\begin{align}
	(-i)^\ell e^{-i\theta_\ell(p)}{\cal F}_{p,\ell}(r) &\simeq \frac{C_\ell}{(2\ell+1)!!} p^{\ell+1} r^{\ell+1}, \label{eq:FFP at r=a}\\
	-i^\ell e^{-i\theta_\ell(p)}{\cal G}_{p,\ell}(r) &\simeq \frac{(2\ell-1)!!}{C_\ell} p^{-\ell} r^{-\ell} + \frac{1}{(2\ell+1)!!} \frac{1}{C_\ell} z_\ell(p) p^{-\ell} r^{\ell+1}. \label{eq:GFP at r=a}
\end{align}
By using Eqs.~\ref{eq:cotdelta}, \ref{eq:FFP at r=a}, \ref{eq:GFP at r=a}, we can show that
\begin{align}
    & p^{2\ell+1} C_\ell^2 \cot\delta_\ell^{(S,{\rm FP})} - z_\ell(k) \nonumber\\
    \simeq&
    -\frac{[(2\ell+1)!!]^2}{|\int_0^a dr r^{\ell+2}v_\ell(r) |^2} \Biggl[ \frac{2^\delta}{2\mu i}  - \int_0^a dr \int_0^a dr' rr' \frac{(2\ell-1)!!}{(2\ell+1)!!} r_<^{\ell+1} r_>^{-\ell} v_\ell(r) v_\ell(r')  \Biggr].
\end{align}
We can see the RHS is independent on momentum $p$.
This is consistent with the behavior of $k_\ell(p)$ defined in Eq.~\ref{eq:kappa_def} which is discussed in Sec.~\ref{eq:mom dependence kappa}.
Thus, we can see the consistency between our wavefunction Eq.~\ref{eq:matching} and the wavefunction $u_{p,\ell}(r)$ which is given in Eq.~38 in Ref.~\cite{Flores:2024sfy}.

\subsection{Comparison to results of Section~\ref{sec:multistate}}

Whereas in the previous subsection we were primarily interested in momentum dependence, reflecting our treatment in Sec.~\ref{sec:singlestate}, we can also compare the results of Ref.~\cite{Flores:2024sfy} to our expressions written in terms of the short-distance scattering amplitude as in Sec.~\ref{sec:multistate} (while still restricting ourselves to the single-state case).

We can anticipate that the comparison will in general be non-trivial because $f_{s,\ell}$ (as defined in Eq.~\ref{eq:multistatematching}) and $a_\ell(s,p)$ (as defined in Ref.~\cite{Flores:2024sfy}) do not encode exactly the same information; $a_\ell(s,p)$ is defined diagramatically, while $f_{s,\ell}$ captures all interactions occurring at $r < a$. In particular, as discussed in the main text, $f_{s,\ell}$ will generally include diagrams that correspond to evolution of the wavefunction in the potential for $r < a$, leading to elastic scattering and Sommerfeld enhancement of annihilation. We showed in Sec.~\ref{sec:multistate} (with further discussion in Apps.~\ref{app:optical},\ref{app:shortrangeinterp}) that where this short-range potential evolution is already captured by $V_\text{long}(r)$, the effective short-range amplitude $\bar{f}_{s,\ell}$ appearing in the $S$-matrix (as defined in Eq.~\ref{eq:multistatekappa}) subtracts the potential-only term and factors out the short-range Sommerfeld enhancement. Thus we might hope that (in this convention) $\bar{f}_{s,\ell}$ would be directly analogous to $a_\ell(s,p)$. However, this identification of  $\bar{f}_{s,\ell}$ with the non-potential part of the short-range physics is only true at leading order (in $\bar{f}_{s,\ell}$), and we have only proved it for the case where the non-potential interactions are modeled as contact terms; we can generally expect higher-order terms and non-contact interactions (not captured in $V_\text{long}(r)$) to generate a residual $a$-dependence in $\bar{f}_{s,\ell}$. Thus we only expect it to be true that $\bar{f}_{s,\ell}$ and $a_\ell(s,p)$ encode the same information (about the size and momentum dependence of the annihilation amplitude) to leading order, in the case where the potential corrections (for evolution at $r < a$) are small.

Based on this intuition, let us suppose that the following conditions hold, at least approximately:
\begin{itemize}
\item As in the previous subsection, there is a characteristic radius $a$ such that we can assume $v_\ell(r) = 0$ for $r > a$.
\item This value of $a$ can be taken sufficiently small that for $r < a$, the evolution under the potential is negligible for $r < a$, and consequently we can approximate the independent solutions of the Schr\"odinger equation (with the long-range potential) in this region as free-particle solutions (weighted by appropriate normalization factors if they correspond to fixed boundary conditions as $r\rightarrow \infty$; these factors will be set by the non-trivial evolution from $r=a\rightarrow \infty$).
\end{itemize}
We will now show that when these conditions are valid, there is a simple and intuitive relationship between $f_{s,\ell}$ and $a_\ell(s,p)$, under which the prescription for the $S$-matrix from our work agrees with that from Ref.~\cite{Flores:2024sfy} (we expect that the two treatments should also agree under broader validity conditions if care is taken in the mapping from $f_{s,\ell}$ to $a_\ell(s,p)$; our claim is only that these conditions are sufficient to see equivalence and lead to a simple relationship between $f_{s,\ell}$ and $a_\ell(s,p)$).

The first condition is one we have assumed throughout this work; the second is stronger than the assumptions we have made previously, and implies that we can make the approximation $V_\text{long}(r)=0$ for $r < a$, which means setting $\Sigma^a_{0,\ell}=1$ in the notation of Sec.~\ref{sec:multistate}, $f_{s,\ell}=\bar{f}_{s,\ell}$, and $\alpha_{\tilde{G}_\ell}(a)=\alpha_{\tilde{G}_\ell}(0)$. Then the $S$-matrix derived from Eq.~\ref{eq:multistateSmatrix} can be written (in the single-state case) as:
\begin{align} S_\ell = S_{0,\ell} \left( 1 + \frac{2i p C_\ell^2}{\kappa_\ell^{-1} - i p C_\ell^2}\right), \quad \kappa_\ell = f_{s,\ell}^{-1} - \alpha_{\tilde{G}_\ell}(0) p^{1/2 - \ell}.\label{eq:SmatrixFPcompare} \end{align}
With this assumption, our different conventions for $V_\text{long}(r)$ for $r < a$ are equivalent. This assumption has as a corollary that $f_{s,\ell}$ can be accurately computed perturbatively without accounting for potential effects, but note that the converse is not true. In particular, in the wino example of Sec.~\ref{sec:wino} and App.~\ref{app:winoextras}, we show that our perturbative matching procedure gives cross-section predictions with only very small residual $a$-dependence and convention-dependence, but intermediate quantities such as $\alpha_{\tilde{G}_\ell}(a)$ can depend substantially on $a$. Nonetheless, we will proceed with this assumption (negligible evolution under the potential for $r < a$) for the purpose of this appendix. 

The combination of the two conditions means that we can make the following replacements, corresponding to the Ref.~\cite{Flores:2024sfy} expression for the $S$-matrix:
\begin{align} {\cal F}_{p,\ell}(r) v_\ell(r) & \rightarrow v_\ell(r) \left[ \alpha_{\mathcal{F}_\ell}(p) s_\ell(p r) - \beta_{\mathcal{F}_\ell}(p) (c_\ell(p r) + i s_\ell(p r))\right]/\sqrt{p}, \nonumber \\
 {\cal G}_{p,\ell}(r) v_\ell(r) & \rightarrow v_\ell(r) \left[ \alpha_{\mathcal{G}_\ell}(p) s_\ell(p r) - \beta_{\mathcal{G}_\ell}(p) (c_\ell(p r) + i s_\ell(p r))\right]/\sqrt{p}, \end{align}
 where $\alpha_{\mathcal{F}_\ell}(p)$, $\beta_{\mathcal{F}_\ell}(p)$, $\alpha_{\mathcal{G}_\ell}(p)$, $\beta_{\mathcal{G}_\ell}(p)$ are the coefficients appearing in the variable phase method decomposition of ${\cal F}_{p,\ell}(r)$ and ${\cal G}_{p,\ell}(r)$ (similar equations also hold if we replace $v_\ell(r)$ with $v_\ell^*(r)$). The second condition ensures these coefficients can be treated as $r$-independent for $r < a$. The first condition means that for $r > a$, the $v_\ell(r)$ factor ensures both the left- and right-hand sides of the expressions above are zero. Furthermore, the boundary conditions on ${\cal F}_{p,\ell}(r)$ impose that $\beta_{\mathcal{F}_\ell}(p)=0$.

Now in Eq.~\ref{eq:S FP}, the $S$-matrix from $V_\text{long}(r)$ alone is $S_{0,\ell}=e^{2i \theta_\ell(p)}$ by definition, and note that any integral including a $v_\ell(r)$ factor can be extended from $r=a$ to $r=\infty$ without changing the answer (by the first assumption). Then we obtain: 
\begin{align}
	S_\ell^{\rm (FP)} & = S_{0,\ell} \left[ 1 + 2i  \frac{2\mu i}{2^\delta p} p |\alpha_{\mathcal{F}_\ell}(p)|^2 \left| \int_0^\infty dr r^2 j_\ell(p r) v_\ell(r) \right|^2 \times \biggl\{ 1 - i  \frac{2\mu i}{2^\delta p}  \sqrt{p} \alpha_{\mathcal{F}_\ell}(p)^*  \right. \nonumber \\
	& \left.  \left. \times \int_0^\infty dr r \int_0^\infty dr' r'  r_< j_\ell(p r_<) \sqrt{p} r_> v_\ell(r) v_\ell^*(r') \Bigl[ \left( \alpha_{\mathcal{F}_\ell}(p) + i \alpha_{\mathcal{G}_\ell}(p)\right)  j_\ell(p r_>)  \right. \right. \nonumber \\
	& \left.  + i \beta_{\mathcal{G}_\ell}(p) \left( y_\ell(p r_>) - i j_\ell(p r_>) \right)  \Bigr]   \biggr\}^{-1} \right], \end{align}
	where we have used the definitions $s_\ell(x) = x j_\ell(x)$, $c_\ell(x) = - x y_\ell(x)$, and also Eq.~\ref{eq:Gpl}. We see that the only integrals we need to evaluate are:
	\begin{align} (1): & \int^\infty_0 r dr \int^\infty_0 r^\prime dr^\prime r_< r_> j_\ell(p r_<) j_\ell(p r_>)v_\ell(r) v_\ell^*(r^\prime) \nonumber \\
	& = \left| \int^\infty_0 r^2 dr j_\ell(p r) v_\ell(r) \right|^2 \nonumber \\
	& = \frac{2^{1+\delta}}{\pi^2\mu} \left|\int^\infty_0 dk k^{3/2} a_\ell(s,k) \int^\infty_0 r^2 dr j_\ell(k r) j_\ell(p r) \right|^2 \nonumber \\
	& = \frac{2^{-1+\delta}}{\mu p} |a_\ell(s,p)|^2,\end{align}
	where we have used $r r^\prime = r_> r_<$ and $j_\ell(r) j_\ell(r^\prime) = j_\ell(r_>) j_\ell(r_<)$, Eq.~24 of Ref.~\cite{Flores:2024sfy}, and the standard orthogonality relationship $\int^\infty_0 r^2 dr j_\ell(k r) j_\ell(p r) = \frac{\pi}{2} \frac{\delta(p - k)}{p^2}$, and:
	\begin{align} (2): & \int^\infty_0 r dr \int^\infty_0 r^\prime dr^\prime r_< r_> j_\ell(p r_<) [ y_\ell(p r_>) - i j_\ell(p r_>) ] v_\ell(r) v_\ell^*(r^\prime).\end{align}
To perform the second integral, we can equate the two forms for the Green's function $G_{k,\ell}(r,r^\prime)$ in Ref.~\cite{Flores:2024sfy}, corresponding to Eqs.~45 and 50 in that work, and set the potential to zero, so ${\cal F}_{p,\ell}(r)$ and ${\cal G}_{p,\ell}(r)$ reduce to free-particle solutions for all $r$. Specifically, we have $\mathcal{F}_{p,\ell}(r) = i^\ell s_\ell(p r) = i^\ell p r j_\ell(p r)$, $\mathcal{H}_{p,\ell}^{(+)}(r) = -i^{\ell+1} [ c_\ell(p r) + i s_\ell(p r) ] = i^\ell (p r) [ j_\ell(p r) + i y_\ell(p r) ]$. Then we obtain:
\begin{align} j_\ell(p r_<) [ -i j_\ell(p r_>) + y_\ell(p r_>) ] & = - i \mathcal{F}_{p,\ell}^*(r_<) \mathcal{H}^{(+)}_{p,\ell}(r_>)  = \frac{-2}{p \pi} \int^\infty_{0} dq q^2 \frac{j_\ell(q r) j_\ell(q r^\prime)}{q^2 - p^2 - i \epsilon}. \end{align}
Inserting this result into the integral above yields:
\begin{align} (2): & \int^\infty_0 r dr \int^\infty_0 r^\prime dr^\prime r_< r_> j_\ell(p r_<) [ y_\ell(p r_>) - i j_\ell(p r_>) ] v_\ell(r) v_\ell^*(r^\prime) \nonumber \\
& =  \frac{-2}{p \pi}  \int^\infty_{0} \frac{q^2 dq }{q^2 - p^2 - i \epsilon} \left| \int^\infty_0 r^2 dr  j_\ell(q r) v_\ell(r)\right|^2 \nonumber \\
& =\frac{-2^{\delta}}{\mu p \pi}  \int^\infty_{0} \frac{q dq }{q^2 - p^2 - i \epsilon}  |a_\ell(s,q)|^2, \end{align}
where in the last line we have used integral (1) above.

Now that we have evaluated the integrals, we can write the $S$-matrix element as:
\begin{align}
	S_\ell^{\rm (FP)} & = S_{0,\ell} \left[ 1 -  \frac{2 }{ p} |\alpha_{\mathcal{F}_\ell}(p)|^2  |a_\ell(s,p)|^2 \times \left( 1 - i  \frac{2\mu i}{2^\delta } \alpha_{\mathcal{F}_\ell}(p)^* \right.\right.  \nonumber \\
	& \left.  \left. \left[ (\alpha_{\mathcal{F}_\ell}(p) + i \alpha_{\mathcal{G}_\ell}(p)) \frac{2^{-1+\delta}}{\mu p} |a_\ell(s,p)|^2
	-  \frac{2^{\delta} i }{\mu p \pi} \beta_{\mathcal{G}_\ell}(p)   \int^\infty_{0} \frac{q dq }{q^2 - p^2 - i \epsilon}  |a_\ell(s,q)|^2 \right]  \right)^{-1} \right], \nonumber \\
	& = S_{0,\ell} \left[ 1 - \frac{2 |\alpha_{\mathcal{F}_\ell}(p)|^2  |a_\ell(s,p)|^2}{ p + \alpha_{\mathcal{F}_\ell}(p)^* \left[ (\alpha_{\mathcal{F}_\ell}(p) + i \alpha_{\mathcal{G}_\ell}(p))  |a_\ell(s,p)|^2
	-  \frac{2  i }{ \pi} \beta_{\mathcal{G}_\ell}(p)   \int^\infty_{0} \frac{q dq }{q^2 - p^2 - i \epsilon}  |a_\ell(s,q)|^2 \right] } \right] \end{align}
Now the dictionary of Tab.~\ref{tab:notation in FP}, the boundary conditions $\alpha_{F_\ell}(p) = \sqrt{p} C_\ell$, $\beta_{G_\ell}(p) =  -\sqrt{p}/C_\ell$, and the relation $\alpha_{G_\ell}(p) = \alpha_{\tilde{G}_\ell}(p)/(p^\ell C_\ell)$, allow us to write this expression in the form:
\begin{align}  S_\ell^{\rm (FP)} & = S_{0,\ell} \left[ 1 +   \frac{2 i p C_\ell^2 }{ (i |a_\ell(s,p)|^2/p)^{-1} - \frac{2 p}{ \pi}   \frac{1}{ |a_\ell(s,p)|^2}  \int^\infty_{0} \frac{q dq }{q^2 - p^2 - i \epsilon}  |a_\ell(s,q)|^2  -  p^{1/2 - \ell} \alpha_{\tilde{G}_\ell}(p) - i p C_\ell^2} \right].   \end{align}
We see this is exactly the form given in Eq.~\ref{eq:SmatrixFPcompare}, if we can identify:
\begin{equation} f_{s,\ell}^{-1} = (i |a_\ell(s,p)|^2/p)^{-1} - \frac{2 p}{ \pi}   \frac{1}{ |a_\ell(s,p)|^2}  \int^\infty_{0} \frac{q dq }{q^2 - p^2 - i \epsilon}  |a_\ell(s,q)|^2.\end{equation}
As desired, this expression is determined purely by the short-range absorptive physics, not by $V_\text{long}(r)$, and at leading order we obtain $p f_{s,\ell} \approx i |a_\ell(s,p)|^2$; i.e. the short-range amplitude is purely imaginary (consistent with assumptions in Ref.~\cite{Flores:2024sfy}), with magnitude determined by the annihilation rate, as required by the optical theorem.

\section{Calculating the short-distance amplitude $f_{s,\ell}$}
\label{app:optical}

\subsection{Normalization convention, optical theorem, and partial waves}

Since the $f_{s,\ell}$ contribution to $\kappa_\ell$ (see Eq.~\ref{eq:multistateSmatrix}) describes only short-distance physics, it can be calculated using the Born approximation or perturbative QFT (assuming a weakly-coupled theory). For contributions that can be captured by a non-relativistic potential, the Born approximation will often be the simplest approach, but for more general interactions, a field-theoretic approach may be necessary. For convenience, let us review the translation from $f_{s,\ell}$ to the matrix element $i\mathcal{M}$ of QFT; here we follow the notation and conventions of Ref.~\cite{Peskin:1995ev}.

The first Born approximation for scattering of two distinguishable particles in quantum mechanics can be written in the form:
\begin{equation} f_s \approx -\frac{2\mu}{4\pi} \int d^3\vec{r} e^{i\vec{q}\cdot \vec{r}} V(\vec{r}) = -\frac{\mu}{2\pi} \tilde{V}(\vec{q}),\end{equation}
where $\tilde{V}(\vec{q})$ denotes the Fourier transform of the potential, and $\mu$ is the reduced mass. Working at the same (leading) order, Ref.~\cite{Peskin:1995ev} relates the potential to the (relativistically normalized) matrix element via $\frac{1}{(4\mu)^2} (i \mathcal{M}) = -i \tilde{V}(\vec{q})$, where the factors of $4\mu$ arise from the difference between relativistic and non-relativistic normalizations of the states (and we assume here $2\rightarrow 2$ scattering where all the particle masses are equal at leading order). Thus we identify $f_s = \frac{1}{32 \pi \mu} \mathcal{M}$ for distinguishable particles (at least at leading order). For $2\rightarrow 2$ scattering, we can extract the individual partial waves by tracking the dependence of $\mathcal{M}$ on the angle $\theta$ between the incoming and outgoing momenta, and computing:
\begin{align} (f_{s,\ell})_{fi} & = \frac{1}{64 \pi \mu}  \int^1_{-1} d\cos\theta P_{\ell}(\cos\theta) \mathcal{M}(i\rightarrow f)(\theta). \end{align}

Where identical particles are involved, this relationship should be corrected to $f_s = \frac{1}{32 \pi \mu} \mathcal{M}/\sqrt{2^{\delta_a} 2^{\delta_b}}$, where $a$ and $b$ label the initial and final states respectively, and $\delta_X = 1$ for identical particles and 0 for distinguishable particles; this corresponds to working with the ``rescaled partial-wave amplitudes'' discussed in Ref.~\cite{Flores:2024sfy}. In the multi-state case, this agrees with the standard rescaling factors for the ``method-2'' annihilation matrix discussed in Ref.~\cite{Beneke:2014gja}. Here $f_s$ remains the scattering amplitude calculated via quantum mechanics for the case of {\it distinguishable} particles (which is the quantity we generally work with in the main text), but $\mathcal{M}$ is the QFT amplitude calculated using all standard rules. 

In the case where both initial and final states contain identical particles, the factor of 2 (relative to the quantum mechanics calculation with distinguishable particles) is due to the inclusion of ``crossed ladder'' diagrams (as well as ladder diagrams) in the QFT amplitude. In the case where only one of the states contains distinguishable particles, the factor of $\sqrt{2}$ in the QFT amplitude can be viewed as the combined effect of including the crossed diagrams (factor of 2), and the fact that our ``method-2'' potential for the quantum mechanics calculation (e.g.~Eq.~\ref{eq:winopot}) already includes a factor of $\sqrt{2}$ for this transition, relative to what one would expect from the tree-level diagram with all distinguishable particles (factor of $1/\sqrt{2}$). In the case where both incoming and outgoing states consist of distinguishable particles, there is no extra rescaling and the standard conversion factor above applies.

As usual, for the bare annihilation rate, we can further simplify the calculation using the optical theorem. As per Ref.~\cite{Flores:2024sfy}, the optical theorem takes its usual simple form when applied to the rescaled amplitudes that account for the identical-particle factors given above, so we will assume below that the matrix elements $\mathcal{M}$ have been rescaled as necessary (and so can be converted to our quantum-mechanical $f_s$ amplitudes for distinguishable particles by the conversion factor $32\pi \mu$). 

The standard optical theorem result (for real momenta) is:
\begin{equation} -i \left[ \mathcal{M}(a\rightarrow b) - \mathcal{M}^*(b\rightarrow a) \right] = \sum_X \int d\Pi_X \mathcal{M}^*(b\rightarrow X) \mathcal{M}(a\rightarrow X).\end{equation}
Here $a$ and $b$ denote the non-relativistic initial and final states coupled by the potential, and $X$ denotes all possible intermediate states. We can apply this result to each partial wave separately, by replacing $\mathcal{M}(a \rightarrow X) = \sum_\ell P_\ell(\cos\theta) (2\ell+1) \mathcal{M}_\ell(a \rightarrow X)$; the double sum on the RHS will collapse to a single sum upon integration of the $P_\ell(\cos\theta)$ terms over the intermediate-state phase space. 

In more detail, given an intermediate state $X$ let us define a unit vector $\hat{X}$; for a two-particle state this could be the direction of the outgoing momenta in the center-of-mass frame. We can perform the partial-wave decomposition for the $a\rightarrow X$ and $b\rightarrow X$ processes in terms of the angles $\theta_a$, $\theta_b$ defined by $\cos\theta_a=\hat{a} \cdot \hat{X}$, $\cos\theta_b = \hat{b}\cdot \hat{X}$, where $\hat{a}$ and $\hat{b}$ are unit vectors describing the incoming momentum directions of particles in the center-of-mass frame, for states $a$ and $b$ respectively. Then the optical theorem becomes:
\begin{align}&  -i \sum_\ell (2\ell+1) P_\ell(\hat{a} \cdot \hat{b})  \left[ \mathcal{M}_\ell(a\rightarrow b) - \mathcal{M}_\ell^*(b\rightarrow a) \right] \nonumber \\
& = \sum_X \int d\Pi_X \sum_{\ell,\ell^\prime} (2\ell+1) (2\ell^\prime+1) P_{\ell^\prime}(\hat{b} \cdot \hat{X}) P_\ell(\hat{a}\cdot \hat{X})   \mathcal{M}_{\ell^\prime}^*(b\rightarrow X) \mathcal{M}_\ell(a\rightarrow X).\end{align}
Let us write the phase space integral as $d\Pi_X = d\Omega_X d\tilde{\Pi}_X$. Since the individual partial wave amplitudes $\mathcal{M}_{\ell^\prime}(a,b\rightarrow X)$ do not depend on $\hat{X}$, we can perform the angular integral as:
\begin{align}&  -i \sum_\ell (2\ell+1) P_\ell(\hat{a} \cdot \hat{b})  \left[ \mathcal{M}_\ell(a\rightarrow b) - \mathcal{M}_\ell^*(b\rightarrow a) \right] \nonumber \\
& = \sum_X \int d\tilde{\Pi}_X \sum_{\ell,\ell^\prime} (2\ell+1) (2\ell^\prime+1)  \frac{4\pi}{2\ell+1} P_\ell(\hat{a}\cdot \hat{b})  \delta_{\ell,\ell^\prime}  \mathcal{M}_{\ell^\prime}^*(b\rightarrow X) \mathcal{M}_\ell(a\rightarrow X) \nonumber \\
& =  \sum_{\ell} (2\ell+1) P_\ell(\hat{a}\cdot \hat{b})   \sum_X 4\pi \int d\tilde{\Pi}_X \mathcal{M}_{\ell}^*(b\rightarrow X) \mathcal{M}_\ell(a\rightarrow X) \nonumber \\
\Rightarrow &  -i \left[ \mathcal{M}_\ell(a\rightarrow b) - \mathcal{M}_\ell^*(b\rightarrow a) \right] =   \sum_X \int d\Pi_X \mathcal{M}_{\ell}^*(b\rightarrow X) \mathcal{M}_\ell(a\rightarrow X) \end{align}

\subsection{The optical theorem with imaginary momenta}

There is an additional subtlety with regard to the case with imaginary momenta, where for $f_{s,\ell}$ we need the full $N\times N$ matrix, not just its $M\times M$ truncation (physically, this can be understood as the potential providing energy to excite states which would be kinematically forbidden in its absence). It is not immediately clear how to define the QFT matrix element in this case.  If we can factor out the momentum dependence of the amplitudes, so that they are characterized in terms of momentum-independent coefficients, then we can match these momentum-independent coefficients in the regime where all the states are kinematically accessible; we can also use the optical theorem to compute these coefficients. Then we can obtain $f_{s,\ell}$ at lower momentum by restoring the (possibly complex) momentum factors.

Suppose that for a given partial-wave cross section, evaluated at real momenta, we have $\mathcal{M}_\ell(a\rightarrow b) = (p_a p_b/\mu^2)^\ell \hat{\mathcal{M}}_\ell(a\rightarrow b)$ where the hatted amplitude is independent of the external momenta. Similarly, suppose we can write $\mathcal{M}_\ell(a\rightarrow X) = (p_a p_X/\mu^2)^\ell \hat{\mathcal{M}}_\ell(a\rightarrow X)$, where $p_X$ is defined as usual if $X$ is one of the non-relativistic two-particle states coupled by the potential, and otherwise can be left unspecified (or can be e.g.~defined to be equal to $\mu$; we will see that this factor does not matter). Here we want $\hat{\mathcal{M}}_\ell(a\rightarrow X)$ to be independent of any external momenta associated with the non-relativistic two-particle states coupled by the potential.

%If for a given partial wave cross section, evaluated at real momenta, we can write $\mathcal{M}_\ell(a\rightarrow b) = (p_a p_b/\mu^2)^\ell \hat{\mathcal{M}}_\ell(a\rightarrow b)$, and $\mathcal{M}_\ell(a\rightarrow X) = (p_a/\mu)^\ell \hat{\mathcal{M}}_\ell(a\rightarrow X)$, 

Then in the optical theorem, evaluated at real $p_a$, $p_b$, we can cancel out the momentum factors associated with the initial/final states  and obtain:
 \begin{equation} -i \left[ \hat{\mathcal{M}}(a\rightarrow b) - \hat{\mathcal{M}}^*(b\rightarrow a) \right] = \sum_X \int  d\Pi_X |p_X/\mu|^2 \hat{\mathcal{M}}^*(b\rightarrow X) \hat{\mathcal{M}}(a\rightarrow X).\end{equation}

If these equations continue to hold when we extend to the case of imaginary momenta, we can write:
  \begin{align} & -i (p_b^* p_a/\mu^2)^\ell \left[ \hat{\mathcal{M}}(a\rightarrow b) - \hat{\mathcal{M}}^*(b\rightarrow a) \right] = (p_b^* p_a/\mu^2)^\ell \sum_X \int d\Pi_X  |p_X/\mu|^2 \hat{\mathcal{M}}^*(b\rightarrow X) \hat{\mathcal{M}}(a\rightarrow X) \nonumber \\
\Rightarrow & -i \left[ (p_b^*/p_b)^\ell \mathcal{M}(a\rightarrow b) - (p_a/p_a^*)^\ell \mathcal{M}^*(b\rightarrow a) \right] = \sum_X \int d\Pi_X \mathcal{M}^*(b\rightarrow X) \mathcal{M}(a\rightarrow X). \end{align}

 This explains why our expressions for the $S$-matrix element and cross section (e.g.~in Eqs.~\ref{eq:multistatekappa}, \ref{eq:explicitfxsec}) are most simply written in terms of $\bar{f}_{s,\ell}$, which contains a factor of $(\tilde{P}^\dagger \tilde{P}^{-1})^{2\ell}$ relative to $f_{s,\ell}$.To illustrate this point, suppose for now that we work in the convention discussed in the main text where $V_\text{long}(r)=0$ for $r < a$, so the relation between the two takes the simple form $\bar{f}_{s,\ell} = (\tilde{P}^\dagger \tilde{P}^{-1})^{2\ell} f_{s,\ell}$.
 
Then, following the conventions above, let us identify $(f_{s,\ell})_{ba} = \frac{1}{32\pi \mu} \mathcal{M}_{\ell}(a \rightarrow b) =  \frac{1}{32\pi \mu} (p_a p_b/\mu^2)^\ell \hat{\mathcal{M}}_\ell(a\rightarrow b)$, then $(\bar{f}_{s,\ell})_{ba} = \frac{1}{32\pi \mu} \mathcal{M}_{\ell}(a \rightarrow b) (p_b^*/p_b)^\ell$, and we can write from the optical theorem above:
 \begin{align} \frac{2\pi i}{\mu} (\bar{f}_{s,\ell}^\dagger - \bar{f}_{s,\ell})_{ba} & =  \frac{1}{(4 \mu)^2} \sum_X \int d\Pi_X \mathcal{M}_\ell^*(b\rightarrow X) \mathcal{M}_\ell(a\rightarrow X).\label{eq:fsoptical} \end{align}
 In this case, the RHS includes all possible final states, including the kinematically allowed states coupled by the potential; to obtain the inclusive annihilation cross section, we must subtract this scattering term. It can easily be seen that this subtraction is implemented by the $\bar{f}_{s,\ell}^\dagger (\bar{Z}_\ell^\dagger - \bar{Z}_\ell) \bar{f}_{s,\ell}$ term in Eq.~\ref{eq:explicitfxsec} (which as discussed in the main text is independent of the long-range potential for this convention, and depends only on the scattering amplitude into kinematically allowed states). Thus we can replace $\bar{f}_{s,\ell}^\dagger - \bar{f}_{s,\ell} \rightarrow \bar{f}_{s,\ell}^\dagger - \bar{f}_{s,\ell} + \bar{f}_{s,\ell}^\dagger (\bar{Z}_\ell^\dagger - \bar{Z}_\ell) \bar{f}_{s,\ell}$ on the LHS of Eq.~\ref{eq:fsoptical} if we understand the sum on the RHS to include only final states for annihilation, not scattering.  For the diagonal terms ($a=b$), we recognize the familiar expression for the cross section on the RHS (in the absence of Sommerfeld enhancement, and up to the $c_i$ factor for identical particles).
 
 Let us define:
 \begin{align} (\bar{\Gamma}_\ell)_{ba} \equiv \frac{1}{32  i \pi \mu} \sum_X \int d\Pi_X \mathcal{M}_\ell^*(b\rightarrow X) \mathcal{M}_\ell(a\rightarrow X), \label{eq:gamma} \end{align} 
 where the sum on the RHS is taken to include only final states for annihilation (i.e. omitting scattering amongst the DM states). We see that within the convention where $V_\text{long}(r)$ is set to zero within $r=a$, the ``numerator'' term in Eq.~\ref{eq:explicitfxsec} is given to all orders by:
 \begin{equation} (\bar{f}_{s,\ell}^\dagger - \bar{f}_{s,\ell} + \bar{f}_{s,\ell}^\dagger (\bar{Z}_\ell^\dagger - \bar{Z}_\ell) \bar{f}_{s,\ell})_{ba} =  (\bar{\Gamma}_\ell)_{ba}. \end{equation}
This is physically intuitive as we expect the overall (corrected or uncorrected) annihilation cross section should vanish when $\bar{\Gamma}_\ell=0$. The cross section is of course independent of our convention for $V_\text{long}(r)$, this statement is just easiest to see in this convention because of the simple behavior of $\bar{Z}_\ell^\dagger - \bar{Z}_\ell$.

However, for the ``denominator'' term we need the full amplitude $\bar{f}_{s,\ell}$, including possibly Hermitian contributions. At leading order in the coupling governing $\bar{f}_{s,\ell}$, we can write $\bar{f}_{s,\ell}^\dagger - \bar{f}_{s,\ell} \approx \bar{\Gamma}_\ell$, since $\bar{f}_{s,\ell}^\dagger (\bar{Z}_\ell^\dagger - \bar{Z}_\ell) \bar{f}_{s,\ell}$ is a higher-order term (note that $\bar{Z}_\ell^\dagger - \bar{Z}_\ell$ is purely a function of momentum and does not receive any enhancement from the long-range potential). We can then write $\bar{f}_{s,\ell} = \frac{1}{2} (\bar{f}_{s,\ell}^\dagger + \bar{f}_{s,\ell}) + \frac{1}{2} ( \bar{f}_{s,\ell} - \bar{f}_{s,\ell}^\dagger) = \frac{1}{2} (\bar{f}_{s,\ell}^\dagger + \bar{f}_{s,\ell}) - \frac{1}{2} \bar{\Gamma}_\ell$,  i.e.~the anti-Hermitian contribution to $\bar{f}_{s,\ell}$ can be approximated as $-(1/2)\bar{\Gamma}_\ell$.
% The corresponding contribution to $f_{s,\ell}$ is given by $ -(1/2) \tilde{P}^{2\ell} (\tilde{P}^\dagger)^{-2\ell} \bar{\Gamma}_\ell$.

In principle there will also be Hermitian contributions to $\bar{f}_{s,\ell}$. $f_{s,\ell}$ describes the full amplitude corresponding to interactions at $r < a$, and so will include elastic scattering (corresponding to the same interactions that give rise to $V_\text{long}(r)$); this will not contribute at leading order to $\bar{\Gamma}_\ell$, but for the convention where we set $V_\text{long}(r)=0$ for $r < a$, these terms will remain in $\bar{f}_{s,\ell}$ (in conventions where they are included in $V_\text{long}(r)$, they will be subtracted as part of going from $f_{s,\ell}$ to $\bar{f}_{s,\ell}$). We can estimate this contribution by using the Born approximation to calculate the scattering amplitude $f_{b,\ell}$ arising from elastic scattering in the non-relativistic potential at $r < a$ (i.e.~if we did {\it not} set the potential to zero inside $r=a$) -- although there may also be relevant relativistic corrections given the small distance scale / high energy scale -- obtaining $\bar{f}_{s,\ell} \approx - (1/2)\bar{\Gamma}_\ell + (\tilde{P}^\dagger \tilde{P}^{-1})^{2\ell} f_{b,\ell}$. The corresponding matching for the convention-independent full short-range amplitude $f_{s,\ell}$ then becomes $f_{s,\ell} \approx  (\tilde{P}^\dagger \tilde{P}^{-1})^{2\ell}  \left[ - (1/2)\bar{\Gamma}_\ell + f_{b,\ell} \right]$.

\subsection{Improved matching between $f_{s,\ell}$ and $\bar{\Gamma}_\ell$}

The lowest-order estimate for $f_{s,\ell}$ can then be obtained by computing $\bar{\Gamma}_\ell$ at tree-level. An improved estimate can be obtained by including the Sommerfeld enhancement arising from evolution of the wavefunction in the potential within $r < a$, since in the matching for $f_{s,\ell}$ we would like to include all interactions within $r < a$. Let us now move to the convention where we do {\it not} set $V_\text{long}(r)$ to zero for $r < a$, instead using the conventional non-relativistic potential in this regime; we will see that the improved matching takes a particularly simple form in this case. The desired short-range Sommerfeld factor is then equivalent to $\Sigma^a_{0,\ell}$ in the main text. For this ``short-range'' Sommerfeld enhancement, which should be modest, we can work to lowest order in the annihilation amplitude, neglecting the corrections derived in this work (i.e.~effectively we adopt the standard Sommerfeld enhancement prescription in the literature).

Suppose we write $\mathcal{M}(a\rightarrow X) \approx \Sigma_{ca} \mathcal{M}_0(c\rightarrow X)$ where $\Sigma$ represents a Sommerfeld enhancement matrix (for a specific partial wave, although we suppress the partial wave indices in the first line below for convenience). Then we would obtain from Eq.~\ref{eq:gamma}:
\begin{align} (\bar{\Gamma}_\ell)_{ba} & = \Sigma_{db}^* \left[ \frac{1}{32 i \pi \mu} \sum_X \int d\Pi_X \mathcal{M}_0^*(d\rightarrow X)  \mathcal{M}_0(c\rightarrow X) \right] \Sigma_{ca} \nonumber \\
\Rightarrow  (\tilde{P}^\dagger \tilde{P}^{-1})^{2\ell} f_{s,\ell} & \approx  -\frac{1}{2} \Sigma^\dagger \bar{\Gamma}_{0,\ell} \Sigma + (\tilde{P}^\dagger \tilde{P}^{-1})^{2\ell} f_{b,\ell} \nonumber \\
(\bar{\Gamma}_{0,\ell})_{dc} & \equiv \frac{1}{32 i \pi \mu} \sum_X \int d\Pi_X \mathcal{M}_0^*(d\rightarrow X)  \mathcal{M}_0(c\rightarrow X). \label{eq:gamma0} \end{align}
Now we can insert $\Sigma = \Sigma^a_{0,\ell}$ (i.e. the uncorrected Sommerfeld enhancement matrix arising from evolution under $V_\text{long}(r)$ for $r < a$) and use Eq.~\ref{eq:sommerfeldidentityNbyN} to write:
\begin{align}  (\Sigma^a_{0,\ell})^\dagger & = \left[ (\tilde{P}^\dagger)^{-1} S_{b,\ell} \tilde{P}^\dagger  \right]^{-1} (\tilde{P}^\dagger \tilde{P}^{-1})^{2\ell}   \Sigma^{aT}_{0,\ell} (\tilde{P}^{-1} \tilde{P}^\dagger)^{-2\ell} \nonumber \\
\Rightarrow (\tilde{P}^\dagger \tilde{P}^{-1})^{2\ell} f_{s,\ell} & \approx  -\frac{1}{2}  \left[ (\tilde{P}^\dagger)^{-1} S_{b,\ell} \tilde{P}^\dagger  \right]^{-1} (\tilde{P}^\dagger \tilde{P}^{-1})^{2\ell}   \Sigma^{aT}_{0,\ell} (\tilde{P}^{-1} \tilde{P}^\dagger)^{-2\ell} \bar{\Gamma}_{0,\ell} \Sigma^a_{0,\ell} + (\tilde{P}^\dagger \tilde{P}^{-1})^{2\ell} f_{b,\ell},\end{align}
where $S_{b,\ell} = 1 + 2 i \tilde{P} f_{b,\ell} \tilde{P}$. Now since we are working to lowest order in the short-range amplitudes, we can drop the cross-term involving both $f_{b,\ell}$ and $\bar{\Gamma}_{0,\ell}$, i.e. setting $S_{b,\ell}\approx 1$ in the first term. Then we obtain the improved matching relation in terms of $\bar{\Gamma}_{0,\ell}$ constructed from the tree-level annihilation matrix elements:
\begin{align} f_{s,\ell} & \approx  -\frac{1}{2}  \Sigma^{aT}_{0,\ell} (\tilde{P}^{-1} \tilde{P}^\dagger)^{-2\ell} \bar{\Gamma}_{0,\ell} \Sigma^a_{0,\ell} + f_{b,\ell}. \label{eq:improvedmatchingcon1} \end{align}
As a reminder, $f_{s,\ell}$ here is the full short-distance amplitude (which is convention-independent), but $f_{b,\ell}$ and $\Sigma^a_{0,\ell}$ are computed from the standard, $a$-independent, non-relativistic potential for $V_\text{long}(r)$ for $r < a$. Recall that we have defined the convention-dependent effective amplitude $\bar{f}_{s,\ell} = (\tilde{P}^\dagger \tilde{P}^{-1})^{2\ell} (\Sigma^{aT}_{0,\ell})^{-1} (f_{s,\ell} - f_{b,\ell}) (\Sigma^a_{0,\ell})^{-1}$ (Eq.~\ref{eq:multistatekappa}); we see that in the convention with $a$-independent $V_\text{long}(r)$, the improved matching relation simplifies to:
\begin{align} \bar{f}_{s,\ell} = -\frac{1}{2} \bar{\Gamma}_{0,\ell}.  \label{eq:improvedmatchingcon2} \end{align}
This aligns with the argument derived from App.~\ref{app:shortrangeinterp} that $\bar{f}_{s,\ell}$ in this convention can be interpreted as the effective scattering amplitude for a contact interaction governing annihilation.

\subsection{Results for the wino example}
\label{app:winomatrix}

We can now go ahead and use existing results for the wino to write out the components of the bare annihilation matrix, $\bar{\Gamma}_{0,\ell}$.
Refs.~\cite{Hellmann:2013jxa, Beneke:2014gja} provide Wilson coefficients for the pure wino that encode $\sum_X \int d\Pi_X \hat{\mathcal{M}}_\ell^*(b\rightarrow X) \hat{\mathcal{M}}_\ell(a\rightarrow X)$, as also discussed in Ref.~\cite{Asadi:2016ybp}. As discussed above, we are using the ``method-2'' basis of  Ref.~\cite{Beneke:2014gja} to define the two-particle states, and so must rescale the elements of the annihilation matrix that involve identical-particle states. Specifically, for $\ell=0$ we can replace $\frac{1}{(4\mu)^2} \left[ \sum_X \int d\Pi_X \mathcal{M}_\ell^*(b\rightarrow X) \mathcal{M}_\ell(a\rightarrow X) \right]$ with $4 \hat{f}(^1S_0)$ from that work, and then apply the method-2 adjustments, yielding:
\begin{equation} \bar{\Gamma}_{0,0} = \frac{2 \mu}{ i \pi} \frac{\pi \alpha_W^2}{(2\mu)^2} \begin{pmatrix} 1 & \frac{1}{\sqrt{2}} \\ \frac{1}{\sqrt{2}} & \frac{3}{2} \end{pmatrix}  = \frac{ \alpha_W^2}{2 i \mu} \begin{pmatrix} 1 & \frac{1}{\sqrt{2}} \\ \frac{1}{\sqrt{2}} & \frac{3}{2} \end{pmatrix} \end{equation}
Note that we use $4 \hat{f}(^1S_0)$ rather than $\hat{f}(^1S_0)$ because the latter describes the contribution to the spin-averaged cross section from the spin-singlet case (i.e. it includes a factor of 1/4 for the fraction of pairs that are in a spin-singlet configuration), whereas for the quantum mechanics calculation we are interested in spin eigenstates (we do not consider spin-flip interactions). The choice of $\ell=0$ and a $\chi^0\chi^0$ final state implies a spin-singlet state.

For $\ell=1$ we instead consider a spin-triplet state, and should replace $(2\ell+1) \frac{2\pi i}{\mu} \bar{\Gamma}_1$ with $\frac{1}{3} \frac{p_b^* p_a}{(2\mu)^2} \hat{f}(^3P_J)$ (and then apply the method-2 adjustments), yielding:
\begin{equation} \bar{\Gamma}_{0,1} = \frac{\mu}{2 i \pi} \frac{1}{3} \frac{1}{(2 \mu)^2}  \frac{28 \pi \alpha_W^2}{9 (2\mu)^2} \begin{pmatrix} |p_1|^2 & \frac{1}{\sqrt{2}} p_1^* p_2 \\ \frac{1}{\sqrt{2}} p_2^* p_1  & \frac{3}{2} |p_2|^2 \end{pmatrix}   \end{equation}

As a cross-check, this predicts the following contributions to the spin-averaged inclusive cross-sections for annihilation from the $\chi^0\chi^0$ and $\chi^+\chi^-$ states, for the $\ell=0$ spin-singlet and $\ell=1$ spin-triplet states, in the absence of Sommerfeld enhancement:
\begin{align} (\sigma_{0,\text{ann}} v_\text{rel})_1 & = \frac{1}{4} 2  \frac{2 \pi i}{\mu} (\bar{\Gamma}_0)_{11} = \frac{2\pi\alpha_W^2}{(2\mu)^2},  \nonumber \\
(\sigma_{0,\text{ann}} v_\text{rel})_2 & = \frac{1}{4} \frac{2 \pi i}{\mu} (\bar{\Gamma}_0)_{22} = \frac{3}{2} \frac{\pi\alpha_W^2}{(2\mu)^2}.
\nonumber \\
(\sigma_{1,\text{ann}} v_\text{rel})_1 & = \frac{3}{4} \times 2 \times 3 \times \frac{2 \pi i}{\mu} (\bar{\Gamma}_1)_{11} = \left(\frac{|p_1|}{2 \mu}\right)^2  \frac{14 \pi \alpha_W^2}{3 (2\mu)^2} , \nonumber \\
(\sigma_{1,\text{ann}} v_\text{rel})_2 & =\frac{3}{4} \times 3 \times \frac{2 \pi i}{\mu} (\bar{\Gamma}_1)_{22} = \frac{|p_2|^2}{(2 \mu)^2}  \frac{7 \pi \alpha_W^2}{2 (2\mu)^2}   
\end{align}
These results are in agreement with Ref.~\cite{Hellmann:2013jxa}, and the $s$-wave results agree with earlier studies, e.g.~Ref.~\cite{Hisano:2006nn}.

\section{Interpreting the short-distance amplitude}
\label{app:shortrangeinterp}

As discussed in the main text, we have freedom in how to define $V_\text{long}(r)$ for $r < a$. To understand how the short-range scattering amplitude $f_s$ will be affected by the presence of a non-zero $V_\text{long}(r)$ for $r < a$, suppose we have such a potential at short range, and also a contact interaction with scattering amplitude $f_{C,\ell}$. In the limit as $a\rightarrow 0$, we expect $f_{s,\ell} \rightarrow f_{C,\ell}$. For other choices of $a$, we can use the results we have derived for a general long-range potential (taken here to be $V_\text{long}(r)$ for $r < a$ and zero for $r > a$) to relate $f_{s,\ell}$ to $f_{C,\ell}$, working to lowest order in $f_{C,\ell}$ and taking the matching radius to zero for a contact interaction. As in the main text, let $f_{b,\ell}$ be the short-distance scattering amplitude obtained when the contact interaction is set to zero. Let $S_{b,\ell}$ be the $S$-matrix element corresponding to $f_{b,\ell}$ and $S_{s,\ell}$ be the $S$-matrix element corresponding to $f_{s,\ell}$. 

As in the main text, the $N\times N$ short-distance Sommerfeld matrix is defined by $\Sigma^a_{0,\ell} = \tilde{P}\alpha_{b,\ell}(0) \tilde{P}^{-2}$. Following Eq.~\ref{eq:sdecomposition}, we can write the modified $S$-matrix element as:
\begin{align} S_{s,\ell} = S_{b,\ell} + 2 i \tilde{P} f_{1,s} \tilde{P}^2 R_\ell \tilde{P}^{-1} = S_{b,\ell} + 2 i \tilde{P}^{-1} (\tilde{P}^{-1} \alpha_{b,\ell}(0))^T \tilde{P}^{2(\ell+1)} R_{s,\ell} \tilde{P}^{-1} , \end{align}
where the $s$ subscripts indicate we are working with the truncated potential (zero outside $r=a$). If we are only interested in working to lowest order in the short-range interaction, the expression for $R_{s,\ell}$ simplifies to:
\begin{align} R_{s,\ell} = \left[ \tilde{P}^{-1} f_{C,\ell}^{-1} \tilde{P}^{-1} \tilde{P}^{2\ell+1}\right]^{-1} \alpha_{b,\ell}(0),\end{align}
so that in the end we obtain:
\begin{align}S_{s,\ell} &= S_{b,\ell} + 2 i \tilde{P}^{-1} \alpha_{b,\ell}(0)^T  \tilde{P} f_{C,\ell} \tilde{P} \alpha_{b,\ell}(0) \tilde{P}^{-1} = S_{b,\ell} + 2 i \tilde{P}  \Sigma^{aT}_{0,\ell}  f_{C,\ell}  \Sigma^a_{0,\ell} \tilde{P} \nonumber \\
\Rightarrow f_{s,\ell} - f_{b,\ell} & =    \Sigma^{aT}_{0,\ell}  f_{C,\ell}  \Sigma^a_{0,\ell} \nonumber \\
\Rightarrow f_{C,\ell}^{-1} & = \Sigma^a_{0,\ell}  \left( f_{s,\ell} - f_{b,\ell} \right)^{-1} \Sigma^{a T}_{0,\ell} .\label{eq:fCdecomposition} \end{align}

%Similarly, re-deriving the analogues of Eq.~\ref{eq:multistateSmatrix} and Eq.~\ref{eq:multistatekappa} for this case, and expanding to lowest order in the short-range interaction, we obtain:
%\begin{align} S_{s,\ell} & = 1 + 2 i \tilde{P} f_{s,\ell} \tilde{P} = S_{b,\ell} (1 + 2 i \tilde{P}^\dagger (\Sigma^a_{0,\ell})^\dagger \bar{f}_{C,\ell} \Sigma^a_{0,\ell} \tilde{P}),\end{align}
%where $\bar{f}_{C,\ell} \equiv (\tilde{P}^\dagger \tilde{P}^{-1})^{2\ell} f_{C,\ell}$. Using Eq.~\ref{eq:sommerfeldidentityNbyN}:
% \begin{align} 1 + 2 i \tilde{P} f_{s,\ell} \tilde{P} & = S_{b,\ell} + 2 i (\tilde{P}^\dagger \tilde{P}^{-1})^{2\ell-1} \tilde{P} \Sigma^{aT}_{0,\ell} f_{C,\ell} \Sigma^a_{0,\ell} \tilde{P} \nonumber \\
% & = 1 + 2 i \tilde{P} \left[ f_{b,\ell} + (\tilde{P}^\dagger \tilde{P}^{-1})^{2\ell-1} (\Sigma^a_{0,\ell})^T f_{C,\ell} \Sigma^a_{0,\ell}  \right] \tilde{P} \nonumber \\
% \Rightarrow f_{s,\ell} - f_{b,\ell} & = (\Sigma^a_{0,\ell})^T f_{C,\ell} \Sigma^a_{0,\ell} \nonumber \\
% \Rightarrow f_{C,\ell}^{-1} & = \Sigma^a_{0,\ell}  \left( f_{s,\ell} - f_{b,\ell} \right)^{-1}  (\tilde{P}^\dagger \tilde{P}^{-1})^{2\ell-1} \Sigma^{a T}_{0,\ell} . \end{align}

These expressions make intuitive sense: at leading order in the strength of the contact interaction, $f_{s,\ell}$ describes the scattering amplitude for the Sommerfeld-enhanced contact interaction, plus the contribution purely from elastic scattering in $V_\text{long}(r)$. The scattering amplitude for the contact interaction can be recovered from $f_{s,\ell}$ (at least at leading order) by subtracting the contribution $f_{b,\ell}$ that arises purely from $V_\text{long}(r)$, and factoring out the Sommerfeld enhancement matrices from what remains. In Sec.~\ref{sec:wino}, we employ this result to justify matching $\bar{f}_{s,\ell}^{-1} (\tilde{P}^\dagger \tilde{P})^{-2\ell} = \Sigma^a_{0,\ell}  \left( f_{s,\ell} - f_{b,\ell} \right)^{-1} \Sigma^{a T}_{0,\ell}$ to the tree-level perturbative QFT result derived purely from the short-range physics, when we choose a convention such that $V_\text{long}(r)$ is non-zero (and $a$-independent) for $r < a$.

Also note that repeating the derivation of Eq.~\ref{eq:sommerfeldidentity} for this short-range Sommerfeld matrix (with the key difference being that it is $N\times N$ not $N\times M$) yields:
\begin{align} (\tilde{P}^\dagger)^{-1} S_{b,\ell} \tilde{P}^\dagger (\Sigma^a_{0,\ell})^\dagger (\tilde{P}^{-1} \tilde{P}^\dagger)^{2\ell}   =  (\tilde{P}^\dagger \tilde{P}^{-1})^{2\ell}   \Sigma^{aT}_{0,\ell}. \label{eq:sommerfeldidentityNbyN} \end{align}

In more detail, we can consider Wronskians evaluated using the following solutions:
\begin{itemize}
\item $u_{b,\ell}(r)$: as previously, we have $\alpha_{b,\ell}(a)=\tilde{P}$, $\beta_{b,\ell}(a) = -\tilde{P}f_{b,\ell} \tilde{P}^2$, $\beta_{b,\ell}(0)=0$.
\item $u_{b,\ell}^*(r)$ (i.e. the conjugate of the solution above): if we denote the $\alpha$ and $\beta$ coefficients for this solution as $\alpha_{b*,\ell}$ and $\beta_{b*,\ell}$, then we have $\beta_{b*,\ell}(0)=0$, and
 \begin{align} \alpha_{b*,\ell}(0) & = (\tilde{P}^{-1} \tilde{P}^\dagger)^{2\ell+1}\alpha_{b,\ell}(0)^*, \quad \alpha_{b*,\ell}(a) = (\tilde{P}^\dagger \tilde{P}^{-1})^{2\ell +1} (\tilde{P}^\dagger - 2 i (\tilde{P} f_{b,\ell} \tilde{P}^2)^*).\end{align}
 \item A solution that is the equivalent of $\tilde{w}_\ell(r)$ when $V_\text{long}(r)$ is truncated to $r < a$. This solution has $\alpha(a)=0$ (since it is purely outgoing outside $r=a$), and $\beta(a) = -\tilde{P} f_1 \tilde{P}^2$. At $r\rightarrow 0$ it has $\beta(0) = -\tilde{P}^{2\ell+1}$. 
\end{itemize}
When we compute the Wronskian (as defined in Eq.~\ref{eq:vpmwronskian}) between the first and third solutions, and the second and third solutions, we obtain the following equations:
\begin{align} \alpha_{b,\ell}(0)^T \tilde{P}^{2\ell+1} & = \tilde{P}^2 f_1 \tilde{P}^2, \nonumber \\
( \alpha_{b,\ell}(0)^\dagger (\tilde{P}^{-1} \tilde{P}^\dagger)^{2\ell + 1}  \tilde{P}^{2\ell+1}) & = (\tilde{P}^\dagger - 2 i (\tilde{P} f_{s,\ell} \tilde{P}^2)^\dagger)  (\tilde{P}^\dagger \tilde{P}^{-1})^{2\ell+1}  \tilde{P} f_1 \tilde{P}^2 . \end{align}
Combining these equations, and noting $S_{b,\ell} = 1 + 2 i \tilde{P} f_{b,\ell} \tilde{P}$, and $\Sigma^a_{0,\ell} = \tilde{P}\alpha_{b,\ell}(0) \tilde{P}^{-2}$, we obtain:
\begin{align}  \tilde{P}^\dagger (\Sigma^a_{0,\ell})^\dagger (\tilde{P}^{-1} \tilde{P}^\dagger)^{2\ell} =  S_{b,\ell}^\dagger (\tilde{P}^\dagger \tilde{P}^{-1})^{2\ell}  \tilde{P}^\dagger \Sigma^{aT}_{0,\ell}.  \end{align}
Then using the fact that $S_{b,\ell}$ is unitary, we can obtain Eq.~\ref{eq:sommerfeldidentityNbyN} as required:
\begin{align} & S_{b,\ell} \tilde{P}^\dagger (\Sigma^a_{0,\ell})^\dagger (\tilde{P}^{-1} \tilde{P}^\dagger)^{2\ell}   =  (\tilde{P}^\dagger \tilde{P}^{-1})^{2\ell}  \tilde{P}^\dagger \Sigma^{aT}_{0,\ell} \nonumber \\
& \Rightarrow ( \tilde{P}^\dagger )^{-1} S_{b,\ell} \tilde{P}^\dagger (\Sigma^a_{0,\ell})^\dagger (\tilde{P}^{-1} \tilde{P}^\dagger)^{2\ell}   =  (\tilde{P}^\dagger \tilde{P}^{-1})^{2\ell}  \Sigma^{aT}_{0,\ell}. \end{align}

Note that if we truncated the Sommerfeld matrix to $N\times M$ and similarly the $S$-matrix to $M\times M$, we would obtain:
\begin{align} 
S_{b,\ell} P (\Sigma^a_{0,\ell})^\dagger (\tilde{P}^{-1} \tilde{P}^\dagger)^{2\ell}  & =   P \Sigma^{aT}_{0,\ell} \end{align}
This matches the structure of the identity in Eq.~\ref{eq:sommerfeldidentity}, as expected.

\section{Useful results for partial wave decomposition}
\label{app:partial_amps}
\subsection{The Born approximation in the partial-wave decomposition for the wino}

We can approximate the contributions to $f_{s,\ell}$ from the long-range potential using the Born approximation. For the wino example, the matrix elements we are interested in computing are $f_{11} = \langle \chi^{0}\chi^{0} | T | \chi^{0}\chi^{0}\rangle, f_{21} = \langle \chi^{0}\chi^{0} | T | \chi^{+}\chi^{-}\rangle, $ and $f_{22} = \langle \chi^{+}\chi^{-} | T | \chi^{+}\chi^{-}\rangle$. The leading order contribution to $f_{11}$ occurs at $\mathcal{O}(\alpha_{W}^{2})$ whereas the other two matrix elements have contributions starting at $\mathcal{O}(\alpha_{W})$.

Up to $\mathcal{O}(\alpha_{W})$, the total matrix elements are given by

\begin{equation} 
\begin{split}
f_{11}^{(1)} & = 0 \\
f_{21}^{(1)} & = 
\alpha_W \frac{M_\chi}{4\pi} \int \frac{d^3 \vec{r}_0}{r_0} \sqrt{2}e^{i (\vec{p} - \vec{k}^\prime) \cdot \vec{r}_0 - m_W r_0} \\
f_{22}^{(1)} & =  
\alpha_W \frac{M_\chi}{4\pi} \int \frac{d^3 \vec{r}_0}{r_0} e^{i (\vec{p}^\prime - \vec{k}^\prime) \cdot \vec{r}_0 } \left(\sin^2\theta_W + \cos^2\theta_W e^{-m_Z r_0}\right) \\
\end{split}
\end{equation} 
The second order matrix elements we are interested in computing are
\begin{equation}
\begin{split}
f_{11}^{(2)} & = \frac{M_\chi^2}{4\pi} \int d^3\vec{r}_1 d^3\vec{r}_0 G_{k^\prime}(\vec{r}_0,\vec{r}_1) V_{12}(r_0) V_{12}(r_1) e^{i\vec{p}\cdot \vec{r}_0 - i \vec{k}\cdot \vec{r}_1}, \\
f_{12}^{(2)} & = \frac{M_\chi^2}{4\pi} \int d^3\vec{r}_1 d^3\vec{r}_0 G_{k^\prime}(\vec{r}_0,\vec{r}_1) V_{12}(r_0) V_{22}(r_1) e^{i\vec{p}\cdot \vec{r}_0 - i \vec{k}^\prime \cdot \vec{r}_1},  \\
f_{22}^{(2)} & = \frac{M_\chi^2}{4\pi} \int d^3\vec{r}_1 d^3\vec{r}_0 \left[ G_{k}(\vec{r}_0,\vec{r}_1) V_{12}(r_0) V_{12}(r_1) + G_{k^\prime}(\vec{r}_0,\vec{r}_1) V_{22}(r_0) V_{22}(r_1) \right] e^{i\vec{p}^\prime \cdot \vec{r}_0 - i \vec{k}^\prime \cdot \vec{r}_1}. 
\end{split}
\end{equation}
For compactness, we have written the matrix elements in terms of the elements of the potential in Eq.~\ref{eq:winopot}, omitting the constant $2\Delta$ term from $V_{22}$, which we absorb into the unperturbed Hamiltonian. Here $\vec{p}$ and $\vec{p}^\prime$ denote the 3-momentum of one of the incoming particles, whereas $\vec{k}$ and $\vec{k}^\prime$ denote the 3-momentum of one of the outgoing particles, both in the center-of-mass frame; in both cases the prime denotes that the particles are in the excited state, so $(k^\prime)^2 + 2 M_\chi \Delta = k^2$, $(p^\prime)^2 + 2 M_\chi \Delta = p^2$.

\begin{figure}
\centering
\includegraphics[width=\hsize]{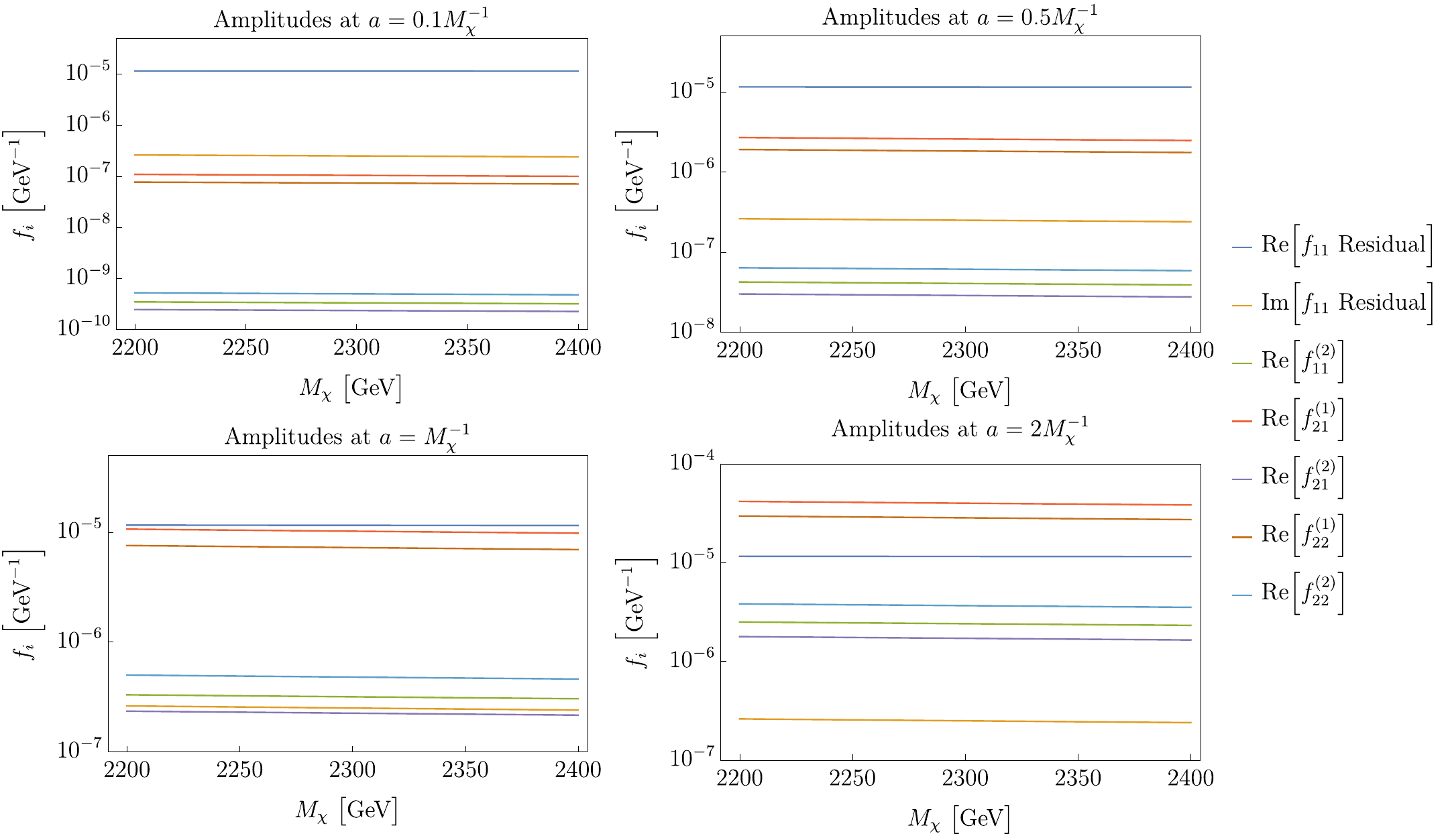}
\caption{Numerical results for the first and second Born approximation, $f_{i}^{(1)}$ and $f_{i}^{(2)}$ respectively, to the various matrix elements $f_{11} = \langle \chi^{0}\chi^{0} | T | \chi^{0}\chi^{0}\rangle, f_{21} = \langle \chi^{0}\chi^{0} | T | \chi^{+}\chi^{-}\rangle, $ and $f_{22} = \langle \chi^{+}\chi^{-} | T | \chi^{+}\chi^{-}\rangle$, and the residual of $f_{11}$ for different values of the matching radius: $a = 0.1 M_{\chi}^{-1}$ (\emph{top left}), $a = 0.5 M_{\chi}^{-1}$ (\emph{top right}), $a = M_{\chi}^{-1}$ (\emph{bottom left}), and $a = 2 M_{\chi}^{-1}$ (\emph{bottom right}).}\label{fig:bornapprox_numerics}
\end{figure}

Furthermore, we have defined $G_k(\vec{r},\vec{r}^\prime) \equiv \frac{e^{i k | \vec{r} - \vec{r}^\prime|}}{4\pi |\vec{r} - \vec{r}^\prime|}$. We see that the general form of these matrix elements is given by
\begin{equation}
f^{(2)} = \frac{M_\chi^2}{4\pi} \int d^3\vec{r}_1 d^3\vec{r}_0 G_{q}(\vec{r}_0,\vec{r}_1) V_{i}(r_0) V_{j}(r_1) e^{i\vec{q}_0 \cdot \vec{r}_0 - i \vec{q}_1 \cdot \vec{r}_1} 
\label{eqn:general_matrix_element}
\end{equation}
Performing a partial wave decomposition on Eq.~\ref{eqn:general_matrix_element} yields 
\begin{equation}
\begin{split}
f^{(2)}_\ell(a) & \approx  \frac{M_\chi^2}{q q_0 q_1} \int^a_0 dr_0 V_{i}(r_0) s_{\ell}(q_0 r_0)  \left[  u_\ell(q r_0) \int^{r_0}_0 dr_1 s_\ell(q r_1) V_{j}(r_1)  s_{\ell}(q_1 r_1) \right. \\
& \left. + s_\ell(q r_0)  \int_{r_0}^a dr_1 u_\ell(q r_1)  V_{j}(r_1)  s_{\ell}(q_1 r_1) \right] 
\end{split}
\label{eqn:partial_wave_second_order}
\end{equation}
and the full second order Born result is recovered by sending $a\to\infty$. We emphasize that Eq.~\ref{eqn:partial_wave_second_order} is a result for general $\ell$. The functions $s_\ell$ and $c_\ell$ are solutions of the radial free particle Schr\"odinger equation, and in terms of the spherical Bessel functions, are given by $s_\ell(kr) \equiv krj_\ell(kr)$ and $c_\ell(kr)\equiv -kry_\ell(kr)$ (as in Eq.~\ref{eq:def of sell and cell}), with $u_\ell(kr) = c_\ell(kr) + is_\ell(kr)$. 
To simplify our numerical calculations presented in Sec.~\ref{sec:wino}, we focus only on the $\ell = 0$ case
and note that $s_0(x) = \sin x \approx x$ and $u_0(x) = e^{ix}$. We also work in the zero-momentum limit so that $k \to 0$. Using these conventions, we compute the first and second Born amplitudes for the three matrix elements of interest and show the results in Fig.~\ref{fig:bornapprox_numerics} for four choices of the matching radius $r = a$. The imaginary parts of the amplitudes vanish for the first and second Born approximation. We also show the results for the real and imaginary parts of the residual of $f_{11}$ which is computed by taking the full one-loop scattering amplitude, rescaling by a factor of $(32\pi M_{\chi})^{-1}$, and then subtracting off the full second order Born approximation.

\begin{figure}
\centering
\includegraphics[width=0.3\textwidth]{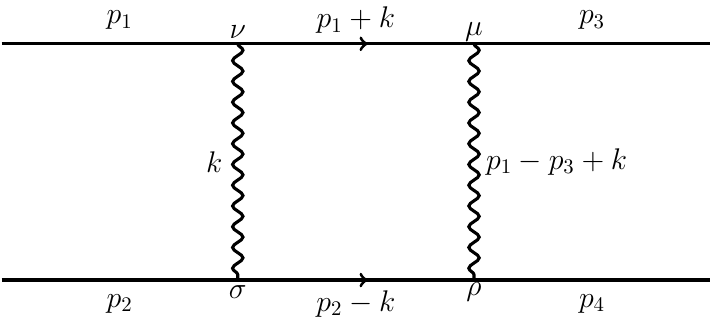}
\includegraphics[width=0.3\textwidth]{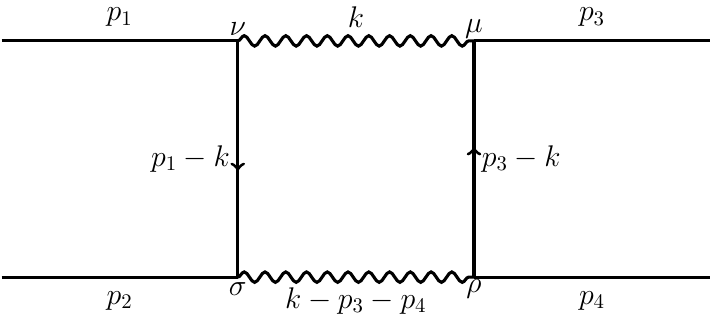}
\includegraphics[width=0.3\textwidth]{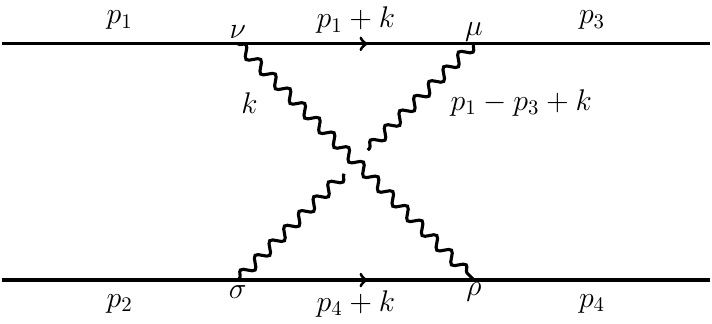}
\includegraphics[width=0.3\textwidth]{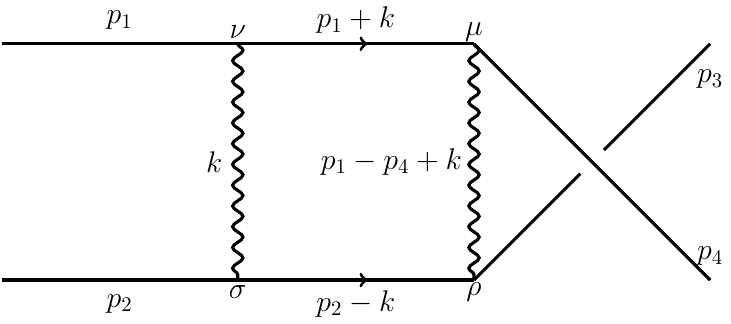}
\includegraphics[width=0.3\textwidth]{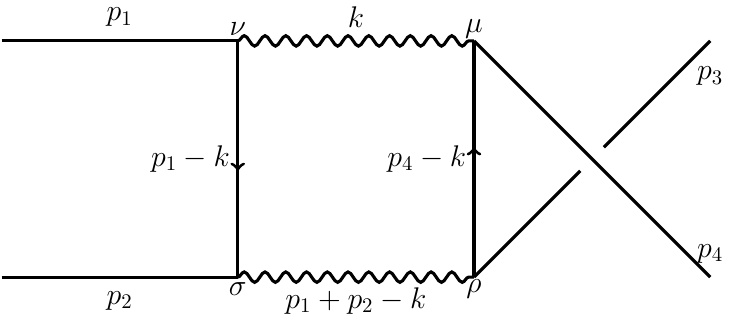}
\includegraphics[width=0.3\textwidth]{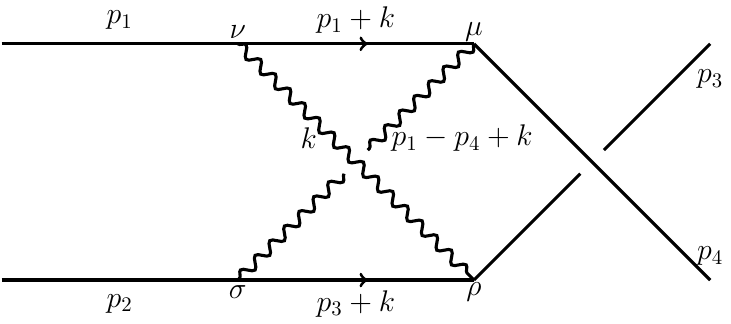}
\caption{Feynman diagram topologies contributing to $\chi^0\chi^0\to\chi^0\chi^0$ scattering. Interchanging $\chi^+ \to \chi^-$ and $W^-\to W^+$ in the loop leads to a doubling of diagrams.}
\label{fig:Feynman_diagrams}
\end{figure}

In the pure wino limit of the MSSM, $\chi^\pm$ and $\chi^0$ only couple off-diagonally via $W^\pm$. As a result, at lowest order, the process $\chi^0\chi^0\to\chi^0\chi^0$ occurs at one-loop level. There are twelve distinct Feynman diagrams which contribute to this process. We show the six topologies which contribute in Fig.~\ref{fig:Feynman_diagrams} and note that there is a doubling due to the interchange of $\chi^+ \to \chi^-$ and $W^-\to W^+$ in the loop. We compute the loop integrals relevant for the amplitude using $\textsc{PackageX}$~\cite{Patel:2015tea,Patel:2016fam} and take the zero-momentum limit which is identical to the $\ell=0$ partial amplitude.

\subsection{Relating helicity and partial-wave amplitudes}

While we chose to focus on the $\ell=0$ case for our wino example, in which case we can obtain the $\ell=0$ component of the full amplitude in the low-velocity limit simply by taking $p\rightarrow 0$, for more general examples we might want to carefully decompose the amplitudes computed from QFT into their partial wave components. In this subsection we review the procedure for extracting the S-matrix elements in the canonical basis given an amplitude, for particles with spin. In particular, it is often easier to compute the amplitude for states of fixed helicity and then convert it to the canonical basis. In the subsequent discussion, we will work this translation out in detail, closely following the discussion in~\cite{deBoer}.

We begin with one particle states with helicity $\lambda$ and momentum $\vec{p}$ which are normalized according to
\begin{equation}
\langle \vec{p}', \lambda'|\vec{p}, \lambda\rangle = \tilde{\delta}(\vec{p}'-\vec{p})\delta_{\lambda'\lambda} = (2\pi)^{3}2E\delta^{3}(\vec{p}'-\vec{p})\delta_{\lambda'\lambda}
\end{equation}
Two particle states are the tensor products of these states and are normalized as 
\begin{equation}
\begin{split}
&|\vec{p}_{a},\lambda_{a},\vec{p}_{b},\lambda_{b}\rangle = |\vec{p}_{a},\lambda_{a}\rangle\otimes |\vec{p}_{b},\lambda_{b}\rangle \\
&\langle \vec{p}_{a}',\lambda_{a}',\vec{p}_{b}',\lambda_{b}' | \vec{p}_{a},\lambda_{a},\vec{p}_{b},\lambda_{b}\rangle = \tilde{\delta}(\vec{p}_{a}'-\vec{p}_{a})\tilde{\delta}(\vec{p}_{b}'-\vec{p}_{b})\delta_{\lambda_{a}'\lambda_{a}}\delta_{\lambda_{b}'\lambda_{b}}
\end{split}
\end{equation}
In the center of momentum frame, the two particle states are defined as
\begin{equation}
|\vec{p}_{a},\lambda_{a},\vec{p}_{b},\lambda_{b}\rangle = \frac{1}{4\pi}\sqrt{\frac{M_{cm}}{|\vec{p}_{1}|}}|\Omega,\lambda_{1},\lambda_{2}\rangle\otimes|P\rangle
\end{equation}
Where $P^{\mu} = p_{1}^{\mu} + p_{2}^{\mu}$ and $M^{2}_{cm} = P^{2}$. Furthermore, $\vec{p}_{1} = -\vec{p}_{2}$ are the momenta of particles a and b in the center of momentum frame. Likewise, $\lambda_{1}$ and $\lambda_{2}$ are the helicities of the particles in the center of momentum frame.

The state $|\Omega,\lambda_{1},\lambda_{2}\rangle$ is defined as
\begin{equation}
|\Omega,\lambda_{1},\lambda_{2}\rangle = \sum_{j, m} N_{j}D^{j}_{m,\lambda}(\Omega)|j, m, \lambda_{1}, \lambda_{2}\rangle = \sum_{j, m} \sqrt{\frac{2j+1}{4\pi}}D^{j}_{m,\lambda}(\Omega)|j, m, \lambda_{1}, \lambda_{2}\rangle
\end{equation}
where $\lambda = \lambda_{1} - \lambda_{2}$. 

The helicity basis states are normalized as
\begin{equation}
\langle j', m', \lambda_{1}', \lambda_{2}' | j, m, \lambda_{1}, \lambda_{2} \rangle = \delta_{jj'}\delta_{mm'}\delta_{\lambda_{1}\lambda_{1}'}\delta_{\lambda_{2}\lambda_{2}'}
\end{equation}

To determine the canonical basis states, we can again start with the states $| \phi = 0, \theta = 0, \lambda_{1}, \lambda_{2}\rangle$. With these choices of $\phi$ and $\theta$, we know that one particle is moving in the $+z$ direction and another is moving in the $-z$ direction. Now, we can rotate this wavefunction. For a single particle, the rotation is defined by
\begin{equation}
\mathcal{R}(\Omega)|s,m\rangle = \sum_{m'}D^{s}_{m',m}(\Omega)|s,m'\rangle
\end{equation}
So, for a two particle state with the particles moving in opposite directions along the $z$ axis, we have
\begin{equation}
|\Omega,\lambda_{1},\lambda_{2}\rangle = \mathcal{R}(\phi,\theta,0)|\phi'=0,\theta'=0,\lambda_{1},\lambda_{2}\rangle = \sum_{m_{1},m_{2}}D^{s_{1}}_{m_{1},\lambda_{1}}D^{s_{2}}_{m_{2},-\lambda_{2}}|\Omega,m_{1},m_{2}\rangle
\end{equation}
Since $D$ is a unitary representation, we also have
\begin{equation}
|\Omega, m_{1}, m_{2}\rangle = \sum_{\lambda_{1},\lambda_{2}}D^{s_{1}*}_{m_{1},\lambda_{1}}D^{s_{2}*}_{m_{2},-\lambda_{2}}|\Omega, \lambda_{1}, \lambda_{2}\rangle
\end{equation}

Adding individual spins $s_{1}$ and $s_{2}$ to a total spin $s$ using Clebsch-Gordan coefficients, we have
\begin{equation}
|\Omega,s,m_{s}\rangle = \sum_{m_{1},m_{2}}\langle s_{1}, m_{1}, s_{2}, m_{2}|s, m_{s}\rangle|\Omega, m_{1}, m_{2} \rangle
\end{equation}
Furthermore, the spherical harmonics are defined as 
\begin{equation}
Y_{l,m_{l}}(\Omega) = \langle\theta,\phi|l,m_{l}\rangle
\end{equation}
Using this, the states can be written as
\begin{equation}
|\Omega,s,m_{s}\rangle = \sum_{l,m_{l}}Y_{l,m_{l}}(\Omega)^{*}|l,m_{l},s,m_{s}\rangle \quad \longleftrightarrow \quad |l,m_{l},s,m_{s}\rangle = \int d\Omega Y_{l,m_{l}}(\Omega)|\Omega,s,m_{s}\rangle
\end{equation}
Finally, we couple spin and orbital angular momentum to form total angular momentum states. Using Clebsch-Gordan coefficients to couple, we have
\begin{equation}
|j,m,l,s\rangle = \sum_{m_{l},m_{s}}\langle l, m_{l}, s, m_{s} | j, m \rangle |l, m_{l}, s, m_{s} \rangle
\end{equation}
Next, writing $|\Omega, m_{1}, m_{2}\rangle$ in terms of $|j, m, \lambda_{1}, \lambda_{2}\rangle$, we have
\begin{equation}
|\Omega, m_{1}, m_{2}\rangle = \sum_{\lambda_{1},\lambda_{2}}\sum_{j,m}\sqrt{\frac{2j+1}{4\pi}}D^{s_{1}*}_{m_{1},\lambda_{1}}D^{s_{2}*}_{m_{2},-\lambda_{2}}D^{j}_{m,\lambda}|j, m, \lambda_{1}, \lambda_{2}\rangle
\end{equation}
Using the following identities 
\begin{equation}
\begin{split}
D^{j_{1}}_{m_{1},m_{1}'}D^{j_{2}}_{m_{2},m_{2}'} &= \sum_{j_{3},m_{3},m_{3}'}\langle j_{1}, m_{1}, j_{2}, m_{2} | j_{3}, m_{3}\rangle\langle j_{1}, m_{1}', j_{2}, m_{2}'|j_{3}, m_{3}'\rangle D^{j_{3}}_{m_{3},m_{3}'} \\
D^{j_{1}}_{m_{1},m_{1}'}D^{j_{3}*}_{m_{3},m_{3}'} &= \sum_{j_{2},m_{2},m_{2}'}\frac{2j_{2} + 1}{2j_{3} + 1}\langle j_{1}, m_{1}, j_{2}, m_{2} | j_{3}, m_{3}\rangle\langle j_{1}, m_{1}', j_{2}, m_{2}'|j_{3}, m_{3}'\rangle D^{j_{2}*}_{m_{2},m_{2}'} \\
\end{split}
\end{equation}
we can simplify the product of three Wigner-D functions and find 
\begin{equation}
\begin{split}
|\Omega, m_{1}, m_{2} \rangle &= \sum_{\lambda_{1}, \lambda_{2}}\sum_{j, m} \sum_{l, m_{l}}\sum_{s, m_{s}}\sqrt{\frac{2l+1}{2j+1}}\langle s, m_{s}, l, m_{l} | j, m \rangle \langle s, \lambda, l, 0 | j, \lambda \rangle \\
&\times \langle s_{1}, m_{1}, s_{2}, m_{2} | s, m_{s} \rangle \langle s_{1}, \lambda_{1}, s_{2}, -\lambda_{2} | s, \lambda \rangle Y_{l, m_{l}}^{*} | j, m, \lambda_{1}, \lambda_{2} \rangle
\end{split}
\end{equation}
Substituting this into the expression for the states $| j, m, l, s \rangle$, we find
\begin{equation}
\begin{split}
|j, m, l, s \rangle &= \sum_{m_{l}, m_{s}}\sum_{m_{1}, m_{2}} \int d\Omega Y_{l,m_{l}}\langle l, m_{l}, s, m_{s} | j, m \rangle \langle s_{1}, m_{1}, s_{2}, m_{2} | s, m_{s} \rangle \\
&\times \sum_{\lambda_{1}, \lambda_{2}}\sum_{j', m'} \sum_{l', m_{l}'}\sum_{s', m_{s}'}\sqrt{\frac{2l'+1}{2j'+1}}\langle s', m_{s}', l', m_{l}' | j', m' \rangle \langle s, \lambda, l', 0 | j', \lambda \rangle \\
&\times \langle s_{1}, m_{1}, s_{2}, m_{2} | s', m_{s}' \rangle \langle s_{1}, \lambda_{1}, s_{2}, -\lambda_{2} | s', \lambda \rangle Y_{l', m_{l}'}^{*} | j', m', \lambda_{1}, \lambda_{2} \rangle
\end{split}
\end{equation}
Integrating the spherical harmonics and using various Clebsch-Gordan identities to simplify these sums, we arrive at 
\begin{equation}
|j, m, l, s \rangle = (-1)^{l+s-j} \sum_{\lambda_{1}, \lambda_{2}} \sqrt{\frac{2l+1}{2j+1}} \langle s, \lambda, l, 0 | j, \lambda \rangle  \langle s_{1}, \lambda_{1}, s_{2}, -\lambda_{2} | s, \lambda \rangle | j, m, \lambda_{1}, \lambda_{2} \rangle
\end{equation}
Next, we will investigate the invariant amplitude which we decompose as follows
\begin{equation}
\mathcal{M}_{fi} = (4\pi)^{2}\frac{M_{cm}}{\sqrt{|\vec{p}_{i}||\vec{p}_{f}|}}\langle \Omega,\lambda_{3},\lambda_{4} | T(M_{cm})|0, 0, \lambda_{1}, \lambda_{2}\rangle
\end{equation}
Enforcing conservation of total angular momentum and writing the amplitude in terms of helicity basis states, we have
\begin{equation}
\mathcal{M}_{fi} = 4\pi\frac{M_{cm}}{\sqrt{|\vec{p}_{i}||\vec{p}_{f}|}}\sum_{j}(2j+1)D^{j*}_{\lambda_{i},\lambda_{f}}(\Omega)\langle j,\lambda_{i},\lambda_{3},\lambda_{4}|T(M_{cm})|j,\lambda_{i},\lambda_{1},\lambda_{2}\rangle
\end{equation}
where $\lambda_{i} = \lambda_{1} - \lambda_{2}$. We also define $\lambda_{f} = \lambda_{3} - \lambda_{4}$ for future reference.

With the invariant amplitude written as a sum over matrix elements in the helicity basis, we can now invert this relationship. Extracting the partial amplitude in the helicity basis, we find
\begin{equation}
\langle j,m,\lambda_{3},\lambda_{4}|T(M_{cm})| j ,m,\lambda_{1},\lambda_{2}\rangle = \frac{\sqrt{|\vec{p}_{i}||\vec{p}_{f}|}}{(4\pi)^{2}M_{cm}}\int d\Omega D^{j}_{m,m^{'}}(\Omega)\mathcal{M}_{fi}
\end{equation}
Finally, we can convert from the helicity basis to the canonical basis using our basis change relations and we find
\begin{equation}
\begin{split}
\langle j, m, \ell', s'|T(M_{cm})|j, m, \ell, s \rangle &= (-1)^{\ell+s-j}(-1)^{\ell'+s'-j}\sum_{\lambda_{1}, \lambda_{2}}\sum_{\lambda_{3}, \lambda_{4}}\sqrt{\frac{2\ell' + 1}{2j + 1}}\sqrt{\frac{2\ell + 1}{2j + 1}} \\
&\times \langle s', \lambda_{f}, \ell', 0 | j, \lambda_{f}\rangle \langle s_{3}, \lambda_{3}, s_{4}, -\lambda_{4} | s', \lambda_{f} \rangle \\
&\times \langle s, \lambda_{i}, \ell, 0 | j, \lambda_{i}\rangle \langle s_{1}, \lambda_{1}, s_{2}, -\lambda_{2} | s, \lambda_{i} \rangle \\
&\times \langle j, m, \lambda_{3}, \lambda_{4} | T(M_{cm}) | j, m, \lambda_{1}, \lambda_{2} \rangle
\end{split}
\end{equation}
Here we have enforced conservation of total angular momentum and defined $\lambda_{i} = \lambda_{1} - \lambda_{2}$ and $\lambda_{f} = \lambda_{3} - \lambda_{4}$ as before.

\section{Supplemental results for the wino example}
\label{app:winoextras}

\begin{figure}
\centering
\includegraphics[width=0.49\hsize]{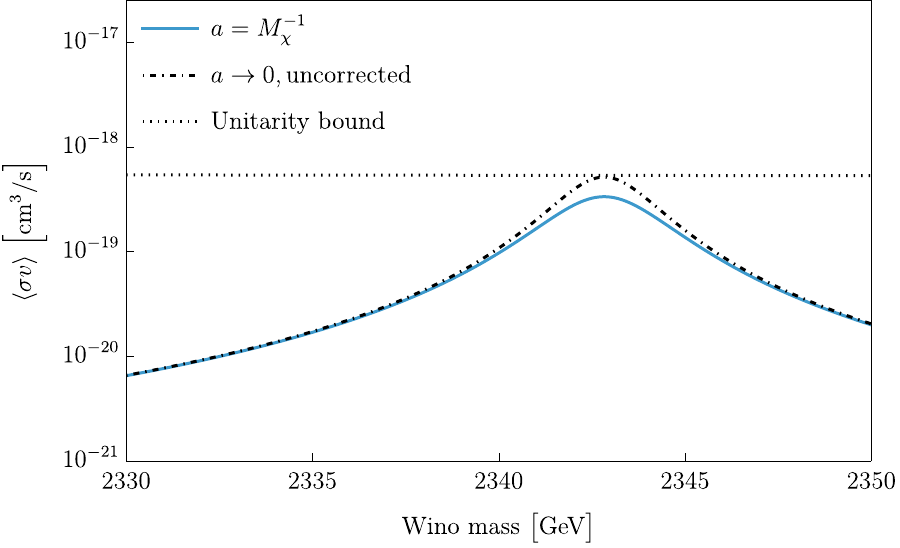}
%xsec_comparison_zoomin_log10vrel=-4_LO.pdf}
\includegraphics[width=0.49\hsize]{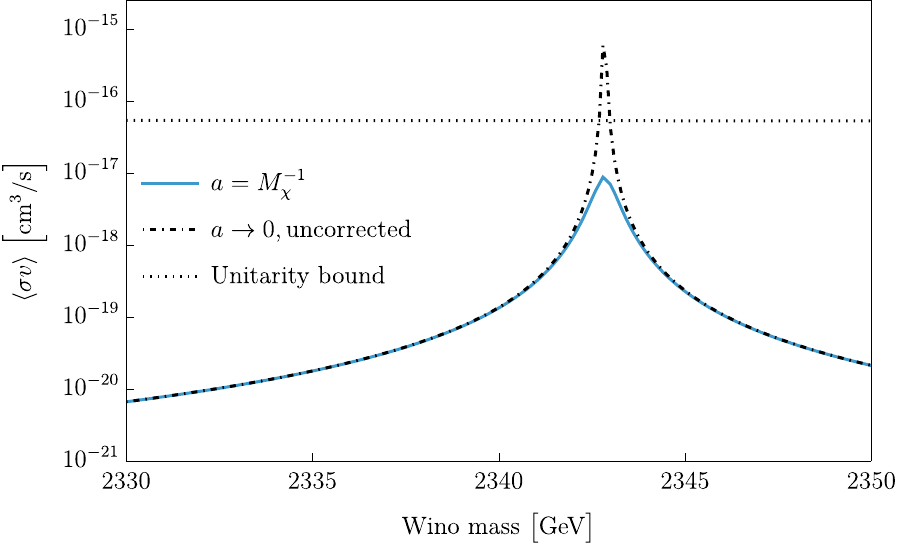}
%xsec_comparison_zoomin_log10vrel=-6_LO.pdf}
\caption{The blue lines show the Sommerfeld-enhanced annihilation cross section for wino dark matter, using the corrected expression derived in this work, for a matching radius of $a = M_{\chi}^{-1}$. The relative velocity between DM particles is taken to be 
$v=10^{-4}$ (\emph{left}) or $v=10^{-6}$ (\emph{right}). We additionally computed the Sommerfeld-enhanced annihilation cross section using matching radii of $a = 0.5 M_{\chi}^{-1}$ and $a = 2M_{\chi}^{-1}$ and find that they are identical to the $a = M_{\chi}^{-1}$ results to one part in $10^5$ for $v=10^{-4}$ and one part in $10^4$ for $v=10^{-6}$. 
The dot-dashed black line shows the result of computing the uncorrected Sommerfeld-enhanced annihilation cross section. The dotted black line indicates the unitarity bound on the annihilation cross section. Results in this figure are computed using the tree-level wino potential.}\label{fig:xsec_wino_LO}
\end{figure}

\subsection{LO vs NLO potential}

In this subsection we show the effect of using Eq.~\ref{eq:winopot} for the wino potential, omitting the corrections described in Eq.~\ref{eq:vnlo}; we also do not include these corrections in $f_{s,\ell}$. We otherwise proceed as described in the main text. Fig.~\ref{fig:xsec_wino_LO} shows the regulated and unregulated cross sections near the resonance point, and can be compared directly with Fig.~\ref{fig:xsec_wino_NLO}. We see that the results are qualitatively very similar; the only major difference is the position of the resonance, which shifts upward in mass by about 140 GeV (from around 2340 GeV to 2480 GeV) when the NLO corrections are included. The shift of the resonance is a known effect, already noted in Refs.~\cite{Beneke:2019qaa,Urban:2021cdu}. However, it is worth noting that the resonance position is quite sensitive to the details of the real potential, and so small real/Hermitian corrections to $f_{s,\ell}$ can also noticeably shift the resonance peak; choosing a convention where $V_\text{long}(r)$ is $a$-dependent will similarly lead to $a$-dependent shifts in the resonances. We will explore these effects further in the following subsections.

%\begin{figure}
%\centering
%\includegraphics[width=0.49\hsize]{fig/xsec_comparison_zoomin_log10vrel=-4_LO.pdf}
%\includegraphics[width=0.49\hsize]{fig/xsec_comparison_zoomin_log10vrel=-6_LO.pdf}
%\caption{Colored lines show the Sommerfeld-enhanced annihilation cross section for wino dark matter, using the corrected expression derived in this work, for different choices of the matching radius $a$. The dot-dashed black line shows the result of computing the uncorrected Sommerfeld-enhanced annihilation cross section. The relative velocity between DM particles is taken to be 
%$v=10^{-4}$ (\emph{left}) or $v=10^{-6}$ (\emph{right}). The dotted black line indicates the unitarity bound on the annihilation cross section. Results in this figure are computed using the tree-level wino potential.}\label{fig:xsec_wino_LO}
%\end{figure}

\subsection{Effect of including one-loop correction to $(f_s)_{11}$}

The NLO potential we employ still has zero $11$ component. We computed the one-loop contribution to the $11$ component for $\ell=0$ through a perturbative QFT calculation; the numerical results for this amplitude are shown in Fig.~\ref{fig:bornapprox_numerics}. We used this calculation to cross-check both (1) that the imaginary part of this amplitude is related to the contribution to $f_{s,\ell}$ from annihilation as described in App.~\ref{app:optical}, and (2) that the real part of this amplitude matches closely to the 2nd Born approximation from the tree-level potential, which we expect to be the dominant contribution, once it is rescaled by a factor $32 \pi M_\chi$ as described in App.~\ref{app:optical}. The residual part of this amplitude (not accounted for by the Born approximation) can then be rescaled appropriately and added to $\hat{f}_s$ as a short-distance correction to the elastic scattering.

\begin{figure}
\centering
\includegraphics[width=0.49\hsize]{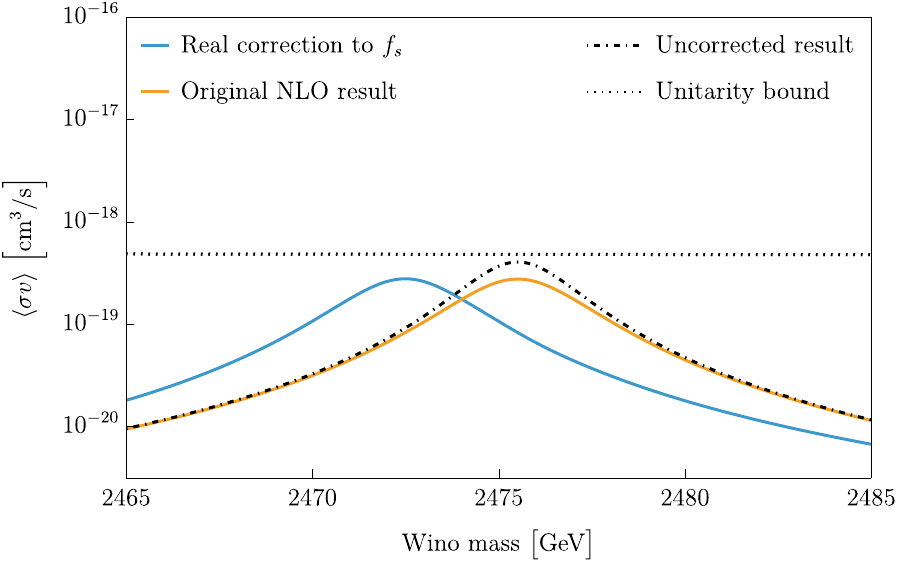}
%xsec_comparison_zoomin_log10vrel=-4_NLO_withresid.pdf}
\includegraphics[width=0.49\hsize]{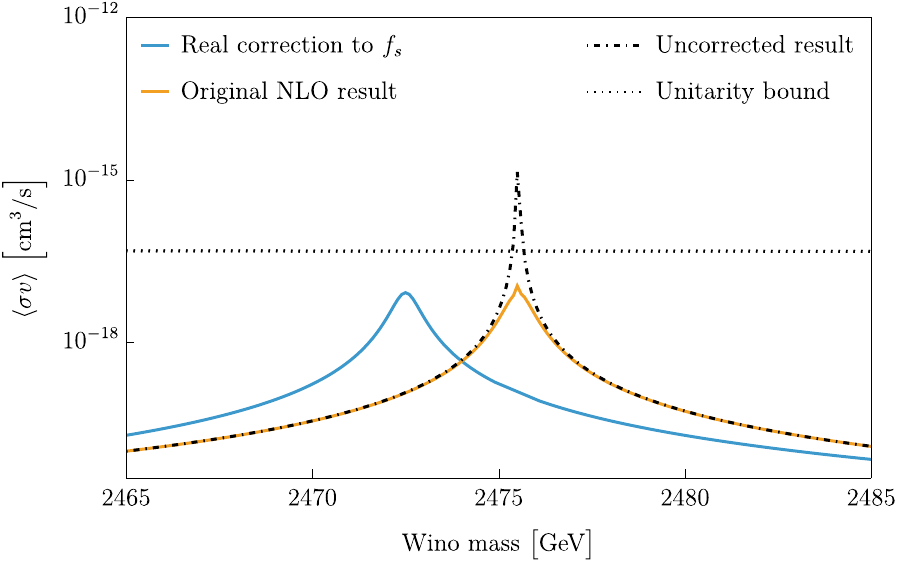}
%xsec_comparison_zoomin_log10vrel=-6_NLO_withresid.pdf}
\caption{The Sommerfeld-enhanced annihilation cross section for wino dark matter, using the corrected expression derived in this work, with (blue line) or without (orange line) including the one-loop correction to the real part of the short-distance scattering amplitude $(f_{s,0})_{11}$. The dot-dashed black line shows the result of computing the uncorrected Sommerfeld-enhanced annihilation cross section, which is independent of the real part of $(f_{s,0})_{11}$. The relative velocity between DM particles is taken to be 
$v=10^{-4}$ (\emph{left}) or $v=10^{-6}$ (\emph{right}). The dotted black line indicates the unitarity bound on the annihilation cross section. Results in this figure are computed using the NLO potential for the wino and taking the matching radius to be $a=1/M_\chi$.}\label{fig:xsec_wino_shortdistreal}
\end{figure}

In Fig.~\ref{fig:xsec_wino_shortdistreal} we show the results of including this term relative to our baseline calculation with the NLO potential, choosing $a=1/M_\chi$ for clarity (since we have already demonstrated the result is largely independent of $a$; we have checked this remains true when the residual term is added). We observe a shift in the resonance position, as expected, but to a much lesser degree than that induced by the NLO correction to the potential ($\sim 3$ GeV vs.~140 GeV). Otherwise the results are very similar -- in particular, the cross section at the peak is similar in the two cases, despite the shift in peak position -- justifying our neglect of this term (and real corrections of similar size to the other components of $f_{s,\ell}$).

\subsection{Results with modified convention for $V_\text{long}(r)$ at $r < a$}

To see how the $a$-independence of the final result emerges in a case where the resonance positions are $a$-dependent, and to demonstrate the convention-independence of the final result, in this section we show results derived by fixing $V_\text{long}(r)=0$ for $r < a$. This means that Eq.~\ref{eq:multistatekappa} takes a relatively simple form, which in our $N=2$, $M=1$, $\ell=0$ case becomes:
\begin{equation} \kappa_0^{-1} = f_{s,0}^{-1}  -  \begin{pmatrix}  \sqrt{p_1} & 0 \\  0 & (p_1^2 - 2 M_\chi \Delta)^{1/4} \end{pmatrix} \alpha_{\tilde{G}_0}(a).\end{equation}

However, we must now include in $f_{s,0}$ contributions from the wino potential for $r < a$, including an elastic scattering term and short-range Sommerfeld correction factors, as discussed in App.~\ref{app:optical} (see in particular Eq.~\ref{eq:improvedmatchingcon1}). Throughout this section, we employ the LO potential of Eq.~\ref{eq:winopot}, and compute the second Born approximation for the contributions to $f_{s,\ell}$ from this potential within the matching radius, as detailed in App.~\ref{app:partial_amps}. We evaluate the short-range Sommerfeld factors by numerically solving the Schr\"odinger equation for the LO potential, which is set to zero outside $r=a$. The results of this section can thus be compared directly to those obtained for the LO case with our previous convention, shown in the first subsection of this appendix.

\begin{figure}
\centering
\includegraphics[width=0.48\hsize]{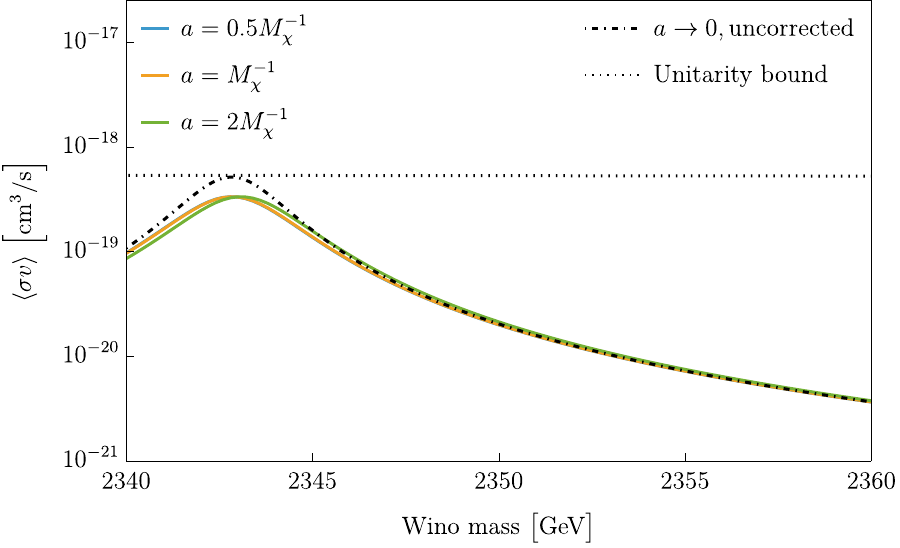}
%xsec_comparison_zoomin_log10vrel=-4_method1.pdf}
\includegraphics[width=0.48\hsize]{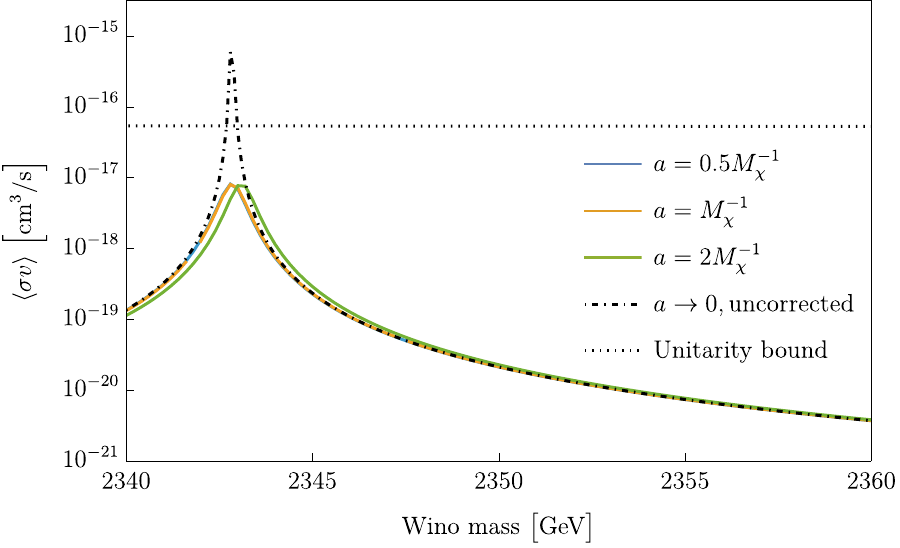}
%xsec_comparison_zoomin_log10vrel=-7_method1.pdf}
\caption{The Sommerfeld-enhanced annihilation cross section for wino dark matter, using the corrected expression derived in this work, for different choices of the matching radius $a$, in the convention where we fix $V_\text{long}(r)=0$ for $r < a$. The dot-dashed black line shows the result of computing the uncorrected Sommerfeld enhancement (i.e. to leading order in the short-range matrix element) with $a\rightarrow 0$ in the evaluation of $\Sigma_{0,\ell}$. The dotted black line indicates the unitarity bound on the annihilation cross section. The relative velocity between DM particles is taken to be $v=10^{-4}$ (\emph{left}), or $v=10^{-6}$ (\emph{right}). All calculations for this figure were performed using the tree-level wino potential.}\label{fig:xsec_wino_method1}
\end{figure}

We plot in Fig.~\ref{fig:xsec_wino_method1} both the full annihilation cross section, and the cross section obtained using the uncorrected Sommerfeld enhancement $\sigma_{\ell,0}$, as a function of the wino mass $M_\chi$, for two different velocities as previously. We observe that the results still converge well for small $a$, although the deviation for $a=2/M_\chi$ is more pronounced than with our previous convention (this is not unexpected, as in this convention we make the additional approximation of treating the potential for $r < a$ perturbatively).

\begin{figure}
\centering
\includegraphics[width=0.48\hsize]{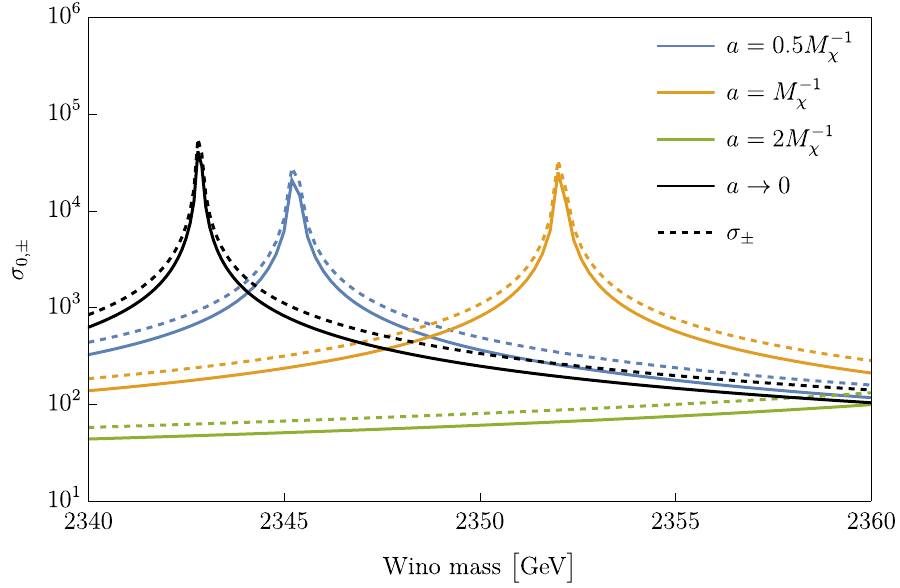}
%sommerfeld_zoomin_log10vrel=-7_method1.pdf}
\includegraphics[width=0.48\hsize]{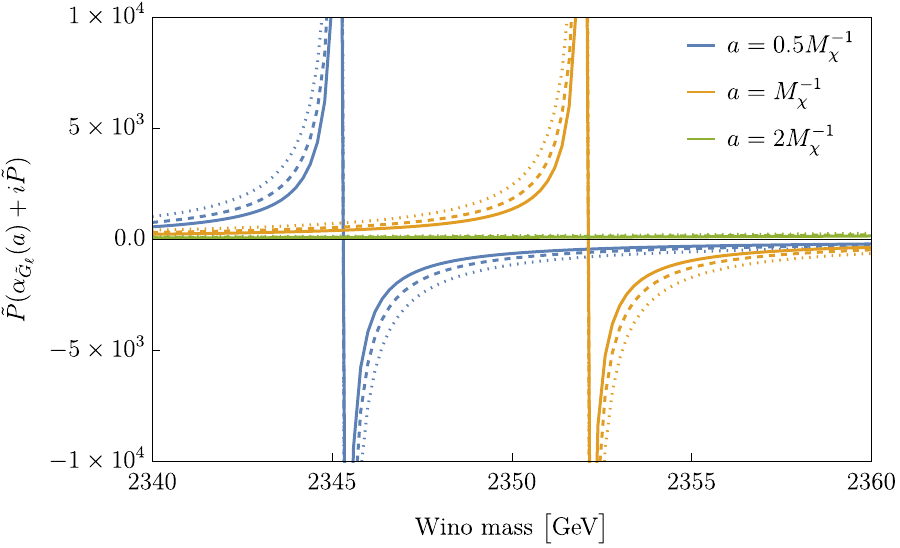}
%alphaG_zoomin_log10vrel=-7_method1.pdf}
\caption{\emph{Left panel:} Uncorrected Sommerfeld factors $\sigma_0$ (solid lines) and $\sigma_\pm$ (dashed lines), as a function of the wino mass, for different choices of $a$, at a relative velocity of $v=10^{-6}$. This calculation uses the convention where $V_\text{long}(r)=0$ for $r < a$. The black lines correspond to the limit $a\rightarrow 0$. Note that the resonance is shifted beyond the right-hand edge of the plot in the case of $a=2/M_\chi$. \emph{Right panel:} components of the real symmetric matrix $\tilde{P} (\alpha_{\tilde{G}_\ell}(a) + i \tilde{P})$, for a relative velocity of $v=10^{-6}$. Solid, dashed, and dotted lines show the values of the $11$, $12$, and $22$ components respectively. All calculations for this figure were performed using the tree-level wino potential.}\label{fig:longrangeparams_wino}
\end{figure}

The ingredients computed from the long-range potential, i.e. the $\sigma_0$, $\sigma_{\pm}$ and $\alpha_{\tilde{G}_\ell}(a)$ factors, are shown as a function of $a$ and $M_\chi$ in Fig.~\ref{fig:longrangeparams_wino}. Note that the resonances in the uncorrected Sommerfeld factors (left panel) are markedly displaced for different choices of $a$, in contrast to the resonances in the corrected cross section (Fig.~\ref{fig:xsec_wino_method1}). Similarly, in the right panel we see that $\alpha_{\tilde{G}_\ell}(a)$ has pronounced features associated with the resonances in the uncorrected Sommerfeld enhancement. This serves as a cross-check on the validity of our method: the inclusion of $a$-dependence in $f_{s,\ell}$ and $\alpha_{\tilde{G}_\ell}(a)$ is correctly canceling the $a$-dependence in $\Sigma_{0,\ell}$, and shifting the resonance peaks (for different choices of $a$) to coincide with each other in the final corrected result.

\subsection{Results with lowest-order matching relation}

\begin{figure}
\centering
\includegraphics[width=0.49\hsize]{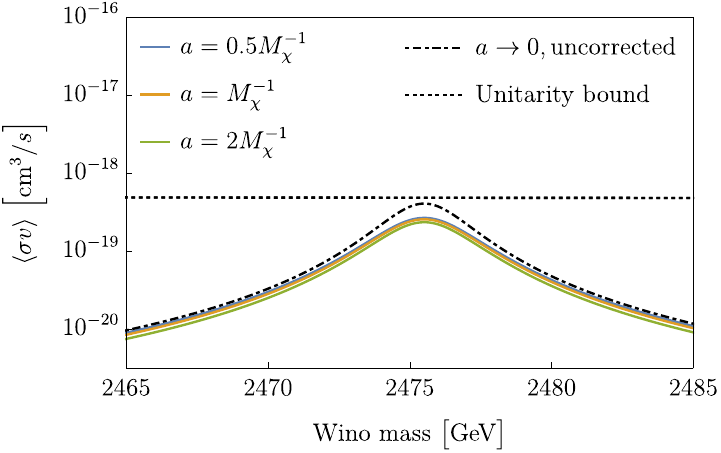}
\includegraphics[width=0.49\hsize]{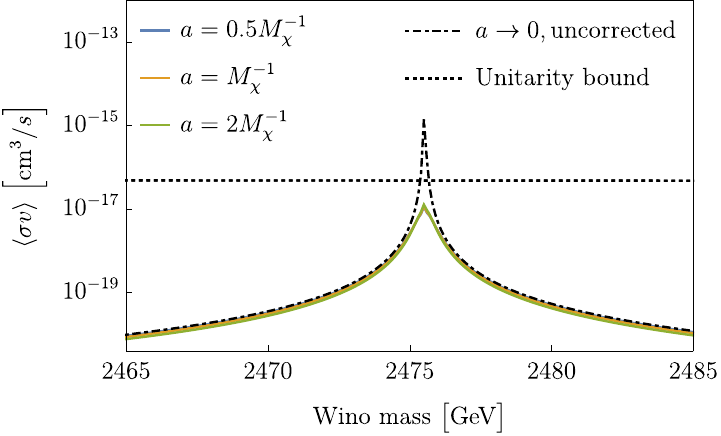}
\caption{Colored lines show the Sommerfeld-enhanced annihilation cross section for wino dark matter, using the corrected expression derived in this work and a simplified tree-level matching relation (see text), for different choices of the matching radius $a$. The dot-dashed black line shows the result of computing the uncorrected Sommerfeld-enhanced annihilation cross section. The relative velocity between DM particles is taken to be 
$v=10^{-4}$ (\emph{left}) or $v=10^{-6}$ (\emph{right}). The dotted black line indicates the unitarity bound on the annihilation cross section.}\label{fig:oldmethod}
\end{figure}

As discussed in App.~\ref{app:optical}, by default we use the improved matching relation given in Eq.~\ref{eq:improvedmatchingcon1} to compute the short-distance amplitude $f_{s,\ell}$. An alternative is to ignore the Sommerfeld factors arising from $r < a$, i.e. approximating $\Sigma^a_{0,\ell} \approx 1$ in that equation, so that $f_{s,\ell}$ is determined solely by the tree-level annihilation cross section (plus the first-order perturbative term from elastic scattering). We show in Fig.~\ref{fig:oldmethod} the analogue of Fig.~\ref{fig:xsec_wino_NLO} where we instead use this ``tree-level matching relation''. Compared to the improved matching relation, we observe a stronger $a$-dependence in the results; this is to be expected, as we anticipate the $a$-dependence to be a consequence of the perturbative treatment of the short-range amplitude (and/or neglect of interactions not encoded in $V_\text{long}(r)$ for $r > a$). The modest size of the effect indicates that the perturbative approximation is quite good.

\subsection{$p$-wave case}

It is straightforward to extend our calculation to the $p$-wave case; we include here the necessary inputs to perform the $p$-wave calculation for the wino. 

Choosing $\ell=1$ for the wino, and choosing a convention where $\sigma_0$ is real and $\sigma_{\pm}$ is purely imaginary, the $p$-wave annihilation cross section is given by:
\begin{align} \sigma_{\text{ann}} v_\text{rel} & = \frac{24 \pi i}{M_\chi} \begin{pmatrix} \sigma_0 & -\sigma_\pm \end{pmatrix}  \left[1 + i  p_1 \kappa_1^\dagger \begin{pmatrix} \sigma_0^2 & -\sigma_0 \sigma_\pm \\ \sigma_0 \sigma_\pm & -\sigma_\pm^2 \end{pmatrix} \right]^{-1} (\kappa_1^\dagger - \kappa_1) \nonumber \\
& \times \left[1 - i p_1 \begin{pmatrix} \sigma_0^2 & - \sigma_0 \sigma_\pm \\ \sigma_0 \sigma_\pm & - \sigma_\pm^2 \end{pmatrix}  \kappa_1 \right]^{-1} \begin{pmatrix} \sigma_0 \\ \sigma_\pm \end{pmatrix}.\label{eq:pwinoxsec} 
\end{align}
Writing $\kappa_1=\begin{pmatrix} \kappa_{11} & \kappa_{12} \\ \kappa_{21} & \kappa_{22}\end{pmatrix}$ and explicitly performing the matrix inversion yields:
\begin{align} \sigma_{\text{ann}} v_\text{rel} & = -\frac{24 \pi i}{M_\chi} \frac{\left[(\kappa_{11} -\kappa_{11}^*) \sigma_0^2 - (\kappa_{22} - \kappa_{22}^*) \sigma_\pm^2 + (\kappa_{12} - \kappa_{21} - \kappa_{21}^* + \kappa_{12}^*) \sigma_0 \sigma_\pm \right]}{\left|i + p_1 \left(\sigma_0^2 \kappa_{11} - \sigma_\pm^2 \kappa_{22} + \sigma_0 \sigma_\pm (\kappa_{12} - \kappa_{21})\right) \right|^2} \nonumber \\
& =  \frac{48 \pi}{M_\chi} \frac{\left[ \sigma_0^2 \text{Im} \kappa_{11}  - \sigma_\pm^2 \text{Im} \kappa_{22}  - i \sigma_0 \sigma_\pm \text{Re}(\kappa_{12} - \kappa_{21})  \right]}{\left|i + p_1 \left(\sigma_0^2 \kappa_{11} - \sigma_\pm^2 \kappa_{22} + \sigma_0 \sigma_\pm (\kappa_{12} - \kappa_{21})\right) \right|^2}. \end{align}
It may be surprising to see a dependence on the real part of the off-diagonal terms in the numerator, but due to the momentum dependence of $\kappa_1$, in the case where the long-range potential can be neglected, the off-diagonal terms in the contribution from the optical theorem are actually real (but the matrix is antisymmetric). Specifically, from the results of App.~\ref{app:optical} we can take the contribution to $\bar{f}_{s,1}$ (which will dominate $\kappa_1$ when the effects of the long-range potential are small) from annihilation to be:
\begin{align} \bar{f}_{s,1} & = -\frac{1}{2}  \frac{\mu}{2 i \pi} \frac{1}{3} \frac{1}{(2 \mu)^2}  \frac{28 \pi \alpha_W^2}{9 (2\mu)^2} \begin{pmatrix} |p_1|^2 & \frac{1}{\sqrt{2}} p_1^* p_2 \\ \frac{1}{\sqrt{2}} p_2^* p_1  & \frac{3}{2} |p_2|^2 \end{pmatrix} \nonumber \\
& =   \frac{7 }{54}  \frac{ \alpha_W^2}{ (2\mu)^3} \begin{pmatrix} i |p_1|^2 & - \frac{1}{\sqrt{2}} |p_1 p_2| \\  \frac{1}{\sqrt{2}} | p_1 p_2| & \frac{3}{2} i |p_2|^2 \end{pmatrix}  .
%& =   \frac{7 i }{54}  \frac{ \alpha_W^2}{ (2\mu)^3} \begin{pmatrix} |p_1|^2 & \frac{1}{\sqrt{2}} p_1^* p_2 \\ - \frac{1}{\sqrt{2}} p_2^* p_1  & - \frac{3}{2} |p_2|^2 \end{pmatrix}  .
\end{align}

\begin{figure}
\centering
\includegraphics[width=0.49\hsize]{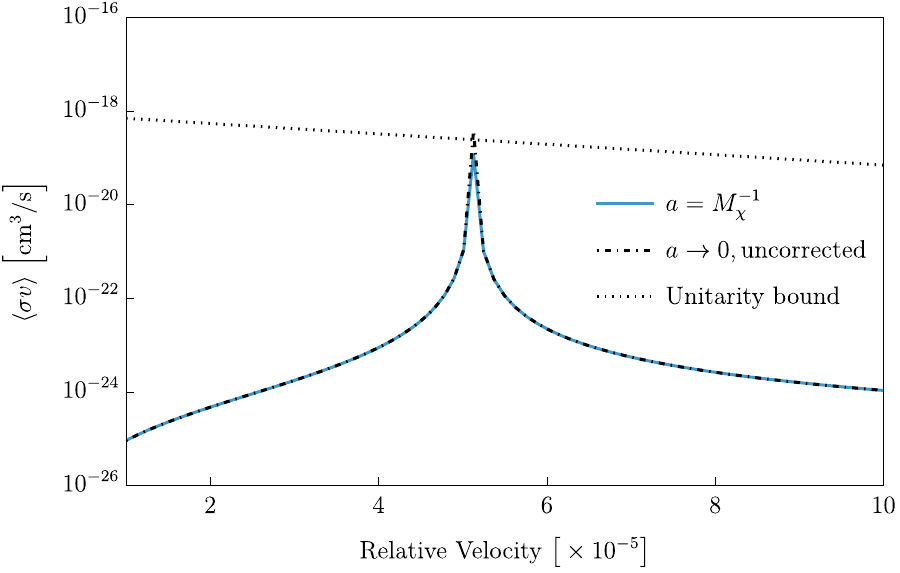}
%xsec_comparison_zoomin_vreldep_l=1_mass=11271_LO.jpg}
\includegraphics[width=0.49\hsize]{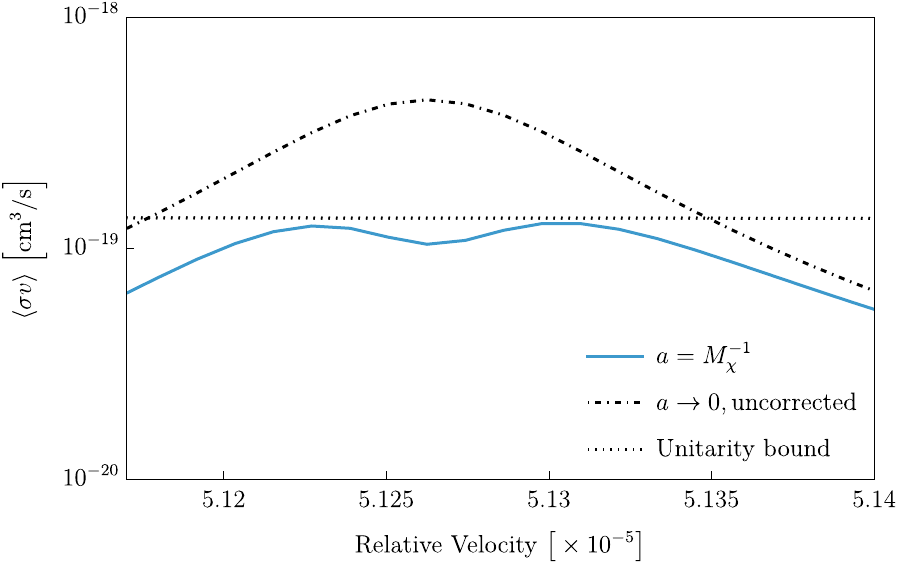}
%xsec_comparison_maxzoomin_vreldep_l=1_mass=11271_LO.jpg}
\caption{The blue lines show the Sommerfeld-enhanced annihilation cross section for wino dark matter, using the corrected expression derived in this work, for a matching radius of $a = M_{\chi}^{-1}$. The DM mass is taken to be 11270.9999 GeV; the \emph{right} panel is a zoom-in of the {\it left} panel showing where the uncorrected cross section crosses the unitarity bound. We additionally computed the Sommerfeld-enhanced annihilation cross section using matching radii of $a = 0.5M_{\chi}^{-1}$ and $a = 2M_{\chi}^{-1}$ and find that they are identical to the $a = M_{\chi}^{-1}$ results to one part in $10^{2}$.
The dot-dashed black line shows the result of computing the uncorrected Sommerfeld-enhanced annihilation cross section. The dotted black line indicates the unitarity bound on the annihilation cross section. Results in this figure are computed using the tree-level wino potential.}\label{fig:pwavepeaks}
\end{figure}

In numerical testing for the wino example, we have generally found that the uncorrected result is a very good approximation for almost all of parameter space. However, one interesting exception is the ``super-resonant Breit-Wigner peaks'' identified by Ref.~\cite{Beneke:2024iev}. In Fig.~\ref{fig:pwavepeaks} we show the mass-velocity region surrounding one such peak, computed with the LO potential for simplicity.\footnote{We thank Tobias Binder for suggesting this analysis.}  We note that the peak location and properties we find are slightly different from those identified in Ref.~\cite{Beneke:2024iev} (e.g. the mass at the resonance is closer to 11.3 TeV rather than 11 TeV), but this may be due to the slightly different electroweak parameters we employ. In this case, the uncorrected cross section only exceeds the unitarity bound for a very narrow range of mass and velocity, and we find that the corrected cross section only significantly modifies the uncorrected result in close proximity to this region.

\section{Generalization of multi-state results to a complex long-range potential}
\label{app:Hermitian}

In Sec.~\ref{sec:multistate} we assumed that $V_\text{long}(r)$ was real and symmetric, which is the case for most commonly-considered examples in the literature. More generally, $V_\text{long}(r)$ can be Hermitian but complex; in this appendix, we discuss how to generalize the results of  Sec.~\ref{sec:multistate} to this case.

For a complex potential, it is not generally true that the real and imaginary parts of a solution to the Schr\"odinger solution are separately solutions, so we cannot decompose $\tilde{w}_\ell(r)$ in terms of the real solutions $\tilde{G}_\ell(r)$ and $\tilde{F}_\ell(r)$. However, we can simply work with $\tilde{w}_\ell(r)$ itself, as the definitions of $w_\ell(r)$ and $\tilde{w}_\ell(r)$ do not rely on the potential being real.

Furthermore, given any two solutions $u(r)$ and $v(r)$, $u^\dagger(r) v^\prime(r) - (u^\dagger(r))^\prime v(r)$ provides a conserved Wronskian, but in the main text we derived other Wronskian relations by conjugating one of the solutions (or breaking one of the solutions into its real and imaginary parts); these relations are no longer valid, so some caution is needed. It is easy to check that the Wronskian relation for $\hat{W}_{ij}$ given in Sec.~\ref{sec:multistate_longrange} is of the form $u^\dagger(r) v^\prime(r) - (u^\dagger(r))^\prime v(r)$, and so remains valid for a complex potential, yielding:
\begin{equation} Q_\ell^\dagger (\tilde{P}^\dagger)^{2(\ell+1)} = (1 - 2 i P^2 f_{0,\ell}^\dagger) P^2 f_{1,\ell} \tilde{P}^2.\end{equation}

Now note that $S_{0,\ell} = 1 + 2 i P f_{0,\ell} P$, so $P S_{0,\ell}^\dagger P^{-1} =  1 - 2 i P^2 f_{0,\ell}^\dagger$, and thus we have $Q_\ell^\dagger (\tilde{P}^\dagger)^{2(\ell+1)} =P S_{0,\ell}^\dagger P f_{1,\ell} \tilde{P}^2$. But as $S_{0,\ell}$ is unitary, this implies
\begin{equation} P f_{1,\ell} \tilde{P}^2  = S_{0,\ell} P^{-1} Q_\ell^\dagger (\tilde{P}^\dagger)^{2(\ell+1)}. \end{equation}

The first equality in Eq.~\ref{eq:sdecomposition} is purely a corollary of writing the full solution as $w_\ell(r) + \tilde{w}_\ell(r) R_\ell$, so we have:
\begin{align} S_\ell & = S_{0,\ell} + 2 i P f_{1,\ell} \tilde{P}^2 R_\ell P^{-1} \nonumber \\
& = S_{0,\ell} \left[1 + 2 i P^{-1} Q_\ell^\dagger (\tilde{P}^\dagger)^{2(\ell+1)} R_\ell P^{-1} \right].\end{align}
This is an alternate (actually somewhat simpler) proof of Eq.~\ref{eq:modifiedS} that does not rely on the potential being real.

Thus writing down the full $S$-matrix element just requires us to determine $R_\ell$. The equivalent of Eq.~\ref{eq:smallralphabeta}, without separating $\tilde{w}_\ell(r)$ into real and imaginary components, is:
\begin{equation}\alpha_{w_\ell}(0) = \tilde{P} Q_\ell, \quad \beta_{w_\ell}(0) = 0, \quad \beta_{\tilde{w}_\ell}(0) = -\tilde{P}^{2\ell+1}.\end{equation}
The matching at $r=a$ gives $\tilde{P} f_{s,\ell} \tilde{P} \alpha_\ell(a) = -\beta_\ell(a)$ as in Eq.~\ref{eq:multistatematchsol}, which becomes:
\begin{align} & \tilde{P} f_{s,\ell} \tilde{P} (\alpha_{w_\ell}(a) + \alpha_{\tilde{w}_\ell}(a) R_\ell)  = -(\beta_{w_\ell}(a) + \beta_{\tilde{w}_\ell}(a) R_\ell) \nonumber \\
\Rightarrow & R_\ell  = - \left[  \beta_{\tilde{w}_\ell}(a)  +  \tilde{P} f_{s,\ell} \tilde{P} \alpha_{\tilde{w}_\ell}(a) \right]^{-1} \left( \tilde{P} f_{s,\ell} \tilde{P} \alpha_{w_\ell}(a) + \beta_{w_\ell}(a) \right) \end{align}
The relation that $\beta_{w_\ell}(a) = -\tilde{P} f_{b,\ell} \tilde{P} \alpha_{w_\ell}(a)$ does not depend on the properties of the potential, and similarly $\tilde{P} Q_\ell = \alpha_{w_\ell}(0) = \alpha_{b,\ell}(0)\tilde{P}^{-1} \alpha_{w_\ell}(a)$ still holds. So we can write $\hat{f}_s = f_s - f_b$ as previously, and thus:
\begin{align} R_\ell  & = - \left[  \beta_{\tilde{w}_\ell}(a)  +  \tilde{P} f_{b,\ell} \tilde{P} \alpha_{\tilde{w}_\ell}(a) + \tilde{P} \hat{f}_{s,\ell} \tilde{P} \alpha_{\tilde{w}_\ell}(a) \right]^{-1} \tilde{P} \hat{f}_{s,\ell}  \tilde{P}  \alpha_{w_\ell}(a) \nonumber \\
&=  - \left[\left(\tilde{P} \hat{f}_{s,\ell}  \tilde{P} \right)^{-1} \left( \beta_{\tilde{w}_\ell}(a)  +  \tilde{P} f_{b,\ell} \tilde{P} \alpha_{\tilde{w}_\ell}(a) \right) + \alpha_{\tilde{w}_\ell}(a) \right]^{-1} \tilde{P} \alpha_{b,\ell}(0)^{-1} \tilde{P} Q_\ell \nonumber \\
& = - \left[\alpha_{b,\ell}(0)\tilde{P}^{-1} \left(\tilde{P} \hat{f}_{s,\ell}  \tilde{P} \right)^{-1} \left( \beta_{\tilde{w}_\ell}(a)  +  \tilde{P} f_{b,\ell} \tilde{P} \alpha_{\tilde{w}_\ell}(a) \right) + \alpha_{b,\ell}(0)\tilde{P}^{-1} \alpha_{\tilde{w}_\ell}(a) \right]^{-1}  \tilde{P} Q_\ell .  \end{align}
We note that the coefficient of the $\hat{f}_s^{-1}$ term in the denominator, the $\alpha_{\tilde{w}_\ell}(a)$ term in the denominator, and the numerator can all be computed solely from the long-range potential and the matching radius. 

Let us now define $\Sigma_{0,\ell} = \tilde{P}^2 Q_\ell P^{-2}$ as previously, and:
\begin{align} - \kappa_\ell^{-1} & \equiv \tilde{P} \left[ \alpha_{b,\ell}(0)\tilde{P}^{-1} \left(\tilde{P} \hat{f}_{s,\ell}  \tilde{P} \right)^{-1} \left( \beta_{\tilde{w}_\ell}(a)  +  \tilde{P} f_{b,\ell} \tilde{P} \alpha_{\tilde{w}_\ell}(a) \right) \right. \nonumber \\
& + \left. \alpha_{b,\ell}(0)\tilde{P}^{-1} \alpha_{\tilde{w}_\ell}(a) \right] (\tilde{P}^\dagger)^{-2\ell} - i \Sigma_{0,\ell} P^2 \Sigma_{0,\ell}^\dagger .\label{eq:hermitiankappa} \end{align}
Then we have:
\begin{equation} R_\ell =(\tilde{P}^\dagger)^{-2\ell}  \left[\kappa_\ell^{-1} - i \Sigma_{0,\ell} P^2 \Sigma_{0,\ell}^\dagger \right]^{-1} \tilde{P}^2 Q_\ell,\end{equation}
and so the $S$-matrix is given by:
\begin{equation} S_\ell = S_{0,\ell} \left(1 + 2 i P \Sigma_{0,\ell}^\dagger  \left[ \kappa_\ell^{-1} - i \Sigma_{0,\ell} P^2 \Sigma_{0,\ell}^\dagger \right]^{-1} \Sigma_{0,\ell} P \right).\end{equation}
This is exactly the form given in Eq.~\ref{eq:multistateSmatrix}, but with a modified/generalized definition for $\kappa_\ell$. Note that $\kappa_\ell^{-1}$ can still be expressed in terms of $\hat{f}_s^{-1}$, a multiplicative factor determined from the long-range potential, and an additive factor determined from the long-range potential. The subsequent discussion in the main text then carries over to this more general case, with the appropriate replacement of $\kappa_\ell$.

For illustration, let us work out $\kappa_\ell$ under the convention where $V_\text{long}(r)=0$ for $r < a$, so that we have $\alpha_{b,\ell}(0)=\tilde{P}$, $f_{b,\ell}=0$, $ \beta_{\tilde{w}_\ell}(a) = -\tilde{P}^{2\ell+1}$,  $\alpha_{\tilde{w}_\ell}(a) = \alpha_{\tilde{w}_\ell}(0)$. Then Eq.~\ref{eq:hermitiankappa} becomes:
\begin{align} \kappa_\ell^{-1} & \equiv   \hat{f}_{s,\ell}^{-1} \tilde{P}^{2\ell} (\tilde{P}^\dagger)^{-2\ell} -  \tilde{P}  \alpha_{\tilde{w}_\ell}(a) (\tilde{P}^\dagger)^{-2\ell}  + i \Sigma_{0,\ell} P^2 \Sigma_{0,\ell}^\dagger. \end{align}
As shown in Eq.~\ref{eq:alphaGextraction}, in the case of a real and symmetric potential $\alpha_{\tilde{G}_\ell}(a) = \alpha_{\tilde{w}_\ell}(a) - i \alpha_{w_\ell}(a)  \Sigma_{0,\ell}^\dagger (\tilde{P}^\dagger)^{2\ell}$, and with the present convention we have $\alpha_{w_\ell}(a) = \tilde{P} Q_\ell$, so we obtain:
\begin{align} \alpha_{\tilde{G}_\ell}(a) & = \alpha_{\tilde{w}_\ell}(a) - i \tilde{P}^{-1} \Sigma_{0,\ell} P^2 \Sigma_{0,\ell}^\dagger (\tilde{P}^\dagger)^{2\ell} \nonumber \\
\Rightarrow \tilde{P} \alpha_{\tilde{G}_\ell}(a) (\tilde{P}^\dagger)^{-2\ell} & = \tilde{P} \alpha_{\tilde{w}_\ell}(a) (\tilde{P}^\dagger)^{-2\ell}  - i \Sigma_{0,\ell} P^2 \Sigma_{0,\ell}^\dagger. \end{align}
Thus we recover our previously derived expression for $\kappa_\ell^{-1}$ in terms of $\alpha_{\tilde{G}_\ell}(a)$ in the case of a real symmetric potential, under this convention, as discussed below Eq.~\ref{eq:multistateSmatrix}.

\end{appendix}

\bibliography{ref}
\bibliographystyle{JHEP}

\end{document}